\documentstyle[12pt]{cernart}
\def \theequation{\thesection.\arabic{equation}}
\def \theequation{\thesubsection.\arabic{equation}}
\def \gsim{\lower.8ex\hbox{$\sim$}\kern-.75em\raise.45ex\hbox{$>$}\;}
\def \lsim{\lower.8ex\hbox{$\sim$}\kern-.8em\raise.45ex\hbox{$<$}\;}
\def \bfk{\mbox{\boldmath $k$}}
\def \bfK{\mbox{\boldmath $K$}}
\def \bfp{\mbox{\boldmath $p$}}
\def \bfbp{\mbox{\boldmath $P$}}
\def \bfP{\mbox{\boldmath $P$}}
\def \bfs{\mbox{\boldmath $s$}}
\def \bfbs{\mbox{\boldmath $S$}}
\def \bfS{\mbox{\boldmath $S$}}
\def \bfo{\mbox{\boldmath $0$}}
\def \bfx{\mbox{\boldmath $x$}}
\def \bfy{\mbox{\boldmath $y$}}
\def \bfA{\mbox{\boldmath $A$}}
\def \bfB{\mbox{\boldmath $B$}}
\def \bfE{\mbox{\boldmath $E$}}
\def \bfG{\mbox{\boldmath $G$}}
\def \bfI{\mbox{\boldmath $I$}}
\def \bfU{\mbox{\boldmath $U$}}
\def \bfpsi{\mbox{\boldmath $\psi$}}
\def \bflambda{\mbox{\boldmath $\lambda$}}
\def \avq{\langle Q^2 \rangle}
\def \gevc{\mbox{\rm (GeV/c)}}
\def \sint{\sum \kern -13.0pt \int}
\def \sla{\kern -5.4pt /}
\def \slas{\kern -6.2pt /}
\def \Cslas{\kern -6.8pt /}
\def \lpartial{\stackrel{\leftarrow}{\partial}}
\def \rpartial{\stackrel{\rightarrow}{\partial}}
\def \lder{\stackrel{\leftarrow}{D}}
\def \rder{\stackrel{\rightarrow}{D}}
\def \ldersl{\stackrel{\leftarrow}{D\Cslas}}
\def \rdersl{\stackrel{\rightarrow}{D\Cslas}}
\begin{document}
\begin{titlepage}
\vskip1cm
\docnum{CERN--TH/7216/94}
\date{April 1994}
\vskip1cm
\title{THE THEORY AND PHENOMENOLOGY OF POLARIZED DEEP INELASTIC SCATTERING}
\begin{Authlist}
{\sf M. Anselmino}
\Instfoot{a1}{Dipartimento di Fisica Teorica, Universit\`a di Torino
and Istituto Nazionale di Fisica Nucleare, Sezione di Torino,
Via P. Giuria 1, 10125 Torino, Italy.}
{\sf A. Efremov}
\footnote{Permanent address: Laboratory of Theoretical Physics, Joint
Institute for Nuclear Research, Dubna, 141980, Russia.}
\Instfoot{a2}{Theory Division, CERN, 1211 Geneva 23, Switzerland.}
{\sf E. Leader}
\Instfoot{a3}{Birkbeck College, University of London, Malet Sreet, London WC1E
7HX, UK.}
\end{Authlist}
\vskip3cm
\collaboration{To appear in Physics Reports}
\end{titlepage}
%
%end of front page
%
\begin{center}
{\bf CONTENTS}
\end{center}

%\vskip 6pt
\noindent
1) {\bf Introduction}

\noindent
{\phantom{1) }} 1.1) Notation

\vskip 6pt
\noindent
2) {\bf Polarized lepton-nucleon deep inelastic scattering -- general
aspects}

\noindent
{\phantom{2) }} 2.1) General formalism in the one photon exchange
approximation

\noindent
{\phantom{2) 2.1) }} 2.1.1) Structure functions and Bjorken scaling

\noindent
{\phantom{2) 2.1) }} 2.1.2) Cross-section differences

\noindent
{\phantom{2) 2.1) }} 2.1.3) Experimental measurement of $g_1$ and $g_2$ on
nucleon targets

\noindent
{\phantom{2) 2.1) 2.1.3) }} a) Longitudinally polarized target

\noindent
{\phantom{2) 2.1) 2.1.3) }} b) Transversely polarized target

\noindent
{\phantom{2) 2.1) 2.1.3) }} c) Combined analysis using both polarizations

\noindent
{\phantom{2) 2.1) 2.1.3) }} d) The problem of extrapolating in $Q^2$

\noindent
{\phantom{2) 2.1) }} 2.1.4) Measurement of $g_{1,2}$ on nuclear targets

\noindent
{\phantom{2) }} 2.2) Neutral and charged current weak interactions initiated
by charged leptons

\noindent
{\phantom{2) 2.1) }} 2.2.1) Weak interaction structure functions and their
measurement

\vskip 6pt
\noindent
3) {\bf The Naive Parton Model in polarized DIS}

\noindent
{\phantom{2) }} 3.1) Projection of $F_{1,2}$ and $g_{1,2}$ from the hadronic
tensor $W^{\mu \nu}$

\noindent
{\phantom{2) }} 3.2) The hadronic tensor and the nucleon structure functions
in the Naive Parton

\noindent
{\phantom{2) 3.2)}} ~Model

\noindent
{\phantom{2) }} 3.3) Intrinsic ${\bf p}_\perp$ and $g_2(x)$

\noindent
{\phantom{2) 2.1) }} 3.3.1) Conflicting Parton Model results for $g_2(x)$

\noindent
{\phantom{2) }} 3.4) Origin of the difficulties with $g_2(x)$

\noindent
{\phantom{2) }} 3.5) Weak interaction structure functions in the Naive Parton
Model

\vskip 6pt
\noindent
4) {\bf Phenomenological analysis of the data on \mbox{\boldmath $g_1(x)$}
and its first moment \mbox{\boldmath $\Gamma_1$}}

\noindent
{\phantom{2) }} 4.1) The SLAC-Yale and EMC data: quark distributions near $x=1$

\noindent
{\phantom{2) }} 4.2) Analysis of data in the framework of the Operator
Product Expansion

\noindent
{\phantom{2) 2.1) }} 4.2.1) Information from hyperon $\beta$-decay

\noindent
{\phantom{2) 2.1) }} 4.2.2) The EMC data on $\Gamma^p_1$ and its consequences

\noindent
{\phantom{2) 2.1) }} 4.2.3) The EMC data: extrapolation to $x$ = 0

\noindent
{\phantom{2) 2.1) }} 4.2.4) The EMC data: higher twist effects

\noindent
{\phantom{2) 2.1) 4.2.4) }} a) The Gerasimov, Drell, Hearn approach

\noindent
{\phantom{2) 2.1) 4.2.4) }} b) The QCD sum rule approach

\noindent
{\phantom{2) 2.1) 4.2.4) }} c) Summary on higher twist effects in the EMC
experiment

\noindent
{\phantom{2) }} 4.3) The EMC result: implications in the Naive Parton Model

\noindent
{\phantom{2) 2.1) }} 4.3.1) The `Spin crisis in the Parton Model'

\noindent
{\phantom{2) 2.1) }} 4.3.2) The angular momentum sum rule

\noindent
{\phantom{2) 2.1) }} 4.3.3) Trouble with the strange quark

\noindent
{\phantom{2) }} 4.4) Analysis of $\Gamma^p_1$ using the QCD Improved
Parton Model

\noindent
{\phantom{2) 2.1) }} 4.4.1) The Operator Product Expansion for $\Gamma^p_1$

\noindent
{\phantom{2) }} 4.5) The new experiments: neutron data and the Bjorken sum rule

\noindent
{\phantom{2) 2.1) }} 4.5.1) The new experiments on deuterium and $^3$He

\noindent
{\phantom{2) 2.1) }} 4.5.2) Tests of the Bjorken sum rule

\vskip 6pt
\noindent
5) {\bf The Operator Product Expansion (OPE)}

\noindent
{\phantom{2) }} 5.1) General structure of the OPE

\noindent
{\phantom{2) }} 5.2) Equations for the moments of $g_{1,2}(x,Q^2)$

\noindent
{\phantom{2) }} 5.3) Is there a connection between $g_1(x)$ and $g_2(x)$?

\noindent
{\phantom{2) }} 5.4) Does the first moment of $g_2(x)$ vanish? (The
Burkhardt-Cottingham sum rule)

\noindent
{\phantom{2) }} 5.5) The Efremov-Leader-Teryaev sum rule

\vskip 6pt
\noindent
6) {\bf \mbox{\boldmath $g_1(x)$}: The axial anomaly and the gluon current
\mbox{\boldmath $K^\mu$}}

\noindent
{\phantom{2) }} 6.1) Is there really a `spin crisis'?

\noindent
{\phantom{2) }} 6.2) On the connection between QCD, the Quark Model and the
Naive Parton Model

\noindent
{\phantom{2) }} 6.3) The axial anomaly

\noindent
{\phantom{2) }} 6.4) The axial gluon current $K^\mu$ and the gluon spin

\vskip 6pt
\noindent
7) {\bf \mbox{\boldmath $g_1(x)$}: Reinterpretation of the measurement of
\mbox{\boldmath $\Gamma^p_1$}}

\noindent
{\phantom{2) }} 7.1) The r\^ole of the anomalous gluon contribution

\noindent
{\phantom{2) }} 7.2) Why the gluon contribution survives as $Q^2 \to \infty$

\noindent
{\phantom{2) }} 7.3) The angular momentum sum rule revisited

\vskip 6pt
\noindent
8) {\bf \mbox{\boldmath $g_1(x)$}: The anomaly in the QCD Field Theoretic
Model}

\noindent
{\phantom{2) }} 8.1) The factorization theorem

\noindent
{\phantom{2) }} 8.2) The gluonic contribution to $g_1(x)$ and the definition
of $\Delta g(x)$

\noindent
{\phantom{2) }} 8.3) Measuring $\Delta g(x)$ via 2-jet production

\noindent
{\phantom{2) }} 8.4) The shape of $g_1^p(x)$ and the anomalous gluon
contribution

\noindent
{\phantom{2) }} 8.5) The anomaly in $\nu p$ elastic scattering and weak
interaction structure functions

\noindent
{\phantom{2) }} 8.6) Operator Product Expansion ${\bf vs.}$ QCD Improved
Parton Model

\vskip 6pt
\noindent
9) {\bf \mbox{\boldmath $g_1(x)$}: Non-perturbative effects in the
interpretation of the data on \mbox{\boldmath $\Gamma^p_1$}}

\noindent
{\phantom{2) }} 9.1) Topological effects and the axial ghost in QCD: a
pedagogical reminder

\noindent
{\phantom{2) }} 9.2) A reminder about the U(1) problem

\noindent
{\phantom{2) }} 9.3) The axial ghost

\noindent
{\phantom{2) }} 9.4) R\^ole of the ghost in the nucleon spin problem

\noindent
{\phantom{2) }} 9.5) Attempts to provide a physical interpretation of the EMC
result

\vskip 6pt
\noindent
10) {\bf \mbox{\boldmath $g_{1,2}(x)$} in the QCD Field Theoretic Model}

\noindent
{\phantom{2) }} 10.1) The QCD Field Theoretic Model

\noindent
{\phantom{2) }} 10.2) $g_1(x)$: Longitudinal polarization

\noindent
{\phantom{2) }} 10.3) $g_2(x)$: Transverse polarization

\noindent
{\phantom{2) }} 10.4) The Naive Parton Model for $g_2(x)$ revisited

\vskip 6pt
\noindent
11) {\bf Future Experiments}

\noindent
{\phantom{2) }} 11.1) Semi-inclusive deep inelastic lepton-hadron reactions

\noindent
{\phantom{2) }} 11.2) Hadron-hadron reactions

\noindent
{\phantom{2) 2.1) }}~~11.2.1) Hard $\gamma$ and Drell-Yan reactions

\noindent
{\phantom{2) 2.1) }}~~11.2.2) Heavy quark production

\noindent
{\phantom{2) }} 11.3) Jet handedness

\vskip 6pt
\noindent
12) {\bf Conclusions}

\vskip 6pt
\noindent
{\bf Appendices}: A) Kinematical relations amongst asymmetries and scaling
functions

\noindent
{\phantom {\bf Appendices:}} B) Current matrix elements in the Parton Model

\noindent
{\phantom {\bf Appendices:}} C) Transverse spin: the restoration of
electromagnetic gauge invariance

\noindent
{\phantom {\bf Appendices:}} D) Distribution and two-parton correlation
functions for transverse spin

\noindent
{\phantom {\bf Appendices:}} E) The Burkhardt-Cottingham and the
Efremov-Leader-Teryaev sum

\noindent
{\phantom {{\bf Appendices:} E)}} rules in the QCD Field Theoretic Model

\vskip 6pt
\noindent
{\bf Acknowledgements}

\vskip 6pt
\noindent {\bf References}

%\vskip 6pt
%\noindent {\bf Figure captions}
%
%end of contents
%
\newpage
\section{\large{Introduction}}
\vskip 6pt

Deep inelastic lepton-hadron scattering (DIS) has played a seminal role in the
development of our present understanding of the sub-structure of elementary
particles. The discovery of Bjorken scaling in the late nineteen-sixties
provided the critical impetus for the idea that elementary particles contain
almost pointlike constituents and for the subsequent invention of the Parton
Model. DIS continued to play an essential role in the long period of
consolidation that followed, in the gradual linking of partons and quarks,
in the discovery of the existence of missing constituents, later identified
as gluons, and in the wonderful confluence of all the different parts of the
picture into a coherent dynamical theory of
quarks and gluons -- Quantum Chromodynamics (QCD).

In more recent times the emphasis has shifted to the detailed study of the
$x$-dependence of the parton distribution functions and to the study of their
$Q^2$-evolution, probably the most direct test of the perturbative aspects
of QCD.

Polarized DIS, involving the collision of a longitudinally polarized
lepton beam on a polarized target (either longitudinally or transversely
polarized) provides a different, complementary and equally important insight
into the structure of the nucleon. Whereas ordinary DIS probes simply the
number density of partons with a fraction $x$ of the momentum of the parent
hadron, polarized DIS can partly answer
the more sophisticated question as to the number density of partons with given
$x$ and given spin polarization, in a hadron of definite polarization, either
along or transverse to the motion of the hadron.

But what is quite extraordinary and unexpected {\it ab~initio} is the
richness and subtlety of the dynamical effects associated with the
polarized case. Whereas the unpolarized scaling functions F$_{1,2}(x)$ have
a simple interpretation in the Naive Parton Model (where the nucleon is
considered as an ensemble of essentially free massless partons) and a
straightforward generalisation in the framework of
perturbative QCD, the spin dependent scaling functions $g_{1,2} (x)$ are much
more subtle, each fascinating in its own way. The function $g_1(x)$ which, at
first sight, seems trivial to deal with in the Naive Parton Model, turns out,
within perturbative QCD, to have an anomalous gluon contribution associated
with
it. In addition the first moment of $g_1(x)$  turns out to be connected with an
essentially non-perturbative aspect of QCD, the axial ghost which is invoked to
resolve the $U(1)$ problem of the mass of the $\eta^\prime$. And $g_2(x)$
turns out not to have any interpretation at all in purely partonic language.

What is also fascinating is the extraordinary interplay of theory and
experiment in the study of $g_1(x)$. For a long time the theory of $g_1(x)$
remained comfortably at the level of the Naive Parton Model. Then, in 1988,
came the disturbing results of the European Muon Collaboration (EMC) [ASH 88,
89], which differed significantly from the naive theoretical predictions. These
results could be argued to imply that the sum of the spins carried by the
quarks
in a proton ($S^{\rm quarks}_z$) was consistent with zero, rather than
with 1/2 as given in the quark model, suggesting a ``spin crisis in the parton
model'' [LEA 88]. This led to an intense scrutiny of the basis of the
theoretical calculation of $g_1(x)$ and the discovery of the anomalous gluon
contribution [EFR 88]. (As so often happens in theoretical physics it turns out
that such an effect had already been studied to some extent in a largely
overlooked paper of 1982 [LAM 82]). So surprising was this discovery that the
calculation was immediately checked by three different groups [ALT 88; CAR~88;
LEA 88a] who all arrived at the same result. (Somewhat fortuitously, as it
turns out, as was demonstrated in [CAR 88]).

In the modified theoretical picture, the quantity $\Delta\Sigma=2S^{quark}_z$,
whose value had to be consistent with zero as a consequence
of the EMC experiment, is replaced by the linear combination (for 3 flavours)
$\Delta\Sigma - (3\alpha_s /2 \pi) \, \Delta g$ where $\Delta g = \int^1_0
dx \, \Delta g(x)$ and $\Delta g(x)$ is the polarized gluon number density.

It has to be stressed that as a consequence of QCD a measurement of the first
moment of $g_1(x)$ {\it does not} measure the $z$-component of the sum of
the quark spins. It measures only the superposition
$$
{1\over 9} \left[ \Delta\Sigma - {3\alpha_s \over 2 \pi} \Delta g \right]
$$
and this linear combination can be made small by a cancellation between quark
and gluon contributions. Thus the EMC results {\it ceases
to imply that $\Delta\Sigma$ is small.}

The function $g_2 (x)$, on the other hand, does not have any simple
interpretation in the Naive Parton Model and it is a triumph of perturbative
QCD that one can derive a sensible, gauge invariant result for it in the
QCD Field Theoretic Model [EFR 84].

In this review we concentrate almost exclusively on the polarized case. A good
survey of the unpolarized case can be found in [ALT 82] or, at a more
introductory level, in [LEA 94]. Our treatment attempts to cover both the
theory and the phenomenology of the subject and is meant to be reasonably
self-contained.

On the phenomenological side we outline the formalism for discussing DIS in
terms of one photon exchange. We briefly mention the extension to weak
interactions (both neutral and charged current) initiated by charged leptons.
We introduce the scaling functions $g_{1,2}(x)$
and consider how data on them can be extracted from experiments on
cross-section asymmetries using both longitudinal and transversely polarized
targets. We consider both hydrogen and nuclear targets and we also address the
problem of how to extract data at fixed $Q^2$ over the whole range of $x$, as
needed for the testing of sum rules.

Since the data taken in different $x$-bins correspond to different ranges
of $Q^2$, the production of a data set for all $x$ at some fixed $Q^2$ is a
non trivial ``experimental'' problem. There is some evidence that the Compton
asymmetry $A_1(x,Q^2)$ is almost independent of $Q^2$ and the extraction of
$g_1(x,Q^2)$ at the desired $Q^2$ is carried out by {\it assuming $A_1$
independent of} $Q^2$. Now it is reasonable to believe that $A_1(x,Q^2)$
will vary only slowly with $Q^2$, but there is no justification for
believing it to be exactly constant. Nonetheless it is experimental
practice NOT to include an error in $g_1(x,Q^2)$ to reflect the uncertainty
in the variation of $A_1$ with $Q^2$. This is, to say the least, quite
unjustifiable. A possible strategy for carrying out the extrapolation
in $Q^2$ is discussed in Section 2.1.3 d).

An extensive phenomenological analysis of all the data on $g_1(x, Q^2)$ and
its first moment $\Gamma_1(Q^2)$ is given including the very recent data from
experiments on deuterium and $^3$He targets. The implications of the data are
considered in the framework of the Naive Parton Model and of the Operator
Product Expansion (OPE). Aspects discussed include: information from hyperon
$\beta$-decay; the extrapolation to $x = 0$; higher twist effects.

We argue, in agreement with several recent analyses, that there is absolutely
no evidence for a failure of the very fundamental Bjorken sum rule.

On the theoretical side we distinguish between three approaches of varying
degrees of sophistication:

i) In the Naive Parton Model the nucleon is viewed as an assembly of
free, parallel moving quarks. The treatment is {\it probabilistic} and its
essence is summarized by the relation connecting the virtual photon-hadron
cross-section $\sigma^{\gamma^* H}$ to a convolution of the $\gamma^*$-quark
cross-sections and the number density of quarks in the hadron, $f_{q/H}(x)$:
$$
\sigma^{\gamma^* H} = \int dx^\prime \, \sigma^{\gamma^* q} (x/x^\prime)
f_{q/H}(x^\prime) \,.
$$

ii) In the QCD Improved Parton Model one supplements the Naive Parton Model
with QCD-controlled, probabilistic, Altarelli-Parisi splitting functions.
The description remains probabilistic and the main effect is the
replacement of number densities $f_{q/H}(x)$ by $Q^2$-dependent densities
$f_{q/H}(x,Q^2)$.

iii) In the QCD Field Theoretic Model one starts with genuine QCD Feynman
diagrams. We call it a ``Model'' because the diagrams have to be divided
into a hard scattering part, which is calculated strictly according to
perturbative QCD, and a soft part which we cannot calculate and which has
to be modelled. One makes certain reasonable sounding assumptions about
the behaviour of this non-perturbative part. The Model is, in general,
{\it not probabilistic} and involves hadronic matrix elements of products
of quark and gluon interacting fields at different space-time points. In
those cases where the result is probabilistic one can usually recover
the Naive Parton Model upon treating the fields as free fields, or, with
a less drastic approximation, the QCD Improved Parton Model.

The Operator Product Expansion (which deals with matrix elements of local
operators), where it can be justified leads to results perfectly compatible
with the Field Theoretic Model, but the latter has a much wider field of
applicability and can be utilised in situations where use of the OPE cannot
be justified.

The calculation of $g_{1,2}(x)$ in the Naive Parton Model is covered in some
detail and the source of the difficulties with $g_2(x)$ is, we hope, very
clearly delineated. We introduce and discuss the general structure of the
Operator Product Expansion and give expressions for the n-th
moments of $g_{1,2}(x, Q^2)$. We are careful to state the range of $n$ for
which the results can be proved and this leads to an examination of various
claims in the literature about relationships between $g_1$ and $g_2$, about
the first moment of $g_2(x)$ ($\int dx~g_2(x)=0$: the Burkhardt-Cottingham
sum rule) and about the second moment of $g_1(x) + 2g_2  (x)$
($\int dx~x[g_1(x) + 2g_2(x)] = 0$: the Efremov-Leader-Teryaev sum rule).

We explain that neither of these sum rules follows directly from the OPE and
that their validity rests upon an assumption about the invertability of certain
Fourier transforms (equivalently, about the behaviour as $x\to 0$) and that
this invertability does not occur if $g_2(x)$ has the small $x$
behaviour $1/x^2$ suggested by Regge Theory, so that the above
integrals diverge.

The axial anomaly and the gluon axial current $K^\mu$ are discussed in some
detail. We study the connection between $K^\mu$ and the gluon spin operator and
derive the anomalous gluon contribution to $g_1(x)$. There is some
disagreement in the literature regarding the kernel relating the polarized
gluon
number density $\Delta g(x)$ to $g_1(x)$, and we give arguments in favour of
a particular choice.

There are also arguments in the literature against the whole concept of an
anomalous gluon contribution, on the grounds that $K^\mu$ is not a gauge
invariant operator. Here we would like to remind readers that in gauge
theories (even in the Abelian case of QED) it often happens that the operators
representing fundamental physical quantities like momentum are
{\it not} gauge-invariant. This lack of gauge invariance is
innocuous and irrelevant provided that the {\it expectation values}
of those operators between physical states {\it are}
gauge invariant. Ultimately, the gauge invariance of $\Delta g(x)$ is guaranted
by showing how it is related to the physically measurable cross-section for
2-jet production.

There are also claims in the literature that the anomalous contribution to
$g_1$ vanishes if the calculation is carried out with strictly massless
gluons. We point out that this is irrelevant. In the QCD Field Theoretic Model
there is, in principle, an integral over the $k^2$ of the gluon and the point
$k^2$ = 0 is of zero measure. Only in a Parton Model calculation with
gluon mass fixed at $k^2$ = 0 would there be any problem.

The whole question of the calculation of $g_{1,2}$ and of the anomaly is
finally approached from the point of view of the QCD Field Theoretic Model.
Contrary to some assertions in the literature we show that there is perfect
consistency between this approach and the Operator Product Expansion.

We contrast the calculation of $g_2(x)$ and $g_1(x)$ in QCD Field Theoretic
Model, showing that the Feynman diagrams which have a simple quark-parton
interpretation lead to an unacceptable result for $g_2(x)$, {\it i.e.} a result
which does not respect gauge invariance. A detailed demonstration is given of
how the  unwanted gauge dependent terms are cancelled when one includes
diagrams corresponding to quark-gluon correlations in the nucleon, a concept
which is outside the framework of the Parton Model.

We provide a brief introduction to non-perturbative and topological effects
in QCD, to the $U(1)$ problem, to the axial ghost and to the isosinglet,
Generalized Goldberger-Treiman relation (GGT). A connection is established
between the total quark spin and the coupling of the $\eta^\prime$ to nucleons
and this suggests that the sum of the quark spins is, in fact, close to the
canonical value of 1/2, in agreement with the interpretation of the EMC data
utilising an anonalous gluon contribution to $g_1(x)$.

A dominant feature throughout is the polarized gluon distribution $\Delta
g(x)$.
It is essential that this important element in the internal structure of
the nucleon should be measured independently, in as many other reactions as
possible. We end with a brief survey of reactions in which this might be
feasible.

Finally, a word about notation. It may seem pedantic to squabble about
choices of notation. But when a subject threatens to become incomprehensible
as a result of an injection of illogical, confusing notation, it is necessary
to draw attention to the danger. Our Section on notation, consequently, will
comment on this issue.

\subsection{Notation}
\vskip 6pt

The Field Theoretic approach to hard scattering was developed in a pioneering
paper by Ralston and Soper in 1979 [RAL 79]. They introduced a precise,
logical, clear and informative notation for the hadronic matrix elements
of the bilocal operators that appear.

On the other hand the spin dependent scaling functions $g_{1,2}(x)$ were
introduced by Bjorken in 1966 [BJO 66]. They too have a precise and clear
meaning in the expression for the measured spin-dependent cross-section.

A complete relabelling of the Ralston-Soper matrix elements [JAF 91] ought
to require serious justification (especially the use of $g_1$ for one of
them when it is only the leading twist approximation to Bjorken $g_1$) and
such justification has never been offered. We therefore follow the
Ralston-Soper notation.

Concerning $\gamma$ matrices and spinors we adopt the conventions of Bjorken
and Drell [BJO 65] except that our spinors are normalised so that for a
fermion with four-momentum
$p^\mu = (E, \mbox{\boldmath $p$})$
$$
u^\dagger u = 2 E,\quad v^\dagger v = 2E,
\quad \bar{u} u = 2 M,\quad \bar{v} v = -2 M
$$
for both massive and massless particles.

One has then
$$
\bar{u}(p)\gamma^\mu u(p) = 2 p^\mu\qquad\bar{v}(p)\gamma^\mu v(p) = 2p^\mu \,.
$$

We use
$$
\gamma_5 = i~\gamma^0\gamma^1\gamma^2\gamma^3
$$
and our convariant spin vectors $S^\mu$ are normalised to $S_\mu S^\mu = -1\,.$

For a fermion of mass $M$, then,
$$
\bar{u}(p, S)\gamma^\mu\gamma_5 u(p,S) =
-\bar{v} (p,S)\gamma^\mu\gamma_5 v(p,S) = 2M S^\mu \,.
$$

For massless fermions of helicity $\lambda = \pm 1/2$, or of chirality
$2\lambda = \pm~1$,
$$
\bar{u}(p,\lambda)\gamma^\mu \gamma_5 u(p,\lambda) =
-\bar{v} (p,\lambda) \gamma^\mu \gamma_5 v(\rho,\lambda) =
\lim_{M\to 0}~2MS^\mu (\lambda) = 4\lambda~p^\mu \,.
$$

All states are normalized so that
$$
\langle P^\prime | P \rangle = (2\pi)^3~2 E~\delta^3
(\mbox{\boldmath $p$}^\prime - \mbox{\boldmath $p$})\,.
$$

We use
$$
\varepsilon_{0123} = + 1~.
$$

Finally, because `transversity' has a quite specific meaning [DAL 56; KOT 70]
in terms of a phase convention {\it etc.}, we avoid its use and simply talk
of longitudinal and transverse polarization states.
%
%end of Section 1
%
\setcounter{section}{1}
\section{\large{Polarized lepton-nucleon Deep Inelastic Scattering - General
aspect}}
\vskip 6pt
\subsection{General formalism in one photon exchange approximation}
\vskip 6pt
\setcounter{equation}{0}

Consider the inelastic scattering of polarized leptons on polarized nucleons.
We denote by $m$ the lepton mass,  $ k~(k^\prime$) the initial (final)
lepton four-momentum and $s~(s^\prime)$ its covariant spin four-vector, such
that $s \cdot  k$ = 0 $(s^\prime \cdot  k^\prime = 0)$ and $s \cdot s =
- 1$ $(s^\prime \cdot s^\prime = -1)$; the
nucleon mass is $M$ and the nucleon four-momentum and spin four-vector are,
respectively, $P$ and $S$. Assuming one photon exchange, see Fig. 2.1, the
differential cross-section for detecting the final polarized lepton in the
solid angle $d\Omega$ and in the final energy range $(E^\prime,~E^\prime +
dE^\prime)$ in the laboratory frame, $P = (M, \bfo), ~k = (E, \bfk),
{}~k^\prime = (E^\prime, {\bfk}^\prime)$, can be written as [LEA 85]
\begin{equation}
{d^2\sigma\over d\Omega~dE^\prime} = {\alpha^2\over 2 M q^4}~{E^\prime\over
E}~L_{\mu\nu}~W^{\mu\nu} \,,
\end{equation}
where $q =  k- k^\prime$ and $\alpha$ is the fine structure constant.

In Eq. (2.1.1) the leptonic tensor $L_{\mu\nu}$ is given by
\begin{equation}
L_{\mu\nu} ( k, s; k^\prime, s^\prime) = [\bar{u} (k^\prime, s^\prime)
{}~\gamma_{\mu}~u(k,s)]^\ast
{}~[\bar{u}(k^\prime, s^\prime)~\gamma_{\nu}~u(k,s)]
\end{equation}
and can be split into symmetric $(S)$ and antisymmetric $(A)$ parts under
$\mu,\nu$ interchange:
\begin{eqnarray}
L_{\mu\nu}(k, s; k^\prime, s^\prime)&=& L^{(S)}_{\mu\nu}~(k;k^\prime) +
iL^{(A)}_{\mu\nu}~(k,s;k^\prime) \\
&+&L^{\prime~(S)}_{\mu\nu}~(k,s;k^\prime,s^\prime) +
iL^{\prime~(A)}_{\mu\nu}~(k;k^\prime, s^\prime)\nonumber
\end{eqnarray}
where
\begin{eqnarray}
L^{(S)}_{\mu\nu}~(k;k^\prime) &=& k_\mu k^\prime_\nu + k^\prime_\mu k_\nu -
g_{\mu\nu}~(k\cdot k^\prime - m^2)\\
L^{(A)}_{\mu\nu}~(k,s;k^\prime) &=&
m~\varepsilon_{\mu\nu\alpha\beta}~s^\alpha~(k-k^\prime)^{\beta}\\
L^{\prime~(S)}_{\mu\nu}(k,s;k^\prime,s^\prime)&=&(k\cdot s^\prime)~
(k^\prime_\mu s_\nu +
s_{\mu}k^\prime_\nu-g_{\mu\nu}~k^\prime\cdot s) \nonumber\\
&-&(k\cdot k^\prime - m^2)~(s_\mu s^\prime_\nu + s^\prime_\mu s_\nu -
g_{\mu\nu}~s\cdot s^\prime) \\
&+&(k^\prime \cdot s)(s^\prime_\mu k_\nu + k_\mu s^\prime_\nu) -
(s \cdot s^\prime)(k_\mu k^\prime_\nu + k^\prime_\mu k_\nu)\nonumber\\
L^{\prime~(A)}_{\mu\nu}~(k;k^\prime,s^\prime) &=&
m~\varepsilon_{\mu\nu\alpha\beta}
{}~s^{\prime\alpha} (k-k^\prime)^{\beta} \,.
\end{eqnarray}

On summing Eq. (2.1.3) over $s^\prime$ and averaging over $s$ one recovers
the usual unpolarized leptonic tensor, 2$L^{(S)}_{\mu\nu}$. On summing over
$s^\prime$, one obtains $2L^{(S)}_{\mu\nu} + 2iL^{(A)}_{\mu\nu}$.

The unknown hadronic tensor $W_{\mu\nu}$ is similarly defined in terms of
four structure functions as [DRE 64; CAR 72; HEY 72].
\begin{equation}
W_{\mu\nu}(q; P,S) = W^{(S)}_{\mu\nu}(q;P) + i~W_{\mu\nu}^{(A)}(q; P, S)
\end{equation}
with
\begin{eqnarray}
{1\over 2M}~W^{(S)}_{\mu\nu}(q;P) &=& \left(-g_{\mu\nu} + {q_\mu q_\nu\over
q^2}\right)~W_1(P\cdot q, q^2) \nonumber\\
&+&\left[\left(P_\mu -{P\cdot q\over q^2}~q_\mu\right)\left(P_\nu -
{P\cdot q\over q^2}~q_\nu\right)\right]{W_2(P\cdot q, q^2)\over M^2}\\
{1\over 2M}~W^{(A)}_{\mu\nu} (q;P,S) &=&
\varepsilon_{\mu\nu\alpha\beta}~q^\alpha\Biggl\{M S^\beta G_1 (P\cdot q, q^2)
\nonumber\\
&+&[(P\cdot q) S^\beta - (S\cdot q) P^\beta]~{G_2 (P \cdot q, q^2)\over M}
\Biggl\} \,.
\end{eqnarray}

{}From Eqs. (2.1.1, 3, 8) one has
\begin{eqnarray}
{d^2\sigma\over d\Omega~dE^\prime}&=&{\alpha^2\over 2M q^4}~{E^\prime\over E}~
\left[ L^{(S)}_{\mu\nu}~W^{\mu\nu(S)} +
L^{\prime~(S)}_{\mu\nu}~W^{\mu\nu (S)} \right.\nonumber\\
&-&\left. L^{(A)}_{\mu\nu}~W^{\mu\nu(A)} - L^{\prime~(A)}_{\mu\nu}~
W^{\mu\nu(A)}\right] \,.
\end{eqnarray}

The individual terms in square brackets can be separately studied by
considering
cross-sections  or differences between cross-sections with particular initial
and final polarizations [ANS 79]. Each of these terms is, at least in
principle, a measurable quantity which is either a function of the two
spin-averaged structure functions $W_1$
and $W_2$ (terms containing $W^{(S)}_{\mu\nu})$ or of the two spin-dependent
structure functions $G_1$ and $G_2$ (terms containing $W^{(A)}_{\mu\nu})$. For
example, the usual
unpolarized cross-section is proportional to $L^{(S)}_{\mu\nu}~W^{\mu\nu(S)}$
\begin{eqnarray}
{d^2\sigma^{unp}\over d\Omega~dE^\prime}\left(k, P;k^\prime\right) &=&{1\over
4}\sum_{s,s^\prime,S}~{d^2\sigma\over d\Omega~d E^\prime}~(k, s, P, S;
k^\prime, s^\prime) \nonumber\\
&=&{\alpha^2\over 2M q^4}~{E^\prime\over E}~2L_{\mu\nu}^{(S)}~W^{\mu\nu(S)}\,,
\end{eqnarray}
while differences of cross-sections with opposite target spins single out the
$L_{\mu\nu}^{(A)}~W^{\mu\nu(A)}$ term:
$$
\sum_{s^\prime}\left[{d^2\sigma\over d\Omega~dE^\prime}~(k, s, P, - S;
k^\prime, s^\prime)
-{d^2\sigma\over d\Omega ~dE^\prime}~(k, s, P, S; k^\prime, s^\prime)\right]
$$
\begin{equation}
={\alpha^2\over 2Mq^4}~{E^\prime\over E}~4L_{\mu\nu}^{(A)}~W^{\mu\nu(A)}\,.
\end{equation}

\subsubsection{Structure functions and Bjorken scaling}
\vskip 6pt

The cross-section for the inelastic scattering of unpolarized leptons on
unpolarized nucleons, in the laboratory frame, can be written explicitely
using Eqs. (2.1.4, 9, 12) and neglecting the lepton mass, as
\begin{equation}
{d^2\sigma^{unp}\over d\Omega~dE^\prime} = {4\alpha^2 E^{\prime 2}\over q^4}
\Biggl[2W_1 \sin^2 {\theta\over 2} + W_2 \cos^2{\theta\over 2}\Biggl]
\end{equation}
where $\theta$ is the LAB scattering angle of the lepton. Its measurement
supplies information on the unpolarized structure functions
$W_1(P\cdot q, q^2)$ and $W_2 (P\cdot q, q^2).$

In the Bjorken limit, or Deep Inelastic Scattering (DIS) regime,

\begin{equation}
-q^2 = Q^2 \to \infty\quad\quad
\nu = E - E^\prime \to \infty \quad\quad x = {Q^2\over 2 P\cdot q} =
{Q^2\over 2M\nu} \,, \>{\mbox{\rm fixed}}
\end{equation}
they are known to approximately scale:
\begin{eqnarray}
\lim_{Bj}~MW_1(P\cdot q, Q^2) &=& F_1 (x)\nonumber\\
\lim_{Bj}~\nu W_2(P\cdot q, Q^2) &=& F_2 (x) \,,
\end{eqnarray}
where $F_{1,2}$ vary very slowly with $Q^2$ at fixed $x$.

Similarly, from Eqs. (2.1.5, 10, 13), one has
$$\sum_{s^\prime}\left[{d^2\sigma\over d\Omega~dE^\prime}(k, s, P, S; k^\prime,
s^\prime) - {d^2\sigma\over d\Omega~d E^\prime} (k, s, P - S; k^\prime,
s^\prime) \right]\equiv$$
\begin{equation}
\equiv {d^2\sigma^{s,S}\over d\Omega~d E^\prime} -
{d^2 \sigma^{s, -S}\over d\Omega~dE^\prime}~~=
\end{equation}
$$={8m \alpha^2 E^\prime\over q^4E}\left\{\left[(q\cdot S) (q\cdot s) +
Q^2 (s\cdot S)\right] MG_1 + Q^2 \Bigl[
(s\cdot S)(P\cdot q)-(q\cdot S)(P\cdot s)\Bigr]{G_2\over M}\right\}$$
which yields information on the polarized structure functions
$G_1 (P\cdot q, q^2)$ and $G_2 (P\cdot q, q^2)$. They also, in the Bjorken
limit, are expected to scale approximately:
\begin{eqnarray}
\lim_{Bj}~{(P\cdot q)^2\over \nu}~G_1(P\cdot q, Q^2)&=& g_1 (x)\\
\lim_{Bj}~\nu ~(P\cdot q)~G_2 (P\cdot q, q^2) &=& g_2 (x) \,. \nonumber
\end{eqnarray}
In terms of $g_{1,2}$ the expression (2.1.10) for $W^{(A)}_{\mu\nu}$ becomes
\begin{equation}
W^{(A)}_{\mu\nu }(q; P, s) = {2M\over P\cdot q}~\varepsilon_{\mu v\alpha\beta}~
q^\alpha\Biggl\{ S^\beta g_1 (x, Q^2)+ \left[S^\beta
- {(S\cdot q) \, P^\beta\over (P\cdot q)}\right] g_2 (x, Q^2)\Biggl\}\,.
\end{equation}

\subsubsection{Cross-section differences}
\vskip 6pt
Aiming at gathering information on the polarized structure functions G$_1$
and G$_2$ we look at Eq. (2.1.17) and specialize it to particular spin
configurations of the incoming leptons and target nucleons.

Let us consider first the case of longitudinally polarized leptons, that is
initial leptons with spin along $(\to)$ or opposite $(\gets)$ the direction
of motion, while the nucleons {\it at rest} are polarized along
$(S)$ or opposite $(-S)$ an arbitrary direction $\hat{\bfbs}$.
We then have
\begin{eqnarray}
s^\mu_{\to} &=& - \, s^\mu_{\gets} = {1\over m}~(|\bfk|,
{}~\hat{\bfk}E)\qquad
\hat{\bfk} = {\bfk \over |\bfk|}\\
S^\mu &=&  (0, \hat{\bfbs}) \,. \nonumber
\end{eqnarray}

We take the $z$-axis along the incoming lepton direction and define
(see Fig. 2.2)
\begin{eqnarray}
k^{\mu}&=&(E,0,0,|\bfk|) \simeq E (1,0,0,1)\nonumber\\
k^{\prime \mu}&=&(E^\prime, {\bfk^\prime})
\simeq E^\prime (1,{\hat{\bfk}}^\prime)\nonumber\\
&=&E^\prime (1, \sin\theta\cos\varphi,~\sin\theta\sin\varphi,~\cos\theta)\\
{\hat{\bfbs}} &=& (\sin\alpha\cos\beta,~\sin\alpha\sin\beta,~\cos\alpha)\,.
\nonumber
\end{eqnarray}

Using Eqs. (2.1.20, 21) into Eq. (2.1.17) one obtains (at leading order in
$m/E$)
%\newpage
\begin{eqnarray}
{d^2\sigma^{\to,S}\over d\Omega~dE^\prime}&-&{d^2\sigma^{\to,-S}\over
d\Omega~dE^\prime} =
-{4\alpha^2\over Q^2}~{E^\prime\over E}\\
&\times&\Biggl\{[E \cos \alpha + E^\prime \cos \Theta] MG_1
+ 2EE^\prime [ \cos \Theta - \cos \alpha]~G_2\Biggl\} \,. \nonumber
\end{eqnarray}
$\alpha$ is the polar angle of the nucleon spin direction,
{\it i.e.} the angle between ${\hat{\bfk}}$ and $\hat{\bfbs}$,
and $\Theta$ is the angle between the outgoing lepton direction,
${\hat{\bfk}}^\prime$, and $\hat{\bfbs}$ :
\begin{eqnarray}
\cos\Theta&=&\sin\theta \cos\varphi \sin\alpha \cos\beta \nonumber\\
&+&\sin\theta \sin\varphi \sin\alpha \sin\beta + \cos\theta \cos\alpha \\
&=& \sin\theta \sin\alpha \cos\phi + \cos\theta \cos\alpha\nonumber
\end{eqnarray}
where $\phi = \beta - \varphi$ is the azimuthal angle between the
$(\hat{\bfk},\hat{\bfk}^\prime)$ scattering plane and the $(\hat{\bfk},
\hat{\bfbs})$ polarization plane (see Figs. 2.2 and 2.3).

For particular values of $\alpha$ one recovers familiar results
[CAR 72; HEY 72]. For nucleons polarized along $(\Rightarrow)$ or opposite
($\Leftarrow$) the initial lepton
direction of motion one has $\alpha$ = 0, $\Theta = \theta$, and
Eq. (2.1.22) gives
\begin{equation}
{d^2\sigma^{\begin{array}{c}\hspace*{-0.2cm}\to\vspace*{-0.3cm}\\
\hspace*{-0.2cm}\Rightarrow\end{array}}\over d\Omega~dE^\prime} -
{d^2\sigma^{\begin{array}{c}\hspace*{-0.2cm}\to\vspace*{-0.3cm}\\
\hspace*{-0.2cm}\Leftarrow\end{array}}\over
d\Omega~dE^\prime} =
-{4\alpha^2\over Q^2}~{E^\prime\over E}\Biggl[(E + E^\prime\cos\theta)MG_1 -
Q^2G_2\Biggl] \,.
\end{equation}

If the nucleons are transversely polarized, that is the nucleon spin is
perpendicular to the direction of the incoming lepton, then $\alpha =
\pi/2$ and Eqs. (2.1.22, 23) yield
\begin{equation}
{d^2 \sigma^{\to\Uparrow}\over d\Omega~dE^\prime} -
{d^2 \sigma^{\to\Downarrow}\over d\Omega~dE^\prime} =
- {4 \alpha^2\over Q^2}~{E^{\prime 2}\over E}~\sin\theta \cos\phi ~(MG_1 +
2EG_2)\,.
\end{equation}

Notice that if the nucleon spin is perpendicular to the scattering plane
$(\alpha = \phi = \pi/2)$, then the cross-section difference in
Eq. (2.1.25) is zero; such a difference has its maximum absolute value when
$\phi = 0$ or $\pi$, that is when the nucleon
spin vector, perpendicular to $\hat{\bfk}$, lies in the scattering plane.

In the above we considered longitudinally polarized leptons. Note that it is
impractical to deal with transversely polarized leptons. Because, for
transversely polarized incoming leptons, that is leptons with spin
perpendicular to the direction of motion, we have
\begin{equation}
s = (0, \hat{\bfs})\,,
\end{equation}
with the unit vector $\hat{\bfs}$ orthogonal to $\hat{\bfk},~\hat{\bfs}
\cdot \hat{\bfk} = 0$. In such a case, contrary to what happens for
longitudinally polarized leptons, Eq. (2.1.20), there is no factor $E/m$ to
cancel the factor $m/E$ which appears in the cross-section differences
(2.1.17), and the latter turn out to be vanishingly small in
the large energy limit $(m/E \to 0)$.

Useful information on the unpolarized structure functions  $W_1$  and $W_2$
could be obtained by looking at initial-final lepton spin asymmetries
[ANS 79]; such a technique, however, requires the measurement of the scattered
lepton polarization, which is probably too difficult to contemplate at present.

\subsubsection{Experimental measurement of \mbox{\boldmath $g_1$} and
\mbox{\boldmath $g_2$} on nucleon targets}
\vskip 6pt
Measurements of cross-section differences with particular lepton and nucleon
spin configurations provide information on the polarized structure functions
$G_1$ and $G_2$ or on the {\it scaling functions} $g_1$ and $g_2$, Eq.
(2.1.18).

A single difference measurement, however, only yields information on a
combination of $G_1$ and $G_2$, rather than on the separate structure
functions. Extracting from the data values of $G_1$ or $G_2$ alone still
requires some further approximate procedure.

a) {\it Longitudinally polarized target}

In the performed experiments [ALG 78; BAU 83, 88; ASH 88, 89; ADE 93, 94; ANT
93] the quantity actually measured is the longitudinal spin-spin asymmetry in
$\ell p \to \ell X$,
\begin{equation}
A_{\parallel} \equiv{d\sigma^{\begin{array}{c}\hspace*{-0.2cm}\to
\vspace*{-0.3cm}\\
\hspace*{-0.2cm}\Leftarrow\end{array}} - d\sigma^{\begin{array}{c}
\hspace*{-0.2cm}\to\vspace*{-0.3cm}\\
\hspace*{-0.2cm}\Rightarrow\end{array}}\over d\sigma^{\begin{array}{c}
\hspace*{-0.2cm}\to\vspace*{-0.3cm}\\
\hspace*{-0.2cm}\Rightarrow\end{array}} + d\sigma^{\begin{array}{c}
\hspace*{-0.2cm}\to\vspace*{-0.3cm}\\
\hspace*{-0.2cm}\Leftarrow\end{array}}}~,
\end{equation}
where $d\sigma$ is short for $d^2\sigma/(d\Omega~dE^\prime)$ and the
denominator is simply twice the unpolarized cross-sections. From Eqs.
(2.1.14) and (2.1.24) we derive \begin{equation}
A_{\parallel} = {Q^2~[(E+E^\prime\cos\theta)MG_1 - Q^2G_2]\over
2EE^\prime~[2 W_1 \sin^2 (\theta/2) + W_2\cos^2 (\theta/2)]} \, \cdot
\end{equation}

For reasons to be explained shortly the asymmetry $A_{\parallel}$ is usually
expressed in terms of virtual Compton scattering asymmetries $A_{1,2}$ [LEA
85],
\begin{equation}
A_{\parallel} = D(A_1 + \eta A_2) \,,
\end{equation}
where $D$ and $\eta$ are known coefficients given in Appendix A. The analysis
of the data then proceeds through subsequent approximate steps, explained in
detail in Appendix A, which lead to the expressions
\begin{equation}
A_{\parallel} \approx DA_1
\end{equation}
and
\begin{equation}
g_1 (x) \approx {A_{\parallel}\over D}~{F_2(x)\over 2 x [ 1 + R (x)]}~,
\end{equation}
where $F_2(x)$ is the unpolarized scaling structure function, Eq. (2.1.16),
and $R$ is the ratio of the longitudinal to transverse cross-section,
\begin{equation}
R = {W_2\over W_1} \Biggl(1 + {\nu^2\over Q^2}\Biggl) - 1 \,.
\end{equation}

All the approximations involved in the above simplifications can be shown
to be quite harmless when measuring $g_1$ [ASH 88, 89]. Indeed not only is
$\eta$ small in (2.1.29) but
one can show that
\begin{equation}
|A_2| \le \sqrt{R}
\end{equation}
and $R$ is known to be small.

If the goal is simply the knowledge of $g_1$ and $g_2$ on an event-by-event
basis then there is a somewhat more direct way of utilising the data.

{}From (2.1.14, 18 and 28) we have
\begin{equation}
{M \nu Q^2E\over 2 \alpha^2E^\prime(E+E^\prime \cos\theta)}~{d^2
\sigma^{unp}\over d
\Omega~dE^\prime}~A_{\parallel} = g_1 - {2xM\over E +
E^\prime \cos\theta}~g_2
\end{equation}
which can be written more concisely as

\begin{equation}
g_1 - \kappa g_2 = 2 K~d\sigma^{unp}~A_{\parallel}
\end{equation}
with
$$\kappa = {2xM\over E + E^\prime \cos\theta} \approx {xM\over E -
Q^2/(4 Mx)}$$
\begin{equation}
K = {M \nu Q^2 E\over 4 \alpha^2 E^\prime (E + E^\prime \cos\theta)} =
{EE^\prime~\cos^2
(\theta/2) \over 2x\sigma_{Mott}~(E + E^\prime \cos\theta)}
\end{equation}
where
$$\sigma_{Mott} = \Biggl[{\alpha \cos(\theta/2)\over 2E \sin^2(\theta/2)}
\Biggr]^2 \cdot$$

The RHS of Eq. (2.1.35) is constructed directly from experiment, with no
need of further data analyses in order to extract $F_2$ and $R$, as required
in Eq. (2.1.31).

As we said previously the single measurement of $A_{\parallel}$ (and
$d\sigma^{unp}$) gives us information on the combination $g_1 - \kappa g_2$,
rather than on $g_1$ or $g_2$ alone. The usual argument [ALG 78; BAU 83, 88],
however, is that the $g_2$ term in Eq. (2.1.35) can be safely neglected
because of the kinematical coefficient $\kappa$
which, in the large energy limit, is very small, as can be seen from
Eqs. (2.1.36). This is indeed confirmed by a more careful analysis of the
$g_2$ term [LEA 88]. We can thus conclude that the measurement of the
quantities on the RHS of Eq. (2.1.35) provides essentially a direct
measurement of the polarized scaling structure function $g_1$.

If one requires data on $g_1 (x, Q^2)$ on an event-by-event basis, (2.1.35)
is the most direct way to use the experimental data on $A_{\parallel}$.
However, it is of great interest to assemble results for
$g_1(x, Q^2)$ over the entire $x$-range $0 \le x \le 1$
{\it at the same $Q^2$}. Experimentally this is impossible.
The kinematics of the experiment dictates that
smaller $x$ will correspond to a smaller range of accessible $Q^2$.

The experimentalist is thus forced to extrapolate the data in $Q^2$ at
fixed $x$, and the question is then: which quantities vary most smoothly and
slowly in $Q^2$? It is an experimental fact that, where it has been studied,
$A_{\parallel}(x, Q^2)/D$ varies only
slowly with $Q^2$. (This question is discussed further in Section 4.5).
For this reason experimentalists prefer to express their measurements in
terms of data on $A_{\parallel} (x, Q^2)$ via (2.1.31). It is then
assumed that $A_{\parallel}(x, Q^2)/D$ is essentially independent of
$Q^2$ and the value of $g_1(x, \langle Q^2 \rangle)$ quoted, for an
experiment with mean $Q^2$ equal to $\langle Q^2 \rangle$, is really
\begin{equation}
g_1(x, \langle Q^2 \rangle) \equiv \Biggl({A_{\parallel}(x)\over D}\Biggr)
{F_2(x, \langle Q^2 \rangle)\over 2x[ 1+ R(x, \langle Q^2 \rangle)]} \, \cdot
\end{equation}

It is important to bear this in mind when testing sum rules which hold at
fixed $Q^2$. A possible strategy for improving the $Q^2$ extrapolation
[EFR 94a] is discussed in Section 2.1.3 d).

As mentioned above, all the approximations leading to (2.1.31) are safe if
one is trying to evaluate $g_1(x)$. In the next Section we consider the
perpendicular asymmetry $A_\perp$ which is used to measure a combination
of $g_1$ and $g_2$ with the aim of extracting $g_2$. Since $g_2$ is expected
to be much smaller than $g_1$ some care may
be necessary in utilising an approximate version of $g_1 (x)$.

b) {\it Transversely polarized target}

If we want to have information on $g_2$ we must then consider other spin-spin
asymmetries. By scattering longitudinally polarized leptons on transversely
polarized nucleons, one can measure the quantity
\begin{equation}
A_\perp \equiv {d\sigma^{\to\Downarrow} - d\sigma^{\to\Uparrow}\over
d\sigma^{\to\Uparrow} + d\sigma ^{\to\Downarrow}}
\end{equation}
where, again, $d\sigma$ is short for $d^2\sigma/(d\Omega~dE^\prime)$ and the
denominator is twice the unpolarized cross-section $d\sigma^{unp}$.

{}From Eqs. (2.1.14) and (2.1.25) one obtains
\begin{equation}
A_\perp = {Q^2 \sin\theta (MG_1 + 2EG_2)\over 2E \, [2W_1
\sin^2(\theta/2) + W_2 \cos^2(\theta/2)]} \, \cos\phi
\end{equation}
where $\phi$ is the difference between the azimuthal angles of $\hat{\bfbs}$
and
$\hat{\bfk}^\prime,~\phi = \beta - \varphi$ (see Figs. 2.2, 3). For
sake of clarity we show in Fig. 2.4 the spin configuration leading to
Eq. (2.1.39), for $\phi = 0$. In the following we shall always assume
$\phi$ = 0; if necessary one could integrate Eq.
(2.1.39) over the $\phi$ range covered by the experimental apparatus.

One can repeat for $g_2$ the same direct procedure followed for
$A_\parallel$, Eqs. (2.1.34--36), and can use Eqs. (2.1.14, 18 and 39)
(with $\phi = 0$) to write
\begin{equation}
g_2 + {\nu\over 2E}~g_1 = \Biggl({\nu\over E}\Biggr)K^\prime~d\sigma^{unp}
{}~A_\perp
\end{equation}
where
\begin{equation}
K^\prime = {Q^2EM\nu\over 4\alpha^2 E^{\prime 2}\sin\theta} =
{E\cos^2(\theta/2)\over 2 x \sigma_{Mott}~\sin\theta}
\end{equation}
and $\sigma_{Mott}$ was given in Eqs. (2.1.36).

A measurement of $A_\perp$, {\it i.e.} the RHS of (2.1.40), provides
direct information
on the structure function combination $g_2 + \nu /(2E)~g_1$.

In this case the coefficient of $g_1$ is not negligible and to isolate
$g_2$ one must
feed in one's knowledge of $g_1$ obtained from the $A_\parallel$ measurement.

The $A_\perp$ type experiments are much harder than the $A_\parallel$
type so it is perhaps worth noting that one can obtain an estimate
of $g_2(x)$ from the $A_\parallel$
experiments if performed at several different beam energies [LEA 88].

At fixed values of $Q^2$ and $x$ the only dependence on the beam energy
$E$ in the LHS of (2.1.35) is in the coefficient of $g_2(x)$.

A study of the energy variation of the RHS of (2.1.34) thus allows one, in
principle, to learn about $g_2(x)$. Unfortunately, because the coefficient
of $g_2(x)$ is so small, it is not clear whether this method is practicable
or not.

In a similar way the measurement of $A_\perp$ at different beam energies
$E$, allows the isolation of $g_2(x)$ from (2.1.40). For example, measurements
at $E = E_1$ and $E_2$ yield
\begin{equation}
(E_1 - E_2)~g_2(x) = [\nu K^\prime~d\sigma^{unp}~A_\perp]_{E=E_1} -
[\nu k^\prime~d\sigma^{unp}~A_\perp]_{E=E_2} \,.
\end{equation}

Equations (2.1.40, 41) provide the most direct access to $g_2$ on an
event-by-event basis, provided one's knowledge of $g_1$ is accurate enough.

However, as with $g_1$, if data is desired at fixed $Q^2$ over the entire
range of $x$ a different strategy is required which we now describe.

c) {\it Combined analysis using $A_\parallel$ and $A_\perp$}

{}From the longitudinal polarization data one can extrapolate $A_\parallel/D$
smoothly in $Q^2$ at fixed $x$. Combining this with the perpendicular
polarization data where it is measured one can construct (see Appendix A for
definitions of kinematical coefficients)
\begin{equation}
A^\prime \equiv {\sqrt{Q^2}\over 2 M}~A_2 = {\sqrt{Q^2}\over 2M (1 + \xi
\eta)}\Biggl\{\xi {A_\parallel\over D} + {A_\perp\over d}\Biggl\}
\end{equation}
at the values of the $A_\perp$ experiment. But $A^\prime$ should itself vary
slowly with $Q^2$ because one has
\begin{equation}
g_1 (x, Q^2) + g_2 (x, Q^2) = {F_2 (x, Q^2)\over 2 x^2 [1 + R(x, Q^2)]}~
A^\prime (x, Q^2)
\end{equation}
so that extrapolation in $Q^2$ at fixed $x$ should be smooth. Via (2.1.44)
one can thus estimate $g_1 + g_2$ over the entire $x, Q^2$ range.

Again, with $A^\prime$ extrapolated to the relevant $Q^2$, one can obtain
an improved evaluation of $g_1(x, Q^2)$ via
\begin{equation}
g_1(x, Q^2) = {F_2 (x, Q^2)\over 2 x[1 + R(x, Q^2)]} \cdot {1\over 1 +
4 M^2 x^2 / Q^2 }\cdot \Biggl\{{A_\parallel\over D}
-{2M\over \sqrt{Q^2}}\Biggl(\eta - {2M x\over \sqrt{Q^2}}\Biggl)
A^\prime\Biggl\}\,.
\end{equation}

The formulae (2.1.44) and (2.1.45) are exact. No approximations have been made.
And they are expressed in terms of the functions
$A^\prime (x, Q^2)$ and $A_\parallel/D$ which
should both be slowly varying functions of $Q^2$ at fixed $x$.

d) {\it The problem of extrapolating in $Q^2$}

We mentioned in a) above that
\begin{equation}
A_1(x,Q^2) = {A_\parallel(x,Q^2) \over D}
\end{equation}
is taken to be independent of $Q^2$ in the experimental evaluation of
$g_1(x, \langle Q^2 \rangle)$ via (2.1.37). There is no allowance made
for the error in extrapolating from the measured region of $Q^2$ for the
$x$ involved to the required value $\langle Q^2 \rangle$. A simple linear
parametrization $a + b\,Q^2$ or $a + b\ln Q^2$ for $A_1$ will not work because
a best fit will yield a very small value of $b$, but with very large errors,
leading to unrealistic error estimates on $g_1(x, \langle Q^2 \rangle)$.

There is no rigorous theoretical solution to this experimental problem,
but the following approximate procedure should lead to an improved
estimate of $g_1(x, \langle Q^2 \rangle)$ and its error [EFR 94a].

Let $x_1, \, x_2, \dots$ be the values of $x$ for the data bins and let
$Q_1^2, \, Q_2^2, \dots$ be the mean values of $Q^2$ for each $x$-bin.

We define a {\it zeroth order approximation} to $g_1(x, \langle Q^2 \rangle)$
for each $x_i$ via (2.1.37), {\it i.e.}
\begin{equation}
g_1^{(0)}(x_i, \langle Q^2 \rangle) \equiv A_1(x_i, Q_i^2)
F_1(x_i, \langle Q^2 \rangle)
\end{equation}
where we have written
\begin{equation}
F_1 = {F_2 \over 2x[1+R]}
\end{equation}
for brevity. We define an improved estimate of $g_1(x, \langle Q^2 \rangle)$
for each $x_i$ via
\begin{equation}
g_1(x_i, \langle Q^2 \rangle) \equiv A_1(x_i, \langle Q^2 \rangle)
F_1(x_i, \langle Q^2 \rangle)
\end{equation}
where $A_1(x_i, \langle Q^2 \rangle)$ is obtained using
\begin{equation}
A_1(x_i, \langle Q^2 \rangle) \approx A_1(x_i, Q_i^2)
+ b(x_i, Q_i^2) \, \ln(\langle Q^2 \rangle/Q_i^2)
\end{equation}
and $b(x_i, Q_i^2)$ is estimated from
\begin{eqnarray}
b(x_i, Q_i^2) &=& {1\over F_1(x_i, Q_i^2)} \left. {\partial g_1(x_i, Q^2)
\over \partial \ln Q^2} \right|_{Q^2=Q_i^2} \\\nonumber
&-&{A_1(x_i, Q_i^2) \over F_1(x_i, Q_i^2)}\left. {\partial F_1(x_i, Q^2)
\over \partial \ln Q^2} \right|_{Q^2=Q_i^2} \, \cdot
\end{eqnarray}

All the terms on the RHS of (2.1.51) are known from experiment except for
the derivative of $g_1$. The latter can be approximately calculated from the
evolution equation provided we use the experimental fact that the flavour
singlet part of $g_1$ is much smaller than the non-singlet part [see (4.2.22)].
Thus we treat the evolution as if it were purely non-singlet, {\it i.e.}
we compute
\begin{equation}
\left. {\partial g_1(x_i, Q^2) \over \partial \ln Q^2} \right|_{Q^2=Q_i^2}
\approx {\alpha_s(Q_i^2) \over 2\pi} \int_{x_i}^1 {dy\over y}~g_1(y,Q_i^2)
\, \Delta P_{qq} \left( {x_i\over y} \right) \,,
\end{equation}
where $\Delta P_{qq}$ is the non-singlet polarized splitting function.
In fact, for the non-singlet case, polarized and unpolarized splitting
functions are equal [ALT 77],
\begin{equation}
\Delta P_{qq}(x)=P_{qq}(x)={4\over 3} \left( {1+x^2 \over 1-x} \right)_+
\,\cdot
\end{equation}

Finally, in the convolution integral in (2.1.52), we approximate $g_1$ by
its known zeroth order estimate (2.1.47). Thus we suggest the approximate
formula
\begin{equation}
\left. {\partial g_1(x_i, Q^2) \over \partial \ln Q^2} \right|_{Q^2=Q_i^2}
\approx {\alpha_s(Q_i^2) \over 2\pi} \int_{x_i}^1 {dy\over y}~
g_1^{(0)}(y,Q_i^2) \, \Delta P_{qq} \left( {x_i\over y} \right) \,.
\end{equation}

This will, of course, provide an estimate for the value of $b$ in (2.1.51)
and for the error on it.

\subsubsection{Measurement of \mbox{\boldmath $g_{1,2}$} on nuclear targets}
\vskip 6pt
We consider spin 1 targets, like deuterium, or spin 1/2 targets like $^3$He.
The
asymmetries considered are the analogues of (2.1.27 and 38) and are defined
for nucleus $A$ by
\begin{equation}
A^A_\parallel={d\sigma_A^{\begin{array}{c}\hspace*{-0.2cm}\to
\vspace*{-0.3cm}\\
\hspace*{-0.2cm}\Leftarrow\end{array}} - d\sigma_A^{\begin{array}{c}
\hspace*{-0.2cm}\to\vspace*{-0.3cm}\\
\hspace*{-0.2cm}\Rightarrow\end{array}}\over d\sigma_A^{\begin{array}{c}
\hspace*{-0.2cm}\to\vspace*{-0.3cm}\\
\hspace*{-0.2cm}\Leftarrow\end{array}} +
d\sigma_A^{\begin{array}{c}\hspace*{-0.2cm}\to\vspace*{-0.3cm}\\
\hspace*{-0.2cm}\Rightarrow\end{array}}}
\end{equation}
\begin{equation}
A^A_\perp={d\sigma_A^{\to\Downarrow} - d\sigma_A^{\to\Uparrow}\over
d\sigma_A^{\to\Downarrow} + d\sigma_A^{\to\Downarrow}}
\end{equation}
where $\sigma^\Rightarrow_A,~\sigma^\Leftarrow_A$ means $J_z = \pm1/2$ for a
spin 1/2 target, but $J_z = \pm 1$ for a spin one target longitudinally
polarized; and similarly for transverse polarization.

We deal only with the case where the constituents of the nucleons are assumed
to
contribute independently to the scattering. This means we neglect shadowing
and Fermi motion.

For {\it unpolarized} scattering on deuterium this is tantamount to
taking
\begin{equation}
\sigma_d = \sigma_p + \sigma_n \,,
\end{equation}
a perfectly reasonable approximation for $\sigma_d$ but one which can be
misleading when differences of cross-section are being studied. For example
the Gottfried sum rule requires the combination $\sigma_p - \sigma_n$ for
which the approximation $2\sigma_p  - \sigma_d$ may be dangerous because the
subtraction of comparable quantities magnifies the error [EPE 93].

In a similar way, for the testing of the Bjorken sum rule we shall require
the quantity $g_1^p - g^n_1$, and one might worry about obtaining
$g^n_1$ from nuclear data on the basis of independent scattering. Happily
because $g_1^p$ and $g^n_1$ are expected
to be quite different in magnitude this should be quite reliable [EPE 92].

On the basis of {\it independent scattering} on the $Z$ protons and
$N$ neutrons in the nucleus the nuclear cross-section difference is given by
\begin{equation}
d\sigma^\Rightarrow_A -
d\sigma^\Leftarrow_A = 2~[Zd\sigma_p~{\cal P}_p~A^p_\parallel +
N d\sigma_n~{\cal P}_n~A_\parallel^n] \,,
\end{equation}
where $d\sigma_{p,n}$ are the unpolarized nucleon cross-sections,
$A_\parallel^{p,n}$ the nucleon longitudinal asymmetries and ${\cal P}_{p,n}$
the longitudinal polarization of the nucleons in the nuclear state with
$J_z = 1/2 $ or 1.

For the asymmetry defined in (2.1.55) one then has
\begin{equation}
A^A_\parallel = f_p {\cal P}_p~A^p_\parallel + f_n {\cal P}_n~A^n_\parallel
\end{equation}
where
\begin{equation}
f_p = {Zd\sigma_p\over Zd\sigma_p + Nd\sigma_n} \qquad f_n = {Nd\sigma_n\over
Zd\sigma_p + Nd\sigma_n}
\end{equation}
are the fractions of events originating on protons and neutrons respectively.

An analogous result holds for $A^A_\perp$.

Equation (2.1.59) is the fundamental formula which should be used to extract
$A^n_\parallel$ from $A^A_\parallel$ and a knowledge of $A^p_\parallel$. This
is
especially true when one does not have data on $A^A_\parallel$ and
$A^p_\parallel$ at the same $\langle Q^2 \rangle$ because it appears that the
asymmetries themselves show very little $Q^2$ dependence.

For deuterium, because of the $D$-state admixture one has
\begin{equation}
{\cal P}^d_p = {\cal P}^d_n = (1 -{3\over 2}~\omega_D)
\end{equation}
where $\omega_D$ = 0.058 is the $D$-state probability.

For the fractions $f_{p,n}$ one has, approximately,
\begin{equation}
f^d_p = {F^p_2 / (1 + R^p)\over 2F^d_2 / (1 + R^d)}\qquad f^d_n =
{F^n_2 / (1 + R^n)\over 2F^d_2 / (1 + R^d)}
\end{equation}
where $F^d_2$ is the deuteron $F_2$ per nucleon.

Often one defines $g^d_1$ {\it per nucleon} via (2.1.30, 31)
\begin{equation}
g^d_1(x,Q^2) \equiv{A^d_\parallel\over D}~{F^d_2(x,Q^2)\over 2 x
[1 + R^d(x,Q^2)]}
\end{equation}
so that (2.1.59) becomes
\begin{equation}
g_1^d~(x,Q^2) = {(1 - {3\over 2} \omega_D)\over 2} \Biggl\{ g^p_1 (x,Q^2) +
g_1^n (x,Q^2)\Biggl\}\,.
\end{equation}

In the above (2.1.55, 56) we have defined asymmetries for 100\% polarized
targets.

For spin 1/2 targets with degree of longitudinal polarization ${\cal P}$ the
generalization of these equations is quite straightforward:
\begin{equation}
A_\parallel = {1\over {\cal P}}\Biggl\{{d\sigma^\to (- {\cal P}) -
d\sigma^\to ({\cal P})\over
d\sigma^\to (- {\cal P}) + d\sigma^\to ({\cal P})}\Biggl\} =
{1\over {\cal P}}~{d\sigma^\to (- {\cal P}) - d\sigma^\to ({\cal P})
\over 2 \, d\sigma}
\end{equation}
where $d\sigma$ is the unpolarized cross-section. A similar result holds
for $A_\perp$.

For spin 1 targets the result is a little more subtle. If $p_+,~p_-,~p_0$
are the probabilities of finding states with $J_z = 1,~-1,~0$ in the target,
then the degree of polarization is [BOU 80]
\begin{equation}
{\cal P} = p_+ - p_-
\end{equation}
and the alignment is
\begin{equation}
{\cal A} = 1 - 3p_0\,.
\end{equation}

Then one has
\begin{equation}
d\sigma^\to (-{\cal P}) - d\sigma^\to ({\cal P}) =
{\cal P} \Big\{ d\sigma^{\begin{array}{c}
\hspace*{-0.2cm}\to\vspace*{-0.3cm}\\
\hspace*{-0.2cm}\Leftarrow\end{array}} - d\sigma^{\begin{array}{c}
\hspace*{-0.2cm}\to\vspace*{-0.3cm}\\
\hspace*{-0.2cm}\Rightarrow\end{array}}\Big\} \,,
\end{equation}
but
\begin{equation}
d\sigma^\to (-{\cal P}) + d\sigma^\to ({\cal P}) =
2d\sigma + {{\cal A} \over 3} [d\sigma_+ + d\sigma_- - 2\,d\sigma_0]
\end{equation}
where $d\sigma$ is the unpolarized cross-section for the spin 1 target.

It follows that for a spin 1 target
\begin{equation}
A_\parallel = {1\over {\cal P}}~{d\sigma^\to (-{\cal P}) -
d\sigma^\to ({\cal P})\over 2\,d\sigma +
({\cal A} / 3) [d\sigma_+ + d\sigma_- - 2\,d\sigma_0]}
\end{equation}
and this is strictly only equal to
\begin{equation}
{1\over {\cal P}}~{d\sigma^\to (-{\cal P}) - d\sigma^\to ({\cal P})
\over 2\,d\sigma}
\end{equation}
if the alignment ${\cal A}$ is known to be zero.

Of course for the 100\% polarized target corresponding to (2.1.55), $p_0 = 0$
and the alignment has its maximum value ${\cal A} = 1$. However, even for
${\cal A} \neq 0$, the term in square brackets in (2.1.69, 70) vanishes in the
approximation of independent scattering from the nucleons. Since the results
(2.1.59, 60) are based on this approximation it is
consistent in these equations to ignore the term proportional to ${\cal A}$.

It is worth noting that one could, in principle, check the assumption of
independent scattering by experimentally testing (2.1.69).

For $^3$He the wave-function is almost entirely an $S$-state with the two
protons having opposite spins. Thus all the spin is carried by the neutron
but there is some mixing in
the wave-function [WOL 89; CIO 93] and one estimates
\begin{equation}
{\cal P}^{^3 He}_n = (87\pm 2)\%\qquad {\cal P}^{^3 He}_p = (-2.5\pm0.3)\% \,.
\end{equation}

For the fractions $f_{p,n}$ one takes the approximation
\begin{eqnarray}
f^{^3 He}_n&=&{F^n_2/(1+R^{n})\over 3F_2^{^3 He}/(1 + R^{^3 He})}\nonumber\\
f^{^3 He}_p&=&{2F^p_2/(1+R^{p})\over 3F_2^{^3 He}/(1 + R^{^3 He})}
\end{eqnarray}
where $F^{^3 He}_2$ is the helium-3 $F_2$ per nucleon.

Defining $g^{^3 He}_1$ {\it per nucleon} via (2.1.31),
\begin{equation}
g^{^3 He}_1~(x,Q^2) = {A^{^3 He}_\parallel\over D}~{F^{^3He}_2(x,Q^2)\over
2x[1 + R^{^3 He}(x,Q^2)]} \,,
\end{equation}
one obtains
\begin{equation}
g^{^3 He}_1~(x,Q^2)={1\over 3} \left[ (0.87 \pm 0.02)~g^n_1(x,Q^2)
- (0.050\pm 0.006)~g^p_1(x,Q^2) \right] \,.
\end{equation}

Note that the above is again based on completely independent scattering from
the
constituent nucleons. This might be reasonable at high $Q^2$ but it is not
likely to be a good approximation in the E142 experiment which has
$\langle Q^2 \rangle$ = 2 (GeV/c)$^2$ and which will be discussed in
Section 4.5.
Unfortunately we do not have any simple prescription for improving the
analysis.

\subsection{Neutral and charged current weak interactions initiated by charged
leptons}
\vskip 6pt
\setcounter{equation}{0}
Consider now the deep inelastic scattering of longitudinally polarized leptons
on polarized nucleons at very high energies, taking into account weak
interactions of both the neutral and charged current type. We do not study
neutrino initiated processes, although the extension of our results to such a
case would be trivial, because of the technical difficulties in polarizing the
large nucleon targets needed for neutrino scattering, which make such
experiments quite unrealistic. Useful information can be found in [NAS 71;
DER 73; NIK 73; AHM 76; KAU 77; JOS 77; CAH 78; BAR 79; CRA 83; LAM 89;
VOG 91; JEN 91; RAV 92; MAT 92, 92b; XIA 93; DEF 93].

Let us start from neutral current processes, $\ell N \to \ell X$,
which, at lowest perturbative order, proceed via the exchange of one photon or
one $Z^0$ boson, Figs. 2.1, 5a. The cross-section then receives contributions
from a purely electromagnetic, a purely weak and an interference term, so that
Eq. (2.1.1) becomes
\begin{equation}
{d^2\sigma_{nc}\over d\Omega~dE^\prime}={\alpha^2\over 2 Mq^4}~{E^\prime\over
E}~\sum_{i= \gamma,\gamma Z,Z}~L^i_{\mu\nu}~W^{\mu\nu}_i~\eta^i\,.
\end{equation}

On summing over the final lepton spins and neglecting terms proportional to
$m/E$ or $m/E^\prime$, the leptonic tensors $L^i_{\mu\nu}$ are given, for
negatively charged leptons $(e^-, \mu^-)$, by
\begin{eqnarray}
L^\gamma_{\mu\nu}&=&\sum_{s^\prime}~[\bar{u}(k^\prime,s^\prime)~\gamma_\mu
{}~u(k,\lambda)]^\ast~[\bar{u}(k^\prime,s^\prime)~\gamma_\nu~u(k,\lambda)]
\nonumber\\
&=& 2~[k_\mu k^\prime_\nu + k^\prime_\mu k_\nu - k \cdot k^\prime g_{\mu\nu}-
2i\lambda~\varepsilon_{\mu\nu\alpha\beta}~k^\alpha k^{\prime\beta}]\\
L^{\gamma Z}_{\mu\nu} &=&\sum_{s^\prime}~[\bar{u}(k^\prime,s^\prime)~
\gamma_\mu(g_{_V} - g_{_A}\gamma_5)~u(k,\lambda)]^\ast
{}~[\bar{u}(k^\prime,s^\prime)~\gamma_\nu~u(k,\lambda)] \nonumber\\
&=& (g_{_V}-2\lambda g_{_A})~L^\gamma_{\mu\nu}\\
L^Z_{\mu\nu}&=&\sum_{s^\prime}~[\bar{u}(k^\prime,s^\prime)~\gamma_\mu
(g_{_V} - g_{_A} \gamma_5)~u(k,\lambda)]^\ast
{}~[\bar{u}(k^\prime, s^\prime)~\gamma_\nu (g_{_V}-g_{_A} \gamma_5)~u(k,
\lambda)] \nonumber\\
&=& (g_{_V} - 2\lambda g_{_A})^2~L^\gamma_{\mu\nu}
\end{eqnarray}
where $2\lambda = \pm 1$ denotes twice the helicity of the initial lepton.
$L^\gamma_{\mu\nu}$ agrees with Eqs. (2.1.3-5), upon remembering that for a
fast moving lepton with helicity $\lambda$ one has $s^\mu \approx
2\lambda k^\mu /m$. For positively charged leptons $(e^+,\mu^+)$ one
should simply replace, in Eqs. (2.2.3, 4), $g_A$ by $-g_A$. In our notation
\begin{equation}
g_{_V} = -{1\over 2} + 2 \sin^2 \theta_W \qquad  g_{_A} = - {1\over 2}\,.
\end{equation}

The factors $\eta^i$ in Eq. (2.2.1) collect some kinematical factors, coupling
constants and the relative weights of different propagators, namely:
\begin{eqnarray}
\eta^\gamma&=& 1 \nonumber\\
\eta^{\gamma Z}&=&\Biggl({GM^2_Z\over 2\sqrt{2}\pi\alpha}\Biggl)\Biggl(
{Q^2\over Q^2 + M_Z^2}\Biggl)\\
\eta^Z&=& {(\eta^{\gamma Z})}^2\nonumber
\end{eqnarray}
where $G$ is the Fermi coupling constant and $M_Z$ is the $Z_0$ mass. Notice
that $GM^2_Z/(2\sqrt{2}\pi \alpha)$ $= (4 \sin^2\theta_W \cos^2 \theta_W)^{-1}
\simeq 4/3$.

Finally the hadronic tensor $W^{\mu\nu}_i$ defines the coupling of the
electro-weak current to the nucleon; exploiting Lorentz and CP invariance it
can be expressed in terms of 8 independent structure functions as [BAR 79]
\begin{eqnarray}
{1\over 2M}~W^i_{\mu\nu}& = &- {g_{\mu\nu}\over M}~F^i_1 + {P_\mu P_\nu\over
M (P\cdot q)}~F^i_2 \nonumber\\
&+& i~{\varepsilon_{\mu\nu\alpha\beta}\over 2 P\cdot q}\Biggl[
{P^{\alpha} q^{\beta}\over M}~F^i_3 +
2 q^\alpha~S^\beta g^i_1 - 4 x P^\alpha S^\beta~g^i_2\Biggl] \\
&-&{P_\mu S_\nu + S_\mu P_\nu\over 2 P\cdot q}~g^i_3 + {S\cdot q\over (P\cdot
q)^2}~P_\mu P_\nu~g^i_4+{S\cdot q\over P\cdot q}~g_{\mu\nu}~g^i_5 \,.\nonumber
\end{eqnarray}

Notice that terms proportional to $q^\mu$ or $q^\nu$ can be dropped in the
definition of $W^i_{\mu\nu}$ because they give no contribution
(in the $m/E \to 0$ limit) when contracted with $L^{\mu\nu}_i$. Other
definitions of the hadronic tensor appearing in the literature differ from
ours because of these terms. In particular the coefficient of
$g_2$ could be written in a more familiar way using the identity [HEI 73]
\begin{eqnarray}
\varepsilon_{\mu\nu\alpha\beta}~P^\alpha S^\beta&=&
{\varepsilon_{\mu\nu\alpha\beta}\over 2 x P\cdot q} \, [(q\cdot S) q^\alpha
P^\beta - (P\cdot q) q^\alpha S^\beta] \nonumber\\
&-& (q_\mu\varepsilon_{\nu\alpha\beta\gamma} -
q_\nu \varepsilon_{\mu\alpha\beta\gamma})~{P^\alpha q^\beta S^\gamma
\over 2xP\cdot q}
\end{eqnarray}
and dropping the last two terms.

The structure functions $F^i_j (P\cdot q, q^2)$ and $g^i_j~(P\cdot q, q^2)$ are
expected to scale approximately in the Bjorken limit and to depend only on $x =
Q^2/(2P\cdot q)$. The $F^i_j$ are the unpolarized structure functions and the
$g^i_j$ the polarized ones: when averaging over the nucleon spin one sums
$W_{\mu\nu}(P,q,S)$ to $W_{\mu\nu}(P,q,-S)$ and all terms proportional to
$g^i_j$ cancel out. In Eq. (2.2.7) we have allowed for parity violation and
indeed $W^i_{\mu\nu}$ is a mixture of second rank tensors and pseudotensors.
In case of pure electromagnetic interactions $(i
= \gamma)$ parity is conserved and one has
\begin{equation}
F^\gamma_3 = g^\gamma_3 = g^\gamma_4 = g^\gamma_5 = 0~.
\end{equation}

In this case Eq. (2.2.7), upon using Eq. (2.2.8) and up to irrelevant terms
proportional to $q^\mu$ or $q^\nu$, reproduces Eqs. (2.1.8-10, 16, 18).
$F_3,~g_3,~g_4$ and $g_5$ only contribute to parity violating interactions
and are often referred to as the parity violating structure functions.

{}From Eqs. (2.2.1-4) and (2.2.6, 7) one obtains explicit expressions of the
cross-sections; some of them can be found in [ANS 93].

In case of charged current interactions, $\ell^\mp N \to \nu (\bar{\nu}) X$,
the process is, at leading order, mediated by the exchange of a $W$ meson and
it resembles the $Z_0$ contribution to the neutral current process, with the
assignments $g_{_V} = g_{_A} = 1$. In fact Eq. (2.2.1) now reads
\begin{equation}
{d^2\sigma_{cc}\over d\Omega~dE^\prime} = {\alpha^2\over 2 M
q^4}~{E^\prime\over
E}~L^W_{\mu\nu}~W^{\mu\nu}_W~\eta^W
\end{equation}
with $W^{\mu\nu}_W$ given by Eq. (2.2.7) in which the label $i$ is now $W$.
For a negatively charged lepton $\ell^-$ (which couples to a $W^-)$,
\begin{eqnarray}
L^{W^-}_{\mu\nu}&=& \sum_{s^\prime}~[\bar{u} (k^\prime,s^\prime)~\gamma_\mu
(1 - \gamma_5)~u(k,\lambda)]^\ast
{}~[\bar{u}(k^\prime,s^\prime)~\gamma_\nu (1-\gamma_5)~u(k,\lambda)]\nonumber\\
&=& (1 - 2\lambda)^2~L^\gamma_{\mu\nu}~.
\end{eqnarray}
Equation (2.2.11) shows that fast $\ell^-$ leptons only couple to a $W$ if they
have a negative helicity $(\lambda = -1/2)$. Equation (2.2.10) is completed by
\begin{equation}
\eta^W = {1\over 2}\Biggl({GM^2_W\over 4\pi\alpha}~{Q^2\over Q^2 + M^2_W}
\Biggl)^2
\end{equation}
where $M_W$ is the $W$ mass. For a positively charged lepton $\ell^+$ one
simply
changes, as usual, the sign of the axial coupling $\gamma^\mu\gamma^5$, that
is one replaces $\lambda$ with $-\lambda$ in Eq. (2.2.11) to obtain
$L^{W^+}_{\mu\nu} = (1 + 2\lambda)^2~L^\gamma_{\mu\nu}$.

\subsubsection{Weak interaction structure functions and their measurement}
\vskip 6pt
In order to obtain information on the polarized structure functions observed
in weak interaction deep inelastic scattering we should study, in analogy to
Eq. (2.1.22), differences of cross-sections with opposite nucleon spin.
Rather than considering here the most general case, which could be derived
from Eqs. (2.2.1-7) or (2.2.10-12) and (2.1.21), we look at particular nucleon
spin configurations, namely longitudinal or
transverse ones, like in Eqs. (2.1.24) and (2.1.25) respectively.

To do so we switch to the more convenient variables $x$ and $y \equiv \nu/E$,
using
\begin{equation}
{d^2\sigma\over d\Omega~dE^\prime} = {E^\prime\over yME}~{d^3\sigma
\over dx~dy~d\varphi}\, \cdot
\end{equation}
We then notice that in the high energy region one can neglect terms
proportional to $M/E$ and that, from Eq. (2.2.6) and for $Q^2$ values up to
$\sim 10^3$ GeV$^2$, one has
\begin{equation}
\eta^Z \ll \eta^{\gamma Z} \ll \eta^\gamma \,.
\end{equation}

Taking all this into account and adopting the same symbols to denote the
nucleon spin as
in Eqs. (2.1.24) and (2.1.25), one obtains for neutral current reactions:
\begin{equation}
{d^2\sigma^{\begin{array}{c}\hspace*{-0.2cm}\to\vspace*{-0.3cm}\\
\hspace*{-0.2cm}\Rightarrow\end{array}}_{nc}\over dx~dy} - {d^2
\sigma^{\begin{array}{c}\hspace*{-0.2cm}\to\vspace*{-0.3cm}\\
\hspace*{-0.2cm}\Leftarrow\end{array}}_{nc}\over dx~dy} \approx - 16\pi
ME{\alpha^2\over Q^4}~xy(2-y)~g^\gamma_1
\end{equation}
\begin{equation}
{d^2\sigma^{\to\Uparrow}_{nc}\over dx~dy~d\varphi} -
{d^2\sigma^{\to\Downarrow}_{nc}\over
dx~dy~d\varphi}x\approx - 8M{\alpha^2\over Q^4}\cos \phi~
\sqrt{2xy (1-y)ME}~x[y~g^\gamma_1 + 2~g^\gamma_2]
\end{equation}
and for charged current reactions initiated by $\ell^\mp$
\begin{eqnarray}
{d^2\sigma^{\begin{array}{c}\hspace*{-0.2cm}\to\vspace*{-0.3cm}\\
\hspace*{-0.2cm}\Rightarrow\end{array}}_{cc}\over dx~dy} - {d^2
\sigma^{\begin{array}{c}\hspace*{-0.2cm}\to\vspace*{-0.3cm}\\
\hspace*{-0.2cm}\Leftarrow\end{array}}_{cc}\over dx~dy}
&\approx& 64 \pi ME{\alpha^2\over Q^4}~\eta^W \\
&\times& \Biggl[ \pm x y (2 - y)~g_1^{W^\mp} + (1 - y) (g^{W^\mp}_3 -
g^{W^\mp}_4) + x y^2~g^{W\mp}_5 \Biggr] \nonumber
\end{eqnarray}
\begin{eqnarray}
{d^3\sigma^{\to^\Uparrow}_{cc}\over dx~dy~d\varphi}-
{d^3\sigma_{cc}^{\to\Downarrow} \over dx~dy~d\varphi}
&\approx& 32 M {\alpha^2\over Q^4} \eta^W \cos\phi~\sqrt{2xy(1-y)ME} \\
&\times& \Biggl[ \pm xy~g_1^{W^\mp} \pm 2 x~g_2^{W^\mp}
+ {1\over 2}~g_3^{W^\mp} + {1 -y\over y}~g^{W^\mp}_4 - xy~g_5^{W\mp}\Biggr]\,.
\nonumber
\end{eqnarray}

Equations (2.2.15, 16) show that in our kinematical range the neutral current
processes are still dominated by one photon exchange and thus agree with
Eqs. (2.1.24, 25) respectively. The charged current cross-sections instead,
although much smaller than the neutral current ones due to the factor
$\eta^W$, Eq. (2.2.12), depend also on the parity violating polarized
structure functions $g^{W^\mp}_{3,4,5}$. (Recall that the $\pm$ signs
and the $W^\mp$ labels in Eqs. (2.2.17, 18) refer to negatively or positively
charged leptons, $\ell^\mp$.)

One could also extract information on the parity violating structure functions
in neutral current processes by looking at appropriate combinations of
cross-sections [ANS 93]. Ideally, it would be very helpful to perform large
$Q^2$ experiments with both $\ell^-$ and $\ell^+$ leptons with positive
$(\to)$ and negative $(\leftarrow)$ helicities. In fact, always within the
$Q^2$ range of validity of Eq. (2.2.14) and noticing, from Eq. (2.2.5), that
$g_{_V} \simeq$ 0.04 whereas $g_{_A} = - 0.5$, one has:
\begin{eqnarray}
\Biggl({d^2\sigma^{\begin{array}{c}\hspace*{-0.2cm}\to\vspace*{-0.3cm}\\
\hspace*{-0.2cm}\Rightarrow\end{array}}_{nc}\over dx~dy} -
{d^2\sigma^{\begin{array}{c}\hspace*{-0.2cm}\to\vspace*{-0.3cm}\\
\hspace*{-0.2cm}\Leftarrow\end{array}}_{nc}\over dx~dy}\Biggl)_{\ell^-}&+&
\Biggl({d^2\sigma^{\begin{array}{c}\hspace*{-0.2cm}\to\vspace*{-0.3cm}\\
\hspace*{-0.2cm}\Rightarrow\end{array}}_{nc}\over dx~dy} -
{d^2\sigma^{\begin{array}{c}\hspace*{-0.2cm}\to\vspace*{-0.3cm}\\
\hspace*{-0.2cm}\Leftarrow\end{array}}_{nc}\over
dx~dy}\Biggl)_{\ell^+} \nonumber\\
%% FOLLOWING LINE CANNOT BE BROKEN BEFORE 80 CHAR
+\Biggl({d^2\sigma^{\begin{array}{c}\hspace*{-0.2cm}\leftarrow\vspace*{-0.3cm}\\
\hspace*{-0.2cm}\Rightarrow\end{array}}_{nc}\over dx~dy} -
{d^2\sigma^{\begin{array}{c}\hspace*{-0.2cm}\leftarrow\vspace*{-0.3cm}\\
\hspace*{-0.2cm}\Leftarrow\end{array}}_{nc}\over dx~dy}\Biggl)_{\ell^-}&+&
\Biggl({d^2\sigma^{\begin{array}{c}\hspace*{-0.2cm}\leftarrow\vspace*{-0.3cm}\\
\hspace*{-0.2cm}\Rightarrow\end{array}}_{nc}\over dx~dy} -
{d^2\sigma^{\begin{array}{c}\hspace*{-0.2cm}\leftarrow\vspace*{-0.3cm}\\
\hspace*{-0.2cm}\Leftarrow\end{array}}_{nc}\over
dx~dy}\Biggl)_{\ell^+} \approx
\end{eqnarray}
$$\approx 64 \pi M E{\alpha^2\over Q^4}~\left\{ (1 - y) \left[ g_{_V}
\eta^{\gamma Z}(g^{\gamma Z}_3 - g^{\gamma Z}_4)
+ g^2_{_A}\eta^Z(g^Z_3 - g^Z_4)\right] \right.$$
$$+\left. xy^2 \left[ g_{_V} \eta^{\gamma Z}
{}~g^{\gamma Z}_5 + g^2_A \eta^Z~g^Z_5 \right] \right\}$$

\begin{eqnarray}
\Biggl({d^2\sigma^{\begin{array}{c}\hspace*{-0.2cm}\to\vspace*{-0.3cm}\\
\hspace*{-0.2cm}\Rightarrow\end{array}}_{nc}\over dx~dy} -
{d^2\sigma^{\begin{array}{c}\hspace*{-0.2cm}\to\vspace*{-0.3cm}\\
\hspace*{-0.2cm}\Leftarrow\end{array}}_{nc}\over dx~dy}\Biggl)_{\ell^-}&-&
\Biggl({d^2\sigma^{\begin{array}{c}\hspace*{-0.2cm}\to\vspace*{-0.3cm}\\
\hspace*{-0.2cm}\Rightarrow\end{array}}_{nc}\over dx~dy} -
{d^2\sigma^{\begin{array}{c}\hspace*{-0.2cm}\to\vspace*{-0.3cm}\\
\hspace*{-0.2cm}\Leftarrow\end{array}}_{nc}\over
dx~dy}\Biggl)_{\ell^+} \nonumber\\
%% FOLLOWING LINE CANNOT BE BROKEN BEFORE 80 CHAR
+\Biggl({d^2\sigma^{\begin{array}{c}\hspace*{-0.2cm}\leftarrow\vspace*{-0.3cm}\\
\hspace*{-0.2cm}\Rightarrow\end{array}}_{nc}\over dx~dy} -
{d^2\sigma^{\begin{array}{c}\hspace*{-0.2cm}\leftarrow\vspace*{-0.3cm}\\
\hspace*{-0.2cm}\Leftarrow\end{array}}_{nc}\over dx~dy}\Biggl)_{\ell^-}&-&
\Biggl({d^2\sigma^{\begin{array}{c}\hspace*{-0.2cm}\leftarrow\vspace*{-0.3cm}\\
\hspace*{-0.2cm}\Rightarrow\end{array}}_{nc}\over dx~dy} -
{d^2\sigma^{\begin{array}{c}\hspace*{-0.2cm}\leftarrow\vspace*{-0.3cm}\\
\hspace*{-0.2cm}\Leftarrow\end{array}}_{nc}\over
dx~dy}\Biggl)_{\ell^+} \approx
\end{eqnarray}
$$
\approx 64 \pi M E~{\alpha^2\over Q^4}~xy(2-y)~g_{_A} \eta^{\gamma Z}~
g^{\gamma Z}_1
$$
\begin{eqnarray}
\Biggl({d^2\sigma^{\begin{array}{c}\hspace*{-0.2cm}\to\vspace*{-0.3cm}\\
\hspace*{-0.2cm}\Rightarrow\end{array}}_{nc}\over dx~dy} -
{d^2\sigma^{\begin{array}{c}\hspace*{-0.2cm}\to\vspace*{-0.3cm}\\
\hspace*{-0.2cm}\Leftarrow\end{array}}_{nc}\over dx~dy}\Biggl)_{\ell^-}&-&
\Biggl({d^2\sigma^{\begin{array}{c}\hspace*{-0.2cm}\to\vspace*{-0.3cm}\\
\hspace*{-0.2cm}\Rightarrow\end{array}}_{nc}\over dx~dy} -
{d^2\sigma^{\begin{array}{c}\hspace*{-0.2cm}\to\vspace*{-0.3cm}\\
\hspace*{-0.2cm}\Leftarrow\end{array}}_{nc}\over
dx~dy}\Biggl)_{\ell^+} \nonumber \\
%% FOLLOWING LINE CANNOT BE BROKEN BEFORE 80 CHAR
-\Biggl({d^2\sigma^{\begin{array}{c}\hspace*{-0.2cm}\leftarrow\vspace*{-0.3cm}\\
\hspace*{-0.2cm}\Rightarrow\end{array}}_{nc}\over dx~dy} -
{d^2\sigma^{\begin{array}{c}\hspace*{-0.2cm}\leftarrow\vspace*{-0.3cm}\\
\hspace*{-0.2cm}\Leftarrow\end{array}}_{nc}\over dx~dy}\Biggl)_{\ell^-}&+&
\Biggl({d^2\sigma^{\begin{array}{c}\hspace*{-0.2cm}\leftarrow\vspace*{-0.3cm}\\
\hspace*{-0.2cm}\Rightarrow\end{array}}_{nc}\over dx~dy} -
{d^2\sigma^{\begin{array}{c}\hspace*{-0.2cm}\leftarrow\vspace*{-0.3cm}\\
\hspace*{-0.2cm}\Leftarrow\end{array}}_{nc}\over
dx~dy}\Biggl)_{\ell^+} \approx
\end{eqnarray}
$$
\approx - 64 \pi M E{\alpha^2\over Q^4} \left\{ (1-y)~g_{_A} \eta^{\gamma Z}
(g^{\gamma Z}_3 - g^{\gamma Z}_4)
+ x y^2~g_{_A} \eta^{\gamma Z}~g^{\gamma Z}_5 \right\}
$$
where $\ell^-$ and $\ell^+$ label negative and positive charge leptons
respectively. Other interesting combinations of cross-sections can be found
in [ANS 93]. Notice that in Eqs. (2.2.19 and 20) a sum over the $\ell^-$ and
$\ell^+$ helicities is performed, which amounts to considering
{\it unpolarized} leptons.

In case only $\ell^-$ unpolarized lepton beams are available one can still
obtain information on the parity violating polarized structure functions,
by scattering the leptons off longitudinally polarized nucleons and
measuring [BIL 75]:
\[
%\begin{equation}
{1\over 2}
\Biggl({d^2\sigma^{\begin{array}{c}\hspace*{-0.2cm}\to\vspace*{-0.3cm}\\
\hspace*{-0.2cm}\Rightarrow\end{array}}_{nc}\over dx~dy} -
{d^2\sigma^{\begin{array}{c}\hspace*{-0.2cm}\to\vspace*{-0.3cm}\\
\hspace*{-0.2cm}\Leftarrow\end{array}}_{nc}\over dx~dy}\Biggl) +
{1\over 2}
\Biggl({d^2\sigma^{\begin{array}{c}\hspace*{-0.2cm}\leftarrow\vspace*{-0.3cm}\\
\hspace*{-0.2cm}\Rightarrow\end{array}}_{nc}\over dx~dy} -
{d^2\sigma^{\begin{array}{c}\hspace*{-0.2cm}\leftarrow\vspace*{-0.3cm}\\
\hspace*{-0.2cm}\Leftarrow\end{array}}_{nc}\over
dx~dy}\Biggl) \equiv {d^2\sigma^\Rightarrow \over dx~dy} -
{d^2\sigma^\Leftarrow \over dx~dy} \nonumber
\]
%\end{equation}
\begin{eqnarray}
&\approx& 16 \pi M E{\alpha^2\over Q^4}~\left\{ (1 - y) \left[ g_{_V}
\eta^{\gamma Z}(g^{\gamma Z}_3 - g^{\gamma Z}_4)
+ g^2_{_A}\eta^Z(g^Z_3 - g^Z_4)\right] \right. \\
&+& \left. xy^2 \left[ g_{_V} \eta^{\gamma Z}
{}~g^{\gamma Z}_5 + g^2_{_A} \eta^Z~g^Z_5 \right] + xy(2-y) g_{_A}
\eta^{\gamma Z} g_1^{\gamma Z} \right\} \,. \nonumber
\end{eqnarray}

Eq. (2.2.22) holds also for unpolarized $\ell^+$ leptons, provided $g_{_A}$
is replaced by $ - g_{_A}$, so that on summing and subtracting the $\ell^-$
and the $\ell^+$ contributions one recovers respectively Eqs. (2.2.19) and
(2.2.20). Of course, the above measurements (2.2.19-22) are difficult ones
in that they single out non leading parts of the cross-sections. However,
they may be feasable and would provide further information
on the structure of the nucleon.
%
%end of Section 2
%
\setcounter{section}{2}
\section{\large{The Naive Parton Model in polarized DIS}}
\vskip 6pt
\subsection{Projection of \mbox{\boldmath $F_1, F_2, g_1$} and
\mbox{\boldmath $g_2$} from the hadronic tensor \mbox{\boldmath $W_{\mu\nu}$}}
\vskip 6pt
\setcounter{equation}{0}

In Section 2 we have written the most general hadronic tensor $W_{\mu\nu}$,
Eqs. (2.1.8-10), which describes the unknown coupling of the virtual photon
to the composite nucleon in terms of the four structure functions
$F_1 = M W_1, ~F_2 = \nu W_2, ~g_1 = M^2 \nu G_1$ and $g_2 = M \nu^2 G_2$;
in the large $Q^2$ Bjorken limit, Eqs. (2.1.16, 18), these are supposed to
scale, that is to depend only on the Bjorken variable $x = Q^2/2M\nu$.
By performing DIS experiments one learns about the structure functions.

The hadronic tensor $W_{\mu\nu}$ can also be computed from models of the
nucleon, as we shall soon illustrate using the Parton Model. In such a case
it is convenient to have a technique which easily allows one to extract the
four different structure functions from a knowledge of $W_{\mu\nu} (N)$.

Let us start from the unpolarized case. Defining
\begin{eqnarray}
P_1^{\alpha\beta} &\equiv & \frac{1}{4} \left[\frac{1}{a} P^\alpha
P^\beta-g^{\alpha\beta}\right]\\
P_2^{\alpha\beta} &\equiv& \frac{3P\cdot q}{4a} \left[\frac{P^\alpha P^\beta}
{a} - \frac{1}{3} \, g^{\alpha\beta}\right]
\end{eqnarray}
with
\begin{equation}
a = \frac{P\cdot q}{2x} + M^2
\end{equation}
one can see, from Eqs. (3.1.1-3) and (2.1.8-10), that
\begin{equation}
P_1^{\alpha\beta} W_{\alpha\beta} (N) = F_1
\end{equation}
and
\begin{equation}
P_2^{\alpha\beta} W_{\alpha\beta} (N) = F_2~.
\end{equation}

Similarly, in the polarized case one can introduce the projectors
\begin{eqnarray}
P_3^{\alpha\beta} &\equiv & \frac{(P\cdot q)^2}{bM^2 (q\cdot S)} \left[
(q\cdot S) S_\lambda + q_\lambda \right] P_\eta
\varepsilon^{\alpha\beta\lambda\eta}\\
P_4^{\alpha\beta} &\equiv& \frac{1}{b} \left\{\left[\frac{(P\cdot q)^2}{M^2}
+ 2 (P\cdot q) x \right] S_\lambda + (q\cdot S) q_\lambda\right\}
P_\eta \varepsilon^{\alpha\beta\lambda\eta}
\end{eqnarray}
with
\begin{equation}
b = -4M \left[\frac{(P\cdot q)^2}{M^2} + 2(P\cdot q) x - (q\cdot S)^2\right]
\end{equation}
such that, from Eqs. (3.1.6-8) and (2.1.8-10),
\begin{eqnarray}
P_3^{\alpha\beta} W_{\alpha\beta} (N) &=& g_2\\
P_4^{\alpha\beta} W_{\alpha\beta} (N) &=& g_1+g_2 \,.
\end{eqnarray}

Equations (3.1.1--10) can be used in any reference frame and for arbitrary
directions of the nucleon spin four-vector; however, some care must be taken
in particular cases in which a limiting procedure is required.

\subsection{The hadronic tensor and the nucleon structure functions in the
Naive Parton Model}
\vskip 6pt
\setcounter{equation}{0}
In the simplest version of the Parton Model the nucleon is considered to be
made of collinear, free constituents, each carrying a fraction $x^\prime$ of
the nucleon four-momentum. The lepton-nucleon DIS is then described as the
incoherent sum of all lepton-constituent quark interactions and the hadronic
tensor $W_{\mu\nu} (N)$, Eq. (2.1.8), is given in terms of
the elementary quark tensor $w_{\mu\nu}$ by [LEA 85]
\begin{eqnarray}
W_{\mu\nu} (q;P,S) &=& W^{(S)}_{\mu\nu} (q;P) + i W^{(A)}_{\mu\nu} (q;P,S)
\nonumber\\
&=&
\sum_{q,s} e^2_q \, \frac{1}{2P\cdot q} \int^1_0 \frac{dx^\prime}{x^\prime}~
\delta (x^\prime-x)~n_q(x^\prime,s;S) ~w_{\mu\nu} (x^\prime,q,s) \,,
\end{eqnarray}
where $n_q (x^\prime,s;S)$ is the number density of quarks $q$ with charge
$e_q$, four-momentum fraction $x^\prime$ and spin $s$ inside a nucleon with
spin $S$ and four-momentum $P$; the $\Sigma_q$ runs over quarks and antiquarks;
$x$ is the Bjorken variable (2.1.15) and the quark tensor $w_{\mu\nu} (x,q,s)$
is the same as the leptonic tensor $L_{\mu\nu}$, Eqs. (2.1.2-7), with the
replacements $k^\mu\to x P^\mu, ~k^{\prime \mu} \to xP^\mu + q^\mu$ and a sum
over the unobserved final quark spin $(s^\prime)$ performed. That is:
\begin{equation}
w_{\mu\nu} (x,q,s) = w^{(S)}_{\mu\nu} (x,q)  + i w^{(A)}_{\mu\nu} (x,q,s)
\end{equation}
with
\begin{eqnarray}
w^{(S)}_{\mu\nu} (x,q) &=& 2\,[2x^2 P_\mu P_\nu + xP_\mu q_\nu + xq_\mu P_\nu
- x(P\cdot q) g^{\mu\nu}]\\
w_{\mu\nu}^{(A)} (x,q,s) &=& -2m_q ~ \varepsilon_{\mu\nu\alpha\beta}
{}~s^\alpha q^\beta
\end{eqnarray}
and the quark mass must for consistency be taken to be $m_q = xM$, before
and after the interaction with the virtual photon. We will further comment
on this point in Section~3.4.

{}From Eqs. (3.1.1-5) and (3.2.1-4) one obtains the well known Naive Parton
Model
predictions for the unpolarized nucleon structure functions:
\begin{eqnarray}
F_1 (x) &=& \frac{1}{2} \sum_q e^2_q ~q(x) \\
F_2 (x) &=& x \sum_q e^2_q ~q(x) = 2x F_1 (x) \,,
\end{eqnarray}
where the unpolarized quark number densities $q(x)$ are defined as
\begin{equation}
q(x) = \sum_s n_q (x,s;S)\,.
\end{equation}

Similarly, from Eqs. (3.1.6-10) and (3.2.1-4), the polarized nucleon structure
functions are obtained:
\begin{eqnarray}
g_1(x) &=& \frac{1}{2} \sum_q e^2_q ~\Delta q(x,S) \\
g_2(x) &=& 0
\end{eqnarray}
where
\begin{equation}
\Delta q(x,S) = n_q (x,S;S) - n_q (x,-S;S)
\end{equation}
is the difference between the number density of quarks with spin parallel to
the nucleon spin ($s=S$) and those with spin anti-parallel ($s=-S$). In the
Parton Model with free non-interacting quarks of mass $m_q= xM$, this quantity
cannot depend on the choice of $S$, {\it i.e.} $\Delta q (x,S) = \Delta q(x)$;
this follows because in such ultra simple case one is allowed to Lorentz
transform to the nucleon rest frame, where the quarks too are at rest and
where the parton distributions $n_q (x, \pm S,S)$, with $S = (0, \hat{\bfbs})$,
cannot depend on the spin quantization direction $\hat{\bfbs}$ [ANS 92].

\subsection{Intrinsic \mbox{\boldmath $p_\perp$} and \mbox{\boldmath $g_2
(x)$}}
\vskip 6pt
\setcounter{equation}{0}
The polarized structure function $g_2 (x)$, Eq. (3.2.9), is zero in the Naive
Parton Model where each parton carries a fraction $x$ of the nucleon
four-momentum, $p^\mu = x P^\mu$. However, non zero values of $g_2$ can be
obtained by allowing the quarks to have an intrinsic Fermi motion inside the
nucleon, $p^\mu = (E_q, p_x, p_y, x^\prime P_z$) (in a nucleon moving along
the $\hat z$-axis).

Equation (3.2.1) then modifies to [LEA 85]
\begin{equation}
W_{\mu\nu} (q;P,S) = \sum_{q,s} e^2_q \int d^3\bfp \left( \frac{P_0}{E_q}
\right) \delta (2p\cdot q-Q^2) \, n_q (\bfp,s;S) \, w_{\mu\nu} (q;p,s)
\end{equation}
which holds in the infinite momentum frame ($|P_z|\to \infty$) and where $n_q
(\bfp,s;S$) is the number density of quarks $q$ with charge $e_q$, energy
$E_q$,
three-momentum $\bfp$ and spin $s$ in a nucleon with spin $S$ and momentum $P$.
It is related to the usual quark distribution functions $q(x)$ and $\Delta
q(x,s;S)$, Eqs. (3.2.7, 10), by:
\begin{equation}
\sum_s \int d^2 \bfp_\perp \, n_q (\bfp,s;S) \, P_z = q(x^\prime)
\end{equation}
\begin{equation}
\int d^2\bfp_\perp \, [n_q (\bfp,s;S) - n_q (\bfp,-s;S)] P_z =
\int d^2\bfp_\perp \, \Delta q (\bfp,s;S) \, P_z = \Delta q (x^\prime,s;S)
\end{equation}
where we have put $d^3 \bfp = d^2 \bfp_\perp \, dx^\prime P_z$.

Equations (3.1.9) and (3.3.1) yield
\begin{equation}
g_2 (x) = \sum_{q,s} e^2_q \int d^3\bfp \left( \frac{P_0}{E_q} \right)
\delta (2p\cdot q-Q^2) ~n_q (\bfp,s;S) ~ P_3^{\alpha\beta}
w_{\alpha\beta}^{(A)} (q;p,s)~,
\end{equation}
and insertion of the explicit expression of $P_3^{\alpha\beta}$,
Eqs. (3.1.6, 8) and $w^{(A)}_{\alpha\beta}$, Eq. (3.2.4), yields
\begin{eqnarray}
g_2 (x) &=& \sum_q e^2_q \, \frac{4m_q (P\cdot q)^2}{bM^2 (q\cdot S)} \int
d^3\bfp \left( \frac{P_0}{E_q} \right) \delta (2p\cdot q - Q^2) \\ &\times&
\Delta q (\bfp,s;S) \{ (P\cdot q) [(q\cdot S)(s\cdot S) + q\cdot s] -
(P\cdot s) [(q\cdot S)^2+q^2]\} \nonumber
\end{eqnarray}
with $b$ as given in Eq. (3.1.8) and the quark mass $m_q=\sqrt{E_q^2-\bfp^2}$.

In the nucleon rest frame let $S^\mu = (0,\hat{\bfbs})$, where $\hat{\bfbs}$
is not perpendicular to the $\hat{z}$-axis (along which the nucleon moves in
the infinite momentum frame); specify the quark spin by choosing
$s^\mu = (0, \hat{\bfbs})$ in the quark rest frame (which, for $\bfp_\perp
\neq 0$, is not the nucleon rest frame). Then Eq. (3.3.5), at leading order
in $M/P_z$, reads [LEA 88]
\begin{equation}
g_2 (x) = \frac{1}{2} \sum_q e^2_q \left(1-\frac {xM}{m_q}\right)
\frac{m_q}{xM}~\Delta q(x,s;S)
\end{equation}
where we have used Eq. (3.3.3).

As the left hand side cannot depend on $s$ and $S$, Eq. (3.3.6) appears
contradictory. However, this is not so since in a fast frame where $p_\perp$
is negligible compared to $p_z$, both $s^\mu$ and $S^\mu$, to leading order,
are parallel to $P^\mu$ so that $\Delta q$ is
the helicity distribution and we have
\begin{equation}
g_2 (x)=\frac{1}{2} \sum_q e^2_q \left( \frac{m_q}{xM}-1\right) \Delta q (x)\,
\end{equation}
where $\Delta q (x)$ is the difference between the number density of quarks
with the same helicity as the nucleon and those with opposite helicity. When
neglecting the intrinsic $p_\perp$ of quarks one has $m_q=xM$, so that from
Eqs. (3.3.6) and (3.3.7) one recovers the previous result $g_2(x)=0$.
Thus the introduction of intrinsic $p_\perp$ allows a non zero value of
$g_2(x)$ in the Parton Model, but we see that such value shows an extreme
sensitivity to the quark mass, Eq. (3.3.7).

We shall argue, as has been emphasized in [JAF 90], that this throws doubt on
all purely Parton Model calculations of $g_2$.

\subsubsection{Conflicting Parton Model results for \mbox{\boldmath $g_2 (x)$}}
\vskip 6pt
Before discussing the difficulties of the Parton Model with $g_2(x)$, and in
order to better realize that indeed there are difficulties, we briefly
summarize some contradictory statements existing in the literature.

Some authors [IOF 84] claim that
\begin{equation}
g_1 (x) + g_2 (x) = 0~,
\end{equation}
whereas Feynman [FEY 72] has
\begin{equation}
g_1 (x) + g_2 (x) = \frac{1}{2} \sum_q e^2_q ~\Delta q_{_T} (x)~,
\end{equation}
where $\Delta q_{_T}$ is given by Eq. (3.2.10) with the spin $S$ perpendicular
to the nucleon momentum. Moreover, on summing Eqs. (3.2.8) and (3.3.7), one
obtains, in terms of helicity densities [LEA 88],
\begin{equation}
g_1 (x) + g_2 (x) = \frac{1}{2} \, \sum_q e^2_q \frac{m_q}{xM} ~\Delta q(x)\,.
\end{equation}
Finally in [JAC 89] it is argued, using a covariant Parton Model formulation,
that
\begin{equation}
g_1 (x) + g_2 (x) = \int^1_x \frac{dy}{y}\, g_1 (y)\,.
\end{equation}

These last authors exploit the experimental data on $g_1 (x)$ to give
predictions for $xg_2 (x)$ [see Fig. 3.1].

In [ANS 92] it is shown that Eq. (3.3.8) is not correct and that Eqs. (3.3.9)
and (3.3.10) are compatible when neglecting the intrinsic $p_\perp$ of quarks.
In section 5.3, we explain why (3.3.11) is incorrect because it neglects
twist-3 terms. Nevertheless (3.3.11) may not be
a bad approximation, so it should be tested experimentally.

Finally in Fig. 3.2 we show a bag model prediction for $g_2(x)$ [JAF 91].

\subsection{Origin of the difficulties with \mbox{\boldmath $g_2(x)$}}
\vskip 6pt
\setcounter{equation}{0}
As we have seen the Parton Model does not lead to clearcut results for
the spin dependent structure function $g_2$. Different authors obtain
different results and these are generally incompatible with each other.

Perhaps the most general way to understand this difficulty is to recall that
the Parton Model is fundamentally an impulse approximation in which binding
effects ({\it i.e.} the virtuality) of the struck parton are unimportant for
the large transverse momentum reactions under consideration. However, in some
cases what one is measuring is not just a cross-section but an asymmetry,
{\it i.e.} a difference of cross-sections, and it may happen that
in this difference the dominant contributions cancel out leaving a result
which does depend upon the binding energy or virtuality. Just such effects
occur in $g_2$. Whenever this happens the result is bound to be unreliable.

We can see this quite directly by reconsidering the antisymmetric part of
Eq. (3.3.1), valid in the impulse approximation:
\begin{equation}
W^{(A)}_{\mu\nu} (q;P,S) = \sum_s e^2_q \int d^3 \bfp \left(\frac{P_0}{E_q}
\right) \delta (2p\cdot q-Q^2) \, n_q (\bfp,s;S) ~w^{(A)}_{\mu\nu} (q;p,s)
\end{equation}
where we ignore the sum over the flavours $q$, irrelevant to this issue.

Now consider the calculation of $w_{\mu\nu}^{(A)}$ describing the interaction
of the hard photon with the quark, as depicted in Fig. 3.3. The final state
quark is a `free' quark and is on mass-hell: $(p^\prime)^2=m^2_q$.
In the impulse approximation also the initial quark is considered to be free
and to have the same mass. But to see the danger of this assertion, let us
put $p^2=m^2$, where for the moment we allow $m^2 \neq m^2_q$ to represent
the fact that the initial quark is really a bound quark. Aside from this we
treat the incoming quark as free, {\it i.e.} its wave function is taken as the
usual free-particle Dirac spinor  $u (p,s)$ for a particle of mass $m$.

One then finds
\begin{equation}
w_{\mu\nu} (q;p,s) = \frac{1}{2} \, Tr \, [(1+\gamma_5 s\sla)
(p\sla + m) \, \gamma_\mu \, (p\sla + q\sla + m_q) \, \gamma_\nu]
\end{equation}
from which one obtains
\begin{equation}
w^{(A)}_{\mu\nu} (q;p,s) = 2 \, \varepsilon_{\mu\nu\alpha\beta}(m_qs^\alpha)
\left[\left(1-\frac{m}{m_q}\right) p^\beta -\frac{m}{m_q}q^\beta\right]\,.
\end{equation}

Equation (3.4.3) is extremely revealing. We see immediately that for a general
$s^\mu$ the result is not gauge invariant ($q^\mu w^{(A)}_{\mu\nu} \neq 0)$
unless $m=m_q$, in which case we recover the Naive Parton Model, Eq. (3.2.4).
Moreover, the offending term, when $m \neq m_q$, is not small in an infinite
momentum frame (where the impulse approximation is supposed to be most
justifiable) even if $(m-m_q)$ is small.

However, in the special case of longitudinal $(L)$ polarization, if the quark
has high momentum so that $m_q/p_z \ll 1$, the product $(m_q s_L^\beta) \to
\pm p^\beta$ and the gauge non-invariant term vanishes because of the
antisymmetric $\varepsilon_{\mu\nu\alpha\beta}$. For the calculation of
$g_1(x)$ it is the longitudinal polarization that is relevant and there is no
unreasonable sensitivity to whether or not
$m$ equals $m_q$. Moreover, since the limit $\lim_{m_q\to 0} ~(m_q S_L^\beta)$
is finite there is also no crucial sensitivity to whether one works with
massive or massless quarks. Thus $g_1(x)$ can be calculated unambigously in
the Parton Model.

Quite the opposite happens for $g_2 (x)$ where the transverse spin is relevant.
There is an extreme sensitivity to whether or not $m$ equals $m_q$ and one
cannot expect to make a reliable calculation of $g_2(x)$ in the Parton Model.

One can put the case even more forcefully. The whole point of quarks is that,
in their pointlike interaction with a hard photon, they produce large
momentum transfer reactions which we are trying to generate for the
photon-hadron interactions. But even if we define the model by insisting that
$m=m_q$, comparing the expression (3.4.3) with the general structure of
$W^{(A)}_{\mu\nu}$ for a spin 1/2 particle (2.1.19), we see
that
\begin{equation}
g_2 (x)|_{quark} = 0 \,.
\end{equation}

Thus the hard photon-free quark interaction does not possess the cross-section
asymmetry which we are seeking to explain in the photon-hadron interaction.
It is clearly unrealistic therefore to try to produce such an asymmetry from
quark-partons.

\subsection{Weak interaction structure functions in the Naive Parton Model}
\vskip 6pt
\setcounter{equation}{0}
We have seen in Section 2.2 that, when taking into account weak interactions,
the most general form of the hadronic tensor, Eq. (2.2.7), contains 8
independent structure functions, 3 unpolarized and 5 polarized ones. On the
other hand, in the Naive Parton
Model, the hadronic tensor is given by (see Eq. (3.2.1))
\begin{equation}
W^i_{\mu\nu} (q;P,S) = \sum_q \frac{1}{2(P\cdot q)x}~[n_q (x,S;S)
w^{i,q}_{\mu\nu} (x,q,S) + n_q (x,-S;S) w^{i,q}_{\mu\nu} (x,q,-S)]
\end{equation}
where we have explicitely performed the sum over $s=\pm S$. The index $i$
denotes the different kinds of contributions ($i=\gamma,\gamma Z, Z, W)$
and the $\sum_q$ runs over quarks and antiquarks. The $w_{\mu\nu}^{i,q}$ are
the (flavour dependent) quark tensors, given by:
\begin{eqnarray}
w^{\gamma,q}_{\mu\nu} (x,q,s) &=& \sum_{s^\prime} e^2_q ~[\bar{u}(p^\prime,
s^\prime) \gamma_\mu u(p,s)]^* \, [\bar{u} (p^\prime,s^\prime)\gamma_\nu
u(p,s)]\nonumber\\
w^{\gamma Z,q}_{\mu\nu} (x,q,s) &=& \sum_{s^\prime} e_q ~[\bar{u}(p^\prime,
s^\prime) \gamma_\mu (g_{_V}-g_{_A}\gamma_5)_q u(p,s)]^* \,
[\bar{u} (p^\prime, s^\prime) \gamma_\nu u(p,s)]\nonumber\\
&+& \sum_{s^\prime} e_q ~[\bar{u}(p^\prime,s^\prime) \gamma_\mu u(p,s)]^*\,
[\bar{u}(p^\prime,s^\prime) \gamma_\nu (g_{_V}-g_{_A}\gamma_5) u(p,s)]\\
w^{Z,q}_{\mu\nu} (x,q,s) &=& \sum_{s^\prime} \, [\bar{u} (p^\prime,s^\prime)
\gamma_\mu (g_{_V}-g_{_A}\gamma_5)_q u(p,s)]^* \, [\bar{u}(p^\prime,s^\prime)
\gamma_\nu (g_{_V}-g_{_A} \gamma_5)_q u(p,s)]\nonumber\\
w^{W,q}_{\mu\nu} (x,q,s) &=& \sum_{s^\prime, q^\prime} \, [\bar{u}(p^\prime,
s^\prime) \gamma_\mu (1-\gamma_5) u(p,s)]^* \, [\bar{u}(p^\prime,s^\prime)
\gamma_\nu(1-\gamma_5) u(p,s)] |(V)_{qq^\prime}|^2\nonumber
\end{eqnarray}
where $p=xP, p^\prime = p+q$, $s$ and $s^\prime$ are respectively the momentum
and spin four-vectors of the initial and final quarks. $(g_{_V})_q$ and
$(g_{_A})_q$ are the vector and axial couplings of the quark of flavour $q$ to
the $Z_0$:
\begin{eqnarray}
(g_{_V})_{u,c} &=& \frac{1}{2}-\frac{4}{3} \sin^2\theta_W
\qquad (g_{_A})_{u,c} = \frac{1}{2} \nonumber\\
(g_{_V})_{d,s} &=& -\frac{1}{2}+\frac{2}{3} \sin^2\theta_W
{}~\quad (g_{_A})_{d,s} = -\frac{1}{2} \,.
\end{eqnarray}

In case of charged current, negatively charged leptons couple to $u$-type
quarks (or $\bar{d}$-type antiquarks) and positively charged leptons couple
to $d$-type quarks (or $\bar{u}$-type antiquarks); one has also to take into
account the proper Cabibbo--Kobayashi--Maskawa matrix elements
$(V)_{qq^\prime}$ occuring in the
transition coupling from a flavour $q$ to a flavour $q^\prime$. However, if we
consider the contribution of four flavours ($u,d,s$ and $c$), one always has
$\sum_{q^\prime}|(V)_{qq^\prime}|^2 = \cos^2\theta_c+\sin^2\theta_c=1$,
where $\theta_c$ is the Cabibbo angle. Equations (3.5.2) and (3.5.3) hold for
quarks; for antiquarks one should only replace $\gamma_5$ with
$-\gamma_5$ in Eqs. (3.5.2) leaving all other expressions unchanged.

Explicit expressions for Eqs. (3.5.2) can be obtained from the general form
\begin{eqnarray}
w_{\mu\nu} (x,q,s) &=& \sum_{s^\prime} \, [\bar{u}(p^\prime,s^\prime)
\gamma_\mu(v_1-a_1\gamma_5) u(p,s)]^* \, [\bar{u}(p^\prime, s^\prime)
\gamma_\nu (v_2-a_2 \gamma_5) u(p,s)] \nonumber\\
&=& 2 (a_1a_2+v_1v_2) [2p_\mu p_\nu-p\cdot q~g_{\mu\nu}]- 4a_1a_2 \,
m^2_q \, g_{\mu\nu} \\
&-& 2v_1a_2\,m_q [2p_\mu s_\nu-s\cdot q~g_{\mu\nu}]-2a_1v_2\,m_q
[2s_\mu p_\nu-s\cdot q ~g_{\mu\nu}] \nonumber \\
&+& 2i\varepsilon_{\mu\nu\alpha\beta} [(v_1a_2+a_1v_2)p^\alpha q^\beta
+2a_1a_2\,m_q p^\alpha s^\beta +(a_1a_2+v_1v_2)m_q \,q^\alpha s^\beta]\nonumber
\end{eqnarray}
by properly fixing the values of $v_{1,2}$ and $a_{1,2}$; in Eq. (3.5.4) we
have dropped, as usual, terms proportional to $q^\mu$ or $q^\nu$ which, when
contracted with the leptonic
tensor $L^{\mu\nu}$ give negligible contributions proportional to $m/E$.

Equations (3.5.1-4) give the Parton Model prediction of the hadronic tensor
$W^i_{\mu\nu}$; by comparing it with the general expression (2.2.7) one
obtains the naive quark-parton model results for the nucleon structure
functions. For completeness we list all of them here, starting from the
electromagnetic case ($i=\gamma)$, for which we recover Eqs. (3.2.5, 6):
\begin{eqnarray}
F_1^\gamma &=& \frac{1}{2} \sum_f e^2_{q_f} (q_f+\bar{q}_f) \qquad
F_2^\gamma = 2xF_1^\gamma \nonumber\\
g_1^\gamma &=& \frac{1}{2} \sum_f e^2_{q_f} (\Delta q_f + \Delta\bar{q}_f)
\qquad g_2^\gamma = 0
\end{eqnarray}
where $q_f =u,d,s,c$; $u$ stands for the number density of quarks $u$ and
so on. The interference contribution $(i=\gamma Z)$ is:
\begin{eqnarray}
F_1^{\gamma Z} &=& \sum_f e_{q_f} (g_{_V})_{q_f} (q_f + \bar{q}_f)
\qquad F_2^{\gamma Z} = 2xF_1^{\gamma Z} \nonumber\\
F_3^{\gamma Z} &=& 2\sum_f e_{q_f} (g_{_A})_{q_f} (q_f-\bar{q}_f)\nonumber\\
g_1^{\gamma Z} &=& \sum_f e_{q_f} (g_{_V})_{q_f}(\Delta q_f +\Delta\bar{q}_f)\\
g_2^{\gamma Z} &=& g_4^{\gamma Z} = 0\nonumber\\
g_3^{\gamma Z} &=& 2x \sum_f e_{q_f} (g_{_A})_{q_f}
(\Delta q_f - \Delta\bar{q}_f) = 2xg_5^{\gamma Z}\nonumber
\end{eqnarray}
and the purely weak interaction $(i=Z)$ leads to:
\begin{eqnarray}
F_1^Z &=& \frac{1}{2} \sum_f (g^2_{_V}+g^2_{_A})_{q_f} (q_f+\bar{q}_f)
\qquad F_2^Z =2xF_1^Z \nonumber\\
F_3^Z &=& 2\sum_f (g_{_V}g_{_A})_{q_f} (q_f-\bar{q}_f)\nonumber\\
g^Z_1 &=& \frac{1}{2} \sum_f (g^2_{_V}+g^2_{_A})_{q_f}
(\Delta q_f + \Delta \bar{q}_f)\\
g_2^Z &=& -\frac{1}{2} \sum_f (g^2_{_A})_{q_f} (\Delta q_f+\Delta \bar{q}_f)
\nonumber\\
g^Z_3 &=& 2x\sum_f (g_{_V}g_{_A})_{q_f} (\Delta q_f - \Delta \bar{q}_f)
= 2xg^Z_5\nonumber\\
g_4^Z &=& 0 \,.\nonumber
\end{eqnarray}
In case of charged current $(i=W)$, on performing explicitely the $\sum_f$, one
obtains, for $\ell^- N\to \nu X$ processes:
\begin{eqnarray}
F_1^{W^-} &=& u+c+\bar{d}+\bar{s} \qquad F_2^{W^-} = 2xF_1^{W^-}\nonumber\\
F_3^{W^-} &=& 2(u+c-\bar{d}-\bar{s})\nonumber\\
g_1^{W^-} &=& (\Delta u+\Delta c+\Delta\bar{d}+\Delta\bar{s}) = -2g_2^{W^-}\\
g_3^{W^-} &=& 2x (\Delta u+\Delta c-\Delta\bar{d}-\Delta\bar{s}) =
2xg_5^{W^-}\nonumber\\
g_4^{W^-} &=& 0 \,.\nonumber
\end{eqnarray}
$\ell^+ N\to \bar{\nu}X$ processes probe different quark flavours and one
obtains the corresponding expressions of the structure functions $F_j^{W^+}$
and $g_j^{W^+}$ by the flavour interchanges $d \leftrightarrow u$ and
$s \leftrightarrow c$ in the above Eqs. (3.5.8).

Notice that in the naive quark-parton model the structure functions $g^i_4$ are
always zero and one finds $g^i_3 = 2xg_5^i$ for any $i=\gamma, \gamma Z, Z, W$.
The functions $g^i_2$ are nonzero only for pure weak interactions (both in
neutral and charged current processes). It is interesting to note that, in
case of neutral current, the integral of $g^Z_2$ is directly proportional to
the total spin carried by the quarks and antiquarks, as can be seen from
Eqs. (3.5.7) and (3.5.3). Particular combinations of the structure functions
can single out interesting quark and quark spin information; some examples
are discussed in [ANS 93].
%
%end of Section 3
%
\setcounter{section}{3}
\section{\large{Phenomenological analysis of the data on
\mbox{\boldmath $g_1(x)$} and its first moment \mbox{\boldmath $\Gamma_1$}}}
\vskip 6pt
\setcounter{equation}{0}

As explained in Section 2.1.3 the measurement of the cross-section
asymmetry using a longitudinally polarized lepton beam on a longitudinally
polarized hadron target may be interpreted as essentially a measurement of
the spin dependent structure function $g_1(x_, Q^2)$. Experimental data on
$g_1^p$ were first obtained at SLAC [ALG 78] as early as 1978 using an
electron beam. Further information came from the SLAC-Yale group
[BAU 83] in 1983, with fascinating implications about the internal
structure of the proton. And then, more recently, very startling results
were obtained by the European Muon Collaboration (EMC) who scattered a
longitudinally polarized muon beam of energy 100--200 GeV on a longitudinally
polarized hydrogen target at CERN [ASH 88, 89]. The unexpectedly low asymmetry
found by the EMC led to what was termed a ``spin crisis in the Parton Model''
[LEA 88] and raised serious questions as to how the spin of the proton
is built up from the spins of its constituents. It also seemed, for a while,
to imply the first failure of the Parton Model \footnote{It is often forgotten
that there are in a sense other failures of the NPM, namely the experimental
observations of large transverse polarization in inclusive $\Lambda$-production
at high $p_T$ first observed in 1976 and of left-right asymmetry in high $p_T$
inclusive pion production on a transversely polarized target (see {\it e.g.}
[HEL~91]).}.
\setcounter{footnote}{0}

The EMC result catalysed a great deal of theoretical research and the
``crisis'' is now believed to be resolved as a consequence of the discovery of
a deep and beautiful connection between the observable measured by the EMC
and the axial anomaly [EFR 88]. Nonetheless the situation is not yet fully
settled and there are subtle non-perturbative issues and ambiguities in
limiting procedures which will be discussed in Section 6.

Two new experiments are under way to check the EMC  measurement of
$g_1^p$ and to measure the analogous neutron function $g_1^n$. First results
have now been presented [ADE 93, 94; ANT 93] and we shall discuss their
implications in Section 4.5. Our discussion in this Section will largely rely
on the EMC experiment, since this is the experiment at largest $\langle Q^2
\rangle$ and should thus be least sensitive to higher twist contributions.
It is important to realize that even if the EMC results turn out to be
incorrect, the theoretical discoveries catalysed by them remain valid and
interesting. In fact at the time of going to press, the SMC has just issued
results at the same $\langle Q^2 \rangle$ as the EMC experiment [ADA 94]
and they are in substantial agreement with those of EMC. We will
comment briefly on the new measurement in Section 4.2.2.

We shall begin by analysing  the data in the simple Parton Model. We
shall find ourselves in difficulty and will therefore invoke the more
sophisticated operator product expansion with its QCD corrections. This will
be seen to offer no significant help and we will be forced to conclude that
the standard approach is in trouble. A possible resolution of these
difficulties in terms of an anomalous gluonic contribution will be discussed
in Section 7, where we deal with the first moment of $g_1^p(x)$ and in
Section 8.4 where we consider the $x$-dependence.

\subsection{The SLAC-Yale and EMC data:
quark distributions near \mbox{\boldmath $x = 1$}}
\vskip 6pt
\setcounter{equation}{0}

As derived in Section 3.2, in the Naive Parton Model one has
\begin{equation}
g_1^p (x)={1\over 2}\Biggl\{{4\over 9}~[\Delta u (x)+\Delta\bar u(x)]
+{1\over 9}[\Delta d(x)+\Delta \bar{d} (x)] + {1\over 9}[\Delta s (x)+
\Delta\bar{s} (x)]\Biggl\}
\end{equation}
where we assume that the contribution of heavy quarks is negligible.

Before discussing the data on $g^p_1 (x)$ itself it will be interesting
to look at the virtual Compton scattering asymmetry $A_1 (x)$ introduced in
Appendix A. Taking the approximate form
\begin{equation}
A_1 (x) \approx {g_1 (x)\over F_1 (x)}
\end{equation}
and using the standard result
\begin{equation}
F_1 (x) = {1\over 2}\Biggl\{{4\over 9}~[u(x)+\bar{u}(x)] + {1\over 9}~
[d(x)+\bar{d}(x)] + {1\over 9}~[s(x)+\bar{s}(x)]\Biggl\}
\end{equation}
one has, using (4.1.1),
\begin{equation}
A_1(x)\approx{4[\Delta u(x)+\Delta\bar u(x)]+\Delta d(x)+
\Delta\bar d(x)+\Delta s(x)+\Delta\bar s(x)\over
4[u(x)+\bar u(x)]+d(x)+\bar d(x)+ s(x)+\bar s(x)} \cdot
\end{equation}

The EMC and early SLAC-Yale data [ALG 78; BAU 83] on $A_1 (x)$ are shown
in Fig. 4.1. The errors are, of course, large, but there is a discernible
trend towards the maximum possible value $A_1 = 1$ as $x \to 1$.

Given that we know empirically that $u(x) \gg d(x),\ s(x),\ \bar q(x)$
as $x \to 1$ and
that one must have $|\Delta q (x)| \le q(x)$ for each quark flavour,
the behaviour $A_1\to 1$ as $x \to 1$ suggests that
\begin{equation}
\Delta u (x)|_{x\to 1} \to u(x)\,.
\end{equation}

This implies that those $u$ quarks which carry a large fraction of the
longitudinally polarized proton momentum are highly polarized along
the direction of the proton spin.

\subsection{Analysis of data in the framework of the  Operator Product
Expansion}
\vskip 6pt
\setcounter{equation}{0}

As will be discussed in Section 5, the operator product expansion gives
results for the moments of $g_{1,2}$ in terms of hadronic matrix elements of
certain operators multiplied by perturbatively calculable coefficient
functions. The general result for the moments of $g_1$ is
given in (5.2.1). If, however, the fields involved in the
electromagnetic currents in (5.1.1) are treated as {\it free quark fields}
the results simplify via (5.1.10) -- in effect the current commutator in
(5.1.1) can be explicitely evalutated -- and one obtains
\begin{equation}
\Gamma^p_1 = \int^1_0 dx~g_1^p(x)={1\over 12}\Biggl\{a_3 +
{1\over \sqrt{3}} \, a_8+{4\over 3} \, a_0\Biggl\} \,,
\end{equation}
where the $a_i$ are [aside from a factor specified below in (4.2.7 and 8)]
hadronic matrix elements of the octet of quark $SU(3)_F$ axial-vector currents
$J^j_{5\mu}\ (j =1,...,8)$ and the flavour singlet axial current
$J^0_{5\mu}$ taken between proton states of definite momentum and spin
direction. (In practice we usually use helicity states). The precise
definitions are given below.

In the following we shall explain how the values of $a_3$ and $a_8$ can
be obtained from data on hyperon $\beta$-decay. The EMC measurement of
$\Gamma^p_1$ can thus, via (4.2.1) be construed as the first ever
measurement of the singlet matrix element $a_0$ and it will
turn out that the value of $a_0$ is unexpectedly small.

We shall show that in the Naive Parton Model
\begin{equation}
a_0= 2S^q_z
\end{equation}
where $S^q_z$ is the component of the total quark spin (carried by all
quarks and antiquarks) in the direction of motion of a proton of helicity
+1/2. That $S^q_z$ turns out to be almost compatible with zero is a great
surprise. Naively it might have been expected to be fairly close to +1/2
implying $a_0 \simeq 1$. When the quark fields are treated as interacting
fields there are small perturbative corrections to (4.2.1) (see Section 4.4.1)
but these have no significant effect on the argument below.

The octet currents are
\begin{equation}
J^j_{5\mu} = {\bar{\bfpsi}} \gamma_\mu\gamma_5 \Biggl({{\bflambda}_j
\over 2}\Biggl) {\bfpsi}\qquad (j=1,2,...,8)
\end{equation}
where the ${\bflambda}_j$ are the usual Gell-Mann matrices and
${\bfpsi}$ is a column vector in flavour space
\begin{equation}
{\bfpsi} = \left(\begin{array}{c}\psi_u\\\psi_d\\\psi_s\end{array}
\right)\,,
\end{equation}
and the flavour singlet current is
\begin{equation}
J^0_{5\mu} = {\bar{\bfpsi}} \gamma_\mu \gamma_5 {\bfpsi}\,.
\end{equation}
(Note that $J^0_{5\mu}$ is sometimes defined like (4.2.3) with a
flavour matrix ${\bflambda}_0/ 2 = (1/\sqrt{6})\,\bfI$.)
The question of the conservation of the non-singlet axial currents will
be discussed in Section 6.1.

The proton states will be labelled by the four-momentum $P^\mu$ and the
covariant spin vector $S^\mu(\lambda)$ corresponding to definite
helicity $\lambda$. Recall that
\begin{equation}
S\cdot P=0\qquad S^2 = -1 \,.
\end{equation}
The forward matrix elements of the $J^j_{5\mu}$ can only be
proportional to $S^\mu$ and the $a_j$ are conventionally defined by
\begin{eqnarray}
\langle P,S|J^j_{5\mu}|P,S \rangle &=& M a_j S_\mu \\
\langle P,S|J^0_{5\mu}|P,S \rangle &=& 2 M a_0 S_\mu \,.
\end{eqnarray}
The relative factor of 2 in (4.2.7 and 8) reflects the  fact that the
$SU(3)$ currents are defined using the generators of the group {\it i.e.}
$\bflambda_j/2$ in (4.2.3).

Analogous to (4.2.3) one introduces an octet of vector currents
\begin{equation}
J^j_\mu = \bar{\bfpsi} \gamma_\mu \Biggl({\bflambda_j\over 2}\Biggl)
\bfpsi  \qquad (j =1,...,8)
\end{equation}
which are {\it conserved currents} to the extent that
$SU(3)_F$ is a symmetry of the strong interactions.

\subsubsection{Information from hyperon \mbox{\boldmath $\beta$}-decay}
\vskip 6pt

Consider now the $\beta$-decays of the spin 1/2 hyperons. If we use the
standard $SU(3)$ labelling [BAI 82] $B_j$ for the hyperons then the
hadronic transition involved is controlled by matrix elements of the
form $\langle B_i|h^\mu_+|B_k\rangle$ where $h^\mu_+$ is the charged hadronic
current  that couples to the $W$ boson in the electroweak Lagrangian. It is
typically of the form of a Kobayashi-Maskawa matrix element multiplied
by some combination of the currents $J^j_\mu$ and $J^j_{5\mu}$.
For example $\Delta Q = \Delta S =1$ transitions
$(\Lambda \to p;~\bar\Sigma \to n;~\Xi^-\to\Lambda)$ are
controlled by the current
\begin{equation}
V_{us}~\bar{u}\gamma^\mu (1 - \gamma_5)s = V_{us}\,[J^4_\mu + i J^5_\mu
- J^4_{5\mu} -iJ^5_{5\mu}] \,.
\end{equation}
If we now assume:
\begin{description}
\item{a)} that the 8 spin 1/2 hyperons form an octet under $SU(3)_F$;
\item{b)} that the currents $J^j_\mu,~J^j_{5\mu}\ (j = 1,...,8)$ transform
as an octet under $SU(3)_F$ with $J^j_\mu,~J^j_{5\mu}$ conserved,
(the conservation of the axial current will be discussed in Section 6.1);
\item{c)} that the momentum transfer and mass differences in the
hadronic transitions are negligible;
\end{description}

\noindent
then all the hyperon $\beta$-decays are described in terms of two
constants, $F$ and $D$ which occur in the matrix elements of the octet
of axial currents [BAI 82]:
\begin{equation}
\langle B_j;P,S| J^i_{5\mu}|B_k;P,S \rangle = 2 M_B S_\mu
\bigl\{-if_{ijk}F+ d_{ijk}D\bigr\}\qquad (i,j,k = 1,...,8)
\end{equation}
where the $f_{ijk}$ and $d_{ijk}$ are the usual $SU(3)_F$ group
constants. The structure of (4.2.11) is dictated by the $SU(3)$
transformation properties of the LHS and is an example of the
Wigner-Eckart theorem. It should be noted that $J^0_{5\mu}$ does not
play any role in the weak interactions.

An analysis in 1983 of hyperon decay data [WA2 83] seemed to be in
good agreement with the above and led to the values
\begin{equation}
F = 0.477 \pm 0.012\,, \qquad D = 0.756 \pm 0.011 \,.
\end{equation}

If now we use the standard $SU(3)_F$ assignments for the baryon octect
[BAI 82] we find from (4.2.11) that
\begin{eqnarray}
a_3 = F + D \\
a_8 = {1\over \sqrt{3}}~(3F -D)\,.
\end{eqnarray}

We see, therefore, that the study of hyperon $\beta$-decay provides us,
via (4.2.12), with the values of $a_3$ and $a_8$, two of the three
parameters occuring in (4.2.1). However there is some doubt as to the
reliability of the values quoted in (4.2.12), and for that reason we
shall adopt a slightly different strategy. Ultimately, however, it will
be seen that the essential result is not sensitive to this issue.

It is straightforward to demonstrate, using only isotopic spin
invariance, that
\begin{equation}
a_3 = g_{_A}\ ,
\end{equation}
where $g_{_A}$ is the axial coupling (= $G_A/G_V$ in the language of the old
Cabibbo theory) which governs neutron $\beta$-decay. We prefer therefore
to use the very accurate measurement of $g_{_A}$ [PDG 92]
\begin{equation}
a_3 = g_{_A} = 1.2573 \pm 0.0028
\end{equation}
rather than the value $F+D=1.233\pm 0.016$ that follows from (4.2.12).

\subsubsection{The EMC data on \mbox{\boldmath $\Gamma^p_1$} and its
consequences}
\vskip 6pt

Our main interest will be in the matrix element of the flavour singlet
axial current, {\it i.e.} in $a_0$. From (4.2.1) we have
\begin{equation}
a_0={3\over 4}\Biggl\{12\Gamma^p_1-a_3-{1\over\sqrt{3}}\,a_8\Biggl\}\ .
\end{equation}

The EMC data on $x g^p_1(x)$ are shown in Fig. 4.2. Of particular
interest, and indeed the source of all the recent theoretical research
in this subject, is the first moment
\begin{equation}
\Gamma^p_1 \equiv \int^1_0 dx~g^p_1(x) \,.
\end{equation}

Since experimental measurements cannot reach $x = 0$ or 1 an
extrapolation of the data is always required in evaluating (4.2.18).
The region $x \to 1$ is harmless but the extrapolation to $x = 0$ is
potentially tricky. The dashed curve in Fig. 4.2 shows the form of the
extrapolation used by the EMC for the value of
$\int^1_x dx^\prime~g_1^p(x^\prime)$ as function of $x$ for $x\to 0$.
It is seen that the extrapolation looks perfectly reasonable (but beware
the logorithmic scale) and it led to the value
$\Gamma^p_1 = 0.126 \pm 0.010 \pm 0.015$.

Recall [see (2.1.31)] that to extract $g_1(x)$ from the measurement of
the asymmetry $A_\parallel$ one has to utilize information on
$F_2(x,Q^2)$ and $R(x,Q^2)$ which are studied in unpolarized DIS.
Since the EMC experiment was carried out there have been improved
determinations of $F_2$ by the NMC group  [AMA 92] and of $R$ at SLAC
[WHI 90]. As a consequence there is a small change and one has [ELL 93]
\begin{equation}
\Gamma^p_1 \, [\avq = 10.7] = 0.128 \pm 0.013 \pm 0.019
\end{equation}
where we have indicated the mean value of $Q^2$ for the experiment.

In Fig. 4.3 we show the new SMC data [ADA 94] on $g_1^p(x)$ at $\avq = 10$
(GeV/c$)^2$ compared with the EMC data. There is excellent
agreement in general where the measurements overlap. The new data, which
extend to smaller value of $x$ give some indication that $g_1^p(x)$ may
be increasing as $x\to 0$, though this trend may not be significant.
The SMC value for the first moment is $\Gamma^p_1 \, [\avq = 10] =
0.136\pm 0.011\pm 0.011$, perfectly consistent with the EMC result (4.2.19).

This result is considerably smaller than the value 0.188 $\pm$ 0.004
expected on the basis of the Ellis-Jaffe sum rule which will be
discussed in Section 4.3. Several attempts were made to try to explain
this discrepancy, initially focusing on the question as to whether
(4.2.19) can be considered as a reliable number to compare with
the theory. If we substitute the value of $\Gamma^p_1$ (4.2.19) into
(4.2.17) we obtain
\begin{equation}
a_0={3\over 4}\Biggl\{(0.279 \pm 0.156 \pm 0.228)-
{1\over 3} (3F - D)\Biggl\} \,.
\end{equation}
The values (4.2.12) yield $(3F - D)/3 = 0.225 \pm 0.013$ which implies
$a_0 = 0.040 \pm 0.117 \pm 0.171$. However a more recent analysis of
hyperon decays obtains [HSU 88]
\begin{equation}
F=0.46\pm 0.01\qquad D=0.79\pm 0.01\qquad{1\over 3}(3F-D)=0.20\pm 0.01
\end{equation}
which implies
\begin{equation}
a_0 = 0.06 \pm 0.12 \pm 0.17 \,.
\end{equation}
As mentioned earlier one might have expected $a_0 \simeq 1$ so that
the value in (4.2.22) is surprisingly small.

This unintuitive result and the disagreement with the Ellis-Jaffe sum
rule inspired several attempts to question the reliability of the EMC
value for $\Gamma^p_1$. Because of its fundamental importance we turn to
this question. We believe that only two issues require serious
consideration: the question of the extrapolation to $x$ = 0 and the
question of higher twist effects.

\subsubsection{The EMC data: extrapolation to \mbox{\boldmath $x = 0$}}
\vskip 6pt

The extrapolation to small $x$ was queried by Close and Roberts [CLO 88]
who stressed that the extraction of $g_1(x)$ from the data, at small
$x$, is dependent upon the assumed behavior as $x \to 0$ of both
$g_1(x)$ and the usual structure function $F_2(x)$, and that, in
particular, the latter may have a more singular behavior than is given by
the usual Regge analysis of small $x$ behavior.

We are convinced that the Regge behavior
\begin{equation}
g_1(x) \sim x^{-\alpha _{a_1}}
\end{equation}
where $\alpha_{a_1}(t)$ is the Regge trajectory of the $a_1$ meson
(previously referred to as the $A_1$),
$\alpha \equiv \alpha(0) \simeq -0.14 \pm 0.20$, is correct, and that
it is not possible to have a contribution to $g_1(x)$ from the
$P \otimes P$ cut [HEI 73]. The reason is that only Regge poles or cuts
with $G(-1)^T\sigma = -1$ can contribute to those virtual Compton
scattering amplitudes that are relevant to $g_1(x)$. (Here $T$ is the
$t$-channel isospin, $G$ is the $G$-parity and $\sigma$ the signature).
However, it is possible to have a three-pomeron cut, but its
contribution relative to $a_1$-exchange is suppressed by both a factor
$(m/Q)\ (\ln\nu)^{-5}$ and a small numerical coefficient. Also we are
reluctant to accept that the non-Regge singular behavior of $F_2(x)$ can
be relevant at the values of $Q^2$ involved in the EMC experiment.
It seems generally accepted that the extrapolation to small $x$  used by
the EMC is reliable and is not the cause of their peculiar result.

It is interesting that there is some support for the validity of the EMC
extrapolation which comes from a totally different source, namely
{\it elastic} $\nu p \to \nu p$. In the latter process the momentum
transfer dependence of the differential cross-section gives information
on the axial form factor $G_A(Q^2)$ [AHR 87] whose value at $Q^2$ = 0 is
directly related to the $a_j$ appearing in DIS. One has
\begin{equation}
G_A(0)={1\over 2}\Biggl\{a_3+{1\over\sqrt 3}\,a_8 -
{1\over 3}\,a_0\Biggl\}\,.
\end{equation}

It is not absolutely straightforward to obtain $G_A (0)$ from the data,
since it is somewhat dependent on the form of the fit assumed for the
$Q^2$ variation in the measured range $0.45\le Q^2 \le 1.05$
(GeV/c$)^2$. Thus, although no extrapolation in $x$
is required, one does have to make an assumption of smooth behaviour in
$Q^2$ as $Q^2 \to 0$. The value of $G_A(0)$ given in [AHR 87] corresponds to
\begin{equation}
a_0 - \sqrt{3}~a_8 = - 0.45 \pm 0.27 \,.
\end{equation}
Using (4.2.14) and (4.2.21) yields
\begin{equation}
a_0 = 0.14 \pm 0.27
\end{equation}
which is perfectly consistent with the EMC value in (4.2.22).

The reader is warned that the $\nu p \to \nu p$ data are sometimes used as
an argument in favour of a large polarized strange quark contribution
$\Delta s$. This interpretation is only valid in the Naive Parton Model
where ${1\over 3} [a_0 - \sqrt{3} \, a_8] = \Delta s$. As will be discussed
in Section 8.5 there is also a gluonic contribution arising from the
axial anomaly.

\subsubsection{The EMC data : higher twist effects}
\vskip 6pt

The possibility that significant higher  twist effects at the values of
$Q^2$ in the EMC experiment might invalidate a comparison of the data
with the Parton Model or with the leading twist operator product
expansion, was studied by Anselmino, Ioffe and Leader [ANS 89] in an
approach based upon compatibility with the Gerasimov, Drell, Hearn sum
rule. A more consistent treatment was later developed by Burkert and
Ioffe [BUR 92, 93].

Higher twist effects have also been approached in a quite different way
by Balitsky, Braun and Kolesnichenko [BAL 90, 93] based upon
the study of higher twist operators and QCD sum rules. The latter
authors find much smaller corrections than in [ANS 89] and in
[BUR 92, 93], but the uncertainties in their approach are intrinsically
difficult to assess. Also there is some question as to the
correctness of the whole approach [IOF 92].

We shall proceed to explain the two methods. Ultimately we feel that
higher twist effects are unimportant in the EMC experiment but are
surely not negligible in the new experiments at
$Q^2$ = 2 and 4.6 (GeV/c$)^2$.

\vskip 4pt
a) {\it The Gerasimov, Drell, Hearn sum rule approach.}

The key point is that $\Gamma^p_1$ is connected to the
Gerasimov-Drell-Hearn (GDH) sum rule [GER 66, DRE 66] for forward
scattering of {\it real} photons on nucleons and that this indicates rapid
$Q^2$-dependences which {\it might} still have consequences at the EMC
$Q^2$-values.

Define
\begin{equation}
I_p(Q^2) \equiv {2M^2\over Q^2}~\Gamma^p_1(Q^2) \,.
\end{equation}
Then at large $Q^2$
\begin{equation}
I_p(Q^2) \to {2M^2\over Q^2}~\Gamma^p_1(Q^2)_{As}
\end{equation}
where $\Gamma^p_1(Q^2)_{As}$ is the asymptotic form given by say the
lowest twist contribution to the operator product expansion and which
is, empirically, positive. But, according to GDH,
\begin{equation}
I_p(0) = -{\kappa^2_p\over 4}
\end{equation}
where $\kappa_p$ is the anomalous magnetic moment of the proton
$(\kappa_p = 1.79)$. Figure 4.4 shows $I_p(Q^2)$ as given by DIS data and
the value at $Q^2~$= 0. (Also shown is $I_p(Q^2)-I_n(Q^2)$ relevant to
the Bjorken sum rule). Since the $Q^2$-dependence of
$\Gamma^p_1(Q^2)_{As}$, as given by perturbative QCD, is slow and
logorithmic, the behaviour of $I(Q^2)$ implies strong higher twist
effects in $\Gamma^p_1(Q^2)$. These were parametrized in [ANS 89] in a
very simple way by putting
\begin{equation}
I_p(Q^2) = 2M^2 \, \Gamma^p_{1,As}\,\Biggl[{1\over Q^2 + \mu^2} -
{c_p~\mu^2\over (Q^2 +\mu^2)^2}\Biggl]
\end{equation}
where
\begin{equation}
c_p=1+{1\over 8}\Biggl({\mu^2\over M^2}\Biggl)~{\kappa^2_p\over
\Gamma^p_{1,As}}
\end{equation}
and $\mu$ is a parameter setting the scale for the $Q^2$ variation. By
vector dominance arguments it was suggested that $\mu^2\simeq m^2_\rho$.
In (4.2.30, 31) $\Gamma^p_{1,As}$ is taken as a constant, the limit of
$\Gamma^p_1(Q^2)_{As}$ as $Q^2 \to \infty$.

Equation (4.2.30) used in (4.2.27) provides a formula for
$\Gamma^p_1(Q^2)$ for all $Q^2$
\begin{equation}
\Gamma^p_1(Q^2) = {\Gamma^p_{1,As}\over 1 + \mu^2/Q^2}
\Biggl[1-{c_p~\mu^2\over Q^2+\mu^2}\Biggl]
\end{equation}
and equating this to the EMC result at $\avq = 10.7$ (GeV/c$)^2$ yields
$\Gamma^p_{1,As} \simeq 1.2~\Gamma^p_{EMC}$, {\it i.e} some 20\% larger
than the EMC result (4.2.19). However the $Q^2$-dependence implied by
(4.2.32) is too rapid and was later shown to contradict the
$Q^2$-dependence of the data by 1.5 standard deviation [ASH 89].

More recently Burkert and Ioffe [BUR 92, 93] have argued that the
contribution from resonance production has a strong $Q^2$-dependence
for small $Q^2$ and then drops rapidly with $Q^2$ and that this should
be substracted out before parametrizing the smooth large $Q^2$
behaviour. The behaviour of the resonance contribution is shown in
Fig. 4.5. Equation (4.2.30) is then modified to
\begin{equation}
I_p(Q^2)  = I^p_{res}(Q^2) + 2M^2\Gamma^p_{1,As}
\Biggl[{1\over Q^2 + \mu^2} - {c_p~\mu^2\over (Q^2 + \mu^2)^2}\Biggl]
\end{equation}
where now
\begin{equation}
c_p = 1 + {1\over 2} \Biggl({\mu^2\over M^2}\Biggl){1\over
\Gamma^p_{1,As}}\Biggl[{\kappa^2_p\over 4} + I^p_{res}(0)\Biggl] \,.
\end{equation}

For the $\Delta$ ($N^*(1238)$) contribution they find [BUR 92]
\begin{equation}
I^p_\Delta (0) = - 0.78
\end{equation}
and including all resonances up to mass 1.8 GeV [BUR 93]
\begin{equation}
I^p_{res} (0) = -1.03 \,.
\end{equation}

Bearing in mind that $\kappa^2_p/4=0.80$ we see that the resonance
contribution has an important effect on the value of $c_p$.

With this revised estimate of $c_p$, and using $\mu = m_\rho$, (4.2.32)
suggests that $\Gamma^p_{1,As}$ is about 8\% larger than the EMC result.
It appears that the $Q^2$-variation implied by (4.2.32) might just be
compatible with the EMC data divided into two bins with $\langle Q^2 \rangle
= 4.8$ and $\langle Q^2 \rangle = 17.2$ (GeV/c$)^2$.

For purposes of comparison with the QCD sum rule approach, let us expand
(4.2.32) in inverse powers of $Q^2$ (we ignore $I_{res}(Q^2)$ at large $Q^2$):
\begin{equation}
\Gamma^p_1(Q^2) = \Gamma^p_{1, As}\Biggl[ 1 - {(1 + c_p)\mu^2\over Q^2}
+\cdots\Biggl] \, \cdot
\end{equation}
The leading term in the higher twist correction is then
\begin{equation}
\delta\Gamma^p_1(Q^2) = - {(1 + c_p) \mu^2~\Gamma^p_{1,As}\over Q^2}
\approx - {0.12~\gevc^2\over Q^2} \,\cdot
\end{equation}

In summary the higher twist corrections, estimated on the basis of the
GDH sum rule, are likely to be less than 8\% for the EMC experiment.

\vskip 4pt
b) {\it The QCD sum rule approach}

In the usual operator product approach explained in Section 5 one keeps
only the operators of lowest twist $\tau$ ($\tau = 2$ for the study of
$g_1(x, Q^2)$). The approach of [BAL 90], based on the QCD sum rule
paper of Shuryak and Vainshtein [SHU 82], is a natural extension in which
one studies also the operators of twist 4 and in which target mass
corrections are taken into account. The problem is that one has to
estimate certain matrix elements of gluon and quark condensates and
certain correlators of quark and gluon operators. The latter is done
assuming pseudoscalar meson dominance.

The leading target mass correction $\delta\Gamma_{1,T}$ to $\Gamma_1^{p,n}$
involves the third moment of $g^{p,n}_1$ and is given by [BAL 90, 93]
\begin{equation}
\delta \Gamma^{p,n}_{1,T} = {2M^2\over 9Q^2}
\int^1_0 dx~x^2\,g_1^{p,n}(x,Q^2)
\end{equation}
which yields
\begin{equation}
\delta\Gamma^p_{1,T} \approx {0.003~M^2\over Q^2}
\end{equation}
which is totally negligible in the EMC experiment.

The higher twist corrections require an estimate of the nucleon matrix
elements of certain twist-4 ($\tau_4$) operators. The contribution to
$\Gamma_1^p$ involves flavour singlet (S) and non-singlet (NS) operators:
\begin{equation}
\delta\Gamma^p_{1,\tau_4} = - {1\over 6 Q^2}\Biggl\{{5\over 3}
\langle\langle O^{S} \rangle\rangle + \langle\langle O^{NS}
\rangle\rangle \Biggl\}
\end{equation}
where  [BAL 90, 93]
\begin{equation}
\langle\langle O \rangle\rangle = {8\over 9}\Biggl[
\langle\langle U \rangle\rangle - {M^2\over 2}
\langle\langle V \rangle\rangle \Biggl]
\end{equation}
and $\langle\langle U \rangle\rangle, \ \langle\langle V \rangle\rangle$
are the reduced matrix elements of operator $U_\mu$
and $V_{\mu\nu,\sigma}$, defined by
\begin{equation}
\langle P, S|U_\mu|P, S \rangle = 2M S_\mu \langle\langle U \rangle\rangle
\end{equation}
and
\begin{equation}
\langle P, S|V_{\mu\nu,\sigma}|P, S\rangle = 2M \langle\langle V
\rangle\rangle \{(S_\mu P_\nu - S_\nu P_\mu)P_\sigma\}_{S_{\nu\sigma}}
\end{equation}
where $S_{\nu\sigma}$ means symmetrization with respect to $\nu,\sigma$.

The actual operators are :
\begin{eqnarray}
U^{S/NS}_\mu &=& g\,[\bar{u}\,\widetilde{\bfG}_{\mu \nu} \gamma^\nu u \pm
\bar{d}\,\widetilde{\bfG}_{\mu\nu}\gamma^\nu d] \\
V^{S/NS}_{\mu \nu,\sigma} &=& g\,\{\bar u \, \widetilde {\bfG}_{\mu\nu}
\gamma_\sigma u \pm\bar{d} \, \widetilde{\bfG}_{\mu \nu}
\gamma_\sigma d \}_{S_{\nu\sigma}}
\end{eqnarray}
where $\widetilde{\bfG}_{\mu \nu}$ is the matrix
\begin{equation}
\widetilde{\bfG}_{\mu \nu} = {1\over 2} \, \varepsilon_{\mu\nu\alpha\beta} \,
G^{\alpha\beta}_a\Biggl({\bflambda_a\over 2}\Biggl)\,.
\end{equation}

It is a non-trivial matter, using sum-rule techniques, to estimate the
nucleon matrix elements of these operators and there is some uncertainty
about the results. Moreover it is claimed [IOF 92] that the handling of
the singlet is incorrect since the axial anomaly (to be discussed in
Section 6) was not taken into account and certain correlators were
estimated using pseudoscalar dominance by $\pi$ and $\eta$ which is
inappropriate to the singlet case. In addition note that there are
errors in [BAL 90] which are corrected in [BAL 93].
In their corrected papers [BAL 93] one has
\begin{eqnarray}
\delta\Gamma^p_{1,\tau_4} &=& - {(0.02 \pm 0.013) \,
\gevc^2 \over Q^2}\nonumber\\
\delta\Gamma^n_{1,\tau_4} &=& - {(0.005 \pm 0.003)\,
\gevc^2 \over Q^2}\, \cdot
\end{eqnarray}

These corrections are much smaller than those obtained from the GDH sum
rule (4.2.38) and are quite negligible for the EMC experiment.

\vskip 4pt
c) {\it Summary on higher twist effects in the EMC experiment}

There is a significant difference between the size of these effects as
calculated on the basis of the GDH sum rule (4.2.38) and on the basis of
QCD sum rules (4.2.48). In both cases, however, the effects for the EMC
experiment are negligible.

We shall thus proceed to analyse the theoretical implications of the EMC
result, which gives a value for $a_0$ compatible with zero. In the
following we shall use the value given in (4.2.22).

\subsection{The EMC result: implications in the Naive Parton Model}
\vskip 6pt
\setcounter{equation}{0}

In Appendix B we show how to express the protonic matrix element of
the axial vector current of a quark field  of definite flavour in terms
of the polarized parton distributions, in the framework of the
Naive Parton Model.

We have
\begin{equation}
\langle P,S|\bar{\psi}_f(0) \gamma_\mu \gamma_5 \psi_f(0)|P,S \rangle
= 2MS_\mu \int^1_0 dx~[\Delta q_f (x) + \Delta \bar{q}_f (x)] \,.
\end{equation}
It follows from (4.2.3, 5, 7 and 8) that
\begin{equation}
a_3=\int^1_0 dx~[\Delta u(x)+\Delta \bar{u}(x)-\Delta d(x)-
\Delta\bar{d} (x)]
\end{equation}
\begin{equation}
a_8={1\over \sqrt{3}} \int^1_0 dx~[\Delta u(x)+\Delta\bar{u}(x)+
\Delta d(x)+\Delta\bar{d}(x)-2\Delta s(x)-2\Delta\bar{s}(x)]
\end{equation}
and
\begin{equation}
a_0 = \Delta\Sigma\equiv \int^1_0 dx~\Delta\Sigma (x)
\end{equation}
where
\begin{equation}
\Delta\Sigma (x)\equiv\Delta u(x)+\Delta\bar{u}(x)+\Delta d(x)+
\Delta\bar{d}(x)+\Delta s(x) + \Delta\bar{s}(x)\,.
\end{equation}
Notice that (4.3.2-4) inserted into (4.2.1) agree with (4.1.1).

If, as is at first sight not an unreasonable assumption, one assumes
that one can neglect the strange quark contribution so that
$a_0 \simeq \sqrt 3\,a_8$ one is led to the Ellis-Jaffe sum rule [ELL 74]
mentioned in Section 4 which gives a much larger value for $\Gamma^p_1$
than found by the EMC:
\begin{equation}
\left( \Gamma_1^p \right)_{EJ} = {1\over 12} \left\{ a_3 +
{5\over \sqrt 3}\, a_8 \right\} \simeq 0.188 \pm 0.004
\end{equation}
where we have used (4.2.14, 16 and 21).

\subsubsection{The ``spin crisis in the Parton Model"}
\vskip 6pt

Let us consider now the physical significance of $\Delta\Sigma(x)$. Since
$q^\pm (x)$ count the number of quarks of momentum fraction $x$ with
spin component $\pm {1\over 2}$ along the direction of motion of the
proton (let us call this the $z$-direction), the total contribution to
$J_z$ coming from a given flavour quark is
\begin{eqnarray}
S_z&=& \int^1_0 dx\Biggl\{\Biggl({1\over 2}\Biggl) q^+(x) +
\Biggl(-{1\over 2}\Biggl) q^-(x) \Biggl\} \nonumber\\
&=&{1\over 2} \int^1_0 dx~\Delta q(x) \,.
\end{eqnarray}
It follows that
\begin{equation}
a_0 = 2 S^{quarks}_z
\end{equation}
where $S^{quarks}_z$ is the contribution to $J_z$ from the spin of all
quarks and antiquarks.

Now in the Naive Parton Model $\bfp_\perp = 0$ and all quarks move parallel
to the parent hadron, {\it i.e.} for a quark of momentum ${\bfp}$,
${\bfp} = x{\bfbp}$. Hence any orbital angular momentum carried by the
quarks is perpendicular to ${\bfbp}$ and thus does not contribute to $J_z$.
In addition it is assumed that the gluons are unpolarized, for reasons
explained in Section 4.3.2. Hence, in the Naive Parton Model, one expects
for a proton of helicity  +1/2 :
\begin{equation}
S^{quarks}_z = J_z = 1/2 \,.
\end{equation}
We stress that this ignores ${\bfp}_\perp$ effects and assumes only quark
and antiquark constituents are polarized.

{}From (4.3.8) and (4.2.22) we get instead of the value 1/2 of (4.3.9)
\begin{equation}
\biggl(S_z^{quarks}\biggr)_{Exp} = 0.03 \pm 0.06 \pm 0.09\ .
\end{equation}
It was this highly unexpected result which was termed the ``spin crisis
in the Parton Model" [LEA 88] .

\subsubsection{The angular momentum sum rule}
\vskip 6pt

In a more general framework, if we allow for the existence of parton
transverse momentum and for the possibility that the gluons are
polarized, (4.3.9) generalizes to the {\it angular momentum sum
rule}
\begin{equation}
J_z = S^{quarks}_z + S^{gluons}_z + L^{partons}_z = 1/2
\end{equation}
and the question then arises as to whether the result (4.3.10) is as
surprising as it at first seems. On the one hand models of the
{\it static} properties of the baryons always have $S$-wave ground
states of quarks only, so in these there is no $S^{gluons}_z$ nor
$L^{partons}_z$. But these models deal with {\it constituent
quarks} whereas the Parton Model involves {\it current quarks}.

Unfortunately nobody knows the precise connection between these so it is
perhaps unjustified to make assertions about partons on the basis of
constituent models. However, Lipkin [LIP 90] has given an argument based
on partons, which emphasizes how peculiar it is that $S^{quarks}_z$ is
so small. We present a simplified version of his treatment. The
key point is that the $SU(6)$ prediction for $g_{_A}$, namely 5/3, is
really quite close to the measured value (4.2.16), so that the proton
wave function cannot be too different from the $SU(6)$ wave function.
Thus if we write for this proton state
\begin{equation}
|p\rangle = \cos\theta ~|SU(6)\rangle + \sin\theta ~|\psi\rangle
\end{equation}
where $|\psi\rangle$ is a state orthogonal to $|SU(6)\rangle$, the latter being
a  state in which the valence partons are in an $SU(6)$ configuration and
anything else present has zero orbital, and zero total spin, angular
momentum, then consideration of the contribution to $g_{_A}$ from various
possible  plausible states $|\psi\rangle$, each having different amounts of
spin carried by the valence quarks and by any other constituents, leads to
the bound $\sin^2\theta \le 3/16$. On the other hand in the $SU(6)$
state, $J_z = S^{quarks}_z = 1/2$, so that requiring that
\begin{equation}
{1\over 2}~\cos^2\theta + \biggl(S^{quarks}_z\biggl)_\psi~\sin^2\theta
\simeq 0
\end{equation}
implies the large value
\begin{equation}
\vert S^{quarks}_z \vert_\psi \gsim 2\,.
\end{equation}
Given the complexity of the state $|\psi\rangle$ that this implies, it is then
hard to understand why the proton is the only stable state with
$J = 1/2,~I = 1/2$.

We shall be able to clarify this question a little more in Section 5
when we consider the surprising feature that QCD induces a
$Q^2$-dependence in $S^{gluons}_z$ and $L^{partons}_z$. This will force
us to rethink our interpretation of the physical meaning
of the matrix element of $J^0_{5\mu}$.

\subsubsection{Trouble with the strange quark}
\vskip 6pt

There is another difficulty associated with the results that follow from
analysing the EMC and hyperon data in terms of the Naive Parton Model.

We can use the numerical values (4.2.16, 21 and 22) to solve, via (4.2.14)
and (4.3.2, 3, 4), for the contributions of quarks and antiquarks of a
given flavour. One finds:
\begin{eqnarray}
\Delta u &\equiv& \int^1_0 dx~[\Delta u(x) + \Delta\bar{u}(x)] =
0.79 \pm 0.03 \pm 0.04 \\
\Delta d &\equiv& \int^1_0 dx~[\Delta d(x) + \Delta\bar{d}(x)] =
-0.47 \pm 0.03 \pm 0.04 \\
\Delta s &\equiv& \int^1_0 dx~[\Delta s(x) + \Delta\bar{s}(x)] =
-0.26 \pm 0.06 \pm 0.09 \,.
\end{eqnarray}

But Preparata and Soffer [PRE 88] suggested that one can bound the
strange quark contribution using the fact that, manifestly,
\begin{equation}
|\Delta s(x)| \le s(x)
\end{equation}
and the knowledge of the behaviour of $s(x)$ as determined from deep
inelastic neutrino experiments.

There is an unjustified step in the analysis of [PRE 88] which can be
modified [LEA 88b], and the argument is not rigorous since a specific
form for $\Delta s$ is assumed. Nonetheless the bound is unlikely to be
far from the truth. It yields
\begin{equation}
|\Delta s|=\Biggl|\int^1_0 dx [\Delta s(x)+\Delta\bar{s}(x)]\Biggl|
\le 0.072 \pm 0.030
\end{equation}
which is roughly a factor of three smaller than the value given in
(4.3.17).

We have to conclude that the analysis based on the Naive Parton Model
or the Operator Product Expansion with free fields is in difficulty.

\subsection{Analysis of \mbox{\boldmath $\Gamma_1$} using the QCD Improved
Parton Model}
\vskip 6pt
\setcounter{equation}{0}

The analysis in Sections 4.1-3 was based on the Naive Parton Model which
emerges from the Operator Product Expansion when all fields are treated
as free fields. An immediate
question is whether the difficulties we encountered can be removed by
the inclusion of perturbative QCD corrections. In this Section we shall
use the QCD-improved expressions which will be derived in Section 5 and
demonstrate that the corrections do not help significantly. We shall
also discuss the Bjorken sum rule.

\subsubsection{The operator product expansion for
\mbox{\boldmath $\Gamma^p_1$}}
\vskip 6pt

As is well known, QCD corrections induce a logarithmic $Q^2$-dependence
which breaks Bjorken scaling. As is outlined in Section 5.2 the
expression (4.2.1) is modified to
\begin{equation}
\Gamma^p_1(Q^2)={1\over 12}\Biggl\{\Biggl(a_3+{1\over\sqrt 3}\,a_8\Biggl)
E_{NS}(Q^2)+{4\over 3}\,a_0\,E_S(Q^2)\Biggl\}
\end{equation}
where the coefficient functions $E_{NS}$ and $E_S$ have perturbative
expansions [KOD 79] recently extended to order $\alpha^2_s$ and
$\alpha^3_s$ respectively [LAR 91, 94]
\begin{equation}
E_{NS}(Q^2) = 1 - {\alpha_s\over \pi} -
\left(\begin{array}{c}3.58\\3.25\end{array}\right)\Biggl({\alpha_s\over
\pi}\Biggl)^2 -
\left(\begin{array}{c}20.22\\13.85\end{array}\right)\Biggl({\alpha_s
\over \pi}\Biggl)^3
\end{equation}
\begin{equation}
E_{S}(Q^2) = 1 - \left(\begin{array}{c}0.333\\0.040\end{array}\right)
\Biggl({\alpha_s\over \pi}\Biggl) -
\left(\begin{array}{c}\phantom{-}1.10\\-0.07\end{array}\right)
\Biggl({\alpha_s\over\pi}\Biggl)^2
\end{equation}
where $\alpha_s = \alpha_s(Q^2)$ and where the upper and lower figures
refer to $N_f = 3\mbox{ or } 4$ respectively. Strictly speaking the
coefficients of $a_3$ and $a_8$ are only equal in the case of massless
quarks, which is assumed in the calculation of (4.4.2). These
perturbative corrections will be important in analysing the new data at
lower $Q^2$, to be discussed in Section 4.5.2. In the notation of
Section 5.2 we have $E_{NS} = E^1_{1,1} = E^1_{1,3};\ E_S = E^1_{1,\psi}$.

Once interactions are allowed the operators and currents have to be
renormalized and their matrix elements will, in general, depend upon the
subtraction point or renormalization scale $\mu$. For a conserved
current, however, the matrix elements are independent of the
renormalization scale. These matters were mentioned in Section 4.2 and will
be taken up again in Section 6. The important  point to remember is that
$a_3$ and $a_8$ are independent of $\mu$, whereas $a_0$ depends upon $\mu$.

If we choose $\mu^2 = Q^2$ then, in leading logarithmic approximation
(LLA), the expressions (4.3.2--4) for the $a_j$ in terms of quark
densities $\Delta q(x)$ are simply modified by the replacement
\begin{equation}
\Delta q(x) \to \Delta q(x;Q^2)
\end{equation}
where the $Q^2$-evolution is controlled by the spin dependent
Altarelli-Parisi equations [ALT~77].

For the EMC experiment the mean value of $Q^2$ is about 10.7 (GeV/c$)^2$
which corresponds to $\alpha_s(Q^2) \simeq 0.24$ for 4 flavours. The
leading correction term is of order 8\% in (4.4.2) and even smaller in
(4.4.3 and we cannot expect a dramatic change in the
values quoted in (4.2.22). Indeed one finds
\begin{equation}
a_0 = 0.17 \pm 0.12 \pm 0.17
\end{equation}
to be compared with the naive expectation $a_0 \simeq 1$. The values
quoted in (4.3.15-17) become, for $Q^2 \simeq 10$ (GeV/c$)^2$
\begin{eqnarray}
\Delta u&=& 0.82 \pm 0.03 \pm 0.04\nonumber\\
\Delta d&=& -0.44 \pm 0.03 \pm 0.04\\
\Delta s&=& -0.21 \pm 0.06 \pm 0.09\nonumber
\end{eqnarray}
which are close to the values given in (4.3.15-17) \footnote{Note added in
proof: An analysis of {\it all} the proton data at $Q^2 = 10$ (GeV)$^2$
[ADA 94] alters (4.2.19) to $\Gamma_1^p = 0.142 \pm 0.008 \pm 0.011$ and
consequently (4.4.5) becomes (notice that in [ADA 94] $\alpha_s \simeq 0.23$
is used) $a_0 = 0.27 \pm 0.08 \pm 0.10$, still uncomfortably far from
$a_0 = 1.$}.
\setcounter{footnote}{0}

Thus the straightforward perturbative QCD corrections to (4.2.1) in no way
alleviate the problem, and the value of $\Delta s$ is still
surprisingly large compared with (4.3.19).

\subsection{The new experiments: neutron data and the Bjorken sum rule}
\vskip 6pt
\setcounter{equation}{0}

The operators $J^0_{5\mu}$ and $J^8_{5\mu}$ that give rise to the
terms $a_0$ and $a_8$ in (4.4.1) are invariant under isotopic spin
rotations. Thus in going from proton matrix elements to neutron matrix
elements they remain unchanged. $J^3_{5\mu}$, on the other hand,
transforms like the 3rd component of an isotopic spin triplet and
therefore changes sign. It follows that the Bjorken sum rule
\begin{equation}
\int^1_0 dx~[g^p_1(x,Q^2) - g^n_1 (x,Q^2)] = S_{Bj} (Q^2)
\end{equation}
where $S_{Bj} (Q^2)$ is the theoretical value of the Bjorken sum rule,
given by
\begin{eqnarray}
S_{Bj} (Q^2)&=& {a_3\over 6}\Biggl\{ 1 - {\alpha_s\over \pi} -
\left(\begin{array}{c}3.58\\3.25\end{array}\right)
\Biggl({\alpha_s\over \pi}\Biggl)^2 -
\left(\begin{array}{c}20.22\\13.85\end{array}\right)
\Biggl({\alpha_s\over\pi}\Biggl)^3\ \cdots \Biggl\}\\
&=&(0.2096 \pm 0.0005)\Biggl\{1 - {\alpha_s\over \pi} -
\left(\begin{array}{c}3.58\\3.25\end{array}\right)
\Biggl({\alpha_s\over \pi}\Biggl)^2 -
\left(\begin{array}{c}20.22\\13.85\end{array}\right)
\Biggl({\alpha_s\over\pi}\Biggl)^3\ \cdots\Biggl\}\,,\nonumber
\end{eqnarray}
holds on fundamental grounds. Note that the fact that $a_3$ is
independent of $Q^2$ is linked to the conservation of the non-singlet
axial current and is discussed more fully in Section 6.1. The upper and
lower figures in (4.5.2) refer to $N_f = 3 \mbox{ or }4$ respectively.

The Ellis-Jaffe sum rule value (4.3.6), $\Gamma^p_1 \simeq 0.188$,
would imply, via (4.5.1,\ 2) at $Q^2 = 10.7$ (GeV/c$)^2$
$$
\Gamma^n_1\equiv \int^1_0 dx~g^n_1(x,Q^2) \simeq - 0.0003
$$
whereas the measured EMC value of $\Gamma^p_1$ gives hope of a somewhat
more sizeable neutron result:
$$
\Gamma^n_1 \simeq - 0.06 \qquad \mbox{\rm at} \qquad Q^2 = 10.7
{}~\mbox{\rm (GeV/c)}^2 \,.
$$

\subsubsection{The new experiments on deuterium and \mbox{\boldmath $^3$}He}
\vskip 6pt

The first experiments to measure $g^n_1(x)$ have now begun to report
results. At CERN the Spin Muon Collaboration (SMC) have used a polarized
deuterium target at an average $Q^2$ of $4.6$ (GeV/c$)^2$ [ADE 93]. At
SLAC the E142 group  uses a polarized $^3$He target at $ \avq = 2.0$
(GeV/c$)^2$ [ANT 93]. (For a combined analysis of the data see [ADE 94]).

The SMC data on the deuteron asymmetry $A^d_1(x)$ at
$Q^2=4.6$ (GeV/c$)^2$ are shown in Fig. 4.6, and in Fig. 4.7 the data on
$xg^d_1(x)$. The E142 data on the neutron asymmetry $A^n_1(x)$ and on
$g^n_1(x)$ at $Q^2$ = 2 (GeV/$c)^2$ are shown in Fig. 4.8. The neutron
data were extracted as explained in Section 2.1.4 using the EMC proton
data.

Soon after the appearance of the new results some rather dramatic
claims were made that the very fundamental Bjorken sum rule had been
violated. In retrospect these incorrect assertions arose because of too
naive a comparison of the new data with the EMC proton data at
$10.7$ (GeV/c$)^2$. A more careful analysis is necessary, as emphasized
by Ellis and Karliner [ELL 93] and by Close and Roberts [CLO 93] and
Altarelli, Nason, Ridolfi [ALT 94].

The main issues are the following:

\vskip 4pt
1) The extraction of $g^n_1(x,Q^2)$ from nuclear data at some $\avq$,
as explained in Section 2.1.4, requires a knowledge of $g^p_1 (x,Q^2)$
at the same mean $Q^2$. It is unjustified to naively combine the EMC
data  at $\avq = 10.7$ (GeV/c$)^2$ with the deuterium data at 4.6 and the
$^3$He data at 2.0 (GeV/c$)^2$.

There is no absolutely safe way to overcome this difficulty. It does seem,
however, that the values of the {\it asymmetries} at fixed $x$, show no
significant variation with $Q^2$ where it has been possible to examine
this. On the other hand the variation of $F_2(x,Q^2)$ with $Q^2$ has
been much studied.

In the improved analysis, therefore, $g_1^p(x,Q^2)$ is calculated at the
required $Q^2$ by using
\begin{equation}
g^p_1(x,Q^2)={A^p_\parallel\over D}~{F_2^p(x,Q^2)\over 2x[1+R^p(x,Q^2)]}
\end{equation}
and {\it assuming $A^p_\parallel$ independent of $Q^2$}. The resulting
dependence of $g^p_1(x,Q^2)$ upon the mean $Q^2$ is shown  in Fig. 4.9
taken from [ELL 93]. In (4.5.3) values of $F^p_2$ were taken from the NMC
data [AMA 92] and of $R^p$ from the SLAC data [WHI 90]. These
$g_1^p(x,Q^2)$ yield, [ELL 93],
\begin{eqnarray}
\Gamma^p_1 \,[2~\gevc^2] &=& 0.124 \pm 0.013 \pm 0.019\nonumber\\
\Gamma^p_1 \, [4.6~\gevc^2] &=& 0.125 \pm 0.013 \pm 0.019 \,.
\end{eqnarray}
Figure 4.10 shows the $g_1^n(x,Q^2)$ extracted from the nuclear data on
this basis.

\vskip 4pt
2) The extrapolation of $g_1^n(x,Q^2)$ to $x = 1$ and $x = 0$ in order to
compute the first moment $\Gamma^n_1$ requires some care.

The SMC experiment at $\avq$ = 4.6 (GeV/c)$^2$ covers the range
0.006 $\le x \le$ 0.6. They find
\begin{equation}
\int^{0.6}_{0.006}~g^d_1(x)~dx = 0.024 \pm 0.020 \pm 0.014 \,.
\end{equation}

The extrapolation to $x = 0$ is taken to be of the reasonable form
$g^d_1 \sim x^{-\alpha}$ with $-0.5 \le \alpha \le 0$, in accord with
the discussion in Section 4.2.3 and compatible with Eq. (4.2.23).
This leads to an estimate of $-0.003\pm 0.003$ for the integral from 0 to
0.006.

The extrapolation to $x = 1$ is based on the fact that the asymmetry is
bounded,
$|A| \le 1$, and leads to the estimate 0.002 $\pm$ 0.004 for the integral
from 0.6 to 1.

The result is, {\it per nucleon},
\begin{equation}
\Gamma^d_1\,[Q^2 = 4.6~\gevc^2] = 0.023 \pm 0.020 \pm 0.015 \,.
\end{equation}

Note that using (2.1.55), (4.4.2, 3) and (4.2.14) we have (taking $N_f = 4$)
\begin{equation}
\Gamma^d_1={(1 - 1.5~\omega_D)\over 12}\Biggl\{{1\over 3}~(3F - D)
\Biggl[1-{\alpha_s\over \pi}-3.25\Biggl({\alpha_s\over\pi}\Biggl)^2\Biggl]
+{4\over 3}~a_0\Bigg [1- 0.04 \, {\alpha_s\over \pi}\Biggl]\Biggl\}\,.
\end{equation}

Taking $\alpha_s = 0.28 \pm 0.03$ at $Q^2$ = 4.6 (GeV/$c)^2$ [NAS 93] and
using the values in (4.2.21), one obtains ($\omega_D = 0.058$)
\begin{equation}
a_0 = 0.09 \pm 0.20 \pm 0.15 \,,
\end{equation}
again consistent with zero and perfectly compatible with the value quoted
in (4.4.5).

For $\Gamma^n_1$ [ELL 93] obtain a result very slightly different from the
value quoted by the SMC because of the small change in $\Gamma^p_1$ shown in
(4.5.4), {\it i.e.}
\begin{equation}
\Gamma^n_1 \,[4.6~(\mbox{\rm GeV/c})^2] = - 0.076 \pm 0.037 \pm 0.046 \,.
\end{equation}

The E142 experiment at $\avq$ = 2 (GeV/c)$^2$ covers the range
0.03 $\le x \le$ 0.6. [ELL~93] have re-analysed the extrapolations to
$x = 0$ and $x = 1$ using the same approach as in the SMC case and they
estimate contributions to $\Gamma^n_1$ of $-0.006 \pm 0.006$ for the
integral from 0 to 0.03 and 0.000 $\pm$ 0.003 for the region 0.6 to 1.
The result is
\begin{equation}
\Gamma^n_1 \,[2.0~(\mbox{\rm GeV/c})^2] = - 0.028 \pm 0.006 \pm 0.009
\end{equation}
which differs somewhat from the value $-0.022 \pm 0.011$ quoted by E142.

\subsubsection{Tests of the Bjorken sum rule}
\vskip 6pt

Using the values of $\Gamma^p_1$ given in (4.5.4) and of $\Gamma^n_1$ given
in (4.5.9 and 10) one can now test the Bjorken sum rule, bearing in mind,
however, that we have {\it assumed} that the asymmetry is
$Q^2$-independent for the range involved. Before comparing with the data
it is instructive to see to what extent the perturbative QCD corrections
influence the expected value of the Bjorken sum rule, $S_{B_j}(Q^2)$ given
by (4.5.2). We show below in Table 4.1 the values of $S_{B_j} (Q^2)$ at
various $Q^2$ calculated to order $\alpha_s,~\alpha^2_s$ and $\alpha^3_s$
starting from the central value 0.2096 without corrections. We have used
values of $\alpha_s(Q^2)$ from [NAS~93] which differ somewhat
from the values used in [ELL~93] and also from those used by the SMC group.
We believe these represent a good compromise between the LEP and
unpolarized DIS determinations which differ somewhat.

\begin{center}
{\sf
{\bf Table 4.1} Values of $\alpha_s (Q^2)$ and of $S_{B_j}(Q^2)$ to various
orders in perturbation theory
\vskip0.3cm
\begin{tabular}{|c|c|c|c|c|c|}
\hline
&&&\multicolumn{3}{|c|}{$S_{B_j}(Q^2)$ = 0.2096 to zero order}\\
\cline{4-6}
&&&&&\\
$Q^2$&$N_f$&$\alpha_s(Q^2)$&to ${\cal O}(\alpha_s)$&to ${\cal O}(\alpha^2_s)$
&to ${\cal O}(\alpha^3_s)$\\
in (GeV/c)$^2$&&&&&\\
\hline
2&3&0.354 $\pm$0.04&0.186&0.176&0.170\\
4.6&4&0.283 $\pm$0.03&0.191&0.185&0.183\\
10.7&4&0.238 $\pm$0.03&0.194&0.190&0.189\\
\hline
\end{tabular}
}
\end{center}
\vskip 6pt
We see that at $Q^2 = 2$ the corrections are very important, $\simeq 20\%$,
and that even at $Q^2$ = 10.7 they are not negligible, $\simeq$ 10\%.

We come now to the comparison between the experimental and theoretical values
of the Bjorken sum rule. In allowing for target mass corrections and higher
twist corrections as discusses in Section 4.2.4 we must recall that the GDH
and QCD sum rule approach give quite different estimates for the higher twist
corrections.

In general terms one has
\begin{equation}
\int^1_0 dx~[g^p_1(x,Q^2) - g^n_1(x,Q^2)] = S_{B_j}(Q^2) +
\delta S^{Bj}_{\tau 4} + \delta S^{Bj}_T \,.
\end{equation}

The target mass correction given in (4.2.40) is negligible for the proton
and an estimate for the neutron is much smaller [ELL 93] so we can ignore
$\delta S^{Bj}_T$ even at $Q^2$ = 2.

In the QCD sum rule approach one has, from (4.2.48)
\begin{equation}
\delta S^{Bj}_{\tau 4}\Biggl|_{QCD \, sum \, rules} \approx
-{(0.015 \pm 0.013) \, \gevc^2 \over Q^2} \, \cdot
\end{equation}

In order to estimate the higher twist corrections in the GDH sum rule
approach let us rewrite (4.2.37) as
\begin{equation}
\Gamma^{p,n}_1(Q^2) = \Gamma^{p,n}_{1,As} - {\mu^2\over Q^2}
[2\Gamma^{p,n}_{1,As} +\lambda_{p,n}]
\end{equation}
where
\begin{equation}
\lambda_{p,n} \equiv {1\over 2}\Biggl({\mu^2\over M^2}\Biggl)
\Biggl[{\kappa^2_{p,n}\over 4} +I^{Res}_{p,n}(0)\Biggl] \,.
\end{equation}
Using the values [BUR 93]
\begin{equation}
I^{Res}_p(0) = -1.028 \qquad I^{Res}_n(0) = -0.829
\end{equation}
we find
\begin{eqnarray}
\delta S^{Bj}_{\tau 4}\Biggl|_{GDH\,sum\,rule} &=& -{\mu^2\over Q^2}
\Biggl\{2(\Gamma^p_1 -\Gamma^n_1)_{As} + \lambda_p - \lambda_n\Biggl\}
\nonumber\\
&=& - {0.16 \, \gevc^2 \over Q^2}\,\cdot
\end{eqnarray}

This appears to be a huge correction at small values of $Q^2$. However
it is illegitimate to use (4.5.16) in a region where the resonance
contribution $I_{Res} (Q^2)$ in (4.2.33) has not yet died out. The $(p-n)$
curve in Fig. 4.5 indicates that $I_{Res}$ is not negligible at
$Q^2 = 2$ (GeV/c$)^2$.
Thus we should add a contribution
\begin{equation}
{Q^2\over 2M^2}~I^{p-n}_{Res} (Q^2)\Biggl|_{Q^2 = 2} \simeq 0.05
\end{equation}
to the correction (4.5.16) at $Q^2 = 2$ (GeV/c$)^2$, yielding
\begin{equation}
\delta S^{Bj}\Biggl|_{GDH\,sum\,rule} = - 0.03 \qquad \mbox{\rm at} \qquad
Q^2 = 2 \, \gevc^2 \,.
\end{equation}

At $Q^2 = 4.6$ (GeV/c$)^2$ it would appear to be safe to neglect
$I_{Res}(Q^2)$ and to use just (4.5.16). However the corrections do seem
surprisingly large. We are unable to estimate the errors in (4.5.16 and 18).

In Table 4.2 we compare the experimental value of the LHS of (4.5.11) with
the theoretical value of the RHS using the results of Table 4.1 and with
the two versions of the higher twist corrections.

\begin{center}
{\sf
{\bf Table 4.2} Test of the Bjorken sum rule
\vskip0.3cm
\begin{tabular}{|c|c|c|c|}
\hline
&&\multicolumn{2}{|c|}{$S_{B_j}(Q^2) + \delta S^{B_j}_T +
\delta S^{B_j}_{\tau 4}$}\\
\cline{3-4}
&&&\\
$\avq$&$\int^1_0 dx~[g^p_1(x,Q^2) - g^n_1(x,Q^2)]$&QCD sum rule&GDH sum rule\\
in (GeV/c)$^2$&&&\\
\hline
2&0.152 $\pm$ 0.014 $\pm$ 0.021&0.162 $\pm$0.023&0.14\\
4.6&0.201 $\pm$ 0.039 $\pm$ 0.050&0.180 $\pm$0.017&0.15\\
\hline
\end{tabular}
}
\end{center}
\vskip 6pt
It is seen that there is absolutely no significant evidence for a failure of
the sum rule. Clearly at the low value of $Q^2$ = 2 the higher twist effects
help to make the agreement more impressive, but even without them there is no
real contradiction, given the size of the experimental errors.
%
%end of Section 4
%
\setcounter{section}{4}
\section{\large{The Operator Product Expansion (OPE)}}
\vskip 6pt
\setcounter{equation}{0}

The fundamental understanding of the $Q^2$-behaviour of the moments in
unpolarized DIS came originally from a study of the operator product
expansion. Later it was discovered that the
same results could be obtained in the QCD Improved Parton Model.

When it comes to the polarized case it has been claimed that the two
approaches yield different results. We shall argue that this is not the case
and claim that there is perfect agreement between the two methods.

We shall also clarify the situation with regard to $g_2(x)$ itself, to its
first moment (the Burkhardt--Cottingham sum rule), and to its relationship
with $g_1(x)$ (the Efremov--Leader--Teryaev and the Wandzura--Wilczek sum
rules), about which there are many misconceptions in the literature.

\subsection{General structure of the OPE}
\vskip 6pt
\setcounter{equation}{0}

It is well known, and can easily be deduced from the Feynman diagram Fig. 2.1,
that the hadronic tensor $W^{\mu\nu}$ involved in the expression for the deep
inelastic cross-section [see (2.1.1)] is given by the Fourier transform of the
nucleon matrix elements of the commutator of electromagnetic currents
$J_\mu (x)$:
\begin{equation}
W_{\mu\nu} (q;P,S) = \frac{1}{2\pi} \int d^4x ~e^{iq\cdot x}
\langle P,S|[J_\mu (x), J_\nu(0)]|P,S\rangle
\end{equation}
where $S^\mu$ is the covariant spin vector specifying the nucleon state of
momentum $P^\mu$. It is convenient to introduce an amplitude $T_{\mu\nu}$
which is closely related to the forward $T$-matrix element for Compton
scattering of a virtual photon of 4-momentum $q$ and helicities $\lambda$
and $\lambda^\prime$. In the convention for the $\hat{T}$-operator
$$
\hat{S} = \hat{I} + i (2\pi)^4 ~\delta^4 (P_f-P_i) ~\hat{T}
$$
one has
\begin{equation}
\langle P,S;q,\lambda^\prime|\hat{T}|P,S;q,\lambda\rangle = 4\pi\alpha
{}~\varepsilon^*_\mu(\lambda^\prime) \, T^{\mu\nu} \, \varepsilon_\nu (\lambda)
\end{equation}
where
\begin{equation}
T_{\mu\nu} (q;P,S)  =  i \int d^4x ~e^{iq\cdot x}
\langle P,S|T(J_\mu (x) J_\nu (0))|P,S\rangle
\end{equation}
is given in terms of the time ordered product of the currents.

Both $W_{\mu\nu}$ and $T_{\mu\nu}$ may be split into parts symmetric ($S$)
or antisymmetric ($A$) under $\mu \leftrightarrow \nu$, and one can show that
\begin{equation}
W_{\mu\nu}^{(S,A)} = \frac{1}{\pi} \, \mbox{\rm Im}\,T_{\mu\nu}^{(S,A)}\,.
\end{equation}

The symmetric part is independent of the spin vector $S^\mu$ and plays a
r\^ole only in unpolarized scattering. We shall therefore concentrate on the
antisymmetric part and we follow the treatment of Kodaira and his group
[KOD 79].

The behaviour of $T_{\mu\nu}$ (and therefore $W_{\mu\nu}$) in the deep
inelastic limit is controlled by the behaviour of the product of currents
near the light cone $x^2 = 0$ and can be derived from Wilson's operator
product expansion.

It is important to note that the expressions (5.1.1 and 3), for which the
operator product approach can be utilised, only arise because of the fully
inclusive nature of the deep inelastic reaction being considered.
Indeed the starting point from which (5.1.1) can be derived is the expression
$$
W_{\mu\nu} \propto \sum_X \langle P,S|J_\mu|X\rangle \langle X|J_\nu|P,S\rangle
$$
which appears in the formula for the cross-section. Only if the sum is over
{\it all} final states $|X\rangle$ does this reduce to (5.1.1).

The antisymmetric part of the Fourier transform of the operator product
appearing in (5.1.3) is expanded in terms of local operators $R$ and
coefficient functions $E$ in the form [KOD 79]:
\begin{eqnarray}
i\int d^4x ~e^{iq\cdot x} \, T(J_\mu (x) J_\nu (0)) &=&
-i \sum_{n=1} \frac{[1-(-1)^n]}{2} \left(\frac{2}{Q^2}\right)^n
q_{\mu_1}...q_{\mu_{n-2}} \sum_i \delta_i \,
\bigg\{ \varepsilon_{\mu\nu\lambda\sigma} \, q^\lambda q_{\mu_{n-1}}
\nonumber\\
&\times& E^n_{1,i}(Q^2,g) \, R_{1,i}^{\sigma\mu_1...\mu_{n-1}}\\
&+&(\varepsilon_{\mu\rho\lambda\sigma} \, q_\nu q^\rho -
\varepsilon_{\nu\rho\lambda\sigma} \, q_\mu q^\rho - q^2
\varepsilon_{\mu\nu\lambda\sigma}) \, \frac{n-1}{n} \nonumber\\
&\times& E_{2,i}^n (Q^2,g) \,
R_{2,i}^{\lambda\sigma\mu_{1}...\mu_{n-2}} \bigg\} \nonumber
\end{eqnarray}
where $g$ is the QCD coupling constant and
where $i$ takes on the values 1,2...,8,$\psi$,$G$, the detailed significance
of which will be explained below. The operators with $i=1,...,8$ transform
like an SU(3) flavour octet. Those with labels $\psi$ or $G$ are flavour
singlets.

The factors $\delta_i$ reflect the charge and isotopic spin structure of the
currents. Taking for the electromagnetic current (in units of $e$)
\begin{equation}
J_\mu = \sum_f e_f \bar{\psi}_f \gamma_\mu \psi_f \qquad f= u,d,s
\end{equation}
(a colour sum is, of course, implied), and bearing in mind that we are
ultimately only interested in the current commutator that occurs in
$W_{\mu\nu}$, one may take for the
$\delta_i$
\begin{equation}
\delta_1 = \frac{1}{3} \qquad \delta_8 = \frac{1}{3\sqrt{3}} \qquad
\delta_\psi = \delta_G = \frac{2}{9}
\end{equation}
and all other $\delta_i=0$.

Note that, as will be seen below, there does not exist an $R_2$-type
operator with $n<2$. Nonetheless because of the explicit factor ($n-1$) in
(5.1.5) we have formally written a sum from $n=1$. Also there is no operator
of the $R_{1,G}$ type for $n=1$. It should formally be regarded as
zero for $n=1$ in the sum over $n$.

The leading twist operators occurring are of two types:

\vskip 4pt \noindent
1) The set $R_{1,i}^{\sigma\mu_1...\mu_{n-1}}$ of twist 2.

For $i=1,...,8$ these are the flavour non-singlet operators
\begin{equation}
R^{\sigma\mu_1...\mu_{n-1}}_{1,i} = (i)^{n-1} \left\{ \bar{\bfpsi} \gamma_5
\gamma^\sigma D^{\mu_1}...D^{\mu_{n-1}} \left(\frac{\bflambda_i}{2}\right)
\bfpsi\right\}_S \qquad (n\ge 1)
\end{equation}
where $S$ implies complete symmetrization in the indices
$\sigma,\mu_1,\mu_2,...,\mu_{n-1}$, the $\bflambda_i$ are the Gell--Mann
$SU(3)$ flavour matrices and $D^\mu$ is the usual QCD covariant derivative.
For $i=\psi$ or $G$ we have the flavour singlet operators
\begin{eqnarray}
R_{1,\psi}^{\sigma\mu_1...\mu_{n-1}} &=& (i)^{n-1} \{\bar{\bfpsi} \gamma_5
\gamma^\sigma D^{\mu_1}...D^{\mu_{n-1}}\bfpsi\}_S \qquad (n\ge1)\nonumber\\
R^{\sigma\mu_1...\mu_{n-1}}_{1,G} &=& (i)^{n-1} {\mbox{\rm Tr}}\{
\varepsilon^{\sigma\alpha\beta\gamma} \bfG_{\beta\gamma} D^{\mu_1}...
D^{\mu_{n-2}} \bfG_\alpha^{\mu_{n-1}}\}_S \qquad (n\ge 2)
\end{eqnarray}
where $\bfG_{\mu\nu}$ is the usual matrix (in colour space) form of the gluon
field tensor [see Eq. (6.3.5)].

\vskip 4pt \noindent
2) The set $R_{2,i}^{\sigma\lambda\mu_1...\mu_{n-2}}$ of twist 3.

For $i=1,...,8$ these are the flavour non-singlet operators
\begin{equation}
R_{2,i}^{\lambda\sigma\mu_1...\mu_{n-2}} = (i)^{n-1} \left\{ \bar{\bfpsi}
\gamma_5\gamma^\lambda D^\sigma D^{\mu_1}...D^{\mu_{n-2}}
\left(\frac{\bflambda_i}{2}\right)\bfpsi \right\}_{S^\prime} \qquad (n\ge 2)
\end{equation}
where $S^\prime$ implies antisymmetrization on $\lambda,\sigma$ and
symmetrization on the indices $\mu_1,..,\mu_{n-2}$.

For $i=\psi$ or $G$ we have flavour singlet operators
\begin{eqnarray}
R_{2,\psi}^{\lambda\sigma\mu_1...\mu_{n-2}} &=& (i)^{n-1} \{\bar{\bfpsi}
\gamma_5 \gamma^\lambda D^\sigma D^{\mu_1}...D^{\mu_{n-2}}\bfpsi\}_{S^\prime}
\qquad (n \ge 2) \nonumber\\
R_{2,G}^{\lambda\sigma\mu_1...\mu_{n-2}} &=& (i)^{n-1} {\mbox{\rm Tr}}\{
\varepsilon^{\sigma\alpha\beta\gamma} \bfG_{\beta\gamma}
D^{\mu_1}...D^{\mu_{n-2}}\bfG_\alpha^\lambda\}_{S^\prime} \quad (n\ge 2)\,.
\end{eqnarray}
The above operators are often referred to as $R^n_{1,i}, R^n_{2,i}$ for
brevity.

It should be noted that the above list of operators is a complete set of
operators in the massless quark theory (if $m_q\neq 0$ there exists a further
set of twist 3 operators). We shall not deal with these. They are discussed
in [KOD 79].

The {\it coefficient functions} $E^n_{1,i} (Q^2,g), E^n_{2,i} (Q^2,g)$ are
the Fourier transforms of the singular functions of $x$ that appear in the
Operator Product Expansion and have to be calculated, using perturbative QCD,
as a power series expansion in the coupling $g$. The factors in (5.1.5) are
arranged so that for free fields, {\it i.e.} $g=0$, one has the simple results:
\begin{eqnarray}
E^n_{1,i} (Q^2,g=0) &=& 1 \qquad\qquad E^n_{2,i} (Q^2,g=0) = 1
\qquad i=1,...,8 \nonumber\\
E^n_{1,\psi} (Q^2,g=0) &=& 1\qquad\qquad E^n_{2,\psi} (Q^2,g=0) = 1\\
E^n_{1,G} (Q^2,g=0)  &=& 0 \qquad\qquad E^n_{2,G} (Q^2,g=0) = 0 \,.\nonumber
\end{eqnarray}

Since gluons are electrically neutral it is no surprise to see that the
operators involving the gluon field tensors $G_{\mu\nu}^a$ play no r\^ole
when $g=0$.

In an interacting theory the $E$'s are functions, $E_{1,i}^n (Q^2/\mu^2;g)$
{\it etc.}, of the renormalization scale $\mu^2$ and they satisfy standard
renormalization group equations. The flavour octet operators are
multiplicatively renormalized, but the $G$ and $\psi$ type singlet operators
mix under renormalization, except for the case $n=1$. This is reflected in
the $Q^2$-evolution of the coefficient functions (for details see [KOD 79]).

The proton matrix elements of these operators have the form:
\begin{eqnarray}
\langle P,S|R_{1,i}^{\sigma\mu_1...\mu_{n-1}}|P,S\rangle &=&
\frac{-2Ma^i_n}{n} \{S^\sigma P^{\mu_1}...P^{\mu_{n-1}}\}_{_S}\\
\langle P,S|R_{2,i}^{\lambda\sigma\mu_1...\mu_{n-2}}|P,S\rangle &=&
Md^i_n (S^\sigma P^\lambda-S^\lambda P^\sigma) P^{\mu_1}...P^{\mu_{n-2}} \,,
\end{eqnarray}
where the factors $a^i_n, d^i_n$ -- essentially reduced matrix elements --
reflect the unknown, non-perturbative aspect of the dynamics.
Depending upon which current is involved, they may or may not depend upon
the renormalization scale $\mu^2$, as is discussed later.

The connection with the simpler operators used in Section 4 is as follows.
For $n=1$ we have
\begin{eqnarray}
(R_{1,i})_\sigma &=& \bar{\bfpsi} \gamma_5 \gamma_\sigma
\left(\frac{\bflambda_i}{2}\right) \bfpsi \qquad i=1,...,8\\
&&= -J^i_{5\sigma}~~~\mbox{\rm of ~eq.~(4.2.3)} \nonumber
\end{eqnarray}
so that via (4.2.7) and (5.1.13)
\begin{equation}
a_i = 2a^i_1 \,.
\end{equation}
Also, note that
\begin{eqnarray}
(R_{1,\psi})_\sigma &=& \bar{\bfpsi}\gamma_5\gamma_\sigma \bfpsi\\
&&= -J^0_{5\sigma}~~~\mbox{\rm of~eq.~(4.2.5)}\nonumber
\end{eqnarray}
so that via (4.2.8) and (5.1.13)
\begin{equation}
a_0 = a^\psi_1\,.
\end{equation}

\subsection{Equations for the moments of \mbox{\boldmath $g_{1,2}(x, Q^2)$}}
\vskip 6pt
\setcounter{equation}{0}

The expressions for the moments are obtained from the OPE in the standard
fashion:
\begin{description}
\item{  i)} {\phantom g}$g_{1,2}$ are related to the absorptive parts of
forward scattering amplitudes for virtual Compton scattering;
\item{ ii)} {\phantom i}dispersion relations in the energy $\nu$ are written
for these scattering amplitudes;
\item{iii)} for small enough $\nu$ the latter can be expanded in a power series
in $\nu$ or, equivalently, in $1/x$. Only even or odd powers appear because
of the crossing symmetry of the scattering amplitudes. The coefficients in the
power series are moments of the absorptive parts and thus of the scaling
functions $g_{1,2}$;
\item{iv)} this power series in $1/x$ is matched to the one which emerges from
the OPE when the scalar products occurring in (5.1.5) are multiplied out.
\end{description}

The result is an expression for the moments of $g_{1,2}$ in terms of the
factors $a^i_n, d^i_n$ and the coefficient functions $E^n_{1,i}$ and
$E^n_{2,i}$. One finds [KOD 79] for the $n$-th moment:
\begin{equation}
\int^1_0 dx~x^{n-1} g_1 (x,Q^2)  =  \frac{1}{2} \sum_i \delta_i a^i_n
E^n_{1,i} (Q^2,g) \qquad n=1,3,5 \dots
\end{equation}
and
\begin{equation}
\int^1_0 dx~x^{n-1} g_2 (x,Q^2) = \left(\frac{1-n}{2n}\right) \sum_i \delta_i
\left[ a^i_n E^n_{1,i} (Q^2,g) -d^i_n E^n_{2,i} (Q^2,g)\right]
\qquad n=3,5,7 \dots
\end{equation}

Note firstly that the Operator Product Expansion only gives information about
the {\it odd} moments of $g_{1,2} (x, Q^2)$.

Note also the important feature that (5.2.2) only gives information about the
moments of $g_2$ for $n \ge 3$. In the original literature (5.2.2) is given
as holding also for $n=1$. That this is incorrect has been stressed by
Jaffe [JAF 90], and we support his contention.
Equation (5.2.2) thus gives no information about the first moment
\begin{equation}
\Gamma_2 (Q^2) \equiv \int^1_0 dx~g_2 (x,Q^2)~.
\end{equation}

Let us now concentrate on the first moment, $n=1$, of $g_1$ in Eq. (5.2.1).
Since, from (5.1.9), there does not exist a gluon operator for $n=1$, the
first moment $\Gamma_1 (Q^2)$ of $g_1$ does not receive a direct contribution
from gluonic operators. In fact in the free theory, with $g=0$, there is no
gluonic contribution at all to $g_1(x)$ or $g_2(x)$ as explained after
Eq. (5.1.12), and the constants $a_n^i, d^i_n$ of (5.1.13, 14) are then
independent of $\mu^2$. Now because $\alpha_s (Q^2) \to 0$ as $Q^2\to \infty$
it often happens that results found in the interacting theory reduce to the
structure of the free theory results with just the replacements
$a^i_n \to a^i_n (Q^2), d^i_n \to d^i_n (Q^2)$ equivalent, in the Parton Model,
to the replacements $q(x)\to q(x,Q^2)$. If this were always the case then
we would expect no gluon contribution to $g_1(x,Q^2)$ as $Q^2\to \infty$.
But as already mentioned in Section 4, and as will be discussed in much
greater detail in Sections 6, 7 and 8 in the framework of the Field Theoretic
Parton Model, {\it there is} a contribution to $g_1(x,Q^2)$ from gluons in the
nucleon even in the limit $Q^2\to \infty$. This apparent disagreement between
the two approaches has upset many people, leading them to question
the validity of the Field Theoretic Parton Model treatment. {\it It must be
stressed that there is no disagreement.} The limit $g\to 0$ and the limit
$Q^2\to \infty$ are not equivalent. The Operator Product Expansion will give
a gluonic contribution to $\Gamma_1(Q^2)$ even though no gluonic
operator exists for $n=1$. The point is that the gluonic contribution emerges
from the quark operators when account is taken of the fact that they are
{\it interacting fields} and not free fields. There is, in fact, complete
equivalence between the approaches as is explained in detail in Section 8.6.

When the detailed results [KOD 79] for the coefficient functions calculated
to order $\alpha_s$ are fed into (5.2.1) one obtains the fundamental formula
(4.4.1) which was the basis for the QCD improved analysis of the polarized
DIS data and which leads to the Bjorken sum rule (4.5.1, 2).

\subsection{Is there a connection between \mbox{\boldmath $g_1(x)$} and
\mbox{\boldmath $g_2(x)$}?}
\vskip 6pt
\setcounter{equation}{0}

There is a much-quoted sum rule, known as the Wandzura--Wilczek relation
[WAN 77] which claims to relate $g_2 (x)$ to an integral over $g_1 (x)$.
The operator product results (5.2.1, 2) allow us to see immediately that
the relation cannot be exact.

{}From (5.2.1) and (5.2.2) we have that
\begin{eqnarray}
\int^1_0 dx~x^{n-1} [g_1(x,Q^2)+g_2(x,Q^2)] &=& \frac{1}{2n} \sum_i
\delta_i a^i_n E^n_{1,i} (Q^2,g) \\
&+& \frac{n-1}{2n} \sum_i\delta_i d^i_n E^n_{2,i} (Q^2,g)
\qquad (n=3,5,7...) \nonumber
\end{eqnarray}
where, recall, the contribution of the last term on the RHS is of twist 3.

Now, if we {\it assume} that (5.3.1) is valid for {\it all} integer $n$
-- and we shall see in a moment that this may be a dangerous assumption --
we can use the convolution theorem for Mellin transforms (which relates the
product of the moments of two functions to the moment of
their convolution) and Eq. (5.2.1) to obtain
\begin{equation}
g_1(x,Q^2) +g_2 (x,Q^2) = \int^1_x \frac{dy}{y}~g_1 (y;Q^2) +
[\mbox{\rm twist}~3] \,.
\end{equation}

In [WAN 77] it is argued, on the basis of a model, that the twist 3 term
can be neglected, leaving a direct connection between $g_1$ and $g_2$.
However, this argument is quite unreliable. The selfsame model gives
nonsensical results for $F_{1,2} (x)$ and $g_1(x)$!

Thus, as will be further explained in Section 5.4, the neglect of the
twist-3 term in (5.3.2) is dangerous and the Wandzura--Wilczek relation
should not be trusted.

But even (5.3.2) may be a dangerous expression. For recall that (5.3.1)
does not hold for $n=1$. In extrapolating it to $n=1$ we are tacitly
assuming [see (5.2.2)] that the first
moment of $g_2, \Gamma_2 = \int^1_0 dx~g_2(x)$, vanishes, whereas, as
stressed earlier, the Operator Product Expansion gives no direct information
on $\Gamma_2$ (this assumed vanishing of $\Gamma_2$ is known as the
Burkhardt--Cottingham sum rule [BUR 70] and will be discussed in the next
Section). Thus even (5.3.2) is, {\it a priori}, not reliable.

The correct way to proceed is to simply use (5.2.2) with (5.2.1) for
$n\ge 3$, {\it i.e.}
\begin{equation}
\int^1_0 dx~x^{n-1} \left\{ g_1(x) + \left( \frac{n}{n-1} \right) g_2(x)
\right\} = \frac{1}{2} \sum_i\delta_i d^i_n E^n_{2,i} (Q^2,g)\qquad (n\ge 3)
\end{equation}
where the RHS is a twist 3 contribution.

This is still a useful equation, for once $g_1(x)$ and $g_2(x)$ are measured
it yields information about the twist 3 operator contribution, which, it
turns out, depends upon {\it correlations} amongst the partons in the nucleon
-- a question of great importance because it is just what determines single
spin asymmetries in QCD (see Section 10).

\subsection{Does the first moment of \mbox{\boldmath $g_2(x)$} vanish?}
\vskip 6pt
\setcounter{equation}{0}

We have already mentioned that according to the Burkhardt--Cottingham sum
rule [BUR 70]
\begin{equation}
\Gamma_2 (Q^2) = \int^1_0 dx~g_2(x,Q^2) = 0~.
\end{equation}

We now wish to ask whether or not we can expect this sum rule to hold.
Firstly, as stressed in the previous section, the result does not follow
from the Operator Product Expansion. In fact the OPE gives no information
about the first moment of $g_2$ since (5.2.2) is not valid for $n=1$.

The result (5.4.1) was derived originally by considering the asymptotic
behaviour of a particular virtual Compton helicity amplitude whose absorptive
part is proportional to $g_2$, and rests on the assumption that the asymptotic
behaviour is controlled by low-lying Regge poles -- in this case Regge poles
like the $a_1$ with $\alpha_{a_1}(0)\simeq - 0.14$. But a very careful analysis
by Heimann [HEI 73] showed that there should be contributions from
multi-pomeron and pomeron-Regge cuts thus invalidating the starting
assumption in the Burkhardt--Cottingham derivation.

It should be noted that this same argument invalidates the Wandzura--Wilczek
result with [twist 3] put equal to zero on the RHS of (5.3.2). For if (5.3.2)
were true without the twist 3 term one could deduce (5.4.1) from it directly
by integrating over $x$.

Because of the well known connection between the behaviour of scaling
functions in the limit $x\to0$ and asymptotic Regge behaviour, the coupling to
multi-pomeron cuts implies a highly singular behaviour
\begin{equation}
g_2 (x)  \stackrel {x\to 0}{\sim} \frac{1}{x^2}
\end{equation}
where we have ignored factors of ln$x$. Thus the integral in (5.4.1) might
not even converge!

That said, we should stress that the couplings of Regge poles and cuts is
a subtle question and such a remarkable conclusion as (5.4.2) must be tested
experimentally. Because Regge cuts always arise from non-planar Feynman
diagrams, and since these correspond to twist greater than 2, it might be
that the coefficient of $1/x^2$ in (5.4.2) tends to zero as $Q^2\to
\infty$. It is then not clear at what scale of fixed $Q^2$ one would expect
to observe the divergence experimentally. Alternatively, if $g_2 (x) \to
C(Q^2)/x^2$ as $x \to 0$ with  $C(Q^2) \ll 1$, it is not guaranteed that
the divergence will be seen experimentally if $x$ is not small enough.

Heimann has given a clear illustration of the mechanism which could be
behind the failure of the Burkhardt--Cottingham sum rule. It involves an
illegitimate interchange of double integrals in a Fourier transform.

One can show that $g_2(x)$ is given by a Fourier transform of the following
type
\begin{equation}
g_2 (x) =\int^\infty_{-\infty} d\lambda ~\frac{e^{-i\lambda x}}{2\pi}
\, [\lambda f (\lambda)]
\end{equation}
where the behaviour of $f(\lambda)$ as $\lambda \to 0$ is such that
$\lambda f (\lambda) \to 0$ as $\lambda \to 0$.

Now because of its relation to the absorptive part of a scattering amplitude
one knows that $g_2(x) = 0$ for $|x|>1$ and $g_2(-x) = g_2(x)$. Then
\begin{eqnarray}
\int^1_0 dx~g_2 (x)\cos\mu x &=& \frac{1}{2} \int^\infty_{-\infty}
dx~g_2 (x) e^{i\mu x}\nonumber\\
&=& \frac{1}{2} \int^\infty_{-\infty} dx \int^\infty_{-\infty}
\frac{d\lambda}{2\pi}\,e^{i(\mu-\lambda)x}\,[\lambda f (\lambda)]\,.
\end{eqnarray}
{\it If} the integrals can be interchanged, we have
\begin{eqnarray}
\int^1_0 dx~g_2 (x) \cos\mu x &=& \frac{1}{2} \int_{-\infty}^\infty
d\lambda ~\delta (\mu-\lambda)\,[\lambda f(\lambda)]\nonumber\\
&=& \frac{1}{2} \mu f (\mu)
\end{eqnarray}
so that taking $\mu = 0$
\begin{equation}
\int^1_0 dx~g_2(x) = 0~.
\end{equation}

However it is NOT always legitimate to interchange the integrals, and this
is obvious if the integral on the LHS of (5.4.5) diverges, say for $\mu=0$.
Indeed the Regge behaviour of $g_2(x)$ as $x\to 0$ is precisely an
example of this type.

If one takes
\begin{equation}
f(\lambda) = \lambda^{\alpha-1} \qquad (0<\alpha<1)
\end{equation}
so that indeed $\lambda f (\lambda) \to 0$ as $\lambda \to 0$, one finds
upon taking the Fourier transform (5.4.3), [HEI 73]
\begin{equation}
g_2 (x) \propto \frac{1}{x^{1+\alpha}}
\end{equation}
and the integral of $g_2(x)$ diverges.

\subsection{The Efremov--Leader--Teryaev sum rule}
\vskip 6pt
\setcounter{equation}{0}

If one assumes that (5.3.3) holds also for $n=2$ one obtains
\begin{equation}
\int^1_0 dx~x[g_1(x) +2g_2(x)] = \frac{1}{2} \sum_i\delta_id_2^i
E^2_{2,i}(Q^2,g)~.
\end{equation}
To lowest order in $\alpha_s$, because of (5.1.12) we need keep only the
operators with $i=1,...,8$ and $i=\psi$. In that case, from (5.1.14) and
(5.1.10 or 11) the $d_2$ are of the form
\begin{equation}
d_2 \propto \langle P,S|\bar{\bfpsi} \gamma_5 [n\slas (S\cdot D) -
S\slas (n\cdot D)] \bfpsi|P,S\rangle
\end{equation}
where $n^\mu$ is a four-vector introduced in the Field Theoretic Parton Model
(Section 10) such that $n^2=0$ and $n\cdot P=1$.

It can be shown, in the framework of the Field Theoretic Parton Model, that
the RHS of (5.5.2) vanishes in the chiral limit. This follows upon integrating
eq. (D.20) of Appendix D and using eqs. (D.17, 13 and 6).
Hence one has the sum rule
%due to Efremov and Teryaev
[EFR 84]
\begin{equation}
\int^1_0 dx~x[g_1(x) + 2g_2 (x)] = 0~.
\end{equation}
This is further discussed in Appendix E.

It should be noted, however, that like the Burkhardt--Cottingham sum rule,
this result does not follow strictly from the OPE. Moreover if $g_2(x)$
diverges like, or close to, $1/x^2$ as $x\to 0$ then, as with the
Burkhardt--Cottingham case, the sum rule will fail because of the
divergence of the integral of $xg_2 (x)$.
%
%end of Section 5
%
\setcounter{section}{5}
\section{\large{The axial anomaly and the axial gluon current
\mbox{\boldmath $K^\mu$}}}
\vskip 6pt

As we have seen in the previous Sections the Naive Parton Model interpretation
of the EMC experiment, based upon  treating the flavour singlet axial current
$J^0_{5\mu}$ as effectively the {\it quark} spin-density operator, leads
to a negligible contribution to the proton spin from the quark spins,
$a_0 =2S^{quarks}_z \simeq 0$, in contradiction with the Quark Model. The
fact that the experiment was done at an average $Q^2 \simeq$ 10 (GeV/c$)^2$
and that the Quark Model is usually trusted for static properties of the
nucleon {\it i.e.} for small momentum transfers, $Q^2 \simeq 0$, does
{\it not} mean that the Naive Parton Model and Quark Model results
cannot be compared with each other, as is sometimes claimed, so that there
is a real contradiction. The statement that in the Quark Model we deal with
constituent quarks and not with partons is somewhat irrelevant, since the
states of three constituent quarks can be considered as a sum of states
involving three valence partons in the same quantum state as the constituent
quarks, with all the other partons (gluons, $q\bar q$ pairs {\it etc.}) having
zero total angular momentum, as emphasized in Lipkin's argument in
Section 4.3.2.

Secondly, there appears to be no significant $Q^2$-variation in the
experimental data down to $Q^2 \simeq$ 1~(GeV/c$)^2$. The sharp change in
$\Gamma^p_1(Q^2)$ demanded by the Drell, Hearn, Gerasimov sum rule occurs for
smaller $Q^2$ (as discussed in Section 4.1.1), so cannot be used, in the
region of the deep inelastic data, as an argument for large higher twist
effects.

We must, therefore, return to the foundations of the Parton Model to try
to seek a resolution to the problem and we shall learn, surprisingly, of
subtle effects which force us to alter some aspects of our naive
interpretation of the Parton Model.

\subsection{Is there really a ``spin crisis"?}
\vskip 6pt
\setcounter{equation}{0}

In field-theoretical language, the usual statement that the nucleon
predominantly consists of three quarks in an $SU(6)$ state (with zero orbital
angular momentum) means that the $3\ quarks \to 3\ quarks$ Green function
with these quantum numbers has a pole at the mass of the nucleon, with
$J = 1/2$, with a significant residue.

The fact that the nucleon is not just a 3 quark state means that the
nucleon pole also occurs, albeit with smaller residue, in more complicated
Green functions such as $3q + g \to 3q + g$, perhaps with non-zero orbital
angular momentum. The operator $J^0_{5\mu}$, containing quark
field operators only, {\it is} the quark spin density operator for
free fields and therefore it does measure the total quark spin in the initial
parton state. But since, as we shall see, $J_{5\mu}^0$ is not a conserved
current the expectation value in the nucleon state where the partons interact
strongly  with each other is not the same as the expectation value in the
initial state made up of free partons. Only in the Naive Parton Model, where
we ignore these interactions, can we equate the expectation values in the
nucleon and free parton states. Since everybody knows that only total $J_z$
and not $S_z$ is generally conserved one may wonder why it is even necessary
to emphasize the non-conservation of $J_{5\mu}^0$. The reason is that naively
,
{\it i.e.} from the equations of motion, $J_{5\mu}^0$ {\it is} conserved for
{\it massless} quarks. So the non-conservation is {\it anomalous} if one
works with massless quarks as is often done in the Parton Model.

For $J^3_{5\mu}$ and $J^8_{5\mu}$ the conservation is exact with massless
quarks. It is also exact if $SU(2)$ and $SU(3)$ are exact symmetries. Since
these symmetries are slightly broken the conservation will be slightly
imperfect with quarks of non-zero mass. The currents are then
{\it partially conserved}. We take it that any dependence on the
renormalization scale for $J^3_{5\mu}$ and $J^8_{5\mu}$ can be neglected.

But there is a further consequence of the non-conservation of $J_{5\mu}^0$
which puts into question our interpretation of the expectation value of
$J_{5\mu}^0$ as the mean value of the total parton spin. To see this consider
the protonic matrix element $\langle P |J_{5\mu}^0| P \rangle$
in the Heisenberg picture. We may insert a sum over free parton states
(assuming, as usual, completeness) so that
\begin{equation}
\langle P |J_{5\mu}^0| P \rangle =
\sint |\langle k_1...k_n|P \rangle|^2
{}~\langle k_1... k_n|J_{5\mu}^0|k_1... k_n \rangle \,.
\end{equation}
Strictly speaking the parton states should either be `IN' or `OUT' type
states, but this is irrelevant for our discussion.

Now in an interacting quantum field theory we are forced to renormalize the
fields and operators, with the result that the value of matrix elements
like $\langle k_1... k_n|\hat{O}|k_1... k_n \rangle$ will, in general, depend
on
the renormalization scale $\mu^2$. Only if the operator $\hat{O}$ is conserved
can one show that the matrix elements are independent of $\mu^2$. This is the
case for $J_{5\mu}^3$ and $J_{5\mu}^8$ with massless quarks. Since anything
of real physical content must be independent of $\mu^2$ we see that the
expectation value of non-conserved operator cannot have any simple physical
significance. Thus it is misleading to think of the expectation value of
$J_{5\mu}^0$ as ``the physical spin of the parton" -- it is {\it not} a
fixed number, it depends on the value of $\mu^2$ and it can, in principle,
have any value whatsoever. To emphasize this one should always indicate the
renormalization scale {\it i.e.} write $\langle k_1...k_n|
J_{5\mu}^0|k_1...k_n\rangle_{\mu^2}$. (Sometimes one
refers to this as the {\it spin at scale  $\mu^2$}, but there is no
implication that its value should be of order 1.)

Bearing this in mind we see that Lipkin's argument (Section 4.3.2) loses its
force in the context of an interacting quantum field theory. The spin crisis
is a crisis only in the domain of the Naive Parton Model, as was indeed
claimed in the paper which introduced the concept of a ``crisis" [LEA 88].

\subsection{On the connection between QCD, the Quark Model and the Naive Parton
Model}
\vskip 6pt

Given that the expectation value of $J_{5\mu}^0$ depends upon the
renormalization scale $\mu^2$ we would like to know if there is a value of
$\mu^2$ at which the expectation value agrees with the Quark Model result
{\it i.e.} corresponds to the physical spin of free quarks. We shall suggest
that this happens as $\mu^2 \to 0$, but because the Quark Model is an
effective theory of the non-perturbative regime of QCD we do not have a
rigorous argument. However we shall adduce some arguments from the
perturbative domain to support our contention.

The Quark Model is characterized by the fact that it does not contain gluons;
the dynamics is in the quark degrees of freedom.
Now neither the quark momentum operator nor the gluon momentum operator is
conserved in QCD. Only their sum is. Thus the momentum fractions of a hadron
carried by quarks or by gluons each depend upon the renormalization scale.
For large $\mu^2$, where $\alpha_s(\mu^2)$ is small so that perturbation
theory can be trusted, we know that the momentum fraction carried by the gluons
{\it increases} to the limit 16/25 at $\mu^2 \to\infty$. Thus the
gluons play a smaller r\^ole in the momentum sum rule at smaller $\mu^2$.
Similarly, we shall see in Section 8 that the spin at scale $\mu$ carried by
the gluons increases without limit as $\mu^2 \to\infty$, so that gluons
play a lesser r\^ole in the angular momentum sum rule at lower $\mu^2$.

Both these examples suggest that one is approaching the Quark Model as
$\mu^2\to 0$. Of course the above are perturbative arguments but it seems
reasonable to assume that the trend continues down to non-perturbatively
small values of $\mu^2$. At the other end of the scale we have the partonic
picture which was invented to explain Bjorken scaling which holds as
$Q^2 \to \infty$, which in the present context corresponds to $\mu^2 \to
\infty$. And indeed, since $\alpha_s(\mu^2) \to 0$ as $\mu^2 \to\infty$
one does usually obtain the relationships of the Naive Parton Model in this
limit, but in fact one never recovers Bjorken scaling and one has to utilize
$Q^2$-dependent parton distributions.

The key exception to this rule is the gluonic contribution to the first moment
of $g_1(x,Q^2)$. Here, as will be seen in Section 6.3 and 7.2, we have a QCD
correction proportional to $\alpha_s(Q^2)$ multiplied by the gluon spin at
scale $Q^2$, which increases like ln~$Q^2$. The logarithmic decrease of
$\alpha_s(Q^2)$ as $Q^2 \to \infty$ is just compensated by the increase in
the gluon spin, leaving a finite, non-zero contribution as $Q^2 \to \infty$.
This is quite anomalous and is, indeed, linked directly to the existence of
the axial anomaly in QCD. It is thus a counterexample to the usual rule that
one recovers the relationships of the Naive Parton Model in the limit
$Q^2 \to\infty$.

This discovery catalyzed by the EMC result is of profound importance and,
ironically, will stand, even if it turns out that the EMC result is modified
by future experiments.

\subsection{The axial anomaly}
\vskip 6pt
\setcounter{equation}{0}

Consider the axial current
\begin{equation}
J_{5\mu}^f = \bar{\psi}_f(x) \, \gamma_\mu \gamma_5 \, \psi_f(x)
\end{equation}
made up of quark operators of definite flavour $f$. (An implicit colour sum
is always implied). From the free Dirac equation of motion one finds that
\begin{equation}
\partial^\mu J_{5\mu}^f = 2im_q \bar{\psi}_f(x)\, \gamma_5 \,\psi_f(x)
\end{equation}
where $m_q$ is the mass of the quark of flavour $f$.

In the chiral limit $m_q \to 0$ (6.3.2) appears to imply that $J_{5\mu}^f$
is conserved. If this were really true there would be a symmetry between
left and right-handed quarks, leading to a parity degeneracy of the hadron
spectrum {\it e.g.} there would exist two protons, of opposite parity.
However, the formal argument from the free equations of motion is not
reliable and, as shown originally by Adler, and by Bell and Jackiw [ABJ 69]
(in the context of QED), there is an anomalous contribution arising from the
triangle diagram shown in Fig. 6.1. As a consequence the axial current is
not conserved when $m_q = 0$. One has instead, for the QCD case
\begin{equation}
\partial^\mu J_{5\mu}^f = {\alpha_s\over 4 \pi}\,G^a_{\mu \nu}
\widetilde{G}^{\mu \nu}_a = {\alpha_s\over 2\pi}\,
\mbox{\rm Tr}\,[\bfG_{\mu \nu}~\widetilde{\bfG}^{\mu \nu}]
\end{equation}
where $\widetilde{G}^a_{\mu \nu}$ is the dual field tensor
\begin{equation}
\widetilde{G}^a_{\mu \nu} \equiv{1\over 2}\,\varepsilon_{\mu
\nu\rho\sigma} G^{\rho\sigma}_a
\end{equation}
and where a field vector or tensor without a colour label stands for a
matrix. In this case
\begin{equation}
\bfG_{\mu\nu} \equiv \Biggl({\bflambda_a\over 2}\Biggl) G^a_{\mu \nu} \,.
\end{equation}
%\newpage
%\begin{figure}[H]
%\vspace*{4cm}
%\hspace*{5cm}\special{picture fig.6.1 scaled 950}
%\end{figure}
%\begin{center}
%Figure 6.1 Triangle diagram giving rise to the axial anomaly
%\end{center}

The result  (6.3.3), which emerges from a calculation of the triangle diagram
(Fig. 6.1) using $m_q$ = 0 and the gluon virtuality $k^2 \neq 0$, is really a
particular limit of a highly non-uniform function. If we take
$m_q \neq~0,~k^2\neq 0$ the RHS of (6.3.3) is multiplied by
\begin{equation}
T(m^2_q/k^2) = 1 -{2 m^2_q/k^2\over \sqrt{1 + 4 m^2_q/k^2}}~
\ln\,\Biggl({\sqrt{1+4 m^2_q/k^2} + 1\over \sqrt{1 + 4 m^2_q/k^2} -
1}\Biggl)\,.
\end{equation}
We see that this anomaly corresponds to $T \to 1$ for $(m^2_q/k^2) \to 0$. On
the other hand, for on-shell gluons, $k^2 = 0$, and $m_q \neq 0$, {\it i.e.}
in the limit $(m^2_q/k^2)\to\infty$ the terms cancel, $T \to 0$, and there
is no anomaly. As explained in Section 8.1 the particular case $k^2 = 0$  is
irrelevant and there is no doubt that the anomaly is relevant.

The anomaly induces a {\it pointlike} interaction between $J_{5\mu}^0$
and gluons. That it is pointlike can be seen by taking different gluon
momenta $k_1$ and $k_2$ in Fig. 6.1  and noting that the amplitude does not
depend on the momentum transfer $k_1-k_2$ when $m_q = 0$. Therefore, in
computing the matrix element of $J_{5\mu}^0$ in a hadron state, we will get
a contribution from the gluon components of the hadron as well as the more
obvious contribution from quarks. From Adler's expression [ABJ 69] for the
triangle diagram, modified to QCD, one finds for the forward gluonic matrix
element of the flavour $f$ current (our convention is $\varepsilon_{0123} = 1$)
\begin{eqnarray}
\langle k,\lambda|J_{5\mu}^f|k,\lambda \rangle &=&
{i\alpha_s\over 2\pi}~\varepsilon_{\mu\nu\rho\sigma} \,k^\nu
\varepsilon^{*\rho}(\lambda)\varepsilon^\sigma(\lambda)\,
T(m^2_q/k^2)\nonumber\\
&=&-{\alpha_s\over 2 \pi}\,S^g_\mu (k,\lambda)\,T(m^2_q/k^2)
\end{eqnarray}
where $\lambda$ is the gluon helicity and we may take
\begin{equation}
S_\mu^g(k,\lambda) \approx \lambda k_\mu
\end{equation}
as the covariant spin vector for almost massless gluons.

Using the methods of Section 3 we may then compute the gluonic contributions
to the hadronic expectation value $\langle P,S|J_{5\mu}^0|P,S \rangle$.
In this case the gluons being bound will be slightly off-shell {\it i.e.}
$k^2 \neq 0$, but small. The full triangle contribution involves a sum over
all quark flavours. We take $m_u, m_d$ and $m_s$ to be $\ll k^2$ whereas
$m_c, m_b$ and $m_t$ are $\gg k^2$. The function $T(m^2_q/k^2)$ thus takes
the values:
\begin{eqnarray}
T&=& 1\qquad \mbox{\rm for}\qquad u,~d,~s\nonumber\\
T&=& 0\qquad \mbox{\rm for}\qquad c,~b,~t
\end{eqnarray}
and the gluon contribution is then given by [see (4.2.8)]
\begin{eqnarray}
a^{gluons}_0(Q^2)&=&-3\,{\alpha_s\over 2\pi}~\int^1_0 dx~\Delta
g(x,Q^2)\nonumber\\
&=&-3\,{\alpha_s\over 2\pi}\,\Delta g(Q^2)
\end{eqnarray}
or from (4.2.1)
\begin{equation}
\Gamma^{gluons}_{1p}(Q^2) = -{1\over 3}\,{\alpha_s\over 2\pi}\,\Delta g(Q^2)\,.
\end{equation}
$\Delta g(x,Q^2)$ is the difference between the number density of gluons with
the same helicity as the nucleon and those with opposite helicity; its integral
$\Delta g(Q^2)$ is the total helicity carried by the gluons.
Note that if $N_f$ massless quark flavours contribute in the anomalous
triangle then on the RHS of (6.3.11)
$$1/3 \to N_f \langle e^2_f \rangle /2$$
where $\langle e^2_f \rangle$ is the mean of the squared charges.

Although (6.3.3) was derived perturbatively to order $\alpha_s$, it is believed
to be an exact result. It was shown long ago by Adler [ABJ 69] that the
anomaly is not influenced by higher order corrections at the 2-loop level.
More recently Anselm and Johansen [ANS 89a] showed that the 3-loop diagram
of Fig. 6.2 effectively  multiplies the anomaly result by a cut-off dependent
constant. But this constant is the same as the one shown by Adler to
renormalize $J_{5\mu}^0$ via Fig. 6.3. Consequently the QED version of
(6.3.3) is unchanged.
%\begin{figure}[H]
%\vspace*{6cm}
%\hspace*{6.5cm}\special{picture fig.6.2 scaled 950}
%\end{figure}
%\begin{center}
%Figure 6.2
%\end{center}
%\begin{figure}[H]
%\vspace*{5cm}
%\hspace*{6cm}\special{picture fig.6.3 scaled 950}
%\end{figure}
%\begin{center}
%Figure 6.3
%\end{center}
These results remain true in QCD for the matrix elements of (6.3.3). Further,
it has been argued by Jackiw [JAC 85] that (6.3.3) is true even outside the
perturbative realm of QCD.

If we consider the anomaly contribution to the hadronic matrix element then
%as shown in Fig. 6.4,
it might appear possible to regard the contribution
(6.3.11) as arising from a QCD correction to the quark distribution function.
However this is quite incorrect since the important region of $p^2$ for the
quark lines in the triangle loop integral, must be $p^2 \to \infty$ in order
to produce the pointlike behavior. The triangle really should be regarded as
a point interaction between the current and the gluons. The result (6.3.10)
is of fundamental importance. It tells us that the Naive Parton Model formula
(4.3.4) for $a_0$ (and hence for $\Gamma^p_1$ in terms of the $\Delta q_f$) is
incorrect. We now have, instead,
\begin{equation}
a_0 = \Delta \Sigma -3 \, {\alpha_s\over 2\pi}~\Delta g
\end{equation}

This result will be discussed in some detail in Section 7. But we note,
immediately, that it has the fundamental implication {\it that the small
measured value of $a_0$ does not necessarily imply that $\Delta \Sigma$ is
small}.

\subsection{The axial gluon current \mbox{\boldmath $K^\mu$} and the gluon
spin}
\vskip 6pt
\setcounter{equation}{0}

It is not difficult to show that if one introduces an {\it axial gluon
current}
\begin{eqnarray}
K^\mu&=&{1\over 2}\,\varepsilon^{\mu\nu\rho\sigma} A^a_\nu \Biggl(
G^a_{\rho\sigma}-{g\over 3}\,f_{abc} A^b_\rho A^c_\sigma \Biggl)\nonumber\\
&=&\varepsilon^{\mu \nu\rho\sigma} \mbox{\rm Tr} \, \Biggl\{ \bfA_\nu
\Bigg(\bfG_{\rho\sigma} + {i\over 3}\, g[\bfA_\rho, \bfA_\sigma]\Biggl)
\Biggl\}
\end{eqnarray}
where, as earlier, the matrix $\bfA_\rho={\bflambda_a\over 2} A^a_\rho$,
then
\begin{equation}
\partial_\mu K^\mu = {1\over 2}\,G^a_{\mu\nu}\widetilde{G}^{\mu \nu}_a =
\mbox{\rm Tr}\,(\bfG_{\mu\nu}~\widetilde{\bfG}^{\mu\nu})\,.
\end{equation}

Consequently, if $m_q = 0$, the modified current
\begin{equation}
\widetilde{J}^f_{5\mu} \equiv J_{5\mu}^f - {\alpha_s\over 2\pi}~K_\mu
\end{equation}
is conserved, $\partial^\mu \widetilde{J}^f_{5\mu} = 0$.

The matrix elements of the modified singlet axial current
\begin{equation}
\widetilde{J}^0_{5\mu} \equiv J^0_{5\mu} - N_f\,{\alpha_s\over 2\pi}\,K_\mu
\end{equation}
are independent of the renormalization scale and should correspond with the
value obtained in the Quark Model (no gluons; approximately $SU(6)$ quark wave
function) {\it i.e.} in the analogue of (4.2.8),
\begin{equation}
\langle P,S|\widetilde{J}_{5\mu}^0|P,S \rangle = 2 M \tilde{a}_0 S^\mu \,,
\end{equation}
we expect $\tilde{a}_0$ independent of $Q^2$ and thus $\tilde{a}_0 \simeq 1$.

We remarked in Section 1 that many of the operators corresponding to standard
dynamical observables are in fact non gauge-invariant in a local gauge theory.
In the gauge $A^a_0(x)=0$ the gluon spin operator $\hat{\bfS}^g$ becomes
\begin{equation}
\hat S^g_i = - \varepsilon_{ijk} A^j_a \partial^0A^k_a \,.
\end{equation}
But in this gauge the  cubic term vanishes for the spatial components of the
vector $K^\mu$ and one finds
\begin{equation}
K_i = - \hat S^g_i\qquad (\mbox{\rm gauge}~A^0_a = 0)\,.
\end{equation}

Consider now the hadronic expectation value of $K^\mu$. By the methods of
Section 3, and in accord with  (6.4.7) we find, in the gauge $A_0 =0$,
\begin{equation}
\langle P,S |K^\mu|P,S \rangle = - 2M\,S^\mu (P)\,\Delta g \,.
\end{equation}

Let us now consider the question of the gauge dependence of this relation. The
current $K^\mu$ is not gauge invariant. In an Abelian theory like QED there is
no cubic term and the gauge transformation induced in $K^\mu$ as a
consequence of
\begin{equation}
A_\mu(x) \to A_\mu(x) + \partial_\mu \Lambda (x)
\end{equation}
can be written in the form
\begin{equation}
K_\mu (x) \to K_\mu (x)-{1\over 2} \left[ \partial^\nu\Lambda (x) \right]
\varepsilon_{\mu \nu \rho \sigma} F^{\rho\sigma}(x)
\end{equation}
where we have used the fact that $F^{\mu\nu}$ is gauge invariant in QED.
Although $K_\mu(x)$ changes, its forward matrix elements, or expectation
values do not, since the expectation value of $F^{\rho\sigma}(x)$ vanishes.
The latter follows because in QED $F^{\rho\sigma}$ is related to the
$A_\mu(x)$ entirely via derivatives, and one may utilize
$[\hat P_\mu, f(x)] = -i \partial f / \partial x^\mu$.

In QCD, under
\begin{equation}
\bfA_\mu \to \bfU \bfA_\mu \bfU^{-1} + {i\over g} (\partial_\mu \bfU)
\bfU^{-1}
\end{equation}
one has
\begin{equation}
\bfG_{\mu\nu} \to \bfU \bfG_{\mu v} \bfU^{-1}
\end{equation}
and one ends up with
\begin{eqnarray}
K_\mu&\to& K_\mu +{2i\over g}\,\varepsilon_{\mu\nu\alpha\beta}\,
\partial^\nu \mbox {\rm Tr}(\bfA^\alpha \bfU^{-1} \partial^\beta \bfU)
\nonumber\\
&+&{2\over 3g^2}~\varepsilon_{\mu \nu\alpha\beta}\,
\mbox{\rm Tr}\{\bfU^{-1}(\partial^\nu \bfU) \bfU^{-1}
(\partial^\alpha \bfU) \bfU^{-1}(\partial^\beta \bfU)\}\,.
\end{eqnarray}

The second term in the RHS of (6.4.13) is a total divergence and gives zero
contribution to the expectation value of $K_\mu$. The third term can also be
shown to be a divergence [CRO 83] but in this case cannot be discarded because
of the non-trivial topological structure of QCD as discussed in Section 8.1.
The last term may indeed be ignored for ``small" gauge transformations
{\it i.e.} those continuously connected to the unit transformation $\bfU=\bfI$,
but it cannot be ignored for ``large" (topologically non-trivial) gauge
transformations. We shall return to study the consequence of this in
Sections 8.1 and 8.4.
%
%end of Section 6
%
\setcounter{section}{6}
\section{\large{Reinterpretation of the measurement of
\mbox{\boldmath $\Gamma^p_1$}}}
\vskip 6pt
\subsection{The r\^ole of the anomalous gluon contribution}
\vskip 6pt
\setcounter{equation}{0}

We showed in Section 4, that the measurement of $g^p_1$ can be interpreted as
effectively the measurement of $a_0$ which is proportional to the proton
expectation value of the flavour singlet axial vector current $J^0_{5\mu}$.
We also argued that the value found by the EMC {\it i.e.} $a_0 \simeq 0$ was
most unintuitive in the framework of the Naive Parton Model, since, therein,
it implies $\Delta \Sigma\simeq 0$.

On account of the anomaly, the connection between $a_0$ and $\Delta\Sigma$ is
quite different from what it is in the Naive Parton Model, and we now have
[see (6.3.12)]
\begin{equation}
a_0(Q^2) = \Delta \Sigma - 3\,{\alpha_s(Q^2)\over 2 \pi}\,\Delta g(Q^2)
\end{equation}
instead of (4.3.4).

The anomaly has generated an effective point interaction between the virtual
photon and the gluons and the small value of $a_0$ can now arise as a result
of a cancellation between $\Delta\Sigma$ and the $Q^2$-dependent gluon spin
contribution $(3\alpha_s(Q^2)/2\pi)\,\Delta g(Q^2)$. Why these terms
should roughly cancel will be discusses presently.

Quantitatively if we take  $a_0 \simeq 0.17$ at $Q^2=10$ (GeV/c$)^2$ where
$\alpha_s \simeq 0.24$ we can let the quarks carry, say 60\% of the proton
spin {\it i.e.} choose $\Delta\Sigma$ = 0.6 and (7.1.1) then implies
\begin{equation}
\Delta g \,[Q^2 = 10~\mbox{\rm (GeV/c)}^2] \simeq 3.8 \,.
\end{equation}
This is, at first sight, quite unphysically large
\footnote{Note added in proof: A value of $a_0 \simeq 0.27$ (see footnote
in Section 4.4.1, [ADA 94]) would imply a smaller value
$\Delta g \,[Q^2 = 10~\mbox{\rm (GeV/c)}^2] \simeq 3.0.$}.
\setcounter{footnote}{0}
But it should not be
forgotten that according to the discussion in Section 6.1, this is the spin
carried by the gluons {\it at scale} 10 (GeV/c)$^2$ and its value
could be anything. To get some idea as to whether this value is absurd or not,
we can use the evolution equation to be derived in Section 7.2 to estimate
$\Delta g (Q^2)$ at some lower scale, closer to the Quark Model regime. Going
as far as we dare, we take $Q^2=4\Lambda^2_{\rm QCD}$ where
$\alpha_s \simeq 1$ and find that
\begin{equation}
\Delta g (4 \Lambda^2_{\rm QCD}) \simeq 0.7 \,,
\end{equation}
a not unreasonable value.

In (7.1.1) we have shown no $Q^2$-dependence in $\Delta\Sigma$. This follows
upon using (6.4.3, 5 and 8) to get
\begin{eqnarray}
\widetilde{a}_0&=&(\Delta\Sigma -3\,{\alpha_s\over 2\pi}\,\Delta g)+
3\,{\alpha_s\over 2\pi}\,\Delta g\nonumber\\
&=&\Delta \Sigma
\end{eqnarray}
and the fact that $\widetilde{a}_0$, being related to the conserved current
$\widetilde{J}^0_{5\mu}$ is independent of the renormalization scale $\mu^2$
(or
$Q^2$), as we shall explain below.

Consider the axial charge $\widetilde{Q}_5$ associated with the conserved
current $\widetilde{J}^0_{5\mu}$. We have
\begin{equation}
\widetilde{Q}_5 = \int d^3 {\bfx} \, \widetilde{J}^0_{50}({\bfx},t)\,.
\end{equation}
Let us ask how it would be possible for $\widetilde{Q}_5$ to depend upon the
renormalization scale $\mu^2$, in a massless theory. The only way would be
via the variable $\mu t$ and such a dependence would then induce a dependence
upon $t$. But we know that for a conserved local current the charge is
time-independent. Thus $\widetilde{Q}_5$ must be independent of the
renormalization scale. It follows, as claimed after Eq. (7.1.4), that
$\widetilde{a}_0$ or $\Delta \Sigma$ are independent of $Q^2$.

\subsection{Why the gluon contribution survives as
\mbox{\boldmath $Q^2 \to \infty$}}
\vskip 6pt
\setcounter{equation}{0}
At first sight it seems surprising that the gluonic term in (7.1.1) survives
at large $Q^2$. It  looks like an $\alpha_s$ correction which usually is
expected to disappear as $Q^2 \to \infty$. However, this is misleading since
the gluon spin at scale $Q^2$ behaves just like $[\alpha_s(Q^2)]^{-1}$ as
$Q^2\to\infty$. This can be seen either from the spin-dependent
Altarelli-Parisi evolution equations [ALT 77] or from the anomalous
dimensions of the operators involved. We shall outline the latter approach.

Now the axial current $J^0_{5\mu}$ is multiplicatively renormalized which
implies that the dependence on renormalization scale appears in a
multiplicative factor. Consequently one can write
\begin{equation}
{d\over d\ln Q^2}\Biggl[\Delta\Sigma-{3\alpha_s(Q^2)\over 2\pi}\,\Delta g(Q^2)
\Biggl]=-\gamma(\alpha_s)\Biggl[\Delta\Sigma-{3\alpha_s(Q^2)\over 2\pi}\,
\Delta g(Q^2)\Biggl]
\end{equation}
where $\gamma(\alpha_s)$ is the anomalous dimension of $J^0_{\mu 5}$.

This is the standard form of an anomalous dimension evolution equation. What
is a little unusual is that the power series expansion for $\gamma(\alpha_s)$
begins only in second order [KOD 80] {\it i.e.}
\begin{equation}
\gamma(\alpha_s) = \gamma_2\Biggl({\alpha_s\over 4\pi}\Biggl)^2 + \ldots
\end{equation}
with
\begin{equation}
\gamma_2 = 16 N_f \,.
\end{equation}

The solution to (7.2.1) is then
\begin{equation}
\Biggl[\Delta \Sigma-{3\alpha_s\over 2\pi}\Delta g\Biggl]_{Q^2}=
\Biggl[\Delta\Sigma-{3\alpha_s\over 2\pi}~\Delta g\Biggl]_{Q^2_0}
\times \exp \, \left\{ {\gamma_2\over 4\pi\beta_0}\,[\alpha_s(Q^2) -
\alpha_s(Q^2_0)] \right\}
\end{equation}
where, as usual,
\begin{equation}
\beta_0 = 11 - {2\over 3}\,N_f\,.
\end{equation}

Equation (7.2.4) implies that the quantity
\begin{equation}
\Biggl[\Delta\Sigma -3{\alpha_s(Q^2)\over 2\pi}\Delta g(Q^2)\Biggl]
\times\exp \, \left\{-{\gamma_2\over 4 \pi\beta_0}\,\alpha_s(Q^2)\right\}
\end{equation}
is independent of $Q^2$. Let us put it equal to a constant $C$, where $C$ is
the value of (7.2.6) at some arbitrary chosen value of $Q^2$.

We now take the limit $Q^2 \to \infty$. Since $\Delta \Sigma$ is independent
of $Q^2$ and $\alpha_s(Q^2)\to 0$ we obtain
\begin{equation}
\lim_{Q^2\to\infty}~{3\alpha_s(Q^2)\over 2\pi}~\Delta g(Q^2)=
\Delta\Sigma-C\,.
\end{equation}
Thus it must be that $\Delta g (Q^2)$ increases like
$[\alpha_s(Q^2)]^{-1}$ as $Q^2$ increases.

\subsection{The angular momentum sum rule revisited}
\vskip 6pt
\setcounter{equation}{0}

We have suggested that the results on the first moment in polarized deep
inelastic lepton-hadron scattering can be explained as a cancellation between a
reasonably large quark spin contribution $\Delta\Sigma \simeq$ (0.6 -- 0.7), as
expected intuitively, and the anomalous gluon contribution $\Delta g$. The
above value of $\Delta\Sigma$ is consistent with various relativistic bag
model calculations and, as will be explained in Section 9.3, with the
so-called ``Generalized Goldberger-Treiman relation".

However, in order to accomplish this cancellation, one requires a large gluon
spin $\Delta g$ at $\langle Q^2 \rangle \simeq 10$ (GeV/c)$^2$ {\it i.e.}
$\Delta g \simeq 4$. As we have explained $\Delta g$ grows indefinitely as
$Q^2$ increases according to the usual QCD evolution equations so that such
a large value of $\Delta g$ cannot be ruled out by our expectation that the
gluon contribution is small at very small $Q^2$.

It should be noted that the large $\Delta g$ has important consequences for
the total angular momentum sum rule. For a nucleon with helicity +1/2 moving
along the $z$-axis we must have
\begin{equation}
J_z = S^q_z + S^g_z + L_z = {1\over 2}
\end{equation}
where $L_z$ represents the total orbital angular momentum of all partons.
With $S^q_z \simeq$ 0.7 and independent of $Q^2$, the growing value of
$S^g_z$ has to be compensated by an analogous growth in the magnitude of
$L_z$. A detailed analysis of how this happens has been given by Ratcliffe
[RAT 87].

In the process of evolution of a quark it radiates a gluon in some
preferred helicity state and, in so doing, conserves its own helicity. It can
then radiate again, with the gluon again produced in the same preferred
helicity
state. Thus gluons of some definite helicity are preferentially being radiated
and $\Delta g$ thus increases. But since the total angular momentum is
conserved at each step of the evolution, something has to compensate for the
component of spin carried by the radiated gluon. This, it turns out, is
provided by the {\it orbital} angular momentum of the $qg$-pair in
the process $q\to qg$. Hence as $\Delta g$ increases with each step in the
evolution, so too does $L_z$ with opposite sign.

The angular momentum sum rule (7.3.1) is thus completely in accord with the
QCD growth of both $S^g_z$ and $L_z$ as $Q^2$ increases.
%
%end of Section 7
%
\setcounter{section}{7}
\section{\large{The anomaly in the QCD Field Theoretic Model}}
\vskip 6pt

We would like to try to understand the previous results concerning the r\^ole
of the anomaly in $\Gamma^p_1$ in more intuitive terms, in particular in the
framework of the Feynman diagram approach which provides a more rigorous,
field-theoretic approach than the Parton Model [EFR 88, ALT 88, LEA 88a, CAR
88]. (We refer to this picture as the QCD Field Theoretic Model). We also wish
to study the gluon contribution to $g_1(x)$ itself.

\subsection{The factorization theorem}
\vskip 6pt
\setcounter{equation}{0}

According to the factorization theorem, which is the basis for deriving Parton
Model type results from field-theory, and which has been proved to all orders
in perturbative QCD, the cross-section asymmetry $\Delta\sigma_{\gamma^*p}$
\begin{equation}
\Delta\sigma_{\gamma^*p} \equiv \sigma_{1,1/2} - \sigma_{-1,1/2}
\end{equation}
for the absorption of a virtual photon on a proton of momentum $P$
(the subscripts label helicities), is given by
\begin{eqnarray}
\Delta\sigma_{\gamma^*p}&=&\int^1_0 dx^\prime \, \Biggl\{\sum_f \,
[\Delta q_f(x^\prime)+\Delta\bar{q}_f(x^\prime)]\,\Delta\sigma_{\gamma^* q}
(x^\prime P)\nonumber\\
&+&\Delta g(x^\prime)\,\Delta\sigma_{\gamma^* g}(x^\prime P)\Biggl\}
\end{eqnarray}
where the sum is over flavours and the partonic cross-sections refer to
photo-absorption on quarks or gluons of  momentum $ p= x^\prime P$. In the
following we consider the $\gamma^*$-proton collision either in the Lab frame
with $\gamma^*$ moving along $OZ$ or in an `infinite momentum' frame where the
fast proton and $\gamma^*$ move parallel to $OZ$.

The partonic sub-process $\gamma^* q$ is dominated by the Born diagram in
Fig. 8.1 which physically corresponds to a forward jet with essentially
zero $k_T$ (momentum perpendicular to the photon direction) in the Lab
frame or in an `infinite momentum frame' boosted from the Lab frame along
the photon direction.
%\begin{figure}[H]
%\vspace*{5cm}
%\hspace*{6cm}\special{picture fig.8.1 scaled 950}
%\end{figure}
%\begin{center}
%Figure 8.1
%\end{center}
(Of course there is another jet, in the backward direction in the
frame where the $\gamma^*$ energy is zero, arising from the fragments of the
proton). Integrated over $k_T$ the Born contribution leads to the usual
expression for $g^p_1$ that one obtains in the Naive Parton Model.

The partonic sub-process $\gamma^*g$ is dominated by $q\bar{q}$ pair
production as shown in Fig.~8.2.
%\begin{figure}[H]
%\vspace*{5cm}
%\hspace*{6cm}\special{picture fig.8.2 scaled 950}
%\end{figure}
%\begin{center}
%Figure 8.2
%\end{center}
It is characterized by the formation of two jets with essentially
opposite ${\bfk}_T$ and is distinguishable from the process of Fig. 8.1
provided $k_T$ is  large enough. If one calculates the differential
cross-section asymmetry $d\Delta\sigma_{\gamma^*g}/dk^2_T$, where ${\bfk}_T$
and $-{\bfk}_T$ are the jet momenta perpendicular to $OZ$, one finds an
interesting dependence on the quark mass, as shown in Fig. 8.3, where the
$k_T$-dependence is shown for three choices of quark mass:
$m^2_q \simeq Q^2,~m^2_q \neq 0$ but $\ll Q^2$ and $m^2_q = 0$.
%\begin{figure}[H]
%\vspace*{5cm}
%\hspace*{6cm}\special{picture fig.8.3 scaled 950}
%\end{figure}
%\begin{center}
%Figure 8.3
%\end{center}
For $k_T\gg m_q$ it is negative and for $k_T \ll m_q$ it is positive.
For $m_q=0$ it is thus always negative. Integrated over $k_T$ one gets for
the contribution to the scaling function of the partonic process
$\ell g \to \ell X$, $g_1^{\gamma^* g}(x)$, from Fig. 8.2, for
{\it each quark flavour f} [BAS 93]
\begin{eqnarray}
g_1^{\gamma^*g}(x, Q^2)&=& {e_f^2\over 2} {\alpha_s\over 2\pi}\Bigg\{(2 x-1)
\Biggl[\ln\Biggl( {Q^2(1-x)/x\over x(1 - x) p^2 + m^2_q}\Biggl)-1\Biggl]
\nonumber\\
&+&(1 - x){2m^2_q - x(2x - 1) p^2\over x(1 - x)p^2 + m^2_q}\Biggl\} \,.
\end{eqnarray}
Here $x$ stands for the Bjorken variable of the partonic process,
$x=Q^2/2p\cdot q$. For $m^2_q = 0$ this simplifies to
\begin{equation}
g_1^{\gamma^*g}(x, Q^2)|_{m^2_q = 0} = {e_f^2 \over 2}
{\alpha_s\over 2\pi} (2x-1)[\ln(Q^2/p^2) - 2 (\ln x + 1)] \,.
\end{equation}

We now wish to feed this into the analogue of (8.1.2) in order to obtain the
gluonic contribution to $g^p_1(x)$. In doing so we must remember that in the
QCD Field Theoretic Model the convolution formula (8.1.2) emerges as the
lowest twist approximation to what was originally a Feynman diagram with an
integration $\int d^4 p$. In this integration the point $p^2 =0$ is of zero
measure. We saw in Section 6.3 that the anomaly is present if $p^2 \neq 0$,
but absent if $p^2 = 0$. From the present point of view the case $p^2=0$ is
irrelevant and it is {\it quite clear that the anomaly does contribute to DIS}.

It should also be remembered that in the QCD Improved Parton Model treating
the above as a {\it gluonic} subprocess assumes that for gluons in
the nucleon $p^2$ is small {\it i.e.} $p^2 \le M^2$.

The analogue of (8.1.2) for the contribution of Fig. 8.2 to the scaling
function $g^p_1(x)$ is, for 3 massless flavours,
\begin{equation}
g^p_1(x) = {1\over 3} \int^1_x {dx^\prime\over x^\prime}~\Delta g
(x^\prime)~g^{\gamma^*g}_1(x/x^\prime)\,.
\end{equation}
Taking the first moment of the convolution (8.1.5) we obtain for the
contribution of Fig. 8.2 to $\Gamma_1^p$ (for 3 flavours)
\begin{equation}
\Gamma_1^p = {1\over 3}\Biggl[\int^1_0 dx \, \Delta g(x) \Biggl]
\Biggl[\int^1_0 dx \, g_1^{\gamma^*g}(x) \Biggl]
\end{equation}
and using (8.1.4) one finds exactly the contribution (6.3.11) due to the
anomaly.

\subsection{Study of the gluonic contribution to \mbox{\boldmath $g_1(x)$}
and the definition of \mbox{\boldmath $\Delta g (x)$}}
\vskip 6pt
\setcounter{equation}{0}

However, some caution is necessary about the interpretation of these results
because they are based on perturbation theory, so may not be reliable as
regards
contributions coming from small $k_T$. In the small $k_T$ region, with $p^2$
limited, the momentum $p^\prime$ carried by the quark propagator in Fig. 8.2
is such that $(p^\prime)^2 \lsim {\cal O}(\Lambda^2_{\rm QCD})$ so cannot be
treated perturbatively. Moreover for $(p^\prime)^2$ of this scale, the
contribution should be considered as due to the interaction of the virtual
photon with the quark-parton of momenta $p^\prime$. This feature is a common
one in the QCD Improved Parton Model. It has nothing specific to do with the
polarized case. And the same diagram, Fig. 8.2, of course contributes in the
unpolarized case, where the contribution of small $k_T$ is absorbed into
a re-definition of the quark distribution function $q(x)$, rendering it
$Q^2$-dependent {\it i.e.} one then deals with functions
$q(x, Q^2)$ which evolve as $Q^2$ varies.

In the unpolarized case the situation is uncontroversial. One finds, for the
Born + first order terms, a result whose structure symbolically
is of the form
\begin{eqnarray}
\{1&+&\alpha_s\ln (Q^2/p^2)+\alpha_s f_q(x)\} \otimes q(x)\nonumber\\
&+&\{\alpha_s \ln(Q^2/p^2) + \alpha_s f_g(x)\}\otimes g(x)
\end{eqnarray}
where $q(x)$ and $g(x)$ are unpolarized quark and gluon distribution
functions. In LLA we neglect the terms $\alpha_s f_q(x)$ and
$\alpha_s f_g(x)$, and, correct to order $\alpha_s$, factorize the remaining
terms by introducing a {\it factorization scale} $\mu^2$:
\begin{eqnarray}
\{1 + \alpha_s \ln (Q^2/p^2)\}\,q(x)&=&\{1+\alpha_s\ln(Q^2/\mu^2)+
\alpha_s\ln (\mu^2/p^2)\}\,q(x)\nonumber\\
&\approx&\{1+\alpha_s\ln (Q^2/\mu^2)\}\{1+\alpha_s\ln (\mu^2/p^2)\}\,q(x)\,.
\nonumber
\end{eqnarray}
We also write
$$
\alpha_s\ln (Q^2/p^2)\,g(x)=\{\alpha_s\ln (Q^2/\mu^2) +
\alpha_s \ln (\mu^2/p^2)\}\,g(x)
$$
and then (8.2.1) becomes, to order $\alpha_s$,
$$
\{1+\alpha_s \ln (Q^2/\mu^2)\}\,q(x,\mu^2) + \alpha_s \ln (Q^2/\mu^2)\,g(x)
$$
where
$$
q(x,\mu^2)\equiv [1+\alpha_s\ln (\mu^2/p^2)]\,q(x) +
\alpha_s \ln (\mu^2/p^2)\,g(x) \,.
$$

In this way the contributions sensitive to low $k_T$ and singular in the limit
$p^2 \to 0$ gets absorbed into new, non-perturbative distribution functions
which we cannot calculate, but have to measure.

The leading logarithm $\ln Q^2$ comes specifically from the ultraviolet region
of the $k_T$ integration which kinematically runs up to
\begin{equation}
k^2_T \le {Q^2\over 4}~{1-x\over x}\, \cdot
\end{equation}
It is therefore insensitive to different ways of factorizing or separating the
low and high $k_T$ regions and there is no ambiguity about the definition of
$q(x, \mu^2)$ {\it in LLA}.

The situation is different when working beyond the LLA where one does not
neglect terms of the form $\alpha_sf(x)$ in (8.2.1). They do not uniquely
arise from the large $k_T$ region and it is essentially a matter of taste or
convention whether they or part of them are included in $q(x,\mu^2)$ or left
as a piece of the gluon contribution. For this reason it is important, beyond
the LLA, to indicate what convention is in use. Hence one uses
$q(x,\mu^2)_{\overline{\rm MS}},~q(x,\mu^2)_{\rm DIS}$ {\it etc.}.

In what way is the polarized case different? The key difference is the factor
$(2x-1)$ in expression (8.1.4). For it means that the usual large logarithm
$\ln (Q^2/p^2)$ does not appear in the first moment, and the anomaly
contribution comes from the second term $2(\ln x+ 1)$ in (8.1.4). But this
seems to be like the terms $\alpha_s f(x)$ in (8.2.1) and therefore
inherently ambiguous. There are many papers on this question. For access to
these see {\it e.g.} [BAS 91]. We believe that there is a reasonably persuasive
argument in favour of a particular resolution of the ambiguity. Our approach
is similar in spirit to that of Ross and Roberts [ROS 90] and Carlitz,
Collins, Mueller [CAR 88] (se also [MAN 90a]).

To study the r\^ole of $k_T$ let us introduce a lower cut-off $\Lambda$ in
the $k_T$ integration {\it i.e.} in evaluating the diagram of Fig. 8.2 we
keep $k_T \ge \Lambda$. The result which replaces (8.1.3) is, for
$Q^2 \gg \Lambda^2$,
\begin{eqnarray}
g_1^{\gamma^* g}(x, Q^2)_{k_T\ge \Lambda}&=& {e_f^2 \over 2}{\alpha\over 2\pi}
\Biggl\{(2x-1)\Biggl[\ln\Biggl({Q^2 (1-x)/x)\over x(1-x)p^2+m^2_q+
\Lambda^2}\Biggl) - 1\Biggl]\nonumber\\
&+&(1-x){2m^2_q-x (2x-1)p^2\over x(1-x)p^2+m^2_q+\Lambda^2}\Biggl\} \,\cdot
\end{eqnarray}
The first term in square brackets can be written
$$
\ln\Biggl({Q^2\over p^2}~{1 - x\over x}\Biggl) - \ln \Biggl[x(1 - x) +
{m^2_q\over p^2} + {\Lambda^2\over p^2}\Biggl]
$$
and we see that the entire term $\ln\{(Q^2/p^2)~[(1 - x)/x]\}$ and not
just $\ln(Q^2/p^2)$ is insensitive to the value of $\Lambda$ and therefore
can be associated with the region of large $k_T$. This is no surprise since
the structure $Q^2(1 - x)/x$ just reflects the upper limit of the $k_T$
integration (8.2.2). It thus seems clear that the $\ln[(1-x)/x]$ emerging
from this term is a high $k_T$ effect and surely belongs to the gluon
partonic sub-process, in agreement with the anomaly analysis.

Returning to the case $m_q=0,~\Lambda =0$, we thus rewrite (8.1.4), for each
flavour $f$, as
\begin{equation}
g^{\gamma^*g}_1(x,Q^2) = {e_f^2 \over 2} {\alpha_s\over 2\pi}~(2x-1)
\Biggl\{\ln\Biggl({Q^2\over p^2}~{1-x\over x}\Biggl)-\ln x(1-x)-2\Biggl\}~,
\end{equation}
introduce the factorization scale $\mu^2$:
\begin{eqnarray}
g_1^{\gamma^*g}(x,Q^2) &=& {e_f^2 \over 2} {\alpha_s\over 2 \pi} (2x-1)
\Biggl\{\ln\Biggl({Q^2\over\mu^2}~{1-x\over x}\Biggl)+\ln (\mu^2/p^2)
\nonumber\\
&-& \ln x(1-x) - 2\Biggl\}~,
\end{eqnarray}
and associate the first term with the gluonic contribution. The rest of the
terms are then absorbed into the quark distribution
$\Delta q_f+\Delta\bar{q}_f$.

Finally then, choosing $\mu^2 = Q^2$, the gluonic contribution from Fig. 8.2
to $g^p_1(x)$ is, via (8.1.5)
\begin{equation}
g^p_1(x)|_{\rm gluonic}={1\over 3}\,{\alpha_s\over 2 \pi} \int^1_x
{dx^\prime\over x^\prime}\Biggl\{ \Biggl( 2{x\over x^\prime} - 1 \Biggr)
\ln \Biggl({1 - x/x^\prime\over
x/x^\prime}\Biggl)\Biggl\}~\Delta g(x^\prime)
\end{equation}
and the modified quark contribution is
\begin{equation}
g^p_1(x, Q^2)|_{\rm quark~singlet} = {1\over 9}\,\Delta \Sigma (x, Q^2)
\end{equation}
where the contribution of Fig. 8.2 yields
\begin{eqnarray}
\Delta \Sigma (x, Q^2)& =& \Delta \Sigma (x)+{\alpha_s\over 2\pi}
\int^1_x {dx^\prime\over x^\prime}~(2{x\over x^\prime}-1) \bigl[\ln(Q^2/p^2)
\nonumber\\
&-&\ln {x\over x^\prime}\left( 1 - {x\over x^\prime} \right) - 2] \,
\Delta g (x^\prime) \,.
\end{eqnarray}
Of course, the first moment
\begin{equation}
\Delta \Sigma = \int^1_0 dx\,\Sigma(x, Q^2)
\end{equation}
remains independent of $Q^2$ in accordance with the discussion in
Section 7.2. Clearly, the full definition of $\Delta\Sigma (x, Q^2)$ to
LLA must also include the contribution from Fig. 8.4.

%\begin{figure}
%\vspace*{3cm}
%\end{figure}
%\begin{center}
%Figure 8.4
%\end{center}

\subsection{Measuring \mbox{\boldmath{$\Delta g(x)$}} via 2-jet production}
\vskip 6pt
\setcounter{equation}{0}

It is important to note that the definition of $\Delta g(x)$ adopted above
associates it with the physical process of the electroproduction of two jets
with large transverse momentum ${\bfk}_T$ and $-{\bfk}_T$ respectively.
Measurement of this semi-inclusive process thus gives direct access to
$\Delta g(x)$. It is determined by a physical measurement and thus,
manifestly, must be gauge invariant. (The formal question of gauge
invariance is discussed in Section 9.3).

We have for the contribution to $g_1(x, Q^2)$ coming from the two-jet events
\begin{equation}
\gamma^* N \to \mbox{\rm jet}({\bfk}_T) + \mbox{\rm jet}(-{\bfk}_T)+X
\end{equation}
\begin{equation}
g^{\rm 2-jet}_1 (x,Q^2)|_{k_T\ge \Lambda} = {1\over3}
\int^1_x {dx^\prime\over x^\prime}
\,g^{\gamma^*g}_1(x/x^\prime, Q^2)|_{k_T \ge \Lambda}~\Delta g(x^\prime)
\end{equation}
where $g_1^{\gamma^*g}$ was given in (8.2.3) and the lower cut off on $k_T$
is given by
\begin{equation}
\Lambda^2 \gg m^2_q,~p^2
\end{equation}
where $p^2$ is the `virtuality' of the gluon.

\subsection{The shape of \mbox{\boldmath $g^p_1(x)$} and the anomalous gluon
contribution}
\vskip 6pt
\setcounter{equation}{0}

We have seen that the anomalous gluonic contribution to the first moment
$\Gamma^p_1$ permits an explanation of the EMC experiment in which the total
spin carried by the quarks can be relatively large provided that
$\Delta g=\int^1_0 dx~\Delta g (x)$ is big enough. For example
$S_z^{\rm quarks} \simeq 0.6 \, J_z^{\rm nucleon}$ requires $\Delta g
\simeq 4$ at $Q^2 \simeq 11$ (GeV/c)$^2$.

We consider now the $x$-dependence of $g_1^p(x)$ given by (we ignore
perturbative QCD corrections)
\begin{eqnarray}
g^p_1(x)&=&{1\over 2}\Biggl\{{4\over 9}[\Delta u(x) + \Delta\bar{u}(x)] +
{1\over 9}[\Delta d (x)+\Delta\bar{d}(x)]\nonumber\\
&+&{1\over 9}[\Delta s(x)+\Delta\bar{s}(x)]\Biggl\} + g_1^p(x)|_{\rm gluonic}
\end{eqnarray}
where $g_1^p(x)_{\rm gluonic}$ is given in (8.2.6).

We saw in Section 4 that {\it without} a gluonic contribution the EMC
experiment forced us to have $\Delta\Sigma\simeq 0$ and this was achieved
by having a surprising large negative $\Delta s$ contribution.

It is interesting to contemplate the other extreme {\it i.e.} to take
$\Delta s(x) \simeq 0$. Assuming a roughly $SU(3)$ invariant description of
the sea quarks (this is not a good approximation for the unpolarized
distributions) $g_1^p(x)$ will in this case depend only upon the valence
quark distributions $\Delta u_{\rm v}(x)$ and $\Delta d_{\rm v}(x)$ and on
$\Delta g(x)$. The normalization
of the valence quark distribution is fixed by the values of F and D (see
Sections 4.2 and 4.3) and their shape is constrained by the requirement
\begin{equation}
|\Delta u_v (x)| \le u_v (x) \qquad |\Delta d_v (x)| \le d_v(x) \,.
\end{equation}

Furthermore at $x\simeq 0.3$, where $x g_1^p(x)$ has its maximum value (see
Fig. 4.2) there is essentially no contribution from $\Delta g(x)$, so that
there is a further constraint on $\Delta u_v(x)$, $\Delta d_v(x)$ in this
region. It then turns out that one requires a significant negative
contribution near $x = 0.1$ to fit the data in Fig. 4.2. The question then
is: can a large gluonic term provide such a contribution? It is found
empirically [ROS 90] that this is not possible as is demonstrated in Fig. 8.5.
One can trace this failure to the feature that the factor in parenthesis in
(8.2.6) has a double zero at $x^\prime = 2 x$. Given that $\Delta g(x)$ is
bounded by
\begin{equation}
|\Delta g(x)|\le g(x)
\end{equation}
and that $g(x)$ dies out rapidly as $x$ increases, the region of overlap in
the convolution in (8.2.6) where both the kernel and $\Delta g (x)$ are big is
a very small one and in effect the integration only samples $\Delta g
(x^\prime)$ for a small portion of the region $x\le x^\prime \le 1$.

Of course there is no reason to suppose a total absence of a strange quark
contribution. The problem with the EMC analysis without the anomalous gluon
contribution was that it demanded a surprisingly large negative $\Delta s(x)$.
Figure 8.6 taken from [ROS 90] shows that a reasonable compromise is possible.
The fit shown corresponds, for F/D = 0.55 (which is approximately the value
used in Section 4), to having
\begin{eqnarray}
\Delta u&=&0.84\qquad \Delta d=-0.41\qquad \Delta s=-0.05 \nonumber\\
\Delta \Sigma &=& 0.38\qquad \Delta g \simeq 3 \,.
\end{eqnarray}

The exact values should not be taken too seriously since they depend upon the
simple, but reasonable parametrization used for the polarized quark
distributions in [ROS 90], and because the small $x$ behavior used does
conform to the expectations from Regge theory. However the fit does indicate
that the  EMC data can be explained with a reasonable strange quark
contribution, a sensible $\Delta\Sigma $ and a moderate $\Delta g$
\footnote{Note added in proof: Presumably with the new proton data [ADA 94]
an even more reasonable $\Delta\Sigma$ and $\Delta g$ would be possible.}.
\setcounter{footnote}{0}

\subsection{The anomaly in \mbox{\boldmath $\nu p$}-elastic scattering and weak
interaction structure functions}
\setcounter{equation}{0}

We mentioned in Section 6.3 that there is no anomaly in the limit $m^2_q/k^2
\to \infty$. Analogously, we see that in the limit of $m^2_q/p^2 \to \infty$
(8.1.3) becomes
\begin{equation}
g_1^{\gamma^*g}(x, Q^2) = {e_f^2\over 2} {\alpha_s\over 2\pi} \left\{ (2 x-1)
\left[ \ln \left( {Q^2\over m_q^2} \, {1-x\over x}\right)-1 \right] +
2(1 - x)\right \}
\end{equation}
and the first moment vanishes. In this case it is natural to consider the
contribution from Fig. 8.2 as a modification of the quark distribution
functions.

There is an interesting application of the r\^ole of quark mass in the
anomaly in {\it elastic} neutrino proton scattering, which is
mediated by $Z_0$ exchange as shown in Fig. 8.7.
%\begin{figure}
%\vspace*{3cm}
%\end{figure}
%\begin{center}
%Figure 8.7
%\end{center}
Since the parity-violating $Z_0$-quark coupling involves $\gamma^\mu\gamma_5$
Fig. 8.7 also depends upon the proton matrix element of the axial vector
current. In the Naive Parton Model, for forward scattering, the contribution
is, in principle, proportional to
\begin{equation}
\delta q \equiv (\Delta u+\Delta\bar{u}-\Delta d-\Delta\bar{d})+
(\Delta c+\Delta\bar{c}-\Delta s-\Delta\bar{s})
\end{equation}
(the sign change for the $I_3 = -1/2$ quarks, compared to the electromagnetic
case, is due to the $Z_0$ couplings).

Experimentally [AHR 87] it seems that $\delta q$ is about 12\% larger than
expected from the $\Delta u,~\Delta d$ contribution (see also the discussion
in Section 4.2.3). This has been interpreted [ELL 88] as evidence for a large
negative $\Delta s$ in agreement with (4.3.17), assuming a negligible
$\Delta c$. However, the anomaly also contributes to this process via the
diagram of Fig. 8.8.
%\begin{figure}
%\vspace*{3cm}
%\end{figure}
%\begin{center}
%Figure 8.8
%\end{center}

By the above discussion there is no contribution from the heavy quarks, but
for each massless quark, in the QCD Field Theoretic Model,
\begin{equation}
\Delta q+\Delta\bar q\to\Delta q+\Delta\bar q-{\alpha_s\over 2\pi}~\Delta g \,.
\end{equation}
Hence, in (8.5.2)
\begin{equation}
\delta q \to \delta q +{\alpha s\over 2\pi}~\Delta g \,.
\end{equation}

In the extreme case $\Delta s=0$ one then requires a rather large
$\Delta g\simeq 5$ to fit the data. As discussed in Section 8.4 such a
scenario leads to difficulties with the $x$-dependence of $g_1^p(x)$. But the
more reasonable choice $\Delta s=-0.05,~g \simeq 3$ is perfectly acceptable.

The weak interaction structure functions discussed in Section 3.5 also receive
an anomalous gluonic contribution via a diagram analogous to that of Fig. 8.2,
with the photon replaced by a $Z_0$ (neutral current) or a by $W^{\pm}$
(charged current). Such contribution leads again, for each massless quark, to
the result (8.5.3) [VOG 91]; for the structure functions measured in charged
current DIS, which involve diagrams with transitions from light to heavy
quarks [see Eqs. (3.5.8)], there are interesting quark mass effects
[VOG 91; MAT 92, 92b].

\subsection{Operator Product Expansion \mbox{\boldmath{$vs.$}} QCD Improved
Parton Model}
\vskip 6pt
\setcounter{equation}{0}

In Section 5 we stressed that only the {\it fully inclusive} electroproduction
process could be expressed in terms of the matrix element of a product
of electromagnetic currents (5.1.1) and hence only in this case could
one utilize the Operator Product Expansion.
Our definition of $\Delta g(x)$ has focused on the semi-inclusive large $k_T$
2-jet process and it might thus seem strange that our results for the first
moment should agree exactly with the anomaly result obtained using the
Operator Product Expansion. But this is simply a `red herring'. We have not
thrown away the part of the cross-section corresponding to small $k_T$. It is
simply absorbed into the redefinition of the quark distribution functions. So
the results of the Operator Product Expansion should agree with the QCD
Improved Parton Model results.

Moreover, if we return to the triangle anomaly discussed in Section 6.3 we
may ask what region of the $k_T$ integration involved in evaluating the
triangle diagram was important for the anomaly. As stressed in [CAR 88] it is
indeed the region of large $k_T$ that is relevant, so there is consistency
between the two methods of deriving the anomaly contribution to the first
moment of $g^p_1(x)$.

These considerations, as emphasized in [ROS 90], draw our attention to an
important issue, namely that whatever cut-off or factorization scheme we use,
we must use it consistently if we wish to compare results from the Operator
Product Expansion with those of the QCD Improved Parton Model. We must also be
careful when dealing with physical quantities that depend on both  polarized
and unpolarized distribution functions.

For example, given that $g_{\pm}(x)$ measures the number density of gluons
polarized along or opposite to the spin direction of the proton, they must be
positive. Thus one expects
\begin{equation}
|\Delta g(x)|=|g_+(x)-g_-(x)|\le|g_+(x)+g_-(x)|=g(x) \,.
\end{equation}
But $g(x)$ is measured in unpolarized scattering and depends upon the
factorization convention utilized there. Clearly  (8.6.1) and analogous
relations for quark distributions can only be expected to hold if the same
cut-off convention is used for $\Delta g(x),~g(x),~\Delta q(x)$ and $q(x)$.
The relevance to the phenomenology of $g_1^p(x)$ was discussed in Section 8.4.

Finally, there is another `red herring' in the literature which needs to be
pointed out. The anomaly emerges from the triangle diagram only if $k_T$ is
integrated up to infinity, as it must be in the Feynman diagram. It is claimed
that, on the contrary, the anomaly emerges when integrating up to
$k^2_T \approx Q^2$ in the QCD Improved Parton Model. The latter statement is
misleading. True, at fixed $x ,~k^{2~{\rm max}}_T \approx Q^2$ but to get the
anomaly we must integrate over $x$ from 0 to 1. The region $x \to 0$
corresponds, via (8.2.2), precisely to $k_T^{\rm max} \to \infty$.

In summary there is perfect agreement between the two approaches to the gluon
contribution in polarized deep inelastic scattering,  provided care is taken
to use consistent factorization or renormalization conventions in both.
%
%end of Section 8
%
\setcounter{section}{8}
\section{\large{Non-perturbative effects in the interpretation of the
measurement of \mbox{\boldmath $\Gamma_1^p$}}}
\vskip 6pt

The previous discussion has focused on the r\^ole of perturbative aspects of
the axial anomaly in the theoretical description of polarized DIS. But the
axial anomaly also has deep and non-trivial non-perturbative consequences in
QCD because of the complicated topological structure of the non-abelian
theory. We first provide a brief survey of the topological issues and than
turn to their r\^ole in DIS. We show that they support the conclusions drawn
from the reinterpretation  of the measurement of $\Gamma^p_1$ including the
effect of the anomaly. Finally we give a brief survey of the open questions
in the understanding of the EMC result.

\subsection{Topological effects and the axial ghost in QCD - a pedagogical
reminder}
\vskip 6pt
\setcounter{equation}{0}

We follow the approach of Dyakonov and Eides [DYA 81]. A more detailed
pedagogical treatment can also be found in [LEA 94]. Consider a
{\it{purely gluonic version of}} QCD, and choose for simplicity the
gauge in which $A^a_0(x) = 0$. While satisfying this gauge condition the
theory is still invariant under arbitrary {\it{time-independent}}
gauge tranformations
\begin{equation}
\bfA_j \to \bfU \bfA_j \bfU^{-1} + {i\over g}~(\partial_j \bfU) \bfU^{-1}
\end{equation}
where, as usual, $\bfA_j$ without a colour label stands for the matrix
$(\bflambda^a/2) A^a_j$ and $\bfU = \bfU({\bfx})$ only and where, for a
consistent theory, one utilises transformation such that $\bfU({\bfx})
\to \bfI$ as $|{\bfx}| \to \infty$.

The fields $A_j^a$ play the r\^ole of generalized coordinates in the quantum
theory and the r\^ole of generalized momenta is played by
\begin{equation}
\pi^j_a(x)\equiv {\partial{\cal{L}}\over\partial\dot{A}^a_j}=-\dot{A}^j_a=
E^a_j
\end{equation}
where ${\bfE}^a$ is the colour-electric field.

The generalized coordinates and momenta satisfy the usual canonical equal
time commutation relations
\begin{equation}
[\pi^a_i({\bfx}, t), A^b_j({\bfy},t)] = -
\delta_{ij}\,\delta_{ab}\,\delta^3({\bfx-\bfy}) \,.
\end{equation}

The Hamiltonian, quite analogous to QED, turns out to be
\begin{equation}
H = \int d^3\bfx~\mbox{\rm Tr} [{\bfE}^2 + {\bfB}^2]
\end{equation}
where the components of the colour-magnetic field ${\bfB}^a$ are defined by
\begin{equation}
B^a_j = {1\over 2} \varepsilon_{jkl}~G^{kl}_a \,.
\end{equation}
Unlike QED, the fields do change under an arbitrary gauge transformation:
\begin{equation}
\bfE_j \to \bfU \bfE_j \bfU^{-1}\qquad \bfB_j \to \bfU \bfB_j \bfU^{-1} \,.
\end{equation}

Bearing in mind that ${\bfE}^2 = \dot{\bfA}^2$ we see that the first term in
(9.1.4) plays the r\^ole of the kinetic energy whereas
$V \equiv \mbox{\rm Tr} [{\bfB}^2]$ represents the potential energy.

The analogue of a wave function in ordinary quantum mechanics is here a
functional $\Psi [{\bfA}(x)]$ of the fields and the stationary states and
energies are found from
\begin{equation}
H\Psi = E\Psi\,.
\end{equation}

Now the gauge transformations $\bfU({\bfx})$ fall into discrete classes
labelled by an {\it{integer}} $n$, the {\it{winding number}}. The class
with $n$ = 0 are called {\it{small}} gauge tranformations and can be built
up from the identity by a series of infinitesimal transformations -- by
smoothly varying the parameters which specify them they can be continously
deformed into the identity transformation. The classes with $n \neq$ 0 are
called {\it{large}} gauge transformations and cannot be continuously
transformed into the identity transformation (for an elementary discussion
see [LEA 94]).

In Section 6.4 we introduced the gluon axial current $K^\mu$ [see (6.4.1)]
which is a function of the fields $A^a_j$. Based on this we can construct a
new variable $X$, a function of the fields $A^a_j$, (often called a
collective coordinate):
\begin{equation}
X \equiv {g^2\over 16\pi^2} \int d^3{\bfx}~K_0(x)
\end{equation}
and such that it changes by the integer $n$ when the fields undergo a gauge
transformation of winding number $n$ but does not change under small gauge
transformation. In fact
\begin{equation}
X \to X + {1\over 24\pi^2}~\int d^3{\bfx}~\varepsilon_{ijk} \mbox{\rm Tr}
\Bigl[ (\partial_i \bfU) \bfU^{-1}(\partial_j \bfU) \bfU^{-1}
(\partial_k \bfU) \bfU^{-1} \Bigr]
\end{equation}
and the change in $X$ can be shown to be just $n$. Thus
\begin{equation}
X \to X + n \,.
\end{equation}

Note that the non-vanishing of the trace term (9.1.9) is a consequence of the
non-Abelian nature of QCD. For normal QED, ({\it i.e.} in 4-dimensional
Minkowski space) on the contrary, $U = e^{i\Lambda ({\bfx})}$ and the trace
term vanishes.

Now we can imagine expressing the potential energy $V$ in terms of $X$ and any
other necessary variables. If then we plot $V~{ \it vs.}~X$ it must have the
same value at all values $X_n = X + n$, since it is, via (9.1.6), invariant
under all gauge transformations, large or small. In other words if we express
the potential in terms of $X$ and the other required variables, $V = V
(X,Y,...),$ then $V$ is a periodic function of $X$ with period 1 as shown in
Fig. 9.1
%\vskip5cm
%\begin{center}
%Figure 9.1
%\end{center}
This periodicity was first noticed by Faddeev [FAD 76], Jackiw and Rebbi
[JAC 76], 't Hooft ['THO 76] and Callan, Dashen and Gross [CAL 76].

Before exploring the consequences it may help to examine a very simple
example of the above phenomenon.

Consider electrodynamics in one time and one space dimension, where $A^\alpha
= (A_0, A_x)$, and define $K_\alpha =
\varepsilon_{\alpha\beta} A^\beta,~(\varepsilon_{01}=-\varepsilon_{10} = 1)$
so that $ K_0 = A_x$.

Let
$$
X = {g\over 2 \pi} \int^\infty_{-\infty}dx~K_0(x) =
{g\over 2\pi} \int^\infty_{-\infty}dx~A_x(x)~.
$$
Then, under the gauge transformation $U = e^{i\phi (x)}$,
$$
A_x \to A_x +{i\over g}~\biggl({\partial\over \partial x}U\biggl)
U^{-1} = A_x -{1\over g}~{d\phi\over dx} \,.
$$
Consequently
$$
X \to X - {1\over 2\pi} \int^\infty_{-\infty} dx~{d\phi \over dx} = X
-{1\over 2\pi}\{ \phi(\infty) - \phi(-\infty)\} \,.
$$
But the requirement $U(x) \to 1$ at spatial infinity means that
$\phi (\infty)$ and $\phi (-\infty)$ must both be multiples of
$2\pi$, but not necessarily the same multiples. Hence $X \to X +$ (an integer
$n$).

Returning to QCD we see that we are dealing with a quantum problem with
periodic potential {\it i.e.} an analogue of an electron in a crystal
lattice. The energy spectrum will thus be a band spectrum. Now, near the
bottom of the lowest energy band the electron behaves like a free particle
with $E=k^2/2m^*$ and $m^*$ an {\it effective mass} which depends on the
details of the shape of the potential. As a consequence the Green's function
$\langle 0| T[x(t)x(0)] |0 \rangle$ has the property that
\begin{equation}
\int dt~e^{i\omega t} \left. \langle 0| T[x(t)x(0)] |0 \rangle
\right|_{\omega \to 0} \longrightarrow {i\over \omega^2m^*}
\end{equation}
like the propagator of a free particle \footnote{This is an immediate
consequence of the equation for the Green's function of a free particle
$$ m{d^2\over dt^2}~\langle 0|T[x(t)x(0)]|0\rangle = i \delta (t).$$}.
\setcounter{footnote}{0}
The electron can move freely in the lattice. Note that these results are
inherently non-perturbative.

Analogously in QCD this means that the vacuum state wave function is not
localized in $X$ at one of the minima $X$ = integer, but that the ground
state is spreat out in $X$ in the form of a Bloch wave
\begin{equation}
|\theta \rangle = \sum_n~e^{in\theta}|n;...\rangle
\end{equation}
where the state $|n;..\rangle$ is approximately localized at the point
$X = n$, and the value of $\theta$ which specifies the lowest energy states
is beyond our power of calculation. The above phenomenon occurs because the
barriers between the different minima of the potential can be penetrated.
In QCD the instantons, which are solutions of the Euclidean version of the
theory, with finite actions, allow one to construct a {\it{semi-classical
approximation}} to $|\theta \rangle$, but the structure (9.1.12) is a general
result, independent of any consideration of instantons. In the following the
vacuum state $|\theta \rangle$ is to be understood as the true QCD vacuum.

Analagously to (9.1.11) the Green's function involving $X(t)X(0)$ will have
a pole at $\omega = 0$. Bearing in mind how $X$ is constructed from $K_0$,
we expect to find a massless pole in the momentum space Feynman propagator
for $K_\mu(x)$, which is the Fourier transform of $\langle 0| T[K_\mu(x)
K_\nu (0)]|0 \rangle$. To avoid a plethora of constants, let us introduce
\begin{equation}
\widetilde{K}_\mu(x) \equiv{N_f\alpha_s\over 2 \pi}~K_\mu (x)
\end{equation}
where $N_f$ is the number of {\it{light}} flavours -- 3 in our case.
We thus expect to have for the $\widetilde{K}_\mu$ propagator:
\begin{equation}
\int d^4x~e^{iq\cdot x} \langle 0| T[\widetilde{K}_\mu(x)\widetilde{K}_\nu
(0)]|0 \rangle = -i\Upsilon^4 {(g_{\mu\nu} - q_\mu q_\nu/q^2)\beta +
q_\mu q_\nu/q^2 \over q^2 + i\epsilon}
\end{equation}
where $\Upsilon$ and $\beta$ are constants.

Some comments are necessary about the above expression. The above form will
hold in any covariant gauge but the value of $\beta$ will be gauge-dependent.
The fact that $\partial_\mu K^\mu(x)$ is gauge-invariant implies that the
longitudinal part of the propagator (9.1.14), {\it i.e.} the part
proportional to $q_\mu q_\nu/q^2$, must be gauge-invariant. Thus the value
of $\Upsilon$ cannot depend upon the gauge and any gauge-dependence associated
with the propagator can only arise from an explicit dependence upon $\beta$.

The massless ghost state thus introduced will be referred to as the `axial
ghost'. The `ghost state' exists in an unphysical sector of the Hilbert
space. (Recall that in quantizing a gauge theory the physical states form
only a subspace of Hilbert space). Finally note that the coupling of the
ghost will be such that the apparent 1/$q^4$ double pole never actually
occurs and, indeed, such that no physical matrix element has a pole at
$q^2 = 0$.

Notice that aside from the normalization factor, the LHS of (9.1.14)
corresponds exactly to the momentum space Feynman propagator for the `field'
$\widetilde{K}_\mu(x)$. We shall introduce the shorthand notation
\begin{equation}
\langle T[AB] \rangle_q \, \equiv \int d^4x~e^{iq\cdot x}
\langle 0|T[A(x)B(0)]|0 \rangle
\end{equation}
for these momentum space propagators.

Equation (9.1.14) implies that in QCD there exists a zero mass axial vector
ghost particle which couples to the axial gluon current $K_\mu$. The
necessity for the ghost pole in QCD was recognised as early as 1975 by Kogut
and Suskind [KOG 75]. The ghost pole was invoked by Veneziano [VEN 79] to
resolve the $U(1)$ problem.

\subsection{A reminder about the \mbox{\boldmath $U(1)$} problem}
\vskip 6pt
\setcounter{equation}{0}

Although we shall not discuss the resolution of the problem we give a brief
reminder as to what exactly the problem is.

As mentioned in Section 6.3 the QCD Lagrangian in the {\it{chiral limit}}
$m_q \to 0$ seems to contain an additional symmetry, namely that it appears
to be invariant under separate flavour transformations on the left and
right-handed parts of the quark fields. Such a symmetry would lead to a
parity doubling of the hadron spectrum if the vacuum were invariant under
chiral transformations and therefore unique. For example, it would imply a
$(J)^P = (1/2)^-,~I = 1/2$ partner for the nucleon. Experimentally the nearest
resonance with these quantum numbers to the nucleon is the $N(1535)$, some 600
MeV heavier than the nucleon. Now the quark mass parameters in the Lagrangian
are believed to be small ($m_u,~m_d < 10$ MeV/c$^2$ and $m_s \simeq 150$
MeV/c$^2$) on the scale of hadron masses, so the chiral limit should be a
reasonable approximation. It is then difficult to believe that the non-zero
quark masses are responsible for the mass gap between the nucleon and the
$N(1535)$.

It is thus assumed that the chiral symmetry is broken principally as a result
of a spontaneous breaking mechanism which gives rise, for example, to a
quark-antiquark condensate. There is no longer a unique vacuum, but a
continuum of states related to each other by chiral rotations and the
symmetry is broken by choosing one of these as the true vacuum state. As a
consequence there should exist a zero mass pseudoscalar Goldstone boson. Its
mass would be strictly zero in the limit $m_q$ = 0 and would be expected to
be non-zero, but small, in the realistic case of small quark masses. There is
one Goldstone boson for each broken generator of the original symmetry.

Since the QCD Lagrangian is, in fact, invariant under $U(3)$ transformations
of flavour there ought to exist nine light pseudoscalars whose masses would
be zero if the $m_q$ were all zero. The obvious pseudoscalar nonet is
($\pi,~K,~\eta,~\eta^\prime)$. In the following we shall treat the $\eta$ as
the eigth member of a pure $SU(3)$ octet, and the $\eta^\prime$ as a pure
$SU(3)$ singlet. (In reality there is a small amount of mixing which we
ignore since it has no bearing on the arguments to follow).

It turns out, however, that the mass of the $\eta^\prime (958)$ is much larger
than expected on theoretical grounds. The other 8 pseudoscalars, associated
with the breaking of chiral $SU(3)_F$, have acceptable masses, so it is only
the Goldstone boson associated with $U(1)_F$ that is in difficulty. It is this
that is referred to as the `$U(1)$ problem'.

\subsection{The axial ghost}
\vskip 6pt
\setcounter{equation}{0}

It turns out that simply introducing an axial ghost pole via (9.1.14) leads to
a result for the $\eta^\prime$ mass which is inconsistent in the chiral limit.
The correct procedure seems to be the following:
\begin{description}
\item{  i)} {\phantom T}The above ghost is considered as a `bare' ghost.
\item{ ii)} {\phantom i}One introduces a quark-antiquark pseudoscalor flavour
singlet field $\eta_0(x)$ corresponding to the `particle'
\begin{equation}
|\eta_0 \rangle \equiv {1\over \sqrt{3}}~|u\bar{u}+d\bar{d}+s\bar{s} \rangle
\end{equation}
which would have corresponded to the $\eta^\prime$ if there were no axial
anomaly. The $\eta_0$ has a mass $m_0$ similar to the other pseudoscalor
mesons.
\item{iii)} We introduce a direct coupling between the ghost field and
$\eta_0(x)$ which results in mixing between the bare ghost and the $\eta_0$
and then study the consequences of the interaction.
\end{description}

To this end we introduce the bare massless ghost field $G^0_\mu(x)$ which
couples directly to $\widetilde{K}_\mu$ so as to induce the mass zero pole in
the $\widetilde{K}_\mu$ propagator (9.1.14). The non-perturbative transition
amplitude for the $\widetilde{K}_\mu$ -- $G_\mu^0$ coupling will be indicated
graphically as in Fig. 9.2. The bare ghost propagator is then given by (9.1.14)
without the factor $\Upsilon^4$. The coupling of the ghost to the $\eta_0$ will
be described by an effective Lagrangian term $-\Delta m~G^0_\mu (x)
\partial^\mu \eta_0(x)$ where $\Delta m$ is a coupling constant that will play
the r\^ole of a mass correction. This induces the diagrammatic vertices shown
in Fig. 9.3.

The bare $G^0_\mu$ and $\eta_0$ now mix and the mass eigenstates have to be
found by diagonalizing their propagators. The simplest method is by summing
the diagrams, Fig. 9.4, generated by the interaction. One finds for the
masses of the renormalized ghost $G_\mu$ and the renormalized physical
$\eta^\prime$
\begin{equation}
m_G = 0\qquad m^2_{\eta^\prime} = m^2_0 + (\Delta m)^2 \,.
\end{equation}
The extra term $(\Delta m)^2$ can be adjusted to raise $m_{\eta^\prime}$ well
above the typical pseudoscalar mass $m_0$ and this resolves the U(1) problem.

We can estimate the size of the parameter $\Upsilon$ involved in the coupling
of $\widetilde{K}_\mu$ and $G^0_\mu$ as follows.

Consider the divergence of the flavour singlet axial current. We have
\begin{equation}
\partial^\mu J^0_{5\mu} = P(x) + \partial^\mu\widetilde{K}_\mu(x)
\end{equation}
where
\begin{equation}
P(x) = 2i \sum_f m_f \, \bar{q}_f \gamma_5 q_f \,.
\end{equation}

The current $J^0_{5\mu} - \widetilde{K}_\mu$ is divergenceless in the chiral
limit and is thus analogous to the flavour non-singlet currents. We thus
define
$f_0$ by
\begin{equation}
\langle 0|J^0_{5\mu}-\widetilde{K}_\mu|\eta_0 \rangle = i\sqrt{3}f_0\,q_\mu
\end{equation}
and expect to have $f_0 \approx f_\pi=135$ MeV. Now one can show that the
parameter $\Upsilon$ is related to $f_0$ and $\Delta m$ via
\begin{equation}
\Upsilon^2 = \sqrt{3} f_0 \, \Delta m \,.
\end{equation}

The value of $m_0$ can be obtained from the magnitude of the quark condensates
$\langle \bar{q}q \rangle$ known from QCD sum rules and this leads,
via (9.3.2) and (9.3.6) to the estimate $\Upsilon \simeq$ 460 MeV [DYA 81].

For the longitudinal part of the complete propagator for $G_\mu$, which
we shall need later, one finds from the diagrams in Fig. 9.4a
\begin{equation}
\langle T[G_\nu G_\mu] \rangle_q = {i\over q^2} \, \left( {q_\mu q_\nu
\over q^2} \right) \, {q^2 - m^2_0\over q^2 - m^2_{\eta^\prime}}
\end{equation}
and for the non diagonal $G_\mu \to \eta^\prime$ propagator given in
Fig. 9.4b
\begin{equation}
\langle T[\eta^\prime G_\mu]\rangle_q = {q_\mu\over q^2}~{\Delta m\over q^2 -
m^2_{\eta^\prime}}\, \cdot
\end{equation}

We see, therefore, that the axial ghost plays an essential r\^ole in
resolving the $U(1)$ problem in QCD. A more sophisticated treatment, allowing
for some breaking of $SU(3)_F$ does not affect the above results significantly,
but does lead to a successful calculation of the mixing between the singlet
$|\eta_0 \rangle$ and the eighth member of the octet $|\eta_8 \rangle$.

Finally we note that there is no danger of the production of the axial ghost
as a physical particle. The cross-section for such production would have the
form
$$
d\sigma_G~\propto {\cal M}_\mu {\cal M}^*_\nu \, \varepsilon^\mu(\lambda)
[\varepsilon^\nu (\lambda)]^* \delta(k^2) d^4k
$$
where $\cal M_\mu$ is the Feynman amplitude and $\varepsilon^\mu(\lambda)$ the
polarization vector for a ghost of helicity $\lambda$ and momentum $k$.
As long as the axial ghost coupling to physical particles is of the
derivative type we
will always have ${\cal M}_\mu \propto k_\mu$. The cross-section will thus
involve $k^\mu k^\nu \, \varepsilon_\mu(\lambda) [\varepsilon_\nu (\lambda)]^*
\, \delta(k^2)$. But for a physical particle $k^\mu\varepsilon_\mu(\lambda)
= 0$ and the cross-section vanishes.

\subsection{R\^ole of the ghost in the nucleon spin problem}
\vskip 6pt
\setcounter{equation}{0}

We showed in Section 6.4 that the nucleon expectation value of $K_\mu$ plays a
direct r\^ole in polarized deep inelastic scattering. In Eq. (6.4.8) we
calculated {\it{perturbatively}} the value of this matrix element. Now,
because of the coupling of $K_\mu$ to the {\it{non-perturbative}} axial
ghost, there appears to be the possibility of, in addition to the RHS of
(6.4.8), a purely non-perturbative contribution. Similar reasoning applies to
matrix elements of $\widetilde{J}^0_{5 \mu}$ [see (6.4.4)] which,
perturbatively, was interpreted as the contribution of the quark spin. We shall
now see how this leads to a new relation for $\Delta \Sigma $.

In any covariant gauge the nucleonic matrix elements of $\widetilde{K}_\mu$
must have the following general form:
\begin{equation}
\langle P^\prime, \lambda^\prime|\widetilde{K}_\mu(0)
|P, \lambda \rangle =\bar{u}_{\lambda^\prime}(P^\prime) \biggl[
\gamma_\mu \widetilde{\cal K}_1(q^2) +
q_\mu \widetilde{\cal K}_2(q^2) \biggr] \gamma_5 u_\lambda (P)
\end{equation}
where $\widetilde{\cal K}_{1,2}~(q^2)$ are scalar form factors,
$q_\mu = P^\prime_\mu - P_\mu$ and $\lambda,~\lambda^\prime$ are helicities.

We shall now show that the ghost contributes only to
$\widetilde{\cal K}_2(q^2)$.
The relevant Feynman diagrams are shown in Fig. 9.5 where the propagators are
the complete propagators shown in Figs. 9.4a, b.

%\vspace*{6cm}
%\begin{center}
%Figure 9.5
%\end{center}

For the interaction of the $\eta_0$ with the nucleon we take a standard
pseudoscalar coupling term in the Lagrangian:
$-i g_{\eta_0N\bar{N}}~\bar{\psi}_N\gamma_5\psi_N\eta_0$ and for the
coupling of $G^0_\mu$ a term $-ig_{G_0N\bar{N}}~G^0_\mu \partial^\mu
(\bar{\psi}_N\gamma_5\psi_N$), chosen so that ghosts cannot be emitted as
physical particles. These give rise to the Feynman vertices shown in Fig. 9.6.
%\vspace*{8cm}
%\begin{center}
%Figure 9.6
%\end{center}

The diagrams of Fig. 9.5 yield a contribution to $\langle P^\prime,
\lambda^\prime|\widetilde{K}_\mu (0)|P, \lambda \rangle$ of
\begin{equation}
{q_\mu\over q^2}~\bar{u}\gamma_5 u~{\Upsilon^2\over m^2_{\eta^\prime}}~
\Bigl(\Delta m \, g_{\eta_0 N\bar{N}} - m^2_0 \, g_{G_0 N\bar{N}}\Bigl) \,.
\end{equation}

Hence, comparing with (9.4.1) the ghost contributes only to
$\widetilde{\cal K}_2$:
\begin{equation}
\widetilde{\cal K}^{ghost}_2 (q^2)={\Upsilon^2\over q^2~m^2_{\eta^\prime}}~
\Bigl(\Delta m \, g_{\eta_0 N\bar{N}}-m^2_0 \, g_{G_0 N\bar{N}}\Bigl) \,,
\end{equation}
{\it i.e.} the ghost induces a pole in $\widetilde{\cal K}_2(q^2)$ at
$q^2 = 0$. Since no other contribution can do this, (9.4.3) becomes an
{\it{exact}} expression for $\widetilde{\cal K}_2(q^2)$ as $q^2 \to 0$.

Note that the gauge-dependent parameter $\beta$ [see (9.1.14)] in the ghost
propagator does {\it{not}} appear in (9.4.3). Thus the value of
$\widetilde{\cal K}_2(q^2)$ as $q^2 \to 0$, as given by (9.4.3), must be
{\it{gauge-invariant}}.

Now the gluonic contribution that was calculated pertubatively in Section 6
occurs only in $\widetilde{\cal K}_1(q^2)$. From (6.4.8) one sees that
\begin{equation}
\widetilde{\cal K}_1(0) = -{3\alpha_s\over 2 \pi}~\Delta g \,.
\end{equation}

We are now able to give the formal argument for the gauge-invariance of
$\Delta g$. We know that $\partial_\mu K^\mu$ is gauge-invariant. But its
forward matrix element just involves a linear combination of
$\widetilde{\cal K}_1(0)$ and $[q^2\widetilde{\cal K}_2(q^2)]_{q^2=0}$
{\it i.e.} a linear combination of $\Delta g$ and other {\it{physical}}
parameters [see (9.4.3)] which do not depend upon the gauge. Thus
$\Delta g$ itself must be independent of the gauge.

Clearly the ghost will also contribute to the modified singlet current
$\widetilde{J}^0_{5\mu}$ defined in (6.4.4). If we put
\begin{equation}
\langle P^\prime,\lambda^\prime|\widetilde{J}^0_{5\mu}|P,\lambda\rangle=
\bar{u}_{\lambda^\prime}(P^\prime)\Bigl[\gamma_\mu \widetilde{G}_1(q^2)+
q_\mu \widetilde{G}_2(q^2)\Bigr]\gamma_5 u_\lambda (P)
\end{equation}
then we will have, for the ghost contribution
\begin{equation}
\widetilde{G}^{ghost}_2(q^2) = - \widetilde{\cal K}^{ghost}_2(q^2)
\end{equation}
and for the perturbative QCD contribution from (6.4.5) and (7.1.4)
\begin{equation}
\widetilde{G}_1(0) = \Delta \Sigma \,.
\end{equation}

We shall now derive a connection between $\Delta\Sigma$, the $\eta^\prime$
decay constant and the $\eta^\prime N\bar{N}$ coupling.

Since in the Parton Model we work with massless quarks, consider the chiral
limit, $m_q \to 0$. Then in (9.4.3), $m^2_{\eta^\prime} \to (\Delta m)^2,~
m^2_0 \to 0$ and using (9.3.6) we have for (9.4.6)
\begin{equation}
\widetilde{G}_2(q^2)=-{\sqrt{3}\over q^2} \, f_0 \, g_{\eta_0N\bar{N}}~+~
\mbox{\rm non~singular~terms}\,.
\end{equation}

Now in this limit $\widetilde{J}^0_{5\mu}$ is conserved. Hence
\begin{equation}
0=\langle P^\prime, \lambda^\prime|\partial^\mu\widetilde{J}^0_{5\mu}
|P, \lambda \rangle = iq^\mu \langle P^\prime, \lambda^\prime|
\widetilde{J}^0_{5\mu} |P,\lambda \rangle \,.
\end{equation}

Using (9.4.5) we have
\begin{equation}
0 = \bar{u}_{\lambda^\prime}(P^\prime) \gamma_5u_\lambda (P) \Bigl[
2M \widetilde{G}_1(q^2) + q^2 \widetilde{G}_2(q^2)\Bigr] \,.
\end{equation}

Hence
\begin{equation}
\widetilde{G}_1 (q^2) = -{q^2\over 2M}~\widetilde{G}_2(q^2)
\end{equation}
and via (9.4.8)
\begin{equation}
\widetilde{G}_1(q^2)={\sqrt{3}f_0~g_{\eta_0N\bar{N}}\over 2M}+{\cal O}(q^2)\,.
\end{equation}

Going to $q^2 = 0$ and using (9.4.7) we obtain [VEN 89, EFR 90]
\begin{equation}
\Delta\Sigma = {\sqrt{3} f_0~g_{\eta_0 N\bar{N}}\over 2M}
\end{equation}
known as a Generalized Golberger-Treiman relation (TGV in [VEN 89]).

Note that the form of (9.4.13) differs from that in [VEN 89]. In [VEN 89]
one starts with the matrix element of $J^0_{5\mu}$ expressed like
(9.4.5) in terms of scalar functions $G_{1,2}(q^2)$:
\begin{equation}
\langle P^\prime, \lambda^\prime|J^0_{5\mu}|P, \lambda \rangle =
\bar{u}_{\lambda^\prime}(P^\prime) \biggl[
\gamma_\mu G_1(q^2) + q_\mu G_2(q^2)\biggr] \gamma_5 u_\lambda(P)
\end{equation}
[note that $G_1(0)$ is then identical to $a_0$ used in Section 4.2 and
Section 6] and one finds
\begin{equation}
G_1(0)={\sqrt{3}\over 2M}~f_{\eta^\prime} (g_{\eta^\prime N\bar{N}}-
\Gamma_{QN\bar{N}}) \,.
\end{equation}

In this form $\Gamma_{QN\bar{N}}$ is the {\it{total}} contribution from
gluonic effects, both from the ghost and from the gluon-parton, and of course
$G_1(0)$ involves not just $\Delta\Sigma$ but $\Delta\Sigma-(3\alpha_s/2\pi)~
\Delta g$. A renormalization group analysis then leads [SHO 92] to the
identification of $\Delta \Sigma$ with the first term on the RHS of (9.4.15).
However, it seems to us that there is some point in separating the ghost and
gluon-parton contributions. The former is totally non-perturbative and
non-Abelian and is absent in QED. The latter is partially perturbative and
would be present also in an Abelian theory. Further, the two contributions
have a significantly different kinematical structure in the matrix element of
$K_\mu$ and so can, at least in principle, be measured separately. Obviously
one learns about $\Delta g$ in deep inelastic scattering; but also in various
high $p_T$ processes, as discussed in Section 11. And it may be possible to
study ghost effects in $NN$ spin dependent
processes in the Regge region, as explained below.

The result (9.4.13) is unchanged if we allow for a non-zero, but flavour
symmetric quark mass. Moreover it can be generalized [EFR 91] to allow for
both $SU(3)_F$ breaking and isospin breaking. The change is {\it{small}}
provided that allowance is made both for breaking of the equality of the quark
masses {\it{and}} for $\pi^0$ -- $\eta$ -- $\eta^\prime$ mixing. If the latter
is neglected one obtains large and misleading corrections, even the wrong sign
for the gluonic contribution [CHE 89]! There is a significant cancellation
between the effects of the two types of symmetry breaking, as is discussed in
detail in [EFR 91].

Now (9.4.13) is interesting because it suggests, in principle, and independent
measurement of $\Delta \Sigma$. The coupling of $\eta^\prime$ to nucleons can
be obtained from studying the r\^ole of $\eta^\prime$-exchange in
nucleon-nucleon elestic scattering. A phase shift analysis in the framework of
a one boson exchange potential model, leads to
$g_{\eta^\prime N\bar{N}} \simeq 7.3$ with, presumably, rather large
uncertainty [DUM 83]. But in this analysis there was no ghost exchange. So the
measured coupling does not really correspond to the $g_{\eta_0 N\bar{N}}$
in (9.4.13). The ghost would generate a contact NN potential and for small $t$,
$|t|\ll m^2_{\eta^\prime}$, what is measured is actually
$\sqrt{g^2_{\eta_0 N\bar{N}} - m^2_{\eta^\prime} \, g^2_{G N\bar{N}}}$ and not
$g_{\eta_0 N\bar{N}}$, but, as we shall argue, it seems likely that
$|m_{\eta^\prime} \, g_{G N\bar{N}}|\ll g_{\eta_0N\bar{N}}$.

Now $g_{\eta^\prime N\bar{N}}$ can also be estimated from
$\eta^\prime \to 2 \gamma$ decay. The decay rate has been calculated using a
triangle diagram made up of the baryon  octet, and assuming, as happens for
$\pi^\circ \to 2 \gamma$, that this gives the same result as a calculation
based on quark triangles [BAG 90]. This yields
$g_{\eta^\prime N\bar{N}} = 6.3 \pm 0.4$.

Both the quoted values are close to the $SU(6)$ value [TOR 84],
$g_{\eta^\prime N\bar{N}} \simeq 6.5$. We see that there will be consistency
between all these values provided that $m_{\eta^\prime} \, g_{G_0 N\bar{N}}$
is small enough.

It may be helpful to note that there is an analogous situation suggesting a
small ghost coupling. In the calculation of the 2-photon matrix element of
$\widetilde{J}^0_{5\mu}$, $\langle \gamma(k_1) \gamma(k_2)|
\widetilde{J}^0_{5\mu}|0 \rangle$
the conservation is broken by a contribution from the QED anomaly {\it i.e.}
there is an additional term $(\alpha = 1/137)$:
\begin{equation}
\partial^\mu \widetilde{J}^0_{5\mu}=\ldots +{\alpha\over 4\pi} N_c \sum_f
e^2_f\,\varepsilon_{\mu\nu\rho\sigma} F^{\mu\nu} F^{\rho\sigma}
\end{equation}
where $N_c$ = number of colours, $e_f$ = quark charge in units of $e$ and
$F^{\mu\nu}$ is the electromagnetic field tensor. Of course, in our previous
strong interaction considerations, we neglected the RHS of (9.4.16)
altogether. Here, because we are dealing with photons, it is this term, and
not the QCD anomaly, that is relevant. Now the above matrix element of
$\widetilde{J}^0_{5\mu}$ is proportional to $g_{\eta^\prime \gamma\gamma}$
(directly related to the rate $\eta^\prime \to 2\gamma$) and earlier studies
[ROE 90] showed consistency with (9.4.16) at the 10\% level. But recently it
was realised [VEN 92] that the ghost should also contribute. This has the
effect of replacing $g_{\eta^\prime\gamma\gamma}$ by
$g_{\eta^\prime\gamma\gamma} - m_{\eta^\prime} \, g_{G\gamma\gamma}$ and the
earlier consistency of (9.4.16) implies that $m_{\eta^\prime} \,
g_{G\gamma\gamma}$ must be $\le 5\%$ of $g_{\eta^\prime\gamma\gamma}$.

The smallness of the ghost coupling to quarks in both the $N\bar{N}$ and
$\gamma\gamma$ case is linked to the success of the OZI approximation. It
would also be expected to be small in the $1/N_c$ expansion [VEN 92], where
for any state $X$,
$$
{m_{\eta^\prime} \, g_{GXX}\over g_{\eta^\prime XX}}\sim {1\over\sqrt{N_c}}
$$
but this does not explain why it is as small as it seems to be.

In summary it thus seems not unreasonable to assume
$$
m_{\eta^\prime} \, g_{G_0 N\bar{N}} \ll g_{\eta_0 N\bar{N}}
$$
and thus
\begin{equation}
g_{\eta_0 N\bar{N}} \simeq 6~-~7 \,.
\end{equation}
Returning to (9.4.13) we take $f_0 \approx f_\pi = 135$ MeV and find
\begin{equation}
\Delta \Sigma \simeq 0.8 \,.
\end{equation}

This supports the interpretation of the EMC result as a cancellation between a
significant $\Delta\Sigma$ and a large gluon contribution as discussed in
Section 7.1. It is in agreement with the naive expectation that the quark spin
contribution is $\simeq 1/2$. At low $Q^2$ this is the dominant
contribution. At high $Q^2$ gluon spin and orbital angular momentum far
outstrip $\Delta\Sigma$. But this should not be a cause for concern.
As stressed in Section 6.1 the proton probed at
large $Q^2$ is very different from the static proton.

It is clearly important to be able to test the validity of the above analysis
by
trying to measure separately $g_{\eta_0N\bar{N}}$ and $g_{G_0 N\bar{N}}$. One
proposal [EFR 91a] is to look for the spin-flip effects caused by the
$\eta^\prime$ and ghost in a situation where other flip mechanisms are absent
or suppressed {\it e.g.} in $A_{NN}$ or $D_{NN}$ in proton-deuteron or
deuteron-deuteron scattering, in the Regge region $|t|\ll s$. The $I=1$
exchanges of $\rho$ and $\pi$ are forbidden and the $I\!\!P,~I\!\!P^\prime,
{}~\omega$ and $f$ exchanges are predominantly non-flip. So $\eta,~\eta^\prime$
and the ghost should be the most important exchanges. The ghost exchange
presumally contributes a fixed pole in the complex-$J$ plane where the $\eta$
and $\eta^\prime$ are associated with moving Regge poles. Consequently the
energy variation of their contributions will be quite different in character
and this should become discernable at high energies.

\subsection{Attempts to provide a physical interpretation of the EMC result}
\vskip 6pt
\setcounter{equation}{0}

The result of the EMC experiment can be summarized as the statement that
[see (6.3.12)]
\begin{equation}
|a_0| = \left| \Delta \Sigma -3 {\alpha_s\over 2\pi} \, \Delta g
\right| \ll 0 \,.
\end{equation}

Now from (6.4.2) we have
\begin{equation}
\langle P^\prime, \lambda^\prime| \partial^\mu\widetilde{K}_\mu
|P, \lambda \rangle = -3{\alpha_s \over \pi} \langle P^\prime,\lambda^\prime
|{\bfE}^a\cdot {\bfB}^a|P,\lambda \rangle
\end{equation}
where ${\bfE}^a$ and ${\bfB}^a$ are the colour-electric and magnetic fields,
and from (9.4.11) the LHS is
\begin{equation}
[2M \widetilde{\cal K}_1(q^2)+q^2\widetilde{\cal K}_2(q^2)] \,
\bar{u}_{\lambda^\prime}(P^\prime) \gamma_5 u_\lambda (P) \,.
\end{equation}

Consider now the term in square brackets as $q^2 \to 0$. From (9.4.4) and
(9.4.6, 8)
\begin{eqnarray}
\lim_{q^2\to 0}~[2M\widetilde{\cal K}_1(q^2)+q^2 \widetilde{\cal K}_2(q^2)]&=
&-2M \, {3\alpha_s\over 2\pi} \, \Delta g + \sqrt{3} f_0 \,
g_{\eta_0 N\bar{N}}\nonumber\\
&=&[\mbox{\rm gluon-parton}] + [\mbox{\rm ghost}]\,.
\end{eqnarray}

Use of the Generalized Goldberger-Treiman relation (9.4.13) then implies, via
(9.5.1), that
\begin{equation}
\left| \lim_{q^2\to 0}~[2M \widetilde{\cal K}_1(q^2) +
q^2\widetilde{\cal K}_2(q^2)] \right| \ll 0 \,.
\end{equation}

In other words there seems to be a cancellation between the
`semi-perturbative' gluon-parton contribution and the non-perturbative `ghost'
contribution in (9.5.4). It remains a question of some interest as to why this
compensation occurs. We consider briefly some of the physical arguments
presented to explain this.

That $a_0 \simeq 0$ was claimed to be quite natural in the Skyrme model
approach to the `spin crisis' [BRO 88]. The $SU(3)$ Skyrme model deals with an
{\it{octet}} of pseudoscalar Goldstone bosons whose self-interactions
are described by the non-linear Skyrme Lagrangian. Quark or gluon degrees of
freedom do not occur explicitly and the baryons appear as topologically
non-trivial soliton solutions of the equations of motion. Since there are no
quark fields it is not absolutely obvious what function of the boson fields
corresponds to the singlet axial current $J^0_{5\mu}$. Nonetheless
arguments can be given for the form of $J^0_{5\mu}$ and in [BRO 88]
it was claimed that $a_0$ is of order $1/N_c$ and, in fact, $a_0 \simeq 0$.
(In these papers $a_0$ is identified with $\Delta\Sigma$, but that is
irrelevant). However, this approach has been criticized by Ryzak [RYZ 89] who
points out that [BRO 88] utilises an effective Lagrangian calculated to order
$1/N_c$ in a $1/N_c$ expansion, yet which does not permit the $\eta^\prime$ to
decay although such decay is precisely a $1/N_c$ effect. In addition, in this
approximation $m_{\eta^\prime}/M\propto (N_c)^{-3/2}$ which is far from
reality. Inclusion of a term to allow $\eta^\prime$ decay would destabilize the
soliton solutions. Ryzak has shown how to overcome this difficulty.
He argues that there are many uncertainties in the estimation of $a_0$ but
concludes that one can have $a_0 = 0.2 \pm 0.1$ in Skyrme
type models. However it is hard to see why the $\eta^\prime$ couples
strongly to the nucleon in these models.

%We conclude that the Skyrme model does not offer a very convincing argument or
%explanation of the fact that $a_0 \simeq 0$.

An entirely different approach to the compensation is discussed in
[FOR 90], where it is argued that the cancellation {\it i.e.} the validity of
(9.5.5) is natural for instanton configurations of the gluon fields when there
is just {\it{one}} flavour. But it is not at all self-evident what
happens for $N_f = 3$.

Yet another proposal [KUH 90] is that there is a link between the so-called
{\it{conformal anomaly}}, which involves $G_{\mu\nu}G^{\mu\nu}$ and the axial
anomaly, which involves $G_{\mu\nu}\widetilde{G}^{\mu\nu}$. By making
an assumption about analyticity in the mass of the regulator fermion fields it
is claimed that one has the following exact result in the chiral limit:
\begin{equation}
a_0 = {-2N_f\over 11 N_c - 2 N_f} = -{2\over 9}
\end{equation}
for 3 light flavours. Such a connection between the anomalies and the relation
(9.5.6) seems quite natural in the framework of supersymmetric QCD where both
the energy-momentum tensor (with its conformal anomaly) and the flavour singlet
current (with its axial anomaly) occur in the {\it{same}} supermultiplet of
currents. But the extension to ordinary QCD is far from obvious, and some
counter-arguments have been given [ANS 92a]. So it is not at all clear that
(9.5.6) can be taken at face value. On the other hand attempts have been made
[DOR 93] to support (9.5.6) on the basis of instanton dominance of the
$\theta$-vacuum.

It has also been suggested [DOR 91] that instanton dominance of the QCD vacuum
could generate large polarized sea quark distributions. In [DOR 91, 93], it
is tacitly assumed that there is no anomalous $\Delta g$ contribution to $a_0$,
so the EMC experiment is interpreted as implying $\Delta\Sigma \simeq$ 0 and
the parameters of the model are adjusted so that the polarized sea cancels the
standard Naive Parton Model polarized valence quark contribution. The
instantons are supposed to generate an effective quark-quark interaction which
mimics Pomeron exchange except that the exchanged object has $I=0,~G=C=1,~P=+1$
and flips helicity. By the usual connection between high energy behaviour in
off-shell Compton scattering and the small-$x$ behaviour of distributions [see
Section 4.1.1] it is argued that for small $x,~[\Delta q(x)+
\Delta\bar{q}(x)]\propto (x \ln^2 x)^{-1}$. This unusually rapid growth at
small $x$ is used to produce the large polarized sea quark contribution.

However, this approach seems to us highly speculative and is governed by the
desire to obtain $\Delta\Sigma \simeq 0$. We have hopefully convinced the
reader that the EMC experiment does {\it{not}} demand $\Delta\Sigma\simeq 0$
and that, moreover, the Generalized Goldberger-Treiman relation is nicely
consistent with the intuitive belief that the value of $\Delta\Sigma$ is not
far from 1.

Finally we comment upon an attempt to resolve the issue by means of a lattice
calculation [MAN 90] of the anomalous gluon contribution to $a_0$. The idea is
to try to measure $\Delta g$ by measuring $\langle P,\lambda|\widetilde{K}_\mu|
P,\lambda \rangle$, or something related to it,
on a lattice. The choice of lattice variables to represent
$\widetilde{K}_\mu$ is a non-trivial task but seems to have been successfully
carried out. The simulation is done on a $6^3 \times 10$ lattice at
$\beta =6/g^2=5.7$, thus corresponding to $\alpha_s\simeq 0.08$.

In [MAN 90] Eq. (6.4.8) {\it{was used to extract}} $\Delta g$, which
is found to be small compared with what is needed to explain the EMC experiment
with a reasonably large $\Delta\Sigma$. Namely it is claimed that
$|\Delta g|\lsim 0.5$.

However, we know from Section 9.4 that (6.4.8) is {\it{not}} the whole
story, and the relation is modified by the ghost contribution. Indeed from
(9.4.1) and (9.4.3) we see that the matrix element of $\widetilde{K}_\mu$ in
the forward direction is quite ambiguous -- the limit $q^2 \to 0$ must be
carefully specified. In fact two different matrix elements were studied,
namely
\begin{equation}
\lim_{P\to 0} \, \langle \bfo,{\bfS}^\prime|\widetilde{\bfK} \cdot
{\bfS}|{\bfP},{\bfS} \rangle \quad \mbox{\rm and} \quad
\lim_{P\to 0}~{\bfS\cdot\hat{\bfP}\over P}
\langle \bfo,\bfS^\prime|\partial_\mu\widetilde{K}^\mu|\bfP,\bfS \rangle
\end{equation}
where $\hat{\bfP}$ is a unit vector along $\bfP$ and $P = |{\bfP}|$.

Using (9.4.1) and (9.5.4) the values found for these matrix elements lead
to the bounds
%\begin{equation}
\begin{eqnarray}
\Bigl|-{3\alpha_s \Delta g\over 2\pi} + {1\over 3}~{\sqrt{3} f_0 \,
g_{\eta_0N\bar{N}}\over 2M}\Bigl| &\lsim& {1\over 20} \\
%\end{equation}
%and
%\begin{equation}
\Bigl|-{3\alpha_s\Delta g\over 2\pi} + {\sqrt{3} f_0 \,
g_{\eta_0N\bar{N}}\over 2M}\Bigl| &\lsim& {5\over 20} \,\cdot
\end{eqnarray}
%\end{equation}

We see that the smallness of the numerical results does {\it{not}} imply
that $\Delta g$ is small. Moreover the smallness of (9.5.9) just confirms the
result (9.5.5) which was obtained from the Generalized Goldberger-Treiman
relation and is thus in complete harmony with the interpretation of the EMC
result where a large $\Delta \Sigma$ is cancelled by a significant gluonic
contribution! Similar comments apply to a more recent numerical study [ALT
94a].

However, there is another difficulty with the interpretation of the lattice
result [EFR 92; MAN 92]. The calculation is done in the so-called
{\it quenched approximation} which means that closed $q\bar{q}$ loops are
not taken into account. It is easy to see diagramatically that this implies
that any $\eta^\prime$ propagator in the lattice calculation of the matrix
elements (9.5.7) is effectively a bare propagator {\it i.e.} the
$\eta^\prime$ mass is not shifted from the value
$m_0$. There will thus be a contribution to (9.5.7) proportional to the bare
$\eta^\prime$ propagator {\it i.e.} proportional to $(m^2_0-q^2)^{-1}$. Bearing
in mind that $m_0$ vanishes in the chiral limit $m_q \to 0$ we see that there
should be a singularity in the quenched lattice calculation of (9.5.7) in the
chiral limit as $q^2 \to 0$. This suggests that the numerical results should
be treated with caution until the quenched approximation can be overcome.

It is interesting to note that an analogous problem would arise in a quenched
lattice calculation of matrix elements of $\partial^\mu J^0_{5\mu}$, which is
central to the Adler-Bardeen relation. In this case the order of taking the
limits $q^2\to 0,~m^2_q\to 0$ determines whether or not a singularity should
occur. Thus, in putting $q^2 = 0$ and then taking the limit $m^2_q \to 0$ one
finds a cancellation of the singularity between the matrix elements of $P$ [see
(9.3.3)] and the anomaly. This was not taken into account in the lattice
calculations. However, taking $m^2_q = 0$ and then latting $q^2 \to 0$ produces
a singularity in $q^2 G_2(q^2)$ of Eq. (9.4.14).
%
%end of Section 9
%
\setcounter{section}{9}
\section{\large{\mbox{\boldmath $g_{1,2}(x)$} in the QCD Field Theoretic
Model}}
\vskip 6pt

In Section 3.2 and 3.3 we discussed various attempts to calculate $g_2(x)$
in the Naive Parton Model and in more sophisticated variants thereof.
We emphasized that these calculations do not lead to coherent results.
Different authors obtain different results and these are usually incompatible
with each other. The basic origin of this difficulty was exposed in
Section 3.4 where we explained that $g_2(x)$ simply cannot be calculated
within the framework of a Parton Model.

In this Section we attack the problem from the much more solid basis of the
twist-3 QCD Field Theoretic Model. We calculate both $g_1(x)$ and $g_2(x)$ so
as to be able to emphasize the differences between the two calculations.

The first `rigorous' derivation of the result for $g_2(x)$ is due to Efremov
and Teryaev [EFR 84]. It is closely connected with the problem of single spin
asymmetries in QCD [HEL 91]. This work was somewhat extended by Ratcliffe
[RAT 86] and by Efremov and Teryaev in a later paper [EFR 87; QIU 91].

\subsection{The QCD Field Theoretic Model}
\vskip 6pt
\setcounter{equation}{0}

Let us now turn to the QCD Field Theoretic Model. In this case one deals with
a Feynman diagram one part of which is handled perturbatively and the other,
being essentially non-perturbative, is parametrized in terms of a small number
of unknown functions which play the r\^ole of the various parton number
densities.

In the simplest diagram only quarks of 4-momentum $k$ from the hadron are
involved and the diagram is split into a `hard' part $E^{\mu\nu}_{\alpha\beta}$
and a `soft' part $\Phi_{\alpha\beta}$ as shown in Fig. 10.1.
%\vspace*{6cm}
%\begin{center}
%Figure 10.1: Separation of $\gamma~hadron\to X$ into soft and hard part.
%\end{center}

Both $E$ and $\Phi$ are absorptive parts for the reactions $\gamma^* q \to X$
and $hadron \to q X$ respectively. There are no explicit propagators for the
quark lines; they are absorbed into $\Phi$.

The mathematical structure which represents the contribution of Fig. 10.1 to
$W^{\mu\nu}$ is
\begin{equation}
W^{\mu \nu} = {1\over 2\pi}\int {d^4 k\over (2\pi)^4}~
E^{\mu \nu}_{\beta\alpha}(q,k)~\Phi_{\alpha\beta}(P,S; k)
\end{equation}
where
\begin{equation}
\Phi_{\alpha\beta} = \int d^4z~e^{ik\cdot z} \langle P,S |\bar{\psi}_\beta (0)
\psi_\alpha (z)|P,S \rangle \,.
\end{equation}
The label $\alpha$ on $\psi_\alpha(z)$ is a spinor label. $\psi_\alpha(z)$ is
a  column vector in colour space, so $\bar{\psi}\psi$ involves a sum over
colour. Colour is irrelevant in $E^{\mu \nu}$.

Using matrix notation for the spinor indices, (10.1.1) becomes
\begin{equation}
W^{\mu\nu}={1\over 2\pi}\int {d^4 k\over (2\pi)^4}~\mbox{\rm Tr}
\Bigl[E^{\mu\nu}(q,k)~\Phi(P,S;k)\Bigr] \,.
\end{equation}

The Parton Model usually emerges upon making the following approximation:
\vskip 4pt

1) The $\gamma^* q$ interaction is given its simplest form, as shown in
Fig. 10.2.
%\vspace*{6cm}
%\begin{center}
%Figure 10.2
%\end{center}

In the following we calculate only with the uncrossed Born diagram. The
result for the crossed diagram is obtained at the end by the replacement
$x_{Bj}\to - x_{Bj}$ in the hadronic matrix elements connected with
$\Phi_{\alpha\beta}$ and is simply to be added to the uncrossed result.
For a quark of flavour $f$, charge $e_f$ (in units of $e$), for the diagram
of Fig. 10.2a,
\begin{equation}
E^{\mu \nu} = -e_f^2 \, \pi \, \gamma^\nu [k\sla + q\sla + m_q]\gamma^\mu
\, \delta[(k + q)^2 - m_q^2]
\end{equation}
where $m_q$ is the quark mass.
Using Fig. 10.2 in Fig. 10.1 gives rise to the familiar `handbag diagram'.
In order to isolate the antisymmetric part of $W^{\mu\nu}$ we make the
replacement in (10.1.4):
\begin{equation}
\gamma^\nu\gamma^\rho\gamma^\mu\to{1\over 2}[\gamma^\nu\gamma^\rho\gamma^\mu
-\gamma^\mu\gamma^\rho\gamma^\nu]=
-i~\varepsilon^{\mu \nu\rho\sigma}~\gamma_\sigma\gamma_5
\end{equation}
\begin{equation}
\gamma^\nu\gamma^\mu\to{1\over 2}[\gamma^\nu\gamma^\mu-\gamma^\mu\gamma^\nu]
=i~\sigma^{\mu v}
\end{equation}
and recalling (2.1.8) find
\begin{eqnarray}
W^{(A)}_{\mu \nu}&=&{e^2_f\over 2} \int{d^4 k\over (2\pi)^4}~
\delta[(k + q)^2-m_q^2]\Bigl\{\varepsilon_{\mu\nu\rho\sigma}(q^\rho+k^\rho)
\,\mbox{\rm Tr}\Bigl[\gamma^\sigma\gamma_5\Phi\Bigr]\nonumber\\
&-&m_q \mbox{\rm Tr}[\sigma_{\mu \nu} \Phi]\Bigr\} \,.
\end{eqnarray}
\vskip 4pt

2) One assumes that the soft matrix element cuts off rapidly for $k^2$
off the mass-shell $k^2 = m_q^2$, and for $k^\mu$ non-collinear with respect
to the hadron momentum $P^\mu$.

This is implemented as follows. Consider a reference frame where the hadron
is moving at high momentum along $OZ$ so that
\begin{equation}
P^\mu = (E, 0, 0, P)\qquad \mbox{\rm with}\qquad E\approx P
\end{equation}
is a `large' four-vector. We introduce a `small' null vector
\begin{equation}
n^\mu = \biggl({1\over P + E}, 0, 0, - {1\over P + E}\biggr)
\end{equation}
such that
\begin{equation}
n \cdot P = 1, \qquad n^2 = 0 \,.
\end{equation}

One can then write for $k^\mu$
\begin{equation}
k^\mu = (k\cdot n)P^\mu + {1\over 2} \left[ {k^2 + {\bfk}^2_T \over (k\cdot n)}
- M^2(k\cdot n) \right]n^\mu+k^\mu_T
\end{equation}
where
\begin{equation}
k^\mu_T = (0, {\bfk}_T, 0) \,.
\end{equation}
In view of the assumption about $\Phi$ we can say that
\begin{equation}
k^\mu \approx (k\cdot n) P^\mu \,.
\end{equation}

It should be noted that some care is necessary in deciding whether the
approximation (10.1.13) is adequate. We shall see that this depends crucially
upon whether we are considering a nucleon with longitudinal ($L$) or with
transverse ($T$) polarization.

\subsection{\mbox{\boldmath $g_1(x)$}: Longitudinal polarization}
\vskip 6pt
\setcounter{equation}{0}

For the study of $g_1$ we consider a nucleon with helicity $\lambda=\pm 1/2$
and it is sufficient to approximate (10.1.7) by putting
\begin{equation}
(q + k)^\rho \approx q^\rho + (k\cdot n) P^\rho
\end{equation}
and dropping the term proportional to the quark mass $m_q$. Then writing
\begin{equation}
q^\rho + (k \cdot n) P^\rho = \int dx~\delta(x - k\cdot n) [q + xP]^\rho
\end{equation}
we can take the integration over $d^4k$ in (10.1.7) through to obtain
\begin{equation}
W^{(A)}_{\mu \nu} = {e^2_f\over 2}~\varepsilon_{\mu\nu\rho\sigma}\int dx~
{\delta(x-x_{\rm Bj})\over 2 P\cdot q}~(q + xP)^\rho A^\sigma (x)
\end{equation}
where (using $S_L$ to denote longitudinal spin)
\begin{eqnarray}
A^\sigma(x)&\equiv&\int {d^4 k\over (2\pi)^4} \, d^4 z~\delta(x-k\cdot n)
e^{ik\cdot z} \langle P,S_L|\bar{\psi}(0)\gamma^\sigma\gamma_5 \psi (z)
|P,S_L \rangle \nonumber\\
&=& \int{d\lambda\over 2\pi}~e^{i\lambda x} \langle P,S_L|\bar{\psi}(0)
\gamma^\sigma\gamma_5\psi(\lambda n)|P,S_L \rangle
\end{eqnarray}
is a pseudo-vector which can only depend upon the {\it vectors}
$P^\mu,~n^\mu$ and $v^\mu \equiv \varepsilon^{\mu\alpha\beta\gamma}~
S_\alpha P_\beta n_\gamma$ and the {\it pseudo-vector} $S^\mu$, and
which must be linear in $S^\mu$. Given that $S\cdot P$ = 0 the only
possibilities are $S^\sigma$ and $(n\cdot S)P^\sigma$. Note that with the
normalization
\begin{equation}
\langle \bfP |{\bfP}^\prime \rangle = (2\pi)^3 \, 2E \,
\delta^3({\bfP}-{\bfP}^\prime)
\end{equation}
$A^\sigma(x)$ has dimensions [$M$].

Recall that for a nucleon with 4-momentum given by (10.1.8)
\begin{equation}
S^\mu(\lambda) = {2\lambda\over M}(P,0,0,E)\qquad S^2 = -1 \,,
\end{equation}
where $\lambda$ is a helicity $(\lambda = \pm {1\over 2})$ so that
\begin{equation}
S^\mu(\lambda) = {2\lambda\over M} (P^\mu - M^2n^\mu)
\end{equation}
and we may take
\begin{equation}
S^\mu (\lambda) \approx {2\lambda\over M} P^\mu \,,
\end{equation}
{\it i.e.} $MS^\mu(\lambda)$ is a `large' vector.

In view of (10.2.8) the structures $(n\cdot S)P^\sigma$ and
$S^\sigma(\lambda)$ are equivalent in leading order and the only possibility
is then (the factor 4 is for later convenience)
\begin{equation}
A^\sigma(x,\lambda) = 4Mh_L(x)\,S^\sigma(\lambda)
\end{equation}
where the dimensionless {\it longitudinal distribution function} is
given by {\footnote{There appears to be a sign error in the definition of
$h_L(x)$ in [COR 92]. Our $h_L(x)=-{1\over 2}~h_L^{COR}$, the factor 1/2
being included so as to obtain (10.2.14). Our sign agrees with [EFR 84].}
\setcounter{footnote}{0}
\begin{equation}
4h_L(x) = {n_\sigma A^\sigma(x) \over 2\lambda}
= \int{d\tau\over 2\pi}~e^{i\tau x} \, \widetilde{h}_L(\tau)
\end{equation}
where
\begin{equation}
\widetilde{h}_L(\tau) = {1\over M(n\cdot S_L)} \langle P,S_L|\bar{\psi}(0)
n\slas \gamma_5\psi(\tau n)|P,S_L \rangle \,.
\end{equation}
Substituting into (10.2.3) and adding the contribution from the crossed Born
diagram, Fig. 10.2b, yields
\begin{equation}
W^{(A)}_{\mu \nu}(L) = e^2_f{M\over P\cdot q}[h_L(x_{\rm Bj}) +
h_L(-x_{\rm Bj})]\varepsilon_{\mu\nu\rho\sigma} q^\rho S^\sigma(\lambda)\,.
\end{equation}

Note that the term $xP^\rho$ in (10.2.3) does not contribute on account of
(10.2.8). Consequently (10.2.12) is gauge invariant
$q^\mu W^{(A)}_{\mu \nu} = 0$. Note that (10.2.8), which holds only for
longitudinal spin, is crucial for the gauge invariance.

Comparing with (2.1.19) in the approximation $S^\mu \propto P^\mu$ we obtain,
for the contribution of a quark of flavour $f$,
\begin{equation}
g_1(x) = {1\over 2} \, e_f^2 \, [h^f_L(x)+h^f_L(-x)] \,.
\end{equation}

If one treats the quark fields in (10.2.10) as free fields and regards the
nucleon as an assemblage of free partons one finds
\begin{equation}
h^f_L(x)=\Delta q_f(x)\qquad h_L^f(-x) = \Delta\bar{q}_f(x)
\end{equation}
so that (10.2.13) reproduces the simple Parton Model result for $g_1(x)$.
Equation (10.2.13) provides a field theoretic generalization of the Parton
Model result. A more intuitive expression is given in Section 10.5. We now
consider how the transverse spin  case differs from the above.

\subsection{\mbox{\boldmath $g_2(x)$}: Transverse polarization}
\vskip 6pt
\setcounter{equation}{0}

In order to see the essential difference between the longitudinal and
transverse spin cases consider again the result (10.2.12). In the CM of the
$\gamma^*$-nucleon collision, as far as magnitudes are concerned, for the
longitudinal case we have
\begin{equation}
|q_\sigma| \sim |MS_\sigma(\lambda)| ={\cal O}(\nu);\qquad P\cdot q=M\nu
\end{equation}
so that for the large components of $W^{(A)}_{\mu\nu}(L)$,
\begin{equation}
|W^{(A)}_{12}(L)| = {\cal O}(\nu/M)
\end{equation}
assuming that $|h_L(x)| ={\cal O}(1)$.

In the transverse spin case the analogue of (10.2.4) can only be proportional
to $S^\sigma_T$ since $n\cdot S_T$ = 0, and will produce a result like
(10.2.12) with $S(\lambda) \to S_T$. Given that $|S_T| ={\cal O}(1)$ one has,
for the `large' components, only
\begin{equation}
|W^{(A)}_{\mu \nu}(T)| ={\cal O}(1) \,.
\end{equation}
This immediately suggests that care must be exercised in neglecting
non-leading terms {\it e.g.} in (10.1.11).

Secondly, note that in (10.2.1) the term $(k\cdot n)P^\mu$ of (10.1.11) did
not contribute because of the fact that, in leading order,
$P_\mu\propto S_\mu(\lambda)$. In the transverse case this will not happen
and the analogue of (10.2.12) will contain a term
$\varepsilon_{\mu\nu\rho\sigma} P^\rho S^\sigma_T$ in $W^{(A)}_{\mu\nu}(T)$
which, (analogously to the Parton Model case (3.4.3) when $m\not =m_q$)
is not gauge invariant.

We must therefore return to (10.1.4) and improve upon the approximation
(10.1.13). However, it will then turn out that  the more complicated non
Parton Model diagram involving gluon exchange, Fig. 10.3, contributes to the
same order. This is not entirely surprising given that the operator product
result (5.2.2) for the moments of $g_2(x)$ involves $A_\mu$.
%\vspace*{6cm}
%\begin{center}
%Figure 10.3: DIS interaction involving quark-gluon correlation.
%\end{center}

Amazingly, as was shown by Efremov and Teryaev [EFR 84], this term just
cancels the unwanted contribution from the $(k\cdot n) P\Cslas$ and the mass
terms of the handbag diagram and the final result is gauge invariant.
Essential in this proof is the use of the equations of motion for the quark
field.

The analysis to show the cancellation is rather complicated and is carried
out in Appendix C. We shall only state the result here. It is the exact
analogue of (10.2.12), namely, including the contribution of the crossed
Born diagram Fig. 10.2b,
\begin{equation}
W^{(A)}_{\mu \nu}(T) = e^2_f \, {M \over P \cdot q}\, [f_T(x_{Bj}) +
f_T(-x_{Bj})] \, \varepsilon_{\mu\nu\rho\sigma} q^\rho S_T^\sigma
\end{equation}
where the analogue of (10.2.9) is
\begin{equation}
A^\sigma(x,T) = 4M f_T(x) \, S^\sigma_T
\end{equation}
with \footnote{Our $f_T(x)=2f_T^{\rm COR}(x)$ for reasons explained after Eqn.
(10.2.10).}
\setcounter{footnote}{0}
\begin{equation}
4f_T(x) = \int {d\tau\over 2\pi}~e^{i\tau x} \widetilde{f}_T(\tau)
\end{equation}
where
\begin{equation}
\widetilde{f}_T(\tau)={1\over M} \langle P,S_T|\bar{\psi}(0)\gamma_5
S\Cslas_T \psi(\tau n)|P,S_T \rangle \,.
\end{equation}
Comparing (10.3.4) with (2.1.19), for the case of transverse polarization,
we obtain for the contribution of a quark of flavour $f$,
\begin{equation}
g_1(x) + g_2 (x) = {e^2_f\over 2}\,[f_T^f(x) + f_T^f(-x)] \,.
\end{equation}

Although the surviving contribution comes from the `handbag' diagram it does
not, in fact, have any simple parton interpretation as is explained briefly
in Section 10.4.

\subsection{The Naive Parton Model for \mbox{\boldmath $g_2(x)$} revisited}
\vskip 6pt
\setcounter{equation}{0}

The result (10.3.8) for $g_1(x) + g_2(x)$ together with the expression
(10.3.6) for $f_T(x)$ which only involves quark field operators might suggest
that there ought to be a formula for $g_1(x) + g_2(x)$ derivable in the Naive
Parton Model. However one can easily see that this is not so.

In the Naive Parton Model we would treat $\psi(z)$ as a free field and the
proton as an assemblage of free partons. The key element in evaluating
(10.3.7) would then be the matrix element.
\begin{equation}
\langle k,s^\prime|\bar{\psi}(0)\gamma_5 S\Cslas_T \psi(\tau n)|k,s \rangle
\end{equation}
taken between free quark states labelled by the covariant spin vectors
$s$ and $s^\prime$. This matrix element will be multiplied by proton wave
functions giving the amplitudes for finding quarks with $(k,s)$ and
$(k,s^\prime)$ respectively. There are essentially two possibilities:

i) If we choose to use quark states of definite helicity
$\lambda,~\lambda^\prime$, for massless quarks the matrix element (10.4.1)
is diagonal {\it i.e.} $\langle k,\lambda^\prime|\ldots|k,\lambda\rangle
\propto \delta_{\lambda^\prime\lambda}$ and the product of wave-functions
becomes their modulus squared, yielding a probabilistic interpretation. But
in this case one finds that (10.4.1) is proportional to
$m_q S_T \cdot s(\lambda)$ which is zero
in the Naive Parton Model with no $k_\perp $. So any non-zero result is only of
order $k_\perp$ and cannot be evaluated without a model of the
$k_T$-distribution, an aspect of the  nucleon structure which is usually
considered to be beyond the Naive Parton Model.

ii) If we choose to use quark states of definite transverse spin {\it e.g.}
states
\begin{equation}
|\uparrow ~\mbox{\rm or}~\downarrow \rangle = {1\over \sqrt 2} \,
\bigg( |+ \rangle \pm i \, |- \rangle \bigg) \,,
\end{equation}
then the matrix element (10.4.1) is zero for massless quarks if
$s^\prime =s$ and is largest for non-diagonal transitions
$|\uparrow \rangle \to |\downarrow \rangle$. Consequently one does not obtain
a probabilistic expression from the nucleon wave-functions. Moreover even
the non-diagonal transitions vanish in the absence of $k_\perp$.

Of course the above arguments do not mean that one cannot do model
calculations of $g_1(x)+g_2(x)$. They simply imply that the results will
not have the traditional Parton Model form and will depend upon specific
assumptions about the $k_\perp$ dependence of the nucleon wave-function.

\subsection{Probabilistic form for \mbox{\boldmath $g_1(x)$} in the Field
Theoretic Model}
\vskip 6pt
\setcounter{equation}{0}

We mentioned in Section 10.2 that the hadronic matrix element $h_L(x)$, in
terms of which $g_1(x)$ is given in (10.2.13), reduces to the usual partonic
$\Delta q(x)$ when the fields are treated as non-interacting free fields.
However even in the case of interacting fields one can rewrite the formulae
(10.2.10, 11) for $h_L(x)$ in such a way as to display manifestly the
probabilistic interpretation.

{}From the analysis of the matrix element $\langle P,S_L| \bar \psi(0)
\gamma^\mu \gamma_5 \psi(\tau n)|P,S_L\rangle$ given in Eq. (E.1) it
is easy to see that to leading order the matrix elements with $\mu=0$ and
$\mu=3$ are equal. From the definition of $n^\mu$ (10.1.9) we can thus
replace the factor $n\slas$ which occurs in the definition of $h_L(x)$ by
\begin{equation}
n\slas = {1\over P+E} (\gamma^0 + \gamma^3) ~ \to ~ {2\over P+E} \, \gamma^0
\approx {\gamma^0 \over P} \,\cdot
\end{equation}

Inside the matrix element (10.2.11) we may then write
\begin{eqnarray}
\bar\psi(0) n\slas \gamma_5 \psi(\tau n) &\to&
{1\over P} \, \bar\psi(0) \gamma^0 \gamma_5 \psi(\tau n) =
{1\over P} \, \psi^\dagger(0) \gamma_5 \psi(\tau n) \nonumber\\
&=& {1\over P} \left\{ \psi^\dagger(0) {1+\gamma_5 \over 2} \psi(\tau n)
- \psi^\dagger(0) {1-\gamma_5 \over 2} \psi(\tau n) \right\} \\
&=& {1\over P} \left\{ \left( {1+\gamma_5 \over 2} \psi(0) \right)^\dagger
\left( {1+\gamma_5 \over 2} \psi(\tau n) \right) \right.\\
&-& \left. \left( {1-\gamma_5 \over 2} \psi(0) \right)^\dagger
\left( {1-\gamma_5 \over 2} \psi(\tau n) \right) \right\}\nonumber
\end{eqnarray}
where, in the last step, we have used
$$
\left( {1 \pm \gamma_5 \over 2} \right)^2 =
\left( {1 \pm \gamma_5 \over 2} \right) \,.
$$

Finally, inserting a sum over a complete set of states, using translational
invariance to shift from $\psi(\tau n)$ to $\psi(0)$, and carrying out
the integration over $\tau$  in (10.2.10), we obtain, taking a nucleon
with helicity  $\lambda = + 1/2$,
\begin{equation}
h_L(x) = q_+(x) - q_-(x)
\end{equation}
where
\begin{equation}
q_\pm(x) \equiv {1\over 4P} \sum_X \left| \langle X |{1 \pm \gamma_5 \over 2}
\,\psi(0) |P,\lambda=1/2 \rangle \right|^2 \,
\delta [n \cdot P_X - (1-x)]\,.
\end{equation}
This displays manifestly the probabilistic nature of $q_\pm(x)$ even in the
fully interactive theory.

The same approach clearly fails for $f_T(x)$ given by (10.3.6, 7), in terms
of which $g_2(x)$ is given by (10.3.8).

Concerning our expression (10.5.5) for $q_{\pm}(x)$, one can derive a more
accurate result, in which one does not make the approximation that the matrix
elements with $\mu=0$ and $\mu=3$ are equal, using the formalism of light-cone
fields [MAN 91a].

Define
\begin{equation}
\gamma^{\pm} = {1\over \sqrt 2}(\gamma^0 \pm \gamma^3)
\end{equation}
and projection operator
\begin{equation}
P_+ = {1\over 2} \gamma^- \gamma^+ \qquad\qquad P_+^2 = P_+
\end{equation}
and ``good'' component of the field
\begin{equation}
\psi_+ \equiv P_+ \psi \,.
\end{equation}

Then a more accurate expression for $q_{\pm}(x)$ is
\begin{equation}
q_\pm(x) = {1\over 4P} \sum_X \left| \langle X |{1 \pm \gamma_5 \over 2}
\,\psi_+(0) |P,\lambda=1/2 \rangle \right|^2 \,
\delta [n \cdot P_X - (1-x)]\,.
\end{equation}
%
%end of Section 10
%
\setcounter{section}{10}
\section{\large{Future experiments}}
\vskip 6pt
\setcounter{equation}{0}

Throughout this paper we have stressed the r\^ole of the axial anomaly in
providing an unexpected gluon contribution to the first moments
$\Gamma_1^{p,n}$
measured in polarized deep inelastic lepton--hadron scattering. As a
consequence of this mechanism the data can be explained with a large spin
contribution $\Delta \Sigma \simeq 0.7$ from the quarks, as expected
intuitively. This is in accord with the Generalized Goldberger--Treiman
relation and with various relativistic bag model calculations. The smallness
of the flavour singlet part of $\Gamma_1^{p}$ is explained as a cancellation
between $\Delta\Sigma$ and the anomalous gluon contribution.

However, in order to accomplish this one requires a large gluon spin
$\Delta g$ at $\langle Q^2\rangle \simeq 10$ (GeV/c)$^2$, {\it i.e.}
$\Delta g \simeq 4$. As we have explained, it is inevitable, within
QCD-evolution, that $\Delta g$ grows indefinitely with $Q^2$, so even if
intuitively we expect $\Delta g$ very small at small $Q^2$ we really cannot
rule out the above value at $\langle Q^2\rangle \simeq 10$ (GeV/c)$^2$.

But, clearly, it is of the utmost importance to test this explanation by
measuring $\Delta g$ in other, independent reactions. If it is as large as
the DIS experiments suggest then there will be measurable  effects in
several other processes. In the following we indicate briefly some of the
possibilities that have been studied thus far [WIN 92, REY 93].

In Section 7.3 we explained how the growth of $\Delta g$ is compensated for by
an analogous growth in $L_z$, the total orbital angular momentum carried by all
partons. The hard processes we are about to discuss, being short range
interactions, are insensitive to $L_z$ and can thus reveal $\Delta g$
directly.

\subsection{Semi-inclusive deep inelastic lepton--hadron reactions}
\vskip 6pt
\setcounter{equation}{0}

Perhaps the most straightforward proposal [CAR 88, MAN 91] is in semi-inclusive
polarized DIS.The idea is to select events with final state hadrons of large
$p_T$ relative to the $\gamma^*$-nucleon collision axis. Such events should be
enriched by two-jet production via the photon--gluon fusion subprocess shown in
Fig. 8.2 and will depend directly on $\Delta g$. The contribution to
$\Gamma^p_1$ coming from this subclass of events should be sensitive to
$\Delta g$ and could even be negative for some choice of $p_T$ cut-off.

Very much in the same spirit one could select on events with charm pair
production [GLU 88] or on $J/\psi$ production at large $p_T$ [KAL 89; SRI 92].
The mechanism is the same as shown in Fig. 8.2. These processes are quite
sensitive to $\Delta g$.

Another proposal is to look at the semi-inclusive production of forward kaons,
which should be sensitive to the strange quarks in the nucleon [CLO 89]. The
idea here is to measure the polarization of the strange quarks in the nucleon,
{\it i.e.} $\Delta s$. This is important because if the anomalous gluon
contribution is NOT the correct mechanism to explain the smallness
of the flavour singlet part of $\Gamma_1^p$ then one can arrange to have
$\Delta\Sigma \simeq 0$ by invoking a surprisingly large and negative
polarized strange quark distribution, as was discussed in Section 4.3.3.
However the interpretation of the experiment might be difficult because
some contribution could arise from $s\bar{s}$ pairs created in the
photon--gluon fusion diagram, Fig. 8.2.

Most of the above measurements should be feasible in the present SMC (CERN)
and SLAC experiments (provided high enough statistics and good enough
particle identification can be achieved) and in the next generation of
experiments now being planned, {\it viz.} HERMES at DESY and HELP at CERN.

\subsection{Hadron--hadron reactions}
\vskip 6pt
\setcounter{equation}{0}

If it is possible to produce high energy collisions of longitudinally polarized
protons or antiprotons on a longitudinally polarized nucleon target then a
whole range of experimental possibilities become feasible for studying
$\Delta g$.

\subsubsection{Hard \mbox{\boldmath $\gamma$} and Drell-Yan reactions}
\vskip 6pt

Several possibilities have been analysed (see [BER 89]).
The classic and most direct experiment would be the polarization
asymmetry in the production cross-section for hard $\gamma$'s at high $p_T$.
The obvious asymmetry
\begin{equation}
A={d\sigma^{\begin{array}{c}\hspace*{-0.2cm}\to\vspace*{-0.3cm}\\
\hspace*{-0.2cm}\to\end{array}} -
d\sigma^{\begin{array}{c}\hspace*{-0.2cm}\to\vspace*{-0.3cm}\\
\hspace*{-0.2cm}\leftarrow\end{array}}\over
d\sigma^{\begin{array}{c}\hspace*{-0.2cm}\to\vspace*{-0.3cm}\\
\hspace*{-0.2cm}\to\end{array}} +
d\sigma^{\begin{array}{c}\hspace*{-0.2cm}\to\vspace*{-0.3cm}\\
\hspace*{-0.2cm}\leftarrow\end{array}}}
\end{equation}
depends strongly on $\Delta g$ and $\Delta q$.

It should also be possible to detect the polarization asymmetry in the
production cross-section for a high $p_T$ jet [RAM 88], but this will be less
sensitive to $\Delta g$ than hard $\gamma$ production because quark initiated
processes play a relatively larger r\^ole.

Experiments of the above type are planned for the RHIC collider now under
construction at Brookhaven but it might also be feasible to do such
measurements using the secondary polarized proton beam colliding with a fixed
polarized target at the Fermilab TEVATRON.

One of the most interesting possibilities for learning about $\Delta g$ in
hadron-hadron collisions involves the production of Drell-Yan lepton pairs at
large $p_T$, {\it i.e.} where the $p_T$ of the {\it pair} is large with respect
to the collision axis. Of course the obvious experiment, analogous to the hard
$\gamma$ case, is the production asymmetry using both polarized beam and
polarized target. But what is especially interesting about the Drell-Yan case,
given the difficulty of providing {\it both} a polarized beam {\it and} a
polarized target at high energies, is that certain asymmetries sensitive to
$\Delta g$ exist even when just the beam or just the target is
{\it longitudinally} polarized.

The problem is that these single spin asymmetries vanish in the usual Born
mechanism for large $p_T$ production shown in Fig. 11.1.
%\vskip3cm
%\begin{center}
%Figure 11.1: Conventional Born mechanism for large $p_T$ Drell-Yan pair
%production.
%\end{center}
A non-zero asymmetry arises from $\gamma-Z^0$ interference for large mass pairs
[LEA 93, 94a] or from interference between the Born diagrams and the one loop
QCD corrections to $qg \to q\gamma$ and $q\bar q \to \gamma g$ [CAR 92].

The most complete information is obtained if the angular distribution of the
lepton in the pair rest frame can be measured [LEA 93, 94a]. Alternatively,
one can measure what might be called the {\it handedness} ($H$) of the lepton
pair. Let ${\bfk}_\pm$ be the laboratory momentum of the $\ell^\pm$ and
${\bfP}$ be the laboratory momentum of the proton polarized beam (or target).
Define
\begin{equation}
X_l \equiv \bfP \cdot ({\bfk}_+ \times {\bfk}_-)
\end{equation}
and let $N(X_l)$ be the number of pairs with given value of $X_l$. Then
\begin{equation}
H \equiv{N(X_l > 0) - N(X_l <0)\over N(X_l > 0) + N(X_l < 0)}
\end{equation}
is sensitive to $\Delta g$.

All such single spin experiments, however, look difficult because of the small
cross-sections. It has also been pointed out that there is a correlation
between the polarization of the produced lepton or direct photon and the
polarization of the beam [CON 91], which depends on $\Delta g$. Such an
experiment might be feasible with muon pairs, but clearly it is very difficult
to monitor the final lepton polarization.

\subsubsection{Heavy quark production}
\vskip 6pt
\setcounter{equation}{0}

Probably the most sensitive experiment to $\Delta g$ and, in many ways, perhaps
the most realistic, is the polarization asymmetry in the production of heavy
quarks using longitudinally polarized beam and target [COR 88]. In this case
the dominant subprocess is the gluon fusion reaction $g + g \to q + \bar{q}$
and the asymmetry will be proportional to $(\Delta g)^2$. Sizable asymmetries
are expected and should be measurable at RHIC.

\subsection{Jet handedness}
\vskip 6pt
\setcounter{equation}{0}

If it were possible to measure the polarization of the  high $p_T$ quark
produced in the partonic reaction
\begin{equation}
q + \vec g \to \vec q + g
\end{equation}
where an unpolarized quark collides with a polarized gluon, then one would
have a very direct measure of the gluon polarization. The advantage of such an
approach is that it requires only an unpolarized beam colliding with a
polarized target. The problem is to measure the final quark's polarization in
the dominant process where it hadronizes into a jet. Thus one has to study the
reaction
\begin{equation}
p + \vec{p} \to jet + X
\end{equation}
and somehow monitor the polarization of the quark which initiated the jet
[STR 92]. (One is effectively trying to measure the longitudinal spin transfer
$D_{LL}$ [BOU 80].)

One can define the {\it handedness} of a jet and show that it is
proportional to the polarization of the initiating parton [NAC 77; EFR 78,
92a].
Namely, analogously to (11.2.2) one can define
\begin{equation}
X_L \equiv {{\bfP}_J \cdot ({\bfp}_1 \times {\bfp}_2)\over |{\bf P}_J|}
\end{equation}
where ${\bfP}_J$ is the jet momentum and ${\bfp}_{1,2}$ are the momenta of two
particles in the jet chosen according to some definite criteria. For example
we could choose the leading and the second leading particles in the jet.

The handedness is then defined in terms of the number of jets with
$X_L > 0$ and $X_L < 0$, {\it i.e.}
\begin{equation}
H_{jet} = {N(X_L>0) - N(X_L<0)\over N(X_L>0) + N(X_L<0)} \,\cdot
\end{equation}
(An analogous quantity can be defined based on using the momenta of the
{\it {three}} most leading particles in the jet).

The essential point is that
\begin{equation}
H_{jet} = \alpha_J \, P_q
\end{equation}
where $P_q$ is the degree of longitudinal polarization of the initiating quark
and $\alpha_J$ is the `jet analysing power', the calculation of which lies in
the realm of non-perturbative QCD. Thus the key issue is the size of
$\alpha_J$.

Miraculously $\alpha_J$ can be measured! The point is that in the electroweak
process
\begin{equation}
e^+e^- \to Z^0 \to q\bar{q} \to 2~jets
\end{equation}
the $q$ and $\bar{q}$ are significantly polarized and their polarization is
known with some confidence from electroweak theory. Thus by measuring
$H_{jet}$ in (11.3.5) we effectively measure $\alpha_J$. One can argue
\footnote{See [NAC 77]: note that $\alpha_J$ corresponds to the quantity
${L-R\over L + R}$ in that paper.}
\setcounter{footnote}{0}
that it is also possible to measure
$\alpha_J$ independently of electroweak theory by studying the correlations
between the handeness of the two jets in (11.3.6).

There is a theoretical estimate that $\alpha_J \simeq$ (5 -- 10)\% [RYS 93,
EFR 94]. If this is confirmed by measurements of (11.3.5) then it will open up
the real possibility of measuring $\Delta g$ at presently existing
accelerators, {\it i.e} at the Fermilab TEVATRON, the CERN SPS and the
Serpukhov UNK.
%
%end of Section 11
%
\setcounter{section}{11}
\section{\large{Conclusions}}
\vskip 6pt

Polarized Deep Inelastic Scattering has proved to be an exciting and
controversial field in which unexpected experimental results have engendered
the discovery of unexpectedly subtle theoretical problems.

Much new data has become available in the past years from the SMC at CERN and
from the E142 at SLAC, including, for the first time, results on the neutron.
And the controversial EMC proton data of 1987 have been confirmed by the SMC.

The principal points that have emerged are:
\begin{description}
\item{1. } The Bjorken sum rule has been tested, and provided sufficient care
is
exercised in dealing with experiments at different mean $Q^2$ and in
incorporating perturbative QCD corrections, there is absolutely no evidence
for any violation of the sum rule. Earlier claims to the contrary were simply
too cavalier in handling the data.
\item{2. } The measurement of the first moment $\Gamma_1^p$ of $g_1^p(x)$ does
not measure the contribution of the quark spins to the spin of the proton.
There
is an anomalous gluon contribution which significantly alters the theoretical
expression for $\Gamma_1^p$ so that the smallness of $\Gamma_1^p$ no longer
implies an almost zero contribution from the quark spins. This explanation
requires a significant polarized gluon density and this feature {\it must} be
tested independently. The polarized gluon number density $\Delta g$ can be
measured in polarized semi-inclusive deep inelastic lepton-hadron interactions,
and in hadron-hadron collisions involving polarized beam and target. Both
hard $\gamma$ and Drell-Yan reactions are sensitive to $\Delta g$, as is the
polarization asymmetry in heavy quark production. Another possible approach is
via ``jet handedness". RHIC seems to be an ideal machine for exploring the
property of $\Delta g$.
\item{3. } While $\Gamma_1^p$ itself no longer determines the contribution of
the quark spins to the spin of the nucleon, there does exist a non-perturbative
Generalized Goldberger-Treiman relation which connects the quark spin to the
coupling of the $\eta^\prime$ to nucleons. This coupling can in principle
be deduced from a detailed phase-shift analysis of elastic nucleon-nucleon
scattering or from some other experiments. Estimates already exist but a more
accurate determination would be of great interest.
\item{4. } It turns out that $g_2(x)$ cannot be calculated in the Naive Parton
Model nor in the QCD Improved Parton Model. Nonetheless an expression can be
obtained for it in a field theoretic approach. The result does not have a
partonic interpretation.
\item{5. } Neither the Burkhardt-Cottingham nor the Efremov-Leader-Teryaev
sum rules follow from the Operator Product Expansion. Indeed if simple Regge
arguments can be trusted, the behaviour at small $x$ of $g_2$ should lead to
divergent integrals in these sum rules. But the argument is not watertight and
every effort should be made to test the sum rules experimentally
\footnote{Note added in proof: The first few data on $g_2^p(x)$ which have just
appeared [ADA 94a] have large errors and do not allow yet a significant test
of the sum rules.}.
\setcounter{footnote}{0}
\end{description}
%
%end of Conclusions
%
%\newpage
\vskip 24pt
\parindent=1cm
\noindent
{\bf Appendix A - Some kinematic relations amongst asymmetries and scaling}

{\bf functions}
\vskip 6pt

The spin--spin asymmetries $A_{\parallel}$ and $A_{\perp}$, Eqs. (2.1.27) and
(2.1.38) respectively, are usually expressed in terms of virtual Compton
scattering asymmetries $A_{1,2}$, as
$$
A_{\parallel} = D (A_1 + \eta  A_2)  \eqno{\mbox{\rm (A.1)}}
$$
and
$$
A_\bot = d\,(A_2-\xi A_1)~, \eqno{\mbox{\rm (A.2)}}
$$
where
$$
A_1 = \frac{M \nu G_1 - Q^2 G_2}{W_1} =
\frac {g_1 - (4M^2 x^2/Q^2) g_2}{F_1}
\eqno{\mbox{\rm (A.3)}}
$$
$$
A_2 =  \sqrt{Q^2} \, \frac{M G_1 + \nu G_2}{W_1} = \frac {2Mx}{\sqrt{Q^2}}
{}~\frac{g_1+g_2}{F_1} \eqno{\mbox{\rm (A.4)}}
$$
$$
D = \frac{E-\epsilon E^\prime}{E(1+\epsilon R)} \qquad \qquad
\eta = \frac{\epsilon \sqrt{Q^2}}{E-\epsilon E^\prime}
\eqno{\mbox{\rm (A.5)}}
$$
$$
d = D\,\sqrt{\frac{2\epsilon}{1+\epsilon}} \qquad\qquad \xi = \eta
\,\frac{1+\epsilon}{2\epsilon}
$$
with
$$
\frac{1}{\epsilon} = 1+2 \left(1+\frac{\nu^2}{Q^2}\right) {\mbox{\rm tg}}^2
\frac{\theta}{2}\eqno{\mbox{\rm (A.6)}}
$$
and, as in Eq. (2.1.32),
$$
R = \frac{W_2}{W_1} \left(1+\frac{\nu^2}{Q^2}\right) -1= \frac{F_2}{2xF_1}
\left(1+\frac{4M^2x^2}{Q^2}\right) -1~.
\eqno{\mbox{\rm (A.7)}}
$$

The analysis of $A_\parallel$ proceeds through subsequent approximations.
One first notices
that both the coefficient $\eta$ and $A_2$, which is bounded by [LEA 85]
$$
|A_2| \le \sqrt{R}~,\eqno{\mbox{\rm (A.8)}}
$$
are small, so that it is reasonably accurate to write
$$
A_\parallel \approx DA_1~.\eqno{\mbox{\rm (A.9)}}
$$
One then neglects the $g_2$ term in $A_1$, Eq. (A.3), and replaces $F_1(x)$ by
$$
F_1 (x) = \frac{F_2(x)}{2x [1+R(x)]} \eqno{\mbox{\rm (A.10)}}
$$
which originates from Eq. (A.7) for $4M^2x^2 \ll Q^2$.

Thus one obtains
$$
g_1 (x) \approx \frac{A_\parallel}{D} \frac{F_2(x)}{2x[1+R(x)]}~,
\eqno{\mbox{\rm (A.11)}}
$$
which expresses $g_1 (x)$ in terms of the measured asymmetry $A_\parallel$,
the unpolarized structure function $F_2 (x)/[2x(1+R)]$ (also measured) and
the known coefficient D. It can be shown that all the above approximations
are actually quite harmless [LEA 88].

The analysis of $A_\perp$ is dicussed in Section 2.1.3.

\vskip 12pt
\noindent
{\bf Appendix B - Current matrix elements in the Parton Model}
\vskip 6pt

In the Standard Model all hadronic currents are expressed in terms of quark
fields $\psi_f(x)$ where $f$ labels the flavour. We require the forward matrix
elements of these currents taken between nucleon states. We shall use the
quark-parton model to evaluate these in the frame $S^\infty$ where the proton
moves along $OZ$ at high speed.

Consider the generic current
$$
J_f = \bar{\psi}_f \Gamma \psi_f\eqno{\mbox{\rm (B.1)}}
$$
where $\Gamma = \gamma^\alpha$ or $\gamma^\alpha \gamma_5$, evaluated at the
space-time origin $x_\mu = 0$. We shall evaluate the forward matrix elements
of $J_f$ in a proton state with covariant spin vector $S_\mu$.

We insert complete sets of quark states, corresponding to partons of
four-momentum $k$ and helicity $\lambda$.

Because the proton is polarized we need to introduce $n_f(\bfk,\lambda;S)
d^3\bfk$ as the number of flavour $f$ quarks with three-momentum in the range
$\bfk \to \bfk+ d^3\bfk$ and with helicity $\lambda$ inside a proton of
four-momentum $P$ and covariant spin vector $S_\mu$. After some
algebra one finds for the {\it forward} hadronic matrix elements,
$$
\langle P,S |J_f|P,S \rangle = \sum_\lambda \int d^3 \bfk \,
\frac{P_0}{\epsilon} \, n_f(\bfk,\lambda;S)
\langle k,\lambda|J_f|k,\lambda \rangle
\eqno{\mbox{\rm (B.2)}}
$$
where $\epsilon$ is the parton energy and where we have utilised the fact that
for fast quarks $\gamma^\alpha$ and $\gamma^\alpha\gamma^5$ conserve helicity.

Let us now define more specifically
$$
J^\alpha_f = \bar{\psi}_f \gamma^\alpha \psi_f \qquad
J^{5\alpha}_{f} = \bar{\psi}_f\gamma^\alpha \gamma_5 \psi_f\,.
\eqno{\mbox{\rm (B.3)}}
$$
Then treating the quark-partons as `free' particles in $S^\infty$ one has
$$
\langle k,\lambda|J^\alpha_f|k,\lambda \rangle =
\bar{u}_\lambda \gamma^\alpha u_\lambda = 2k^\alpha
\eqno{\mbox{\rm (B.4)}}
$$
and
$$
\langle k,\lambda|J^{5\alpha}_{f}|k,\lambda \rangle = \bar{u}_\lambda
\gamma^\alpha \gamma_5 u_\lambda = 2m_f s^\alpha (\lambda)
\eqno{\mbox{\rm (B.5)}}
$$
where $s^\alpha(\lambda)$ is the parton's covariant spin vector corresponding
to a helicity state $\lambda = \pm \frac{1}{2}$,
$$
s^\alpha (\lambda) = \frac{2\lambda}{m_f} (|\bfk|,\epsilon \hat{\bfk})\,.
\eqno{\mbox{\rm (B.6)}}
$$

On grounds of Lorentz covariance the forward nucleonic matrix elements have
a similar structure:
$$
\langle P,S |J^\alpha_f|P,S \rangle = 2v_f P^\alpha \eqno{\mbox{\rm (B.7)}}
$$
and
$$
\langle P,S |J^{5\alpha}_{f}|P,S \rangle = 2M a_f S^\alpha
\eqno{\mbox{\rm (B.8)}}
$$
where $v_f, a_f$ are numbers (Lorentz scalars) that measure the strength of
the vector and axial-vector matrix elements. As explained in [BAI 82] the
numbers $v_f$ are exactly
known when the currents $J^\alpha_f$ are conserved currents.

Substituting (B.4 and 5) into (B.2) and using (B.7 and 8) one finds
$$
P^\alpha v_f = \sum_\lambda \int d^3\bfk \, \frac{P_0}{\epsilon} \,
n_f(\bfk,\lambda;S) \, k^\alpha
\eqno{\mbox{\rm (B.9)}}
$$
and
$$
MS^\alpha a_f = \sum_\lambda \int d^3\bfk \, \frac{P_0}{\epsilon} \,
n_f(\bfk,\lambda;S) \, m_f s^\alpha(\lambda)~.
\eqno{\mbox{\rm (B.10)}}
$$
Now taking in $S^\infty$
$$
k \approx \left( x^\prime P + \frac{m^2 + \bfk^2_T}{2x^\prime P},~\bfk_T,
{}~x^\prime P \right)                             \eqno{\mbox{\rm (B.11)}}
$$
and using it in (B.6), and taking
$$
S^\alpha (\Lambda) = \frac{2\Lambda}{M} \left( P,~0,~0,~-P-\frac{M^2}{2P}
\right)                                             \eqno{\mbox{\rm (B.12)}}
$$
for a proton of helicity $\Lambda = \pm \frac{1}{2}$, we find eventually,
from (B.9 and 10)
$$
v_f = \int^1_0 dx^\prime~[q^P_f(x^\prime) +q^A_f(x^\prime)] =
\int^1_0 dx~q_f (x) \eqno{\mbox{\rm (B.13)}}
$$
and
$$
a_f = \int^1_0 dx^\prime~[q^P_f(x^\prime)-q^A_f (x^\prime)] =
\int^1_0 dx~\Delta q_f (x)\,. \eqno{\mbox{\rm (B.14)}}
$$
where $q_f^P(x)$ and $q_f^A(x)$ are respectively the number densities of quarks
$f$ with spin parallel or antiparallel to the nucleon spin.

In the above we considered only quarks. A similar arguments holds for antiquark
constituents, where the partonic matrix elements are
$$
\langle k,\lambda|J^\alpha_f|k,\lambda \rangle = -\bar{v}_\lambda
\gamma^\alpha v_\lambda = -2k^\alpha
$$
$$
\langle k,\lambda|J^{5\alpha}_{f}|k,\lambda \rangle =
-\bar{v}_\lambda\gamma^\alpha\gamma_5 v_\lambda = 2m_f s^\alpha (\lambda)\,.
\eqno{\mbox{\rm (B.15)}}
$$

Finally then one finds
$$
v_f = \int^1_0 dx~[q_f(x)-\bar{q}_f (x)] \eqno{\mbox{\rm (B.16)}}
$$
$$
a_f = \int^1_0 dx~[\Delta q_f (x) + \Delta \bar{q}_f (x)]~.
\eqno{\mbox{\rm (B.17)}}
$$

The currents occuring in DIS are all just linear combination of the currents
$J_f^\alpha$ and $J_{f}^{5\alpha}$ so that all results quoted in Section 4.3
follow trivially from (B.7, 8, 16 and 17).

\vskip 12pt
\goodbreak
\noindent
{\bf Appendix C - Transverse spin: the restoration of electromagnetic
gauge}

{\bf invariance}
\nobreak
\vskip 6pt

We work in a reference system where the nucleon 4-momentum $P^\mu$ is given by
(10.1.8) and where the null-vector $n^\mu$ is as in (10.1.9).

We shall choose the gauge $A\cdot n = 0$ for simplicity. In the general
treatment [EFR 84] where $E^{\mu\nu} (...,k;...)$ of (10.1.3) does not refer
just to the Born (`handbag') diagram, the technique is to replace (10.2.1) by
$$
(q+k)^\rho=q^\rho+(k\cdot n) P^\rho + \Delta k^\rho\eqno{\mbox{\rm (C.1)}}
$$
and to do a Taylor expansion in the small quantity $\Delta k^\rho$, so that
(10.1.1) will involve $(\partial E^{\mu\nu})/(\partial k_\rho)$ evaluated at
$k^\rho = (q+xP)^\rho$.

The contribution to $W^{\mu\nu}$ from Fig. 10.3 which involves gluon exchange
($G$) is of the form (we show explicitely the colour labels $i,j,a$)
\begin{eqnarray}
W^{\mu \nu}_G&=&{1\over 2\pi}\int{d^4k_1\over (2\pi)^4}\,{d^4k_2\over (2\pi)^4}
{}~E^{\mu\rho\nu}_{\beta\alpha}(k_1,k_2)\,g\, \bigl(\lambda^a/2\bigr)_{ji}
\nonumber\\
&& \times\int d^4z_1\,d^4z_2~e^{ik_1z_1}\,e^{i(k_2-k_1)z_2}
\langle P,S |\bar{\psi}_{\beta,j}(0)A^a_\rho (z_2)\psi_{\alpha,i}(z_1)|
P,S \rangle\,.\nonumber
\end{eqnarray}
\vskip-1.5cm\hfill(C.2)
\vskip1cm

We absorb the colour matrix $(\lambda^a/2)_{ji}$ and the factor $g$ into the
hadronic matrix element to obtain
$$
W_G^{\mu\nu}=\frac{1}{2\pi} \int \frac{d^4k_1}{(2\pi)^4} \,
\frac{d^4k_2}{(2\pi)^4} \,\mbox{\rm Tr}\,\Bigl[ E^{\mu\rho\nu}(k_1,k_2)
\, \Phi^G_\rho (P,S;k_1,k_2)\Bigr]
\eqno{\mbox{\rm (C.3)}}
$$
where
$$
\Phi^G_\rho = \int d^4z_1 d^4z_2~e^{ik_1z_1}~e^{i(k_2-k_1)z_2}
\langle P,S| \bar{\psi}(0) \, gA_\rho (z_2) \, \psi (z_1)|P,S \rangle
\eqno{\mbox{\rm (C.4)}}
$$
and where now $\psi$ is a column vector in spin space and in colour space and
$A_\rho$ is the matrix $A_\rho = (\bflambda^a/2) A^a_\rho.$ The trace in (C.3)
is only over spinor labels.

In the general treatment a key r\^ole is played by the Ward identity
$$
\frac{\partial E^{\mu\nu}}{\partial k_\rho}(q+xP)=E^{\mu\rho\nu} (q+xP, \,
q+xP)
\eqno{\mbox{\rm (C.5)}}
$$
which permits a connection between the leading order part of $W_G^{\mu\nu}$
and the Taylor expansion of $W^{\mu\nu}$ from the quark diagram of Fig. 10.2 .
The reader is referred to [EFR 84] for the general analysis.
Here we shall illustrate the phenomenon by taking the simplest case where
$E^{\mu\nu}$ is given by the Born diagram in Fig. 10.2 and $E^{\mu\rho\nu}$
by the Born diagram shown in Fig. C.1.
%\vskip3cm
%\begin{center}
%Fig. B.1
%\end{center}
Both $E^{\mu\nu}$ and $E^{\mu\rho\nu}$ are independent of colour.

For this simplest case it is easier not to use the Taylor expansion (C.1) in
(10.1.4) but to split $q\sla + k\sla + m_q$ into $q\sla$, which yields the
result (10.3.4), and the contribution from $k\sla + m_q$, which we shall call
$\Delta W_{\mu\nu}^{(A)}$, where
$$
\Delta W^{\mu\nu}_{(A)}=\frac{e_f^2}{2}\int\frac{d^4k}{(2\pi)^4} \, \biggl\{
\varepsilon^{\mu\nu\rho\sigma} \mbox{\rm Tr} \bigl[ \gamma_\sigma \gamma_5
\Phi_\rho \bigr] - m_q \mbox{\rm Tr} \bigl[\sigma^{\mu\nu}\Phi\bigr]
\biggr\}~\delta [(k+q)^2-m_q^2]
\eqno{\mbox{\rm (C.6)}}
$$
where we have used (10.1.5) and (10.1.7) and where
$$
\Phi_\rho=\int d^4z~e^{ikz}\langle P,S |\bar{\psi}(0)i\partial_\rho\psi (z)
|P,S \rangle \,.
\eqno{\mbox{\rm (C.7)}}
$$

To obtain (C.7) we have used the fact that $k_\rho$ can be obtained by
differentiating $e^{ikz}$, plus the usual assumption that the hadronic matrix
element vanishes for $|z|\to \infty$. Note for future reference that by
translational invariance one can show that
$$
\bar{\psi}(0) i\rpartial_\rho \psi(z) \Longleftrightarrow \bar{\psi}(0)
(-i \lpartial_\rho) \psi(z)
\eqno{\mbox{\rm (C.8)}}
$$
are equivalent inside the matrix element in (C.7).

Introducing $I = \int dx~\delta(x-k\cdot n)$, and using (C.8), Eq. (C.6) can
eventually be written as
$$
\Delta W^{\mu\nu}_{(A)} = \frac{e_f^2}{8P\cdot q} \int dx~\delta (x-x_{Bj})
\langle \varepsilon^{\mu\nu\rho\sigma} \gamma_\sigma\gamma_5
(-i\rpartial_\rho + i\lpartial_\rho)
-2m_q \sigma^{\mu\nu}\rangle
\eqno{\mbox{\rm (C.9)}}
$$
where we have introduced the shorthand for any X
$$
\langle X\rangle\equiv\int\frac{d\tau}{2\pi}~e^{i\tau x}
\langle P,S_T|\bar{\psi}(0) X \psi (\tau n)|P,S_T \rangle \,.
                                                    \eqno{\mbox{\rm (C.10)}}
$$

Turning now to the gluon diagram, using the Born approximation in Fig. C.1,
we keep only the leading terms in the quark propagators:
$$
q^\rho+k_1^\rho \approx q^\rho+(k_1\cdot n) P^\rho =
\int dx_1~\delta (x_1 - k_1\cdot n)(q^\rho + x_1P^\rho)
$$
$$
q^\rho+k_2^\rho \approx q^\rho+(k_2\cdot n) P^\rho =
\int dx_2~\delta (x_2 - k_2\cdot n)(q^\rho + x_2P^\rho)\,.
\eqno{\mbox{\rm (C.11)}}
$$
After some manipulation (C.3) becomes
\begin{eqnarray}
W_G^{\mu\nu} &=& \frac{e^2_f}{8P\cdot q} \int dx_1 \, dx_2 \, \langle\langle
\gamma^\nu (q\sla + x_2 P\Cslas) g A\Cslas (\tau_1n)(q\sla + x_1 P\Cslas)
\gamma^\mu \rangle\rangle \nonumber\\
&&\times \mbox{\rm Disc}
\left[\frac{1}{(x_1-x_{\rm Bj})(x_2-x_{\rm Bj})} \left(\frac{i}{\pi P\cdot
q}\right)\right] \nonumber
\end{eqnarray}
\vskip-1.5cm\hfill(C.12)
\vskip0.5cm
\noindent
where, for any $X$,
\begin{eqnarray*}
\langle\langle X(x_1,x_2;\tau_1)\rangle\rangle \equiv \int
\frac{d\tau_1}{2\pi}\frac{d\tau_2}{2\pi} e^{i\tau_2x_2}
e^{i\tau_1(x_1-x_2)} \\ \times \langle P,S_T|\bar{\psi}(0)
X(x_1,x_2,\tau_1n) \psi (\tau_2n)|P,S_T \rangle \,.
\end{eqnarray*}
\vskip-1.5cm\hfill                                                   (C.13)
\vskip1cm
\noindent

Note that if $X$ depends on only one of the variables $x_1$ or $x_2$, one
has the useful results
$$
\int dx_2~\langle\langle X(x_1;\tau_1)\rangle\rangle =
\langle X(x_1;\tau_1)\rangle
\eqno{\mbox{\rm (C.14)}}
$$
$$
\int dx_1~\langle\langle X(x_2;\tau_1)\rangle\rangle =
\langle X(x_2;0)\rangle\,.
\eqno{\mbox{\rm (C.15)}}
$$

Note that although (C.14, 15) look innocuous there is the hidden assumption
that the Fourier transforms in (C.10, C.13) can be inverted. We shall return
to this question in Appendix E.

The discontinuity in (C.12) is most simply evaluated by splitting into partial
fractions. It will turn out to be useful to do this in two ways:
\begin{eqnarray}
\mbox{\rm Disc} \left[ \frac{1}{(x_1-x_{\rm Bj})(x_2-x_{\rm Bj})}
\right] &=& \mbox{\rm Disc} \left\{ \frac{1}{x_2-x_1} \left[
\frac{1}{x_1-x_{\rm Bj}}-\frac{1}{x_2-x_{\rm Bj}} \right] \right\}
\nonumber\\
&& = \frac{-i\pi}{x_2-x_1} [\delta (x_1-x_{\rm Bj})-\delta (x_2-x_{\rm Bj})]
\nonumber
\end{eqnarray}
\vskip-1.5cm\hfill(C.16)
\vskip0.5cm
\noindent
or
\begin{eqnarray}
\mbox{\rm Disc} \left[ \frac{1}{(x_1-x_{\rm Bj})(x_2-x_{\rm Bj})}
\right] &=& \mbox{\rm Disc} \left\{ \frac{1}{x_1+x_2-2x_{\rm Bj}} \left[
\frac{1}{x_1-x_{\rm Bj}} + \frac{1}{x_2-x_{\rm Bj}} \right] \right\}
\nonumber\\
&=& \frac{-i\pi}{x_1+x_2-2x_{\rm Bj}} [\delta
(x_1-x_{\rm Bj})+\delta (x_2-x_{\rm Bj})] \,.\nonumber
\end{eqnarray}
\vskip-1.5cm\hfill(C.17)
\vskip0.5cm

The $\gamma$-matrix identity
$$
\gamma^\alpha \gamma^\rho \gamma^\beta -g^{\alpha\rho} \gamma^\beta =
g^{\rho\beta}\gamma^\alpha-g^{\alpha\beta}\gamma^\rho
-i\varepsilon^{\alpha\rho\beta\sigma} \gamma_\sigma \gamma_5
\eqno{\mbox{\rm (C.18)}}
$$
is helpful in simplifying the $\gamma$-algebra and will be crucial at a later
stage of the analysis.

When using this to simplify $\gamma^\nu (q\sla + x_2 P\Cslas) A\Cslas
(q\sla + x_1 P\Cslas) \gamma^\mu$, and when the piece antisymmetric in $\mu\nu$
is extracted, one can discard all terms of the type $\gamma^\nu (P\Cslas$ or
$q\sla) (P\cdot A$ or $q\cdot A) \gamma^\mu$, since they give a vanishing
contribution in leading order. To see
this, via (10.1.5), such contributions involve terms like
$$
(P^\rho~\mbox{\rm or}~q^\rho)(P_\alpha~\mbox{\rm or}~q_\alpha)
\varepsilon^{\mu\nu\alpha \sigma} \langle P ,S_T|
\bar{\psi} (0) \gamma_\sigma \gamma_5 A_\rho(\tau_1n)\psi(\tau_2 n)
|P ,S_T \rangle \,.
\eqno{\mbox{\rm (C.19)}}
$$
Now the matrix element is a pseudotensor. The leading terms linear in $S_T$
can then only have the forms $P_\sigma S^T_\rho$ and $S^T_\sigma P_\rho$.
But the latter must be absent since in our gauge $A\cdot n = 0$. Thus (C.19)
involves $P\cdot S_T$ or $q \cdot S_T$ which both vanish.

Finally, the antisymmetric part of the $\gamma$-matrix expression in (C.12)
becomes
$$
g (P\cdot q)~\Bigl\{ -i\varepsilon^{\mu\nu\rho\sigma} \gamma_\sigma \gamma_5
A_\rho(\tau n)(x_1+x_2-2x_{\rm Bj})+(x_1+x_2) n\slas [P^\nu A^\mu
(\tau n) - P^\mu A^\nu (\tau n)] \Bigr\}\,.
\eqno{\mbox{\rm (C.20)}}
$$

Combining this judiciously with (C.16 and 17) we get for the antisymmetric
part of (C.12)
\begin{eqnarray}
W^{\mu\nu}_{G(A)} &=& \frac{e^2_f}{8P\cdot q} \int dx_1 \, dx_2 ~ \Bigl\{
g\,\delta (x_1-x_{\rm Bj})\langle\langle -\varepsilon^{\mu\nu\rho\sigma}
\gamma_\sigma \gamma_5 A_\rho(\tau n)\nonumber\\
&-&in\slas [P^\nu A^\mu (\tau n) - P^\mu A^\nu (\tau n)]\rangle\rangle
- g\,\delta(x_2-x_{\rm Bj}) \langle\langle \varepsilon^{\mu\nu\rho\sigma}
\gamma_\sigma \gamma_5 A_\rho(\tau n) \nonumber\\
&-& i n\slas [P^\nu A^\mu(\tau n) -
P^\mu A^\nu(\tau n)]\rangle\rangle \Bigr\}\,.\nonumber
\end{eqnarray}
\vskip-0.9cm\hfill(C.21)
\vskip0.5cm

The trivial dependence on $x_1,x_2$ allows us to use (C.14 and 15) to obtain
\begin{eqnarray}
W^{\mu\nu}_{G(A)} &=& \frac{e^2_f}{8P\cdot q} \int dx~\delta (x-x_{\rm Bj})
\Bigl\{ \langle -\varepsilon^{\mu\nu\rho\sigma} \gamma_\sigma \gamma_5
A_\rho(0)\nonumber\\
&-& i n\slas [P^\nu A^\mu (0) -P^\mu A^\nu (0)]\rangle - \langle
\varepsilon^{\mu\nu\rho\sigma} \gamma_\sigma \gamma_5 A_\rho(\tau n)
\nonumber\\
&-& in\slas [P^\nu A^\mu (\tau n) -P^\mu A^\nu (\tau n)]\rangle
\Bigr\}\,.\nonumber
\end{eqnarray}
\vskip-1.5cm\hfill(C.22)
\vskip1cm

Now note that the structures of $\Delta W^{\mu\nu}_{(A)}$ in (C.9) and
$W^{\mu\nu}_{G(A)}$ in (C.22) are identical, so we may add them to obtain
\begin{eqnarray}
\Delta W^{\mu\nu}_{(A)}+W^{\mu\nu}_{G(A)} &=& \frac{e^2_f}{8P\cdot q} \int dx
{}~\delta(x-x_{\rm Bj}) \Bigl\{ \langle \varepsilon^{\mu\nu\rho\sigma}
\gamma_\sigma\gamma_5 \rder_\rho \!\!(\tau n)\nonumber\\
&-&m_q\sigma^{\mu\nu}-ign\slas [P^\mu A^\nu (\tau n) -
P^\nu A^\mu (\tau n)] + \varepsilon^{\mu\nu\rho\sigma} \gamma_\sigma
\gamma_5 \lder_\rho \!\!(0)\nonumber\\
&-&m_q\sigma^{\mu\nu} + ign\slas [P^\mu A^\nu (0) - P^\nu A^\mu (0)]
\rangle \Bigr\} \nonumber
\end{eqnarray}
\vskip-0.9cm\hfill(C.23)
\vskip0.5cm
\noindent
where we have introduced
\begin{eqnarray}
\rder_\rho \!\!(z) &=& i\rpartial_\rho - g A_\rho (z)\nonumber\\
\lder_\rho \!\!(z) &=& -i\lpartial_\rho -g A_\rho (z)\,.\nonumber
\end{eqnarray}
\vskip-1cm\hfill(C.24)
\vskip0.5cm
\noindent

Consider first the terms involving $n\slas$. We can simplify them by
analysing the structure of the matrix element
$$
\langle P,S_T|\bar{\psi} \gamma^\mu A^\nu \psi|P, S_T\rangle\,.
$$
It is a true tensor which can only be built from the vectors $P^\alpha,
n^\alpha$ and
\begin{eqnarray}
v^\alpha &=& \varepsilon^{\alpha\rho\sigma\lambda} S^T_\rho P_\sigma
n_\lambda \quad\quad v^2=-1\nonumber\\
P\cdot v &=& n\cdot v = S^T\cdot v = 0~, \nonumber
\end{eqnarray}
\vskip-1cm\hfill(C.25)
\vskip0.5cm
\noindent
and the pseudovector $S^T_\alpha$, and which must be linear in $S^T_\alpha$.
The only possible forms are $P^\mu v^\nu$ or $P^\nu v^\mu$ of which the latter
cannot occur because of the gauge condition $A\cdot n = 0$. It follows that
$$
\langle P\Cslas A^\nu \rangle = \langle S\Cslas_T A^\nu \rangle
=\langle v\sla A^\nu \rangle = 0\,.
\eqno{\mbox{\rm (C.26)}}
$$

If therefore we write, to leading order,
$$
\gamma^\mu A^\nu = [n\slas P^\mu + P\Cslas n^\mu - S\Cslas_T S^\mu_T
- v\sla v^\mu]A^\nu
\eqno{\mbox{\rm (C.27)}}
$$
then inside (C.23) only the term $n\slas P^\mu$ contributes.
Inside the matrix element we thus have the equivalence
$$
n\slas P^\mu A^\nu \Longleftrightarrow \gamma^\mu A^\nu\,.
\eqno{\mbox{\rm (C.28)}}
$$
We may thus, in (C.23), make the replacement
$$
n\slas g [P^\mu A^\nu - P^\nu A^\mu] \to \gamma^\mu g A^\nu
-\gamma^\nu g A^\mu\,.
\eqno{\mbox{\rm (C.29)}}
$$
We may also, as a consequence of (C.8) add $0=\langle
\rpartial_\rho + \lpartial_\rho \rangle$ so that {\it e.g.}
$$
\langle i\gamma^\mu g A^\nu(0)-i\gamma^\mu gA^\nu(\tau n)\rangle
= \langle -i\gamma^\mu \lder^\nu\!\!(0) + i\gamma^\mu \rder^\nu\!\!(\tau n)
\rangle\,.
\eqno{\mbox{\rm (C.30)}}
$$

Finally, (C.23) becomes:
\begin{eqnarray}
\Delta W^{\mu\nu}_{(A)}+W^{\mu\nu}_{G(A)} &=& \frac{e^2_f}{8P\cdot q} \int dx
{}~\delta(x-x_{\rm Bj}) \Bigl\{ \langle \varepsilon^{\mu\nu\rho\sigma}
\gamma_\sigma\gamma_5 \rder_\rho\!\!(\tau n) - m_q\sigma^{\mu\nu}\nonumber\\
&+& i \gamma^\mu \rder^\nu \!\!(\tau n)-i\gamma^\nu \rder^\mu \!\!(\tau n)
\rangle + \langle \varepsilon^{\mu\nu\rho\sigma} \gamma_\sigma\gamma_5
\lder_\rho \!\!(0)\nonumber\\
&-& m_q\sigma^{\mu\nu} -i\gamma^\mu \lder^\nu \!\!(0) + i\gamma^\nu \lder^\mu
\!\!(0)\rangle \Bigr\} \,.\nonumber
\end{eqnarray}
\vskip-1cm\hfill(C.31)
\vskip0.5cm
We shall now demonstrate that both matrix elements in (C.31) vanish as a
consequence of the equations of motion
$$
[\rdersl(z)-m_q] \psi (z) = 0 \quad\quad \bar{\psi}(z)
[\ldersl(z)-m_q] = 0 \,.
\eqno{\mbox{\rm (C.32)}}
$$

The starting point is the $\gamma$-matrix identity which follows from (C.18):
$$
0=-\sigma^{\mu\nu} \gamma^\rho + ig^{\nu\rho}\gamma^\mu -ig^{\mu\rho}
\gamma^\nu + \varepsilon^{\mu\nu\rho\sigma} \gamma_\sigma \gamma_5 \,.
\eqno{\mbox{\rm (C.33)}}
$$
Taking the scalar product with $\rder_\rho \!\!(z)$, acting on $\psi (z)$,
and using (C.32) yields
$$
0=[-m_q \sigma^{\mu\nu} + \varepsilon^{\mu\nu\rho\sigma} \gamma_\sigma\gamma_5
\rder_\rho \!\!(z) + i\gamma^\mu \rder^\nu \!\!(z) - i\gamma^\nu
\rder^\mu \!\!(z)] \psi (z)
\eqno{\mbox{\rm (C.34)}}
$$
which is exactly the structure that appears in the first matrix element of
(C.31). A similar argument shows that the second matrix element also vanishes.

Thus the miracle is achieved. The unwanted $xP\Cslas$ and mass terms have
disappeared as a consequence of the equations of motion when the two parton
handbag diagram is combined with the three parton gluon diagram. The result
is (10.3.4).

\vskip 12pt
\noindent
{\bf Appendix D - Distribution and two-parton correlation functions for
transverse}

{\bf spin}
\vskip 6pt

It is of some interest to consider the distribution and correlation functions
associated with the matrix elements of the quark and gluon operators appearing
in Appendix C and to derive certain relationships between them which follow
from the equations of motion. The structure of these correlation functions
will be further studied in Appendix E.

Consider first $\langle \sigma^{\mu\nu}\rangle$ in the notation of (C.10). Its
leading order tensorial structure can only be
$\epsilon^{\mu\nu\alpha\beta} P_\alpha S^T_\beta$. Thus we put
$$
\langle \sigma^{\mu\nu}\rangle = 4h_T (x) \epsilon^{\mu\nu\alpha\beta}
P_\alpha s^T_\beta                                      \eqno{\mbox{\rm (D.1)}}
$$
and find that \footnote{Our $h_T(x) = 1/2 h^{COR}_T (x)$ for reasons
explained after Eq. (10.2.11).}
$$
4h_T(x) = \int \frac{d\tau}{2\pi}~e^{i\tau x}~\widetilde{h}_T (\tau)
                                                     \eqno{\mbox{\rm (D.2)}}
$$
where
$$
\widetilde{h}_T (\tau) = \langle P,S_T|\bar{\psi}(0)n\slas\gamma_5 S\Cslas_T
\psi (\tau n)|P,S_T \rangle \,.                   \eqno{\mbox{\rm (D.3)}}
$$

It is $h_T(x)$ that is the true analogue of $h_L(x)$ of (10.2.10) and which
has a simple parton interpretation in terms of probabilities to find quarks
with transverse spin in a transversely polarized nucleon. As we have seen,
however, in DIS this term appears multiplied by the quark mass [Eq. (10.1.7)]
and in fact eventually cancels out. Nonetheless it is a fundamental
distribution reflecting a different aspect of the structure of the nucleon,
and, as has been emphasized by [COR 92], can be measured
directly in Drell-Yan reactions.

Consider next the matrix element
$\langle\langle \gamma^\mu\rder^\nu \!\!(\tau n)\rangle\rangle$. In leading
order one can only have the tensor structures $P^\mu v^\nu$ or $P^\nu v^\mu$
but one can show that the latter term is not permitted by
time-reversal invariance. So one can put
$$
\langle\langle \gamma^\mu \rder^\nu \!\!(\tau n)\rangle\rangle =
MB^V (x_1,x_2) P^\mu v^\nu \,.                       \eqno{\mbox{\rm (D.4)}}
$$
It follows that
$$
B^V (x_1,x_2) = \int \frac{d\tau_1}{2\pi} \frac{d\tau_2}{2\pi}~
e^{i\tau_2x_2}\,e^{i\tau_1(x_1-x_2)}\,\widetilde{B}^V(\tau_1,\tau_2)
$$
where
$$
\widetilde{B}^V (\tau_1,\tau_2) = -\frac{1}{M} \langle P,S_T |
\bar{\psi}(0) n\slas(v\cdot \rder \!\!(\tau_1 n)\, \psi (\tau_2 n)
|P,S_T \rangle \,.                                  \eqno{\mbox{\rm (D.5)}}
$$

Note that it can be shown [EFR 84] that as a consequence of time reversal
invariance
$$
B^V (x_1,x_2) = -B^V (x_2,x_1)~.                    \eqno{\mbox{\rm (D.6)}}
$$

Now using (C.14) one has
$$
\langle i\gamma^\mu\rder^\nu \!\!(\tau n)-i\gamma^\nu\rder^\mu
\!\!(\tau n) \rangle = iM\int dx_2~B^V (x_1,x_2) \,
\varepsilon^{\mu\nu\alpha\beta} P_\alpha S^T_\beta  \eqno{\mbox{\rm (D.7)}}
$$
where we have used
\begin{eqnarray}
P^\mu v^\nu -P^\nu v^\mu &=& P^\mu \varepsilon^{\nu\alpha\beta\gamma}
S^T_\alpha P_\beta n_\gamma -P^\nu\varepsilon^{\mu\alpha\beta\gamma} S^T_\alpha
P_\beta n_\gamma\nonumber\\
&=& \varepsilon^{\mu\nu\alpha\beta} P_\alpha S^T_\beta \,.\nonumber
\end{eqnarray}
\vskip-1cm\hfill                                                       (D.8)
\vskip0.5cm

Finally, consider
$\langle\langle \gamma^\sigma\gamma_5 \rder^\rho \!\!(\tau n)\rangle\rangle$.
In leading order, one can only have
$$
\langle\langle \gamma^\sigma\gamma_5\rder^\rho \!\!(\tau_1n)\rangle\rangle =
MB^A(x_1,x_2)P^\sigma S^\rho_T + MB^A_T (x_1,x_2) P^\rho S_T^\sigma
                                                      \eqno{\mbox{\rm (D.9)}}
$$
with
$$
B^A (x_1,x_2) = \int \frac{d\tau_1}{2\pi}\frac{d\tau_2}{2\pi}~
e^{i\tau_2x_2}\,e^{i\tau_1(x_1-x_2)}\widetilde{B}^A(\tau_1,\tau_2)
                                                      \eqno{\mbox{\rm (D.10)}}
$$
where
$$
\widetilde{B}^A (\tau_1,\tau_2) = -\frac{1}{M} \langle P,S_T
|\bar{\psi}(0) n\slas\gamma_5 [S^T\cdot\rder\!\!(\tau_1n)]
\psi (\tau_2n)|P,S_T \rangle                      \eqno{\mbox{\rm (D.11)}}
$$
and
$$
B^A_T (x_1,x_2) = \int \frac{d\tau_1}{2\pi} \frac{d\tau_2}{2\pi}~
e^{i\tau_2x_2}\,e^{i\tau_1(x_1-x_2)}\widetilde{B}_T^A(\tau_1,\tau_2)
                                                      \eqno{\mbox{\rm (D.12)}}
$$
where
$$
\widetilde{B}^A_T (\tau_1,\tau_2) = -\frac{1}{M} \langle P,S_T
|\bar{\psi}(0) S\Cslas_T\gamma_5 n
[ n\cdot D(\tau_1n)] \psi (\tau_2n)
|P,S_T \rangle \,.                                    \eqno{\mbox{\rm (D.13)}}
$$
{}From time reversal invariance one has [EFR 84]
$$
B^A (x_1,x_2) = B^A (x_2,x_1)\,.                      \eqno{\mbox{\rm (D.14)}}
$$

Note that in (D.12), because of the gauge condition,
$$
n\cdot\rder \!\!(\tau_1 n) = in^\mu \partial_\mu\,,\eqno{\mbox{\rm (D.15)}}
$$
so that the $\tau_1$ integration can be done yielding
\begin{eqnarray*}
B^A_T (x_1,x_2) = -\frac{i}{M} \delta (x_1-x_2) \int \frac{d\tau_2}{2\pi}~
e^{i\tau_2x_2} \\ \times \langle P,S_T|\bar{\psi}(0)S\Cslas_T \gamma_5
n^\alpha\partial_\alpha \psi (\tau_2n)|P,S_T \rangle \,.
\end{eqnarray*}
                                                     \vskip-1cm\hfill(D.16)
\vskip0.5cm
It is fairly obvious, and is shown in Appendix E, that the derivative
$n\cdot\partial$ acting on $\psi$ is equivalent to $\partial/\partial\tau_2$
which may then be switched so as to act on the exponential, yielding
$$
B^A_T (x_1,x_2) = 4\delta (x_1-x_2) x_2 f_T (x_1)     \eqno{\mbox{\rm (D.17)}}
$$
where $f_T$ is the function appearing in (10.3.6).

Using (C.14) and (D.9) we have
$$
\langle \varepsilon^{\mu\nu\rho\sigma}\gamma_\sigma\gamma_5\rder_\rho \!\!
(\tau_1n)\rangle = M\int dx_2 ~ \{B^A_T (x_1,x_2)-B^A(x_1,x_2)\}~
\varepsilon^{\mu\nu\alpha\beta}P_\alpha S^T_\beta\,.  \eqno{\mbox{\rm (D.18)}}
$$

The vanishing of the matrix element of (C.34) thus implies, using
(D.1, 7 and 17)
$$
\int dx_2~\{ B^A_T(x_1,x_2)-B^A(x_1,x_2)+iB^V (x_1,x_2)\}=
\frac{4m}{M}h_T(x_1)~.                                \eqno{\mbox{\rm (D.19)}}
$$

Taking account of (D.17) this becomes:
$$
\int dx_2~\{iB^V(x_1,x_2)-B^A(x_1,x_2)\}=\frac{4m}{M}h_T (x_1)-4x_1f_T (x_1)~.
                                                      \eqno{\mbox{\rm (D.20)}}
$$
Use of the equations of motion acting to the left upon $\bar{\psi}(0)$
eventually leads to an analogous relationship:
$$
\int dx_1~\{iB^V(x_1,x_2)+B^A(x_1,x_2)\}=\frac{4m}{M}h_T(x_2)+4x_2f_T (x_2)~.
                                                      \eqno{\mbox{\rm (D.21)}}
$$

\vskip 12pt
\noindent
{\bf Appendix E - The Burkhardt--Cottingham and the Efremov--Leader--Teryaev}

{\bf Sum Rules in the QCD Field Theoretic Model}
\vskip 6pt

We shall show how the above sum rules can be derived from a careful study of
the structure of the matrix elements involved in Section 10 and in Appendix D.
The other ingredient is the equation of motion for $\psi (z)$. Also one has
to make assumptions about the invertability of certain Fourier transforms. One
sees very clearly where the weak point is in the `proof' of these sum rules
(see Sections 5.4 and 5.5). In the case of the Efremov--Leader--Teryaev
sum rule, the original derivation [EFR 84] was a little unsatisfactory,
because it appealed to a particular reaction to derive properties inherent to
the nucleon. The present derivation deals only with nucleon matrix elements.

Consider first the forward matrix element of the bilocal operator
$$\bar{\psi} (0) \gamma^\mu \gamma_5 \psi (z)$$ on the light cone $z^2=0$.
Its most general form is
$$
\frac{1}{M} \langle \bar{\psi}(0)\gamma^\mu\gamma_5 \psi (z) \rangle_{P,S} =
A_1 S^\mu + (z\cdot S) A_2 P^\mu +(z\cdot S) A_3 z^\mu   \eqno{\mbox{\rm
(E.1)}}
$$
where $\langle ~... ~ \rangle_{P,S}$ is short for $\langle P,S |~...~|
P,S \rangle$. The scalar functions $A_{1,2,3}$ are functions only of $z\cdot
P$.

{}From (E.1) we deduce
\begin{eqnarray}
\frac{1}{M} \langle \bar{\psi} (0)\gamma^\mu \gamma_5 \partial^\nu\psi (z)
\rangle_{P,S} &=&
A^\prime_1 S^\mu P^\nu +A_2 P^\mu S^\nu+A_3z^\mu S^\nu\nonumber\\
&+& (z\cdot S) [A^\prime_2 P^\mu P^\nu +A_3^\prime z^\mu P^\nu +
A_3 g^{\mu\nu}]\nonumber
\end{eqnarray}
\vskip-1cm\hfill                                                     (E.2)
\vskip0.5cm
where
$$
A^\prime \equiv \frac{dA(z\cdot P)}{d(z\cdot P)} \,\cdot
\eqno{\mbox{\rm (E.3)}}
$$

We now put $z^\mu =\tau n^\mu$ as required in the matrix elements involved
in Section 10 and Appendix D. Then
$$
\frac{1}{M} \langle \bar{\psi} (0)\gamma^\mu\gamma_5\psi (\tau n)
\rangle_{P,S} = A_1
S^\mu + \tau (n\cdot S) [A_2 P^\mu + \tau A_3n^\mu]
\eqno{\mbox{\rm (E.4)}}
$$
where now $A_i = A_i (\tau)$, and
\begin{eqnarray}
\frac{1}{M} \langle \bar{\psi}(0)\gamma^\mu\gamma_5\partial^\nu\psi(\tau n)
\rangle_{P,S} &=& A^\prime_1 S^\mu P^\nu + A_2P^\mu S^\nu +
\tau\{A_3 n^\mu S^\nu\nonumber\\
&+& (n\cdot S)[A^\prime_2 P^\mu P^\nu +\tau A^\prime_3 n^\mu P^\nu +
A_3 g^{\mu\nu}]\} \,. \nonumber
\end{eqnarray}
\vskip-1cm\hfill                                                      (E.5)
\vskip0.5cm
We assume that all scalar functions are such that $\tau A(\tau) \to 0$
as $\tau \to 0$ for all terms occurring in (E.4 and 5). Then at $\tau =0$
we have the simple structures
$$
\frac{1}{M} \langle \bar{\psi}(0)\gamma^\mu\gamma_5\psi (0)\rangle_{P,S}
= A_1(0)S^\mu                                       \eqno{\mbox{\rm(E.6)}}
$$
and
$$
\frac{1}{M} \langle \bar{\psi}(0)\gamma^\mu\gamma_5 \partial^\nu \psi (0)
\rangle_{P,S} = A^\prime_1 (0) S^\mu P^\nu +A_2 (0)P^\mu S^\nu\,.
\eqno{\mbox{\rm (E.7)}}
$$

We shall also require, from (E.5)
$$
\frac{1}{M} \langle \bar{\psi}(0)\gamma_5{\partial\slas}\psi (\tau n)
\rangle_{P,S} = -\tau (n\cdot S)[M^2 A^\prime_2+5A_3+\tau A^\prime_3]
\eqno{\mbox{\rm (E.8)}}
$$
so that at $\tau = 0$
$$
\frac{1}{M} \langle \bar{\psi}(0)\gamma_5{\partial\slas} \psi (0)
\rangle_{P,S} = 0 \,.                   \eqno{\mbox{\rm (E.9)}}
$$

Finally note, from (E.5) that
\begin{eqnarray}
\frac{1}{M} \langle \bar\psi(0)\gamma^\mu\gamma_5n\cdot\partial\psi(\tau n)
\rangle_{P,S} &=& A^\prime_1S^\mu +(n\cdot S) [(A_2+\tau A^\prime_2)P^\mu
+\tau (2A_3 + \tau A^\prime_3)n^\mu \nonumber\\
&=& \frac{1}{M}\frac{d}{d\tau} \langle \bar{\psi}(0)\gamma^\mu\gamma_5
\psi (\tau n)\rangle_{P,S} \nonumber
\end{eqnarray}
\vskip-1.0cm\hfill                                                 (E.10)
\vskip0.5cm
\noindent
a result used in Appendix D.

Consider now the gluonic matrix element
$$1/M ~ \langle \bar{\psi}(0) \gamma^\mu \gamma_5 gA^\nu (z) \psi (z)
\rangle_{P,S}$$
with
$z=\tau n$. Its most general form is
\begin{eqnarray*}
\tau (S\cdot n)~[B_1 P^\mu P^\nu +\tau B_2 P^\mu n^\nu
+\tau B_3 n^\mu P^\nu +\tau^2 B_4n^\mu n^\nu ] \\ +B_5 S^\mu P^\nu +
B^6 P^\mu S^\nu +\tau B_7 S^\mu n^\nu +\tau B_8 n^\mu S^\nu \,.
\end{eqnarray*}
\vskip-1cm\hfill                                                    (E.11)
\vskip0.5cm

The gauge condition $n_\mu A^\mu = 0$ implies that
$$
B_5=0\,, \qquad \tau B_1=-B_6\,, \qquad \tau B_3=-B_8
\eqno{\rm (E.12)}
$$
so that
\begin{eqnarray}
\frac{1}{M} \langle \bar{\psi}(0)\gamma^\mu\gamma_5gA^\nu (\tau n)
\psi (\tau n) \rangle_{P,S} &=&
\tau B_1 [(S\cdot n) P^\mu P^\nu -P^\mu S^\nu]\nonumber\\
+\tau (S\cdot n) [B_2 P^\mu n^\nu &+& \tau B_4 n^\mu n^\nu]\nonumber\\
+\tau^2 B_3 [(S\cdot n)n^\mu P^\nu &-& n^\mu S^\nu]+\tau B_7 S^\mu
n^\nu\,.
\nonumber
\end{eqnarray}
\vskip-1.5cm\hfill                                                     (E.13)
\vskip0.8cm
Notice the crucial feature, that the imposition of the gauge condition,
together with the assumptions about the vanishing of products like
$\tau B(\tau)$ as $\tau \to 0$, leads to the vanishing of (E.13) at
$\tau = 0$ {\it i.e.}
$$
\langle \bar{\psi}(0)\gamma^\mu \gamma_5gA^\nu (0)\psi (0)\rangle_{P,S} = 0\,.
\eqno{\mbox{\rm (E.14)}}
$$
This result will be crucial for deriving the Efremov--Leader--Teryaev sum rule.

Let us now relate some of the above coefficients to the functions occurring in
the discussion of $g_1$ and $g_2$.

{}From (10.2.11) and (E.4) we have
$$
\widetilde{h}_L(\tau)=A_1(\tau)+\tau A_2(\tau) \,.
\eqno{\mbox{\rm (E.15)}}
$$
{}From (10.3.7) and (E.4)
$$
\widetilde{f}_T (\tau) = A_1(\tau)\,.          \eqno{\mbox{\rm (E.16)}}
$$
Then from (10.2.13) and (10.2.10), if the Fourier transforms can be inverted,
\begin{eqnarray}
\int^1_0 dx~g_1(x) &=&\frac{e^2_f}{8}\,\widetilde{h}_L(0)\nonumber\\
&=& \frac{e^2_f}{8} \, A_1 (0)\qquad \mbox{\rm by (E.15)}\,. \nonumber
\end{eqnarray}
\vskip-1.5cm\hfill                                                   (E.17)
\vskip0.7cm
Similarly, from (10.3.8 and 6)
\begin{eqnarray}
\int^1_0 dx~[g_1(x) +g_2 (x)] &=& \frac{e^2_f}{8}\widetilde{f}_T (0)\nonumber\\
&=& \frac{e_f^2}{8} A_1(0)\qquad\mbox{\rm by (E.16)}\,.\nonumber
\end{eqnarray}
\vskip-1.5cm\hfill                                                   (E.18)
\vskip0.7cm
Equations (E.17 and 18) imply the Burkhardt--Cottingham sum rule
$$
\int^1_0 dx~g_2(x) = 0 \,.                       \eqno{\mbox{\rm (E.19)}}
$$

We have already discussed in Section 5.4 why the above derivation may fail
because of the non-invertability of the Fourier transforms. We turn now to the
Efremov--Leader--Teryaev sum rule.

Consider first Eq. (D.20) which followed from the equations of motion.
Integrating over $x_1$, using (D.6), (D.10) and (10.3.6) there results:
$$
\widetilde{B}^A(0,0)= \left. i\frac{d\widetilde{f}_T}{d\tau}
\right|_{\tau = 0}
                                                  \eqno{\mbox{\rm (E.20)}}
$$
where we have taken the quark mass to be zero for simplicity and where we have
taken, on the basis of (10.3.6),
$$
4 x f_T (x)=i\int \frac{d\tau}{2\pi}~e^{i\tau x}~
\frac{d\widetilde{f}_T}{d\tau}(\tau)\,.      \eqno{\mbox{\rm (E.21)}}
$$
Now because of (E.14), from (D.11)
$$
\widetilde{B}^A (0,0) = -\frac{i}{M} \langle \bar{\psi}(0)n\slas\gamma_5
(S_T\cdot\partial )\psi (0) \rangle_{P,S}
$$
so that via (E.7)
$$
\widetilde{B}^A (0,0) = i A_2 (0)\,.                \eqno{\mbox{\rm (E.22)}}
$$
Use of this and (E.16) in (E.20) yields
$$
A_2(0) = \left. \frac{d}{d\tau} A_1 (\tau)
\right|_{\tau=0}=A^\prime_1 (0)\,.
\eqno{\mbox{\rm (E.23)}}
$$
Now by arguments similar to those that lead to (E.21), we have
\begin{eqnarray}
\frac{8}{e^2_f} \int^1_0 dx~xg_1(x) &=& \left. i
\frac{d{\widetilde h}_L(\tau)}{d\tau}\right|_{\tau=0}\nonumber\\
&=& i [A^\prime_1(0)+A_2 (0)]     \qquad \mbox{\rm by (E.15)} \nonumber\\
&=& 2 i A^\prime_1(0)  ~~~~~~~~~~~\qquad \mbox{\rm by (E.23)}\,. \nonumber
\end{eqnarray}
\vskip-1.5cm\hfill                                                  (E.24)
\vskip1.0cm
Similarly we have
\begin{eqnarray}
\frac{8}{e^2_f}~2\int^1_0 dx~x[g_1(x) &+& g_2(x)] = \left. 2i
\frac{d}{d\tau} {\widetilde f}_T(\tau)\right|_{\tau =0} \nonumber \\
&=& 2i A^\prime_1 (0)      \qquad \mbox{\rm by (E.16)} \,.\nonumber
\end{eqnarray}
\vskip-1.5cm\hfill                                                  (E.25)
\vskip1.0cm
Subtracting (E.24) from (E.25) yields the Efremov--Leader--Teryaev sum rule
$$
\int^1_0 dx~x[g_1(x) + 2g_2(x)] = 0 \,.     \eqno{\mbox{\rm (E.26)}}
$$
The same caveats apply to this `proof' as do to the Burkhardt--Cottingham sum
rule.

\vskip 36pt
\goodbreak
\noindent
{\bf Acknowledgements}
\nobreak
\vskip 6pt
The authors are grateful for the hospitality and support of the Theory
Division, CERN, the Physics Department, Birkbeck College and the Dipartimento
di Fisica Teorica, University of Torino. They wish to acknowledge helpful
discussions with A.~Anselm, D.~Diakonov, S.~Forte, P.~Hoodbhoy, V.~Hughes,
B.~Ioffe, S.~Kulagin, G.~Mallot, P.~Mulders, P.~Nason, R.~Sassot, T.~Sloan,
G.~Smirnov, K.~Sridhar, O.~Teryaev and M.~Velasco.

M.A. is grateful to the British Council for financial support. A.V.E. thanks
the Royal Society of London, the U.K. SERC, the International Science
Foundation (Grant RFE 000) and the Russian Foundation for Fundamental
Investigation (Grant 93-02-3811), for financial support. E.L. is grateful to
the U.K. SERC and the INFN for support. He also acknowledges the warm
hospitality of the European Centre for Theoretical Studies, Trento.
%
%end of Appendices
%
\vskip 36pt
%\newpage
\begin{center}
{\Large \bf References}
\end{center}
\vskip 12pt
\begin{description}

\item{\bf [ABJ 69]} S.I. Adler, Phys. Rev. {\bf 177} (1969) 2426;
J.S. Bell and R. Jackiw, N.C. {\bf 51A} (1969) 47.
\item{\bf [ADA 94]} D. Adams {\it et al.}, Phys. Lett. {\bf B329} (1994) 399.
\item{\bf [ADA 94a]} D. Adams {\it et al.}, Phys. Lett. {\bf B336} (1994) 125.
\item{\bf [ADE 93]} B. Adeva {\it et al.}, Phys. Lett. {\bf B302} (1993) 533.
\item{\bf [ADE 94]} B. Adeva {\it et al.}, Phys. Lett. {\bf B320} (1994) 400.
\item{\bf [AHM 76]} M.A. Ahmed and G.G. Ross, Nucl. Phys. {\bf B111} (1976)
441.
\item{\bf [AHR 87]} L.H. Ahrens {\it et al.}, Phys. Rev. {\bf D35} (1987) 785.
\item{\bf [ALG 78]} M.J. Alguard {\it et al.}, Phys. Rev. Lett. {\bf 41}
(1978) 70.
\item{\bf [ALT 77]} G. Altarelli and G. Parisi, Nucl. Phys. {\bf B126} (1977)
298.
\item{\bf [ALT 82]} G. Altarelli, Phys. Rep. {\bf 81} (1982) 1.
\item{\bf [ALT 88]} G. Altarelli and G.G. Ross, Phys. Lett. {\bf B212} (1988)
391.
\item{\bf [ALT 94]} G. Altarelli, P. Nason and G. Ridolfi, Phys. Lett.
{\bf B320} (1994) 152; {\it Erratum,} {\bf B325} (1994) 538.
\item{\bf [ALT 94a]} R. Altmeyer {\it et al.}, Phys. Rev. {\bf D49} (1994)
3087.
\item{\bf [AMA 92]} NMC Collaboration: P. Amaudruz {\it et al.}, Phys. Lett.
{\bf B295} (1992) 159.
\item{\bf [ANS 79]} M. Anselmino, Phys. Rev. {\bf D19} (1979) 2803.
\item{\bf [ANS 89]} M. Anselmino, B.L. Ioffe and E. Leader, Yad. Fiz {\bf 49}
(1989) 214 [Sov. J. Nucl. Phys. {\bf 49} (1989) 136].
\item{\bf [ANS 89a]} A.A. Anselm and A. Johansen, JETP Lett. {\bf 49} (1989)
214.
\item{\bf [ANS 92]} M. Anselmino and E. Leader, Phys. Lett. {\bf B293}
(1992) 216.
\item{\bf [ANS 92a]} A.A. Anselm, Phys. Lett. {\bf B291} (1992) 455.
\item{\bf [ANS 93]} M. Anselmino, P. Gambino and J. Kalinowski,
Z. Phys. {\bf C64} (1994) 267.
\item{\bf [ANT 93]} D.L. Anthony {\it et al.}, Phys. Rev. Lett. {\bf 71}
(1993) 759.
\item{\bf [ASH 88]} J. Ashman {\it et al.}, Phys. Lett. {\bf B206} (1988) 364.
\item{\bf [ASH 89]} J. Ashman {\it et al.}, Nucl. Phys. {\bf B328} (1989) 1.

\item{\bf [BAG 90]} B. Bagchi, A. Lahiri, J. Phys. {\bf G16} (1990) L239.
\item{\bf [BAI 82]} See {\it e.g.} Chapter 4 of D. Bailin, `Weak
Interactions', Sussex University Press (1982). Note that our sign
convention for F and D is opposite to Bailin's.
\item{\bf [BAL 90]} I.I. Balitsky, V.M. Braun and A.V. Kolesnichenko, Phys.
Lett. {\bf B242} (1990) 245.
\item{\bf [BAL 93]} I.I. Balitsky, V.M. Braun and A.V. Kolesnichenho, Penn
State University Preprint, hep-ph 9310316.
\item{\bf [BAR 79]} J.A. Bartelski, Acta Phys. Pol. {\bf B10} (1979) 10; 923.
\item{\bf [BAS 91]} S.D. Bass, B.L. Ioffe, N.N. Nikolaev and A.W. Thomas,
J. Moscow Phys. Soc. {\bf 1} (1991) 317.
\item{\bf [BAS 93]} S.D. Bass, A.W. Thomas, J. Phys. {\bf C19} (1993) 925.
\item{\bf [BAU 83, 88]} G. Baum {\it et al.}, Phys. Rev. Lett. {\bf 51} (1983)
1135; V.W. Hughes {\it et al.}, Phys. Lett. {\bf 212} (1988) 511.
\item{\bf [BER 89]} E. Berger and J. Qiu, Phys. Rev. {\bf D40} (1989) 778
and 3128;
S. Gupta, D. Indumathi and M.V.N. Murthy, Z. Phys. {\bf C42} (1989) 493;
H.Y Cheng and S.N. Lai, Phys. Rev. {\bf D41} (1990) 91;
C. Bourrely, J. Ph. Guillet and J. Soffer, Nucl. Phys. {\bf B361} (1991) 72;
R.M. Godbole, S. Gupta and K. Sridhar, Phys. Lett. {\bf B225} (1991) 120;
A.P. Contogouris {\it et al.}, Phys. Lett. {\bf B304} (1993) 329.
\item{\bf [BIL 75]} S.M. Bilenky, N.A. Dadajan and E.H. Hristowa, Sov. J.
Nucl. Phys. {\bf 21} (1975) 657.
\item{\bf [BJO 65]} J.D. Bjorken and S.C. Drell, `Relativistic Quantum
Fields', New York, Mc Graw-Hill, 1965.
\item{\bf [BJO 66]} J.D. Bjorken, Phys. Rev. {\bf 148} (1966) 1467.
\item{\bf [BOU 80]} C. Bourrely, E. Leader and J. Soffer, Phys. Rep.
{\bf 59} (1980) 95.
\item{\bf [BRO 88]} S. Brodsky, J. Ellis and M. Karliner, Phys. Lett.
{\bf B206} (1988) 309; J. Ellis and M. Karliner, Phys. Lett. {\bf B213}
(1988) 73.
\item{\bf [BUR 70]} H. Burkhardt and W.N. Cottingham, Ann. Phys.
{\bf 56} (1970) 453.
\item{\bf [BUR 92]} V.D. Burkert and B.L. Ioffe, Phys. Lett. {\bf B296}
(1992) 223.
\item{\bf [BUR 93]} V.D. Burkert and B.L. Ioffe, CEBAF preprint,
CEBAF--PR--93--034, October 1993.

\item{\bf [CAH 78]} R.N. Cahn and F.G. Gilman, Phys. Rev. {\bf D17}
(1978) 1313.
\item{\bf [CAL 76]} C. Callan, R. Dashen, D. Gross, Phys. Lett {\bf 63B}
(1976) 334.
\item{\bf [CAR 72]} C.E. Carlson and Wu-ki Tung, Phys. Rev. {\bf D5}
(1972) 721; J.S. Bell and R. Jackiw, N.C. {\bf 51A} (1969) 47.
\item{\bf [CAR 88]} R.D. Carlitz, J.C. Collins and A.H. Mueller, Phys. Lett.
{\bf B214} (1988) 229.
\item{\bf [CAR 92]} R.D. Carlitz and R.S. Willey, Phys. Rev. {\bf D45}
(1992) 2323.
\item{\bf [CHE 89]} T.P. Cheng and L.F. Li, Phys. Rev. Lett. {\bf 62}
(1989) 1441.
\item{\bf [CIO 93]} C. Ciofi degli Atti, S. Scarpetta, E. Pace and G. Salm\`e,
Phys. Rev. {\bf C48} (1993) R968.
\item{\bf [CLO 88]} F.E. Close and R.G. Roberts, Phys. Rev. Lett. {\bf 40}
(1988) 1471.
\item{\bf [CLO 89]} F. Close and L. Milner, ORNL-report 31770 (1989);
L. Frankfurt {\it et al.}, Phys. Lett. {\bf B230} (1989) 141;
A.V. Efremov and V. Alizade, Prepr. JINR E--2--91--465 (1991);
M. Veltri {\it et al.}, Proceedings of `Physics at HERA' (Hamburg, 1991)
Vol. 1, p. 447.
\item{\bf [CLO 93]} F.E. Close and R.G. Roberts, Phys. Lett. {\bf B316}
(1993) 165. See also plenary talk: 13th Intl. Conf. on Particles and Nuclei,
PANIC 93, July 1993, Perugia, Italy.
\item{\bf [CON 91]} A.P. Contogouris and S. Papadopoulos, Phys. Lett
{\bf B260} (1991) 204;
M. Gl\"{u}ck and W. Vogelsang, Phys. Lett {\bf B277} (1992) 515;
P. Matheus and R. Ramachandran, Z. Phys. {\bf C53} (1992) 305.
\item{\bf [COR 88]} J.L. Cortes and B. Pire, Phys. Rev. {\bf D38} (1988)
3586; G. Ballochi {\it et al.}, HELP, CERN/LEPC 89--90 and LEPC/M88 (1989);
M.A. Doncheshi and R.W. Robinett, Phys. Lett. {\bf B248} (1990) 188;
A.P Contogouris, S. Papadopoulos and B. Kamal, Phys. Lett. {\bf B246} (1990)
523;
K. Culter {\it et al.}, HERMES, DESY/PRE 90--1 (1990)
\item{\bf [COR 92]} J.L. Cortes, B. Pire and J.P. Ralston, Z. Phys.
{\bf C55} (1992) 409.
\item{\bf [CRA 83]} N.S. Craigie, K. Hidaka, M. Jacob and F.M. Renard,
Phys. Rep. {\bf 99} (1983) 69.
\item{\bf [CRO 83]} C. Cronstr\"{o}m and J. Mickelsson, J. Math. Phys.
{\bf 24} (1983) 2528. See also [JAC 85].

\item{\bf [DAL 56]} R.H. Dalitz, Int. School of Physics "Enrico Fermi",
Corse, Vol. 33, Academic Press, 1956.
\item{\bf [DEF 93]} D.E. de Florian {\it et al.}, La Plata preprint 93-03
(1993).
\item{\bf [DER 73]} E. Derman, Phys. Rev. {\bf D7} (1973) 2755.
\item{\bf [DOR 91]} A.E. Dorokhov and N.I. Kochelev, Phys. Lett. {\bf B259}
(1991) 335.
\item{\bf [DOR 93]} A.E Dorokhov, N.I Kochelev and A.Yu. Zubov,
Int. J. of Mod. Phys. {\bf A8} (1993) 603.
\item{\bf [DRE 64]} S.D. Drell and J.D. Walecka, Ann. Phys. {\bf 28} (1964)
18.
\item{\bf [DRE 66]} S.D. Drell and A.C. Hearn, Phys. Rev. Lett. {\bf 16}
(1966) 908.
\item{\bf [DUM 83]} O. Dumbrais {\it et al.}, Nucl. Phys. {\bf B216} (1983)
277.
\item{\bf [DYA 81]} D.I. Dyakonov and M.I. Eides, Sov. Phys. JETP {\bf 54}
(1981) 2.

\item{\bf [EFR 78]} A.V. Efremov, Sov. J. Nucl. Phys. {\bf 28} (1978) 83;
R.H. Dalitz, G.R. Goldstein and R. Marshall, Z. Phys. {\bf C42} (1989) 441;
A.V. Efremov, L. Mankiewicz and N.A. T\"{o}rnqvist, Phys. Lett. {\bf B284}
(1992) 394.
\item{\bf [EFR 84]} A.V. Efremov and O.V. Teryaev, Sov. J. Nucl. Phys.
{\bf 39} (1984) 962 [Yad. Fiz. {\bf 39} (1984) 1517].
\item{\bf [EFR 87]} A.V.Efremov and O.V. Teryaev, Phys. Lett. {\bf B200}
(1987) 363.
\item{\bf [EFR 88]} A.V. Efremov and O.V. Teryaev, JINR, Report
No. E2--88--287 (1988); unpublished.
\item{\bf [EFR 90]} A.V. Efremov, J. Soffer and N. T\"{o}rnqvist, Phys. Rev.
Lett. {\bf 64} (1990) 1495;\\
A.V. Efremov, J. Soffer and O.V. Teryaev, Nucl. Phys. {\bf B396} (1990) 97.
\item{\bf [EFR 91]} A.V. Efremov, J. Soffer and N. T\"{o}rnqvist, Phys. Rev.
{\bf D44} (1991) 1369.
\item{\bf [EFR 91a]} A.V. Efremov and V. Karotkijan, Sov. J. Nucl. Phys.
{\bf 54} (1991) 937.
\item{\bf [EFR 92]} A.V. Efremov, `Polarization Dynamics in Nuclear and
Particle Physics' (Proceedings of the 2nd Adriatico Research Conference,
Trieste, 1992), Ed. A.O. Barut {\it et al.}, World Scientific, Singapore,
1993, p. 218.
\item{\bf [EFR 92a]} A.V. Efremov, L. Mankiewicz and N.A. T{\"o}rnqvist,
Phys. Lett. {\bf B284} (1992) 394.
\item{\bf [EFR 94]} A.V. Efremov, `A model for the estimation of jet
handedness', (to be published).
\item{\bf [EFR 94a]} A.V. Efremov and E. Leader, BNL preprint BNL--60734
(1994); submitted to Phys. Rev. Lett.
\item{\bf [ELL 74]} J. Ellis and R.L. Jaffe, Phys. Rev. {\bf D9} (1974) 1444;
{\it Erratum}, {\bf D10} (1974) 1669.
\item{\bf [ELL 88]} J. Ellis and M. Karliner, Phys. Lett. {\bf B213} (1988)
73.
\item{\bf [ELL 93]} J. Ellis and M. Karliner, Phys. Lett. {\bf B313} (1993)
131; See also Plenary talk at PANIC--93.
\item{\bf [EPE 92]} L.N. Epele, H. Fanchiotti, C.A. Garcia Canal and
R. Sassot, Phys. Lett. {\bf 287} (1992) 247.
\item{\bf [EPE 93]} L.N. Epele, H. Fanchiotti, C.A. Garcia Canal, E. Leader
and R. Sassot, Birkbeck College Preprint, October 1993.

\item{\bf [FAD 76]} L.D. Faddeev, `Non-local field theories', Proceedings of
Alushta Conference on Nonlocal and Nonlinear Theories, Dubna, 1976.
\item{\bf [FEY 72]} R.P. Feynman, `Photon-hadron Interactions', Benjamin,
Reading, MA, 1972.
\item{\bf [FOR 90]} S. Forte and E. Shuryak, Nucl. Phys. {\bf B357} (1990)
153; S. Forte, Phys. Lett. {\bf B224} (1989) 189 and Nucl. Phys. {\bf B331}
(1990) 1.

\item{\bf [GER 66]} S. Gerasimov, Sov. J. Nucl. Phys. {\bf 2} (1966) 930.
\item{\bf [GLU 88]} M. Gl\"{u}ck and E. Reya, Z. Phys. {\bf C39} (1988) 569;
J.P. Guillet, Z. Phys. {\bf C39} (1988) 75;
G. Altarelli and W.J. Stirling, Particle World {\bf 1} (1989) 40;
M. Gl\"{u}ck, E. Reya and W. Vogelsang, Nucl. Phys. {\bf B351} (1991) 579.

\item{\bf [HEI 73]} R.L. Heimann, Nucl. Phys. {\bf B64} (1973) 429.
\item{\bf [HEL 91]} See {\it e.g.} K. Heller, 'High Energy Spin Physics'
(Proc. of 9th Int. Symp., Bonn, 1990), Ed K.-N. Althoff and W. Meyer, Springer,
1991, Vol .1, p.97; S. Nurushev, `Frontier of High Energy Spin Physics' (Proc.
of 10 Int. Symp. on High Energy Spin Physics, Nagoya, Japan, 1992), Ed.
Hasegawa
T. {\it et al.}, Yamada Sci. Foundation and Univ. Acad. Press. Inc., 1993,
Tokyo, Japan, p. 311
\item{\bf [HEY 72]} A.J.G. Hey and J.E. Mandula, Phys. Rev. {\bf D5} (1972)
2610.
\item{\bf [HSU 88]} S.Y. Hsuch {\it et al.}, Phys. Rev. {\bf D38} (1988) 2056.

\item{\bf [IOF 84]} B.L. Ioffe, V .A. Khoze and L.N. Lipatov, `Hard processes',
 North-Holland, Amsterdam, 1984.
\item{\bf [IOF 92]} B.L. Ioffe and A.Yu. Khodjamirian, Preprint ITEP--41
(1992).

\item{\bf [JAC 76]} R. Jackiw, G. Rebbi, Phys. Rev. Lett. {\bf 37} (1976)
172.
\item{\bf [JAC 85]} R. Jackiw, `Current algebra and anomalies',
World Scientific, Singapore (1985).
\item{\bf [JAC 89]} J.D. Jackson, G.G. Ross and R.G. Roberts, Phys. Lett.
{\bf B226} (1989) 159.
\item{\bf [JAF 90]} R.L. Jaffe, Comments Nucl. Phys. {\bf 19} (1990) 239.
\item{\bf [JAF 91]} R.L. Jaffe and Xiangdong Ji, Phys. Rev. {\bf D43} (1991)
724.
\item{\bf [JEN 91]} E. Jenkins, Nucl. Phys. {\bf B354} (1991) 24.
\item{\bf [JOS 77]} A.S. Joshipura and P. Roy, Ann. Phys. {\bf 104} (1977) 440.

\item{\bf [KAL 89]} P. Kalyniak, M.K. Sundaresan and P.I.S. Watson,
Phys. Lett. {\bf B216} (1989) 397.
\item{\bf [KAU 77]} J. Kaur, Nucl. Phys. {\bf B128} (1977) 219.
\item{\bf [KOD 79]} J. Kodaira, S. Matsuda, K. Sasaki and T. Uematsu, Nucl.
Phys. {\bf B159} (1979) 99;
J. Kodaira, Nucl. Phys. {\bf B165} (1979) 129;
J. Kodaira, S. Matsuda, T. Muta, T. Uematsu and K. Sasaki, Phys. Rev.
{\bf D20} (1979) 627. Note that our covariant spin vector $S_\mu$ is
1/(2$M$) times the spin vector in the above papers.
\item{\bf [KOD 80]} J. Kodaira, Nucl. Phys. {\bf B165} (1980) 129.
\item{\bf [KOG 75]} J. Kogut and L. Suskind, Phys. Rev. {\bf D11} (1975) 3583.
\item{\bf [KOT 70]} A. Kotanski, Acta Physica Polonica {\bf B1} (1970) 45.
\item{\bf [KUH 90]} J.H. K\"{u}hn and V.I. Zakharov, Phys. Lett. {\bf B252}
(1990) 615.

\item{\bf [LAM 82]} C.S.Lam and B. Li, Phys. Rev. {\bf D28} (1982) 683.
\item{\bf [LAM 89]} B. Lampe, Phys. Lett. {\bf B227} (1989) 469.
\item{\bf [LAR 91]} S.A. Larin and J.A.M. Vermaseren, Phys. Lett. {\bf B259}
(1991) 345.
\item{\bf [LAR 94]} S.A. Larin, Phys. Lett. {\bf B334} (1994) 192.
\item{\bf [LEA 85]} See, {\it e.g}., E. Leader and E. Predazzi, `An
Introduction
to Gauge Theories and the New Physics', Cambridge University Press (1985).
\item{\bf [LEA 88]} E. Leader and M. Anselmino, Z. Phys. {\bf C41} (1988)
239.
\item{\bf [LEA 88a]} E. Leader and M. Anselmino, Santa Barbara Preprint
NSF--88--142, July 1988, unpublished.
\item{\bf [LEA 88b]} E. Leader and P. Ratcliffe, unpublished.
\item{\bf [LEA 93]} E. Leader and K. Sridhar, Phys. Lett. {\bf B311} (1993)
324.
\item{\bf [LEA 94]} E. Leader and E. Predazzi, `An Introduction to Gauge
Theories and Modern Particle Physics', Cambridge University Press, in press.
\item{\bf [LEA 94a]} E. Leader and K. Sridhar, Nucl. Phys. {\bf B419} (1994) 3
\item{\bf [LIP 90]} H. Lipkin, Phys. Lett. {\bf B237} (1990) 130.

\item{\bf [MAN 90]} J.E. Mandula, Phys. Rev. Lett. {\bf 65} (1990) 1403.
\item{\bf [MAN 90a]} L. Mankiewicz and A. Schafer, Phys. Lett. {\bf B242}
(1990) 455.
\item{\bf [MAN 91]} A.V. Manohar, Phys. Lett. {\bf B255} (1991) 579.
\item{\bf [MAN 91a]} L. Mankiewicz, Phys. Rev. {\bf D43} (1991) 733.
\item{\bf [MAN 92]} J.E. Mandula, `Polarization Dynamics in Nuclear and
Particle Physics' (Proceedings of the 2nd Adriatico Research Conference,
Trieste, 1992), Ed. A.O. Barut {\it et al.}, World Scientific, Singapore,
1993, p. 227.

\item{\bf [MAT 92]} P. Mathews and V. Ravindran, Int. J. Mod. Phys.
{\bf A7} (1992) 309.
\item{\bf [MAT 92b]} P. Mathews and V. Ravindran, Phys. Lett. {\bf B278} (1992)
175.

\item{\bf [NAC 77]} O. Nachtmann, Nucl. Phys. {\bf B127} (1977) 314.
\item{\bf [NAS 71]} C. Nash, Nucl. Phys. {\bf B31} (1971) 419.
\item{\bf [NAS 93]} We are very grateful to P. Nason for supplying values of
$\alpha_s(Q^2)$ which are based upon next-to-leading order $Q^2$-dependence
with normalisation to $\alpha_s(M_Z) = 0.116 \pm 0.01$ and exact matching at
heavy quark thresholds. The error at $Q^2 = M^2_Z$ reflects the disagreement
between LEP and DIS values. See S. Catani, Proc. EPS Conf, Marseille, July
1993.
\item{\bf [NIK 73]} N.N. Nikolaev, M.A. Shifman and M.Z. Shmatikov, JETP
{\bf 18} (1973) 39.

\item{\bf [PDG 92]} Particle Data Group, Phys. Rev. {\bf D45} (1992).
\item{\bf [PRE 88]} G. Preparata and J. Soffer, Phys. Rev. Lett. {\bf 61}
(1988) 1167

\item{\bf [QIU 91]} J. Qiu and G. Sterman, Phys. Rev. Lett. {\bf 67} (1991)
2264; Nucl. Phys. {\bf B378} (1992) 52.

\item{\bf [RAL 79]} J. Ralston and D.E. Soper, Nucl. Phys. {\bf B152}
(1979) 109. The reader is warned that there is a term $\gamma_5 [k\sla_A,
{\cal A}\Cslas ]$ where ${\cal A}^\mu = (k_A \cdot S_A) P_A^\mu -
(k_A \cdot P_A) S_A^\mu$, missing from what was meant to be the most general
expression for the tensor structure of $\Phi$. This term is not important
for the analysis in [RAL 79]. (E. Leader, private communication to the authors,
unpublished.)
\item{\bf [RAM 88]} G.P Ramsey, D. Richards and D. Sivers, Phys. Rev.
{\bf D37} (1988) 3140;
Z. Kunst, Phys. Lett {\bf B218} (1989) 243;
G.P. Ramsey and D. Sivers, Phys. Rev. {\bf D43} (1991) 2861;
M.A. Doncheshi, R.W. Robinett and L. Weinhauf, Phys. Rev. {\bf D44}
(1991) 2717;
P. Chiappetta and G. Nardulli, Z. Phys. {\bf C51} (1991) 435.
\item{\bf [RAT 86]} P.G. Ratcliffe, Nucl. Phys. {\bf B264} (1986) 493.
\item{\bf [RAT 87]} P. Ratcliffe, Phys. Lett. {\bf B192} (1987) 180.
\item{\bf [RAV 92]} V. Ravishankar, Nucl. Phys. {\bf B374} (1992) 309.
\item{\bf [REY 93]} E. Reya, `QCD 20 years later', Ed. P.H. Zerwas and
H.A. Kastrup, World Scientific, 1993, Vol. 1, p. 272.
\item{\bf [ROE 90]} N.A. Roe {\it et al.}, (ASP-Collaboration) Phys. Rev.
{\bf D41} (1990) 17; D.I. Diakonov and M.I. Eides, `Elementary Particle
Physics', Lecture at XVI LIYaP Winter School, LIYaP, 1981, p. 123.
\item{\bf [ROS 90]} G.G. Ross and R.G. Roberts, Preprint RAL--90--062,
August 1990; unpublished.
\item{\bf [RYS 93]} M.G. Ryskin, Phys. Lett. {\bf B319} (1993) 346.
\item{\bf [RYZ 89]} Z. Ryzak, Phys. Lett. {\bf B217} (1989) 325; also private
communication.

\item{\bf [SHO 92]} G.M. Shore and G. Veneziano, Nucl. Phys. {\bf B381}
(1992) 23.
\item{\bf [SHU 82]} E.V. Shuryak and A.I. Vainshtein, Nucl. Phys. {\bf B201}
(1982) 142.
\item{\bf [SRI 92]} K. Sridhar and E. Leader, Phys. Lett. {\bf B295} (1992)
283.
\item{\bf [STR 92]} M. Stratmann and W. Vogelsang, Phys. Lett. {\bf B295}
(1992) 277.

\item{\bf [TOR 84]} N. T\"{o}rnqvist, P. Zenczykowski, Phys. Rev. {\bf D29}
(1984) 2139.
\item{\bf ['THO 76]} G. 't Hooft, Phys. Rev. Lett. {\bf 37} (1976) 8;
Phys. Rev. {\bf D14} (1976) 3432.

\item{\bf [VEN 79]} G. Veneziano, Nucl. Phys. {\bf B159} (1979) 213.
\item{\bf [VEN 89]} G. Veneziano, Mod. Phys. Lett. {\bf A4} (1989) 1605;
G.M. Shore and G. Venezia\-no, Phys. Lett {\bf B244} (1990) 75.
\item{\bf [VEN 92]} G. Veneziano and G. M Shore, Nucl. Phys. {\bf B381}
(1992) 3.
\item{\bf [VOG 91]} W. Vogelsang and A. Weber, Nucl. Phys. {\bf B362}
(1991) 3.

\item{\bf [WA2 83]} WA2 Group, Z. Phys. {\bf C21} (1983) 27; Particle Data
Group, Phys. Lett. {\bf 111B} (1982).
\item{\bf [WAN 77]} S. Wandzura and F. Wilczek, Phys. Lett. {\bf B72} (1977)
195.
\item{\bf [WHI 90]} L.W. Whitlow {\it et al.}, Phys. Lett.  {\bf B250} (1990)
193.
\item{\bf [WIN 92]} R. Windmolders, Int. J. Mod. Phys. {\bf A7} (1992) 639.
\item{\bf [WOL 89]} R.M. Woloshyn, Nucl. Phys. {\bf A496} (1989) 749.

\item{\bf [XIA 93]} Xiangdong Ji, Nucl. Phys. {\bf B402} (1993) 217.
\end{description}
%
%end of References
%
\newpage
\begin{center}
{\large \bf Figure captions}
\end{center}
\vskip 12pt

\begin{description}
\item{\bf Fig. 2.1} Feynman diagram for inelastic lepton--nucleon scattering:
$\ell N \to \ell X$.
\vskip 6pt

\item{\bf Fig. 2.2} Definition of the angles
$\alpha, \beta, \theta, \varphi, \phi$ and $\Theta$.
\vskip 6pt

\item{\bf Fig. 2.3} Further definition of the angles
$\alpha, \theta$ and $\phi$.
\vskip 6pt

\item{\bf Fig. 2.4} Spin configuration relevant to Eq. (2.1.39) for the
measurement of $A_\perp$ with $\phi=0$.
\vskip 6pt

\item{\bf Fig. 2.5} a) Lowest order Feynman diagrams for weak interaction
contributions to  $\ell N\to  \ell X$ and b) to $\ell^\mp N \to \nu
(\bar{\nu}) X$.
\vskip 6pt

\item{\bf Fig. 3.1} Predictions for $xg_2(x)$ and $x [g_1 (x) + g_2 (x)]$
based on Eq. (3.3.11) [JAC 89].
\vskip 6pt

\item{\bf Fig. 3.2} Prediction for $g_2(x)$ in the bag model [JAF 91]. Dotted
line = twist-2 contribution; dashed = twist-3; solid = sum.
\vskip 6pt

\item{\bf Fig. 3.3} Interaction of hard photon with quark.
\vskip 6pt

\item{\bf Fig. 4.1} EMC ($\bullet$) [ASH 89] and SLAC--Yale ($\circ$)
[ALG 78; BAU 83] data on $A_1(x)$.
\vskip 6pt

\item{\bf Fig. 4.2} EMC data on $xg_1^p (x)$. Also shown is the extrapolation
to $x=0$ of the integral of $g_1^p(x)$ (see text).
\vskip 6pt

\item{\bf Fig. 4.3} Comparison of the new SMC data on $g^p_1(x)$ at
$\langle Q^2\rangle = 10$ (GeV/c)$^2$ [ADA 94] with EMC data.
\vskip 6pt

\item{\bf Fig. 4.4} Behaviour of $I_p(Q^2)$ (curve 1) and $I_p(Q^2)-I_n (Q^2)$
(curve 2) at large $Q^2$ and as given by the GDH sum rule at $Q^2$ = 0
(cross and triangle respectively, from [ANS 89]).
\vskip 6pt

\item{\bf Fig. 4.5} The resonance contribution to $I(Q^2)$ for proton, neutron
and their difference (from [BUR 93]).
\vskip 6pt

\item{\bf Fig. 4.6} SMC data on the deuteron asymmetry $A_1^d$ at
$Q^2 = 4.6$~(GeV/c)$^2$ (from [ADE 93]).
\vskip 6pt

\item{\bf Fig. 4.7} SMC data on $x g_1^d (x)$ at $Q^2 = 4.6$ (GeV/c)$^2$
(from [ADE 93]).
\vskip 6pt

\item{\bf Fig. 4.8} E142 data on the neutron asymmetry $A^n_1(x)$ and on
$g^n_1(x)$ at $Q^2 = 2$ (GeV/c)$^2$. See text (from [ANT 93]).
\vskip 6pt

\item{\bf Fig. 4.9} $g_1^p (x, Q^2)$ at various $Q^2$ as extracted from the
deuterium and helium-3 data by [ELL 93] (see text). Continuous curve,
$Q^2 = 2$ (GeV/c)$^2$; dotted, $Q^2 = 4.6$; dot-dashed, $Q^2 = 10.7$.
\vskip 6pt

\item{\bf Fig. 4.10} $g^n_1(x,Q^2)$ as extracted by [ELL 93] (see text).
Continuous curve, $Q^2 = 2$ (GeV/c)$^2$; dotted, $Q^2 = 4.6$, dot-dashed,
$Q^2 = 10.7$.
\vskip 6pt

\item{\bf Fig. 6.1} Triangle diagram giving rise to the axial anomaly.
\vskip 6pt

\item
{\bf Fig. 6.2} Diagram giving rise to renormalization of axial anomaly.
\vskip 6pt

\item
{\bf Fig. 6.3} Diagram giving rise to  renormalization of the axial current.
\vskip 6pt

\item{\bf Fig. 8.1} Born diagram for $\gamma^* q \to q$.
\vskip 6pt

\item{\bf Fig. 8.2} Lowest order diagram for $\gamma^* g \to q\bar{q}$.
\vskip 6pt

\item{\bf Fig. 8.3} Dependence of $d\Delta \sigma^{\gamma^*q}/dk_T^2$ on $k_T$
for several choices of the quark mass.
\vskip 6pt

\item{\bf Fig. 8.4} Lowest order diagram for $\gamma^* q \to g~q$.
\vskip 6pt

\item{\bf Fig. 8.5} Result of fitting $g^p_1 (x)$ with $\Delta s = 0$, and
$\Delta g$ constrained by the unpolarized glue. The solid line corresponds to
$F/D = 0.63$ and the dashed line to $F/D$ = 0.55.
\vskip 6pt

\item{\bf Fig. 8.6} Result of fitting $g_1^p (x)$ with $\Delta s \neq 0$,  and
$\Delta g$ constrained by the unpolarized glue. The solid line corresponds
to $F/D = 0.63$ and the dashed line to $F/D = 0.55$.
\vskip 6pt

\item{\bf Fig. 8.7} $Z_0$ exchange contribution to elastic $\nu p \to \nu p$.
\vskip 6pt

\item{\bf Fig. 8.8} Anomaly contribution to elastic $\nu p \to \nu p$.
\vskip 6pt

\item{\bf Fig. 9.1} The QCD potential $V(X,~Y,...)$ as a function of $X$.
\vskip 6pt

\item{\bf Fig. 9.2} Graphical representation of the non-perturbative coupling
of the bare axial ghost to the current $\widetilde{K}_\mu$.
\vskip 6pt

\item{\bf Fig. 9.3} Graphical representation of the coupling between the axial
ghost and the field $\eta_0 (x)$.
\vskip 6pt

\item{\bf Fig. 9.4} Graphical representation of matrix elements a) $\langle
T[G_\mu G_\nu] \rangle$ and b) $\langle T[G_\mu \eta^\prime] \rangle$.
\vskip 6pt

\item{\bf Fig. 9.5} Graphical representation of matrix element $\langle
P^\prime,\lambda^\prime|\widetilde{K}_\mu (0)|P,\lambda\rangle$.
\vskip 6pt

\item{\bf Fig. 9.6} Graphical representation of interaction of $\eta^0$
and ghost $G^0_\mu$ with nucleon.
\vskip 6pt

\item{\bf Fig. 10.1} Separation of $\gamma^* +$ hadron $\to X$ into soft and
hard parts.
\vskip 6pt

\item{\bf Fig. 10.2} a) Born diagram for $\gamma^* q\to \gamma^* q$ and b) the
crossed version.
\vskip 6pt

\item{\bf Fig. 10.3} DIS interaction involving quark-gluon correlation.
\vskip 6pt

\item{\bf Fig. 11.1} Conventional Born mechanism for large $p_T$ Drell-Yan pair
production.
\vskip 6pt

\item{\bf Fig. C.1} Born diagram for $E^{\mu\rho\nu}$. There is also a crossed
diagram analogous to Fig. 10.2b.

\end{description}

\end{document}